\documentclass[twoside,a4paper,10pt]{book} 
\usepackage{amsmath,amssymb,graphicx,psfrag,fancyhdr,epsfig,oldgerm} 
\usepackage{cite,here,exscale}

\usepackage{amsmath,amssymb,graphicx,psfrag,fancyhdr,epsfig,oldgerm} 
\usepackage{cite,here,exscale}

\usepackage[active]{srcltx}
\usepackage{dcolumn}
\usepackage{bm}

\begin{document}

\pagestyle{empty}
  \vspace{2cm}
  \sffamily\upshape\mdseries
  \noindent
  {\Huge \textbf{Linear Algebra with Disordered Sparse Matrices that have Spatial Structure: Theory and Computation}}

  \vspace{9.5cm}

  \noindent
  Thesis for the Dottorato di Ricerca \\
  Department of Physics  \\
  Universit\`a degli Studi di Roma "La Sapienza," Roma

  \vspace{1.5cm}

  \noindent  {\large \textbf{Vincent Edward Sacksteder IV}}

  \vspace{1.5cm}

  \noindent
  Advisor: \parbox[t]{7cm}{Prof.~Giorgio~Parisi \\
    Department of Physics \\
    Universit\`a degli Studi di Roma "La Sapienza"}
  
  \cleardoublepage

\vspace*{-2.0cm} 

\begin{figure}[h]
\begin{center}
\includegraphics[width=2cm]{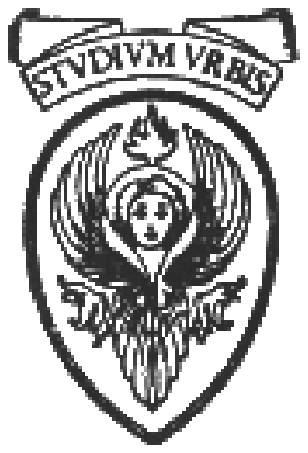}
\end{center}
\end{figure}

\vspace{-0.5cm}

\begin{center}

{\large UNIVERSIT\`{A} DEGLI STUDI DI ROMA\\ ``LA SAPIENZA''}

\end{center}



\vspace{1.5cm}

\begin{center}
\begin{tabular}{c}
{\Huge \textbf{Linear Algebra with Disordered }}\\
\\
{\Huge \textbf{Sparse Matrices that have }}\\
\\
{\Huge \textbf{Spatial Structure: }}\\
\\\
{\Huge \textbf{Theory and Computation}}\\
\end{tabular}
\end{center}

\vspace{3.0cm}

\begin{center}

{\Large Thesis submitted for the degree of} \\

\vspace{0.2cm}

{\Large {\it Doctor Philosophi\ae}} \\

\vspace{0.5cm}

{\large PhD in Physics - XVII cycle}

\end{center}

\vspace{1.5cm}

\begin{tabular}{lp{2cm}r}
{{\bf \Large Candidate}} & & {{\bf \Large Supervisor}}
\\
\\
{\Large Vincent Edward Sacksteder IV} & & {\Large Giorgio Parisi}  
\end{tabular}

\vspace{1.5cm}

\begin{center}

{\normalsize November 2004}

\end{center}

\cleardoublepage

\pagestyle{fancy}

\chapter*{Dedication}
I dedicate this thesis to my father, Vincent Edward Sacksteder III, who shared with me his love for physics and for so many other things, and for me.  One of my favorite science-related memories of him is of long hours during car trips when he would answer my many questions about science, industry, and lots of other subjects.  Another is of him reading science fiction aloud to us.

\medskip 
I would also like to remember my grandpa Richard Anthony Miller, who died on the day that this thesis was completed.  He remains a strength to us all, and I'm very thankful for him and for the way he lived and died.  

\medskip
And I should remember Dr. Phil Peters, who died a year after teaching me graduate level electromagnetism.
\begin{flushright}
Rome, November 7, 2004
\end{flushright}
\chapter*{Acknowledgements}
I want most to thank God: the Father, his son Jesus our Saviour, and the Holy Spirit of love.  I am thankful to them for the last five years which have probably been the happiest years yet in my life, for the past three and a half years living in the beautiful and holy city of Rome and traveling to many other extraordinary places, for good friends and community, for the privilege of both begining and ending this Ph.D, and for letting this Ph.D. be of a quality that I can be very proud of.   But those are just the externals: the deeper reality is love. I am thankful for God's constant intimate friendship, love, and mercy, which are faithful even when I am not faithful, that have been enabling me to change slowly for the better, and that challenge me to share what I can with others.  I believe that each person was created through God's deliberate choice to create that individual person and not another person, and that his choice was motivated by love, by seeing something immensely precious in the person he was choosing, and by wanting to share a friendship and a love with that person.  I am very thankful for the choice that God made about me; for the gift of being alive.

I also want to thank my advisor, Giorgio Parisi.  Without his faithful support I probably would not have been permitted to pursue the research projects which developed into this thesis. He worked to understand what I was doing even when it didn't relate directly to his research, encouraged me, and gave me comments and advice.  He also shared some of his vast scientific knowledge with me, and gave me a first hand experience of working with a great scientist.  

I also need to thank many other scientists.  Dr. Lightman paid me to learn C and to do some scientific programming, and Dr. John Cramer paid me to do some more programming and advised me for three quarters.  The nuclear theorist George Bertsch, who advised me for three quarters in 1994-95, got me started on trying to renormalize the eigenvalue problem, which was the seed of this thesis.  Dr. Bachelet here in Rome informed me of the existence of $O{(N)}$ algorithms, and talks by Dr. Boris Altshuler and Dr. Beenaaker kindled my interest in mesoscopic physics.  Several physicists sat down with me to counsel me about what research direction to pursue for my thesis; I am particularly thankful to Massimiliano Testa, who keeps his door open to me, writes recommendations, and always has a friendly smile.  Many others have spent one on one time answering long sets of questions, most notably numerical analysts Martin Gutknecht and Lars Elden, and physicists Alexander Mirlin, Jacques Verbaarschot, and Kostantin Efetov.  Jacques Verbaarschot recommended to me the paper by Fyodorov that stimulated my development of the new sigma model presented in chapter \ref{NoGradedMatrices}.  In Manila physicists Br. Joe Scheiter, FSC and Fr. Dan McNamara, SJ gave me quite a welcome, as did theoretical physicist Ithnin Abdul Jalil in Kuala Lumpur.  I had the fortune of being able to attend two different courses on numerical analysis by the master Gene Golub.  My introduction to the numerical analysis community was sponsored by Dr. Nicola Mastronardi, who welcomed me at a summer school.   Many scientists on the NA Digest mailing list contributed to my research on reliablity and reproducibility: among the most helpful were doctors Tim Trucano, Steve Stevenson, Stan Scott, Rob Easterling, Marek Gutowski, Alan Karp, Gaston Gonnet, Carol Marians, Brian Gough, Arnold Neumaier, and Alan Calder.  There are many other scientists and staff who have given me their good will, trust, and even friendship along the way, in fact too many to list or even remember them all.  I would, however, like to particularly mention the other physics Ph.D. students (especially my office mates!) here at La Sapienza, as well as Andrea Cavagna, Irene Giardina, and Tomas Jorg down the hall.  May all of you, including the ones I have not listed explicitly, be rewarded and blessed.

Probably my friends and family provided an even greater contribution, allowing me to stay sane and happy and grow personally through the many changes and challenges of these years.  I really can't list you all, but I would like to particularly mention my Dad and Mom, each of my siblings (Rachel, Mary, Bernadette, Sarah, Jacob, Ann, Joanna, Liz, Jess, and Paul), my relatives, Luisa and Sixto Calayag and their extended family, Fr. Aloysius Deeney, Fr. Lawrence D'Souza and family, John Petry, Pascuale Midiri, Alessandro Guccione, Claudia Bassi, Matteo Parlagreco, Ilaria Palombo, Livia Sain, and the Midiri, Dodero, Ghersinti-Sorrentino, Mercanile, and Livier families.  And all the people who came and stayed with me as guests, most notably six of my sisters and brothers!  There are also a number of friends to whom I am enormously indebted but who I am sure would prefer to remain anonymous.  

To you and to all who are unlisted, may God's peace and love be with you continuously, and may you be able to accept the gift.  L'Chaim! (To Life!)

\tableofcontents

\mainmatter

\chapter{\label{Introduction}Introduction}

In this thesis I study - both numerically and theoretically - how to do linear algebra with disordered sparse matrices that have a spatial structure.  Here's what I mean:
\begin{itemize}
\item Spatial Structure: The basis represents a manifold; a volume; a set of $D$ coordinates living in a $D$ dimensional world.  (I will use the word "system" to refer to this volume.) The basis may also include other, non-spatial, coordinates.
\item Sparse: When represented in the position basis, the matrices are nearly diagonal; i.e. the matrix elements $\langle \vec{x} | M | {\vec{x} + \vec{\delta}} \rangle$ are zero if the distance $| \vec{\delta} |$ is not small.  However, it is important that some matrix elements off the diagonal (${| \vec{\delta} |} \neq 0$) be non-zero, in order to describe the system's spatial structure.  These non-zero off-diagonal elements are often called the kinetic part of the matrix.
\item Disordered: The matrix elements vary significantly at spatial scales which are small compared to the size of the system.   One can define the scale of variation $\lambda$ as ${\lambda}^{-1} \equiv {|{\vec{\nabla}}_{\vec{x}}\ln{{\langle \vec{x} | M | {\vec{x} + \vec{\delta}} \rangle}}|}$; being disordered means that $\lambda$ is much smaller (a factor of ten or more) than the linear size of the system. 
\item Linear Algebra: I study the eigenvectors and eigenvalues of these matrices, as well as their Green's function (also known as the resolvent), exponential, logarithm, density matrix (step function), Gaussian function, and Cauchy distribution.  These are all problems from linear algebra.  
\item Numerics: Numerical algorithms for calculating these quantities typically run in $O{(N^{3})}$ time where $N$ is the basis size and is proportional to the system's volume.  Therefore even modern computers are limited to small systems, for instance cubes of width $\sim20$.  In this thesis I examine some approximate algorithms - called $O{(N)}$ algorithms - which allow computation of much larger systems. These algorithms rely on certain assumptions about the physics of the system, and this thesis goes a long way toward understanding what those assumptions are and when they are justified. 
\item Theory: I model these disordered matrices with models where some of the matrix elements are determined by random numbers.  Therefore one has a large set of possible disordered matrices, each member of the set distinguished by a particular choice of random numbers.  I follow the usual theoretical approach of computing averages over this set.  Using various techniques, I compute averages of multiples of the Green's function, and also estimate average eigenfunctions.
\end{itemize}

Why are these problems important, and why are they physics?  Because any physical system with independent degrees of freedom acting at two or more different length scales can be considered to be disordered.  This includes mesoscopic physics where electrons flow large distances through an impure and spatially distorted medium, as well as lattice QCD where neutrons and protons are collective excitations of much smaller quarks and gluons.  But it also includes oil flow through rock, turbulence, protein behavior in cells, heat and light absorption in clouds, and most other important problems.  Any system where renormalization ideas are applicable is a disordered system, though actually most renormalization techniques make an additional requirement that there be a self-similarity between the various scales.    In all of these systems the short-scale physics makes a huge qualitative, not just quantitative, difference in the large-scale physics.

It is in precisely these problems that even the fastest computers are just not up to the job.  Consider a three-dimensional system with just two relevant length scales, one ten times the size of the other.  To treat both length scales accurately one will need to use a grid roughly $100$ points on a side, for a total of a million points in the volume.  Linear algebra tasks that run in $N^{3} = {{10}^{6}}^{3} = {10}^{18}$ time are already well beyond the capability of modern computers.  These $O{(N^{3})}$ tasks can include simulating a system's movements, its response to external forces, or its equilibrium state.  In condensed matter physics simulating movement caused by electronic forces requires evaluating the density matrix function (an $O{(N^{3})}$ problem), in lattice QCD calculating the ground state requires evaluating the logarithm or the inverse ($O{(N^{3})}$ again), and in engineering calculating a structure's response to external forces is also an $O{(N^{3})}$ calculation. As a consequence, many important problems remain well out of reach of numerical study.

Disorder at short distances may drastically change the physics at large distances.  In particular, it can cause a disconnection between points that are far apart, so that what happens at a given point is independent of everything far away from it.   In real life people often make this "nearsightedness" assumption, i.e. that what happens here is largely independent of things far away, but they rarely have rigorous justification.  Disorder can give the nearsightedness assumption solid backing.  A little over ten years ago, some condensed matter physicists began to use this assumption to justify using certain approximate algorithms to solve a problem which would require $O{(N^{3})}$ time if a non-approximate algorithm were used\cite{Yang91, Goedecker99}.  These new approximate algorithms are called $O{(N)}$ algorithms or, alternatively, linear scaling algorithms, because they run in a time which is proportional to the basis size; they are much, much faster than $O{(N^{3})}$.

Almost all $O{(N)}$ algorithms were developed specifically for calculating the density matrix function, which is important in simulating motion of atoms in condensed matter physics. In the context of condensed matter physics it was not clear whether the nearsightedness assumption was correct, and it was unclear whether $O{(N)}$ algorithms could be justified.  Physicists pursued both theoretical and computational strategies to support the $O{(N)}$ algorithms.  On a theoretical level, they justified the nearsightedness assumption not with disorder but instead with the band theory of crystals - a bit strange considering that often they were simulating not-crystalline materials, and also condering that, in the important case of pure metals, band theory does not predict strong localization.  On a computational level, physicists used three kinds of evidence to support $O{(N)}$ approximations: tests that each $O{(N)}$ algorithm converged to a single, unambiguous result, checks that on a qualitative level the simulated atoms did more or less what was expected, and demonstrations that much, much larger systems could be simulated. Within these criteria the $O{(N)}$ algorithms seemed to work well.  There have been very few\cite{Smirnov01, Smirnov02} more rigorous computational checks, and these concerned themselves with the accuracy of the Green's function, which certain $O{(N)}$ algorithms compute as an intermediate step in the process of computing the density matrix.

In my thesis research I did the first quantitative and systematic check of the accuracy of $O{(N)}$ algorithms for calculating the density matrix.  I studied a class of $O{(N)}$ algorithms which I call basis truncation algorithms.  Contrary to my previous expectations, I found that $O{(N)}$ algorithms can be very accurate even in metals with a quite small disorder.  Examining the results in detail, I found that the algorithms' accuracy is determined by the coherence length of the eigenfunctions, and occurs even when the eigenfunctions are extended throughout the volume of the system.  To repeat, even when the eigenfunctions are extended $O{(N)}$ algorithms which ignore any long-distance physics can still provide accurate results.  I also found a formula which accurately predicts the error. These results have been accepted for publication; an expanded version is given here in chapter \ref{DensityMatrixAccuracy}.  This research is significant for anybody who wants to simulate a system with several length scales, since it widens the class of problems where $O{(N)}$ algorithms are applicable, suggests that the coherence length is the best way of determining when these algorithms are applicable, and gives a tested formula for estimating the error.

Many of the same $O{(N)}$ algorithms used to calculate the density matrix can also be applied to computing other matrix functions which are important in many other fields, but these opportunities remain largely unexplored.  I followed up my density matrix research by studying the accuracy of $O{(N)}$ algorithms when computing the inverse, logarithm, exponential, Cauchy (Lorentz) distribution, and Gaussian distribution.  I plan to publish these results, which are described in chapter \ref{FunctionAccuracy}.  They are significant because they again extend the range of applicability of $O{(N)}$ algorithms, this time to the most important matrix functions in physics and the other sciences.

One of the original motivations for this thesis is the challenge of simulating fermions in lattice QCD.  I review this problem and its connections with disordered systems in chapter \ref{LatticeQCD}.  Briefly, modern supercomputers have considerable difficulty in doing "unquenched" calculations; i.e. calculations with dynamical quark degrees of freedom, or, in other words, with fermion loops.  Numerically, the difficulty is that one must repeatedly calculate either a determinant (related to the matrix logarithm) or a matrix inverse.  Both of these problems run in $O{(N^{3})}$ time. My goal is to develop new $O{(N)}$ algorithms for this problem, and in fact this was the motivation for studying the logarithm and the inverse in chapter \ref{FunctionAccuracy}.  

While developing theoretical estimates of the error of $O{(N)}$ algorithms, I had to estimate matrix elements and eigenfunctions in disordered systems.  About twenty-five years ago, Berry introduced a model of incoherent, unlocalized, nonfractal eigenfunctions\cite{Berry77}.  I developed a generalization of his model which is able to describe both localized eigenfunctions and multifractal eigenfunctions.  Multifractal eigenfunctions are expected in disordered systems, as was first noted by Castellani and Peliti\cite{Castellani86}.  I expect to submit this phenomenological model for publication soon, after revisions to reflect some results published over the past few years which hold in the limit of small disorder.  This work is presented in chapter \ref{CoherenceLength}.

I have a lot of experience in the software industry, having worked almost five years for Microsoft.  One of my biggest lessons from this experience is that software can never be trusted until both it and its results are systematically and comprehensively checked and rechecked.  A few years ago I returned to academic life, and was immediately confronted by the fact that the scientific community does not do systematic checking of its software and routinely publishes papers containing the results of computations which are neither documented nor systematically tested, and are neither archived nor reproducible.  According to the standards of the software industry this sort of conduct is very unprofessional.  However this conduct could possibly be partially excused by the fact that the scientific community addresses problems which are somewhat different than those addressed by the software industry. It is important to understand the scientific community and the scientific process before coming to conclusions about correct software practices.

In order to decide what my own software practices should be as a physicist, I decided to go through a process of researching and reviewing the physics community's awareness and management of software unreliability, in other words, bugs.  This work culminated in two things: first, realizing that it is very important to make one's computations reproducible - i.e. to allow both oneself and other parties to easily and reliably reproduce one's own results, if possible simply by installing and running the script or program one has written.  The second result was a set of recommendations about scientific discipline, for both individuals and of organizations.  I wrote this work up into a rough draft of a paper on the reliability of scientific computations and wrote to a scientific mailing list where roughly fourty people gladly gave comments.   A second draft is included here as chapter \ref{Reliability}.  It should be of interest to anyone who is concerned about the reliability of the numerical results (numbers, graphs, or formulas derived using software) which they either publish themselves or read in published scientific papers.  There are very few articles on this topic within the whole scientific community, let alone the physics community, and I am aware of none that have the comprehensive approach which I have taken, with both a review of current practices and a holistic set of recommendations for best practices.

My numerical research into the accuracy of $O{(N)}$ algorithms, documented in chapters \ref{DensityMatrixAccuracy} and \ref{FunctionAccuracy}, was also an exercise in implementing the ideas of my paper on reliability.  I followed most of the best practices I had suggested, in order to see whether they were really such good ideas before submitting the paper for publication.  One very exceptional best practice is that I am making both this Ph.D. thesis and my research papers entirely reproducible; meaning that I supply an automated way for interested parties to re-compute all my results, recreate the graphs shown in my papers, and recompile the papers themselves.  All necessary code and files are freely available on the Internet under the GNU public license\cite{GNUGPL} or other free licenses.  Reproducible papers are an important tool for managing the complexity and unreliability of computation, but they are still only beginning to be used.  There are very few such completely reproducible papers available in the scientific community as a whole, perhaps less than ten.  The lessons I have learned from writing these reproducible papers are included in chapter \ref{Reliability}.

 One important way of checking numerical results is to develop as many analytical predictions as possible, in various limits or special cases, and then to check that the software reproduces the correct result.  Such analytical predictions can be very difficult to obtain if the problem being calculated is complex. In particular, there are currently no analytical predictions that can be used to test the algorithms used to simulate fermions in lattice QCD, except in test cases where gauge fields have very special smooth configurations.  These cases therefore fall outside the region of physical interest, where the gauge field has its own independent degrees of freedom and is therefore disordered.  One of my research goals is to develop analytical predictions about fermionic algorithms in the presence of disorder, which would provide for the first time the ability to check the convergence and accuracy of fermionic algorithms in physically realistic problems.  This would allow me to not only check the existing fermionic algorithms but also compare and contrast them with new $O{(N)}$ algorithms.
 
The overlapping communities of mesoscopic physics and random matrices have developed a variety of field theoretical techniques for making analytical predictions about disordered systems.  In the limit of weak disorder, they are able to make analytical predictions about eigenvalues, eigenfunction self-correlation, correlation between eigenfunctions, products of Green's functions, etc.  Two field theories are available for these calculations: the first is a sigma model, based on a "replica" technique, which involves modelling many species of particles, with all the species having the exact same physical properties.  This replica technique is very flexible and has been applied to a wide range of very hard problems.  Models derived using the replica technique correctly describe physical systems where there really are many species of particles.  However, the replica sigma model is most often used to describe physical systems with only one species of particle.  In this case, the replica model is correct to all orders of perturbation theory, but can be wrong when treating non-perturbative phenomena\cite{Zirnbauer99}.  The second field theory is also a sigma model, but is called a supersymmetric sigma model because it is based on using graded matrices, i.e. matrices containing both Grassman variables and non-Grassman variables. This field theory in terms of graded matrices is derived from the original field theory via a Hubbard-Stratonovich tranformation.  The supersymmetric sigma model is confined to a much smaller set of problems than the replica sigma model, but within its range of applicability it distinguishes itself by correctly handling non-perturbative phenomena.

To date, no work has been done generalizing these models to the problems I'm interested in, namely fermions moving in a disordered potential, with the fermions modeled by a determinant as is normal in quantum field theory.  There is no real conceptual difficulty with this generalization, it's just that no one has realized that it would be useful, partly because of the scientific community's limited interest in testing its computations.  However, a considerable amount of work would be required to generalize the supersymmetric sigma model: one's graded matrices would have more fermions than before, and there is a complicated and tricky step in the supersymmetric method where one factors the supersymmetric matrix into a product of other matrices.  So despite the lack of conceptual difficulty, the actual dirty work of this generalization is a bit daunting.  Generalizing the replica sigma model is much easier, but one can't be sure of the non-perturbative results.
 
So I was preparing to do this generalization, but along the way I realized that there is a third sigma model, which can be used to obtain all the results which are accessible to the supersymmetric sigma model, but does not involve graded matrices.  This realization was stimulated by a recent paper by Fyodorov\cite{Fyodorov02}  which derived the third model in the case of a system with no spatial extent; i.e. one which is a single point in space.  At the end of his paper he wrote "Finally, it is interesting to explore if the Ingham-Siegal integrals and their natural generalizations could provide a serious alternative to the Hubbard-Stratonovich transformation in the whole domain of random matrices and disordered systems."  However, Fyodorov did not begin to follow up this speculation until September 5, 2004, when he submitted a preprint which, in an appendix, briefly describes how how to use his technique to derive a replica sigma model on a lattice \cite{Fyodorov04}.  The sigma model which he derived was first introduced twenty-four years ago, and was the first sigma model derived for a disordered system\cite{Wegner79, Wegner79a, Schafer80, Pruisken82, Jungling80}.  Fyodorov's derivation was exact because he derived the sigma model from a hamiltonian with many identical species of particles.  He did not treat the more physically important case of a system with only one species of particle, which would traditionally require either the replica technique or graded matrices.  However, he did speculate that this was possible, writing: "Another interesting problem is how to include anticommuting degrees of freedom in the above derivation.  Technically this can be done following various methods ...."  Fyodorov then cites both a paper by Zirnbauer which does not include Grassman variables and also unpublished research done by Fyodorov himself.

I did not become aware of Fyodorov's preprint until early October, 2004.  A month earlier, in early September, I began thinking that Fyodorov's zero-dimensional sigma model can be extended to systems which are spatially extended, and that all the results previously obtained with graded matrices can be obtained without them.  This assertion, if true, would have a large influence on mesoscopic theory, as it offers a drastic simplification and also makes the theory much more flexible, so that generalizing, for instance, to the problem of fermion determinants in disordered potentials becomes quite easy.  So I began very carefully working through the details of deriving the new model and reproducing results already derived using the supersymmetric sigma model. This work, contained in chapter \ref{NoGradedMatrices}, is not yet complete, but already it contains a detailed derivation of the new sigma model of extended systems, as well as the most detailed and thorough tutorial in the literature about how to use a field theory to calculate observables in the unitary ensemble of random matrices.  

Earlier I examined the accuracy of a certain class of $O{(N)}$ algorithms, namely basis truncation algorithms, which can perform remarkably well.  However these algorithms are particularly simple minded, and one might expect to find situations where they fail but more sophisticated $O{(N)}$ algorithms succeed.  In chapter \ref{LinearAlgorithms} I review all the $O{(N)}$ algorithms which I am aware of.  I then suggest several new algorithms which I believe are my own contribution.

My interest in $O{(N)}$ algorithms and in disordered systems was stimulated by an attempt many years ago to figure out how to apply renormalization ideas to the $O{(N^{3})}$ problem of diagonalizing a matrix.  The theory of renormalizing linear algebra $O{(N^{3})}$ problems is very simple compared to that of renormalizing field theories, but it is still very hard to figure out how to use this theory non-perturbatively to obtain faster algorithms, because the renormalization itself requires $O{(N^{3})}$ time.  When renormalizing field theories, there is a similar difficulty (i.e. an infinite number of operators are created), which is surmounted by assuming that the renormalized theory is local, and also assuming that the system is self-similar between various length scales, and also using perturbation theory.  Here we are doing numerical calculations, which are usually aimed at obtaining non-perturbative information, so perturbation theory is not an option.  Moreover one wants to avoid making assumptions about self-similarity and to compute the relation between length scales from first principles.   
 
However the assumption that renormalization preserves locality can still be useful, and translates into a sort of near-sightedness principle.  Renormalization is basically a program of averaging neighboring short distance degrees of freedom to produce a smaller set of long distance degrees of freedom.  Locality preservation means that one assumes that the short distance degrees of freedom influence only the long distance degrees of freedom which are close by.  In other words, short distance degrees of freedom that are far apart interact only via long distance degrees of freedom.  This is a key assumption; otherwise the perturbative renormalization program in field theory would fail.
 
 $O{(N)}$ algorithms are simply the result of making the locality assumption which is so basic to renormalization.   $O{(N)}$ nearsightedness states that none of the degrees of freedom in the problem interact at long distances; that the problem can be effectively broken into parts.  The is just the simplest case of renormalization's locality preservation assumption, which I discussed in the last paragraph.  
 
 It is natural to consider physical problems where there are more length scales, and to seek a way of renormalizing away first the smallest length scale, then the next smallest, etc.  Therefore even before my thesis I analyzed a class of $O{(N {{\ln}^{2} N})}$ algorithms which can be viewed as generalizations of $O{(N)}$ algorithms, and are specifically designed to handle multi-scale systems.  In chapter \ref{Multiscale} I explain my new $O{(N {{\ln}^{2} N})}$ algorithms, and the very substantial accelerations that they can produce.  These new algorithms use some of the new $O{(N)}$ algorithms explained in chapter \ref{LinearAlgorithms} as building blocks.  Both chapters are of interest to persons working on numerical computing in problems with multiple length scales.

\chapter{\label{LaypersonsIntroduction}Layperson's Introduction to Chapters \ref{DensityMatrixAccuracy} and \ref{FunctionAccuracy}}
\section{High School Algebra}
Do you remember doing algebra in high school?  There were
variables $x$ and $y$, $a$ and $b$, letters which represented
numerical quantities but whose values were not known.  You were
given equations relating those variables to numbers, for instance
${5} = {{3 x} + 2}$.  Then your teacher asked you to solve the
equation, which meant find out what value the variable had; in
this case you would find ${x} = {1}$.

Perhaps you were also given equations with two variables in them;
${x + y} = {1}$.  It turned out that if you had two variables, you
needed two equations in order to find a solution.  For instance,
if you had a second equation ${y} = {1}$, you could solve for $x$;
${x} = {0}$.

And you may even have been told, but probably not, that you can
actually have as many variables as you like, a whole alphabet, or
thousand or millions of them. But you also need as many equations
as variables before you can find your solution.  Thus you can
imagine having a thousand equations for a thousand variables, and
solving to find the value of each variable.  Of course this would
be a nightmare; doing two unknowns was already a chore in high
school; a thousand would be impossible for mere mortals.  But
computers - champions at busywork - can do this sort of thing.

In algebra you probably also were introduced to inequalities,
graphing, and the quadratic equation.  If you remember the
quadratic equation then you know it was exceptionally complicated
to solve.  This is because all the other equations (and sets of
equations) you saw were linear, which means that variables were
always multiplied by numbers, never by other variables. Nonlinear
equations are considerably harder than linear ones: for instance
nonlinear equations are connected to chaos and unpredictable
behavior and sometimes do not have any solution at all, while
linear equations are much better behaved and easier to solve. Yet
despite their relative simplicity linear equations are interesting
and complex enough that there is a whole mathematical specialty
devoted to them.  My concern in this introduction focuses on
problems formulated in terms of linear equations, which are called linear algebra problems.  One linear algebra problem that is very interesting is the eigenvalue problem,
which is a way of solving many sets of linear equations at the
same time.

\section{Scientific Calculations}
 The reason why I took you on this walk through memory
lane is that most sciences and applied sciences make predictions
by creating large sets of variables and equations, and then
solving them. For instance, suppose an engineer wanted to predict
how a structure (a house, a bridge) will respond to forces on it.
First she would break it down conceptually into pieces:  boards, struts, walls,
cables, whatever.  Second, she would represent each piece with a
variable.  Next, for each piece she would write an equation which
tells - in a mathematical way - how it acts on, and is acted on
by, all the other pieces in the structure; i.e. the forces on it.
Lastly, having written down these equations, our engineer would
solve them, using a computer of course.  This would allow her to
predict how a house would respond to an earthquake, or whether a
bridge would bear a 100 ton load.

The same sort of thing happens in chemistry.  Everything is made
up of atoms like hydrogen and copper, and each atom has one or
more electrons around it.  The quantum rules of how atoms connect
up to make molecules, liquids, and solids are determined by what
their electrons do; sometimes all the electrons will stay close to
the atoms that own them; other times the atoms will share electrons.  The
details of this process determine what molecules and materials are
formed, their colors and consistencies, how chemical reactions
happen; basically everything. In order to make a quantum
calculation of these things, you need at least one variable for
each atom, and more if you want to be more accurate.  You also
need a set of equations saying how each electron is influenced by
the neighboring electrons.  With the variables and equations in
place, you need to solve the equations to find out what really
happens. This sort of procedure has worked very well for calculating
the properties of smallish molecules with up to a few hundred
atoms, and also a lot of crystals because they can be broken down
into their smallest building block, which is often just a few
atoms.

\section{Speed Problems}
There is a problem, though: modern computers often can't solve the
equations in a reasonable amount of time.  It turns out that
solving the eigenvalue problem (and also a lot of other linear algebra problems) 
with a few thousand variables
stretches their limits, especially when (as often happens) you
have to repeatedly change the numbers a bit and then solve again.
The practical consequences are pretty bad. For instance, to
calculate what happens in biology you have to deal with proteins,
which contain thousands or tens of thousands of atoms. And these
proteins are immersed in liquids, which contain many more atoms.
So right now there is no way to do a quantum calculation for most
biological problems, and therefore accuracy is very limited.

The same sort of issue comes up with many other important
scientific and engineering problems. For instance, if you wanted
to predict the behavior of a skyscraper or supertanker you might
want to think about all their beams and struts and plates and
joints, but you can't because there are too many. In this case
what you would have to do is start grouping things together into
units (for instance a set of struts making a truss), and assume
that each unit moves together. This simplification will allow you
to calculate stuff on a computer, but it will also be less
accurate, as you can see if you consider that your computer is no
longer allowing for the possibility that units may twist, bend, or
even break apart.

You may have heard that computers speeds are growing fast.  In
fact their speed is growing exponentially, just like the world
population.  Of course such explosions can't last forever;
eventually a limit is reached, but you could still imagine
(optimistically) that we might not hit the limit for another
twenty years, in which case computer speeds might multiply by a
thousand.  Then it would be interesting to know what the speed
increase would do for our ability to solve equations.

There is a whole mathematical discipline devoted to analyzing
different problems and figuring out how much effort is required to
solve them.   Some problems are so inherently complex that
multiplying your speed (or, alternatively, multiplying the time
you spend) doesn't really do much. A simple example of this would
be trying to understand how to move in a board game like chess,
if you know the rules but don't know how to play the game.
(Actually, almost any problem in real life is like this if you
approach it in a literal, systematic, pencil-pusher sort of way.)
Other problems are much easier, even for pencil pushers: for
example when sorting books into alphabetical order, if you have
twice as much time, then you can sort almost twice as many books.
Solving the eigenvalue problem is an intermediate case, but still
pretty bad: if you have ten times as many variables, you will need
to put in a thousand times as much effort.  Which means that we
can't expect faster computer speeds to help us much with a
lot of important scientific and engineering problems, for example 
 accurate quantum calculations of most biological processes.

You may know from experience that often a hard problem can made
easier by figuring out what is important and what is not, and then
concentrating on the important part.  Moreover, if you have a big
problem, you might try to break it up into smaller problems which
can be solved easily one by one.  "Divided we fall, united we
stand." Unfortunately many linear algebra problems are not friendly to the
ideas of solving one piece at a time, or of figuring out important
things first.  In linear algebra problems, every variable is equally
important, and depends on every other variable.   For example,
they predict that the movement of one beam in a building depends
on what's happening to every other part of the building.  This is
why solving the eigenvalue problem is so hard; everything is
bundled together into the same knot, and you have to find the
solution all at once instead of picking the knot apart bit by bit.

\section{The Distance Strategy and $O{(N)}$ Algorithms}
When confronted with a roadblock, find another way.  In the past
ten years many scientists have begun to think that maybe linear
algebra could be improved upon.  We know from real life that not
every fact is equally important, and in fact we know that a lot of
stuff can be ignored.  Maybe there is some way to modify the
equations so that they sort out important from unimportant.
Of course, you need a strategy for deciding what's important for
calculating the value of a variable. One simple strategy is
distance: if something is close by, it's important; if not, it's
not. Scientists have used this strategy to create new equations
where variables depend only on things that are close by them. For
instance, when using these equations on a molecule,
electrons only depend on what's happening close by, and therefore
you can break a molecule up into pieces which can be calculated
separately, and everything is much quicker.  The new equations - called
$O{(N)}$ algorithms (pronounced "order N algorithms") are much
easier to solve, and nowadays molecules with many thousands of
atoms can be calculated.  This is very exciting: by choosing a very simple 
strategy (which I will call the distance strategy) of making a black and white distinction between near and far, lots of previously impossible problems are suddenly in reach.

But does the distance strategy make sense?   It depends. Imagine a
house with a tiled roof, and that you have the job of predicting
what will happen to one of the roof tiles.  Well, certainly you
need to know about things close to the tile: its support, the
tiles around it, workmen walking on the roof, and any tree
branches falling close by.  And - at least if the house is well
designed - you can ignore things a little farther away: whether
the windows are open or closed, the number of rooms in the house,
and even damage to house: an errant driver, or a tree falling
against another part of the roof.  So, clearly the distance strategy
makes some sense. And it might make even more sense when applied
to other physical problems, like many chemistry calculations.

But there are also circumstances where the distance strategy
doesn't work. The foundation is the part of the house farthest
from our roof tile, and yet if it washes away then the roof will
not stand on its own.  Hundreds of miles away, deep in the ground,
masses of earth slipping past each other can destroy the house in
an earthquake.  And the last week's weather thousands of miles
away can determine the rain, wind, hail, or sun beating on your
tile.

The next two chapters describe some work that I did on checking when these distance-based strategies work, and trying to figure out what decides whether they work or fail.  
I solved some linear algebra problems the correct-but-slow way, and also using an $O{(N)}$ algorithm, and compared the results. I was surprised that the $O{(N)}$ algorithm worked far better than I had expected.

\chapter{\label{DensityMatrixAccuracy}Accuracy of $O{(N)}$ Algorithms for the Density Matrix}
In this chapter I discuss some fast algorithms for evaluating the density
matrix, which is a matrix function; i.e. a function whose argument and value are both matrices.  This is a difficult problem from linear algebra, a field which contains immense challenges that
have stimulated a very active and diffuse research effort ever since
computers were introduced.  When solving a linear algebra problem one must first 
choose a finite basis.
Second, one must ensure stability of the algorithm, or in other
words ensure that it always converges to a result. Third, one must
analyze and understand the accuracy of the algorithm; its result
must be in some sense close to the correct answer. Fourth and
lastly, one must accelerate the computation as much as possible,
whether by choosing an algorithm which is fast or well-suited to
the problem at hand, or by choosing a basis which matches the
physics well and thus can be trucated to a small number of basis
elements. Traditionally, physicists have worked on the choice of
basis - and, to some extent, the choice of algorithm - while the
invention of algorithms, as well as their convergence and
accuracy, have often been left to computer scientists.

One of the biggest challenges is a computational bottleneck: evaluation of a matrix function
generally require $O{(N^{3})}$ time, where $N$ is the basis size
and usually scales linearly or worse with the system volume.  This
prohibits computations of large systems, even with modern
computers. In particular, if a matrix function must be
recalculated at every step of a system's evolution in time, then
calculations with a basis size larger than $O{(1000 - 10000)}$ are
not practical.  For example, quantum simulations of a single protein molecule require re-evaluation of the density matrix function at every time step.  Calculations of protein dynamics, which would involve at least $O{(2000)}$ atoms, are well out of reach\cite{Segall02}; in fact modern computers are limited to $O{(100)}$ atoms.  Larger numbers of atoms can be handled only by smearing them all together into a jelly (jellium), or else by stepping away from quantum physics.  
Moore's law will not solve this problem: a ${10}^{3}$ increase of speed allows only a ${10}^{1}$ increase in system volume.  No exponential growth lasts forever.

\section{\label{OrderN Algorithms} $O{(N)}$ Algorithms}
In the last ten years physicists have taken the lead in developing
a new class of algorithms motivated by a physical insight: a
substance's properties at a point $A$ should not depend on things far away from $A$.  For instance, a point lattice defect should have little effect on electron
orbitals even a few lattice spacings away.  Although this is a familiar idea, perhaps Kohn\cite{Kohn96} is responsible for naming it, calling it the "nearsightedness" principle.  The nearsightedness principle is translated into an expectation that the density matrix function should have a finite range in position space, beyond which its matrix
elements die off exponentially.  Therefore the density matrix contains only $O{(N)}$ non-zero elements, and one can invent algorithms for evaluating the density matrix which run in $O{(N)}$ time.  In 1991 W. Yang introduced the first such algorithm, the Divide and Conquer algorithm, which first diagonalizes the argument of the density matrix and then approximates the density matrix function, all in $O{(N)}$ time\cite{Yang91}. This
stimulated the development of many other $O{(N)}$ algorithms
\cite{Ordejon98, Scuseria99, Goedecker99, Wu02, Bowler02} which have met
considerable success, permitting quantum calculations of systems with tens of thousands of atoms.  

$O{(N)}$ algorithms succeed because they are an application of a
very basic problem solving principle.  Both humans and the
computers we create are fundamentally limited, and therefore our
problem solving must always start by separating the important
information from the unimportant, the wheat from the chaff.
We break this basic problem-solving rule when we try to evaluate a matrix function while using a large basis.  The real problem is probably not the basis size but the conceptual framework of linear algebra, which contains no notion of the relative importance of information, so that a solution's value can be changed by varying any matrix element whatsoever.  $O{(N)}$ algorithms correct this mistake by selectively ignoring certain information.

All $O{(N)}$
algorithms rely on special characteristics of the system under
study, and the question of their validity can be answered only
after having specified that system. To date, theoretical studies
of the validity of these algorithms have occurred almost
exclusively within the conceptual framework of ordered systems,
using ideas of metals, insulators, and band gaps\cite{Baer97,
Baer97b, Goedecker98, IsmailBeigi99, Goedecker99b, Goedecker99,
Zhang01, Koch01, He01, Taraskin02}. In this chapter and the next I carefully
examine the applicability of $O{(N)}$ algorithms to disordered
systems. I calculate the density matrix function in two ways: with an
$O{(N)}$ algorithm, and via diagonalization. Comparison of the two
results provides some new insight into when disorder can make
$O{(N)}$ calculations feasible.

 All $O{(N)}$ algorithms make three basic assumptions:
 \begin{itemize}
\item Existence of a Preferred, Local Basis. It is assumed that
the system is best described in terms of a localized basis set. I
here define a basis as localized if for any basis element ${| \psi
\rangle}$, only a small number of positions $\vec{x}$ satisfy
${\langle \vec{x} | \psi \rangle} \neq 0$.  In particular, plane wave bases are excluded. Note
that crystal calculations are still possible, by using a free
propagator that properly includes the crystal lattice structure.
 This theory, called KKR band theory, was developed by Korringa, Kohn, and Rostoker\cite{Korringa47, Kohn54}.

\item Existence of a Distance Metric.  There must be a way of
computing the physical distance between any two basis elements
${|\psi\rangle}$ and ${|\acute{\psi}\rangle}$.

 \item Locality of both the Matrix Function and its Argument.
 I call a matrix $A$ local if, for every pair
 of basis elements $| \vec{x}
 \rangle$ and $| \vec{y} \rangle$ that are far apart,
  ${{\langle \vec{x} |} A {| \vec{y} \rangle}}
= 0$.  Throughout this chapter and the next I choose a simple criterion for
being far apart: comparison with a radius $R$.
\end{itemize}

 There are a
number of ways for an $O{(N)}$ algorithm to exploit the three
basic assumptions. In this chapter and the next I focus on the class of
algorithms based on basis truncation. This class includes Yang's
Divide and Conquer algorithm\cite{Yang91}, the "Locally
Self-Consistent Multiple Scattering" algorithm\cite{Abrikosov96,
Abrikosov97}, and Goedecker's "Chebychev Fermi Operator
Expansion"\cite{Goedecker94, Goedecker95}, which I will henceforth
call the Goedecker algorithm. Basis truncation algorithms break
the matrix function into spatially separated pieces. Given the
position of a particular piece, the basis is truncated to include
only elements close to that position, and then the matrix function
is calculated within the truncated basis. Thus, for any generic
matrix function $f{(H)}$, a basis truncation algorithm calculates
$ {{\langle \vec{x} |} {f{(H)}} {| \vec{y} \rangle}} = {{\langle
\vec{x} |} {f{(P_{\vec{x}, \vec{y}} H P_{\vec{x}, \vec{y}})}} {|
\vec{y} \rangle}}$, where $P_{\vec{x}, \vec{y}}$ is a projection
operator truncating all basis elements far from $\vec{x}$,
$\vec{y}$. There may be also an additional step of interpolating
results obtained with different $P$'s, but I will ignore this. In
this work I choose $P$ to be independent of the left index
$\vec{x}$, and truncate all basis elements outside a sphere of
radius $R$ centered at $\vec{y}$.

Basis truncation algorithms vary only in their choice of how to
evaluate the function $ {f{(P_{\vec{x}, \vec{y}} H P_{\vec{x},
\vec{y}})}} $. Specifically, Yang's algorithm calculates $f$ in
terms of the argument's eigenvectors, the Goedecker algorithm does
a Chebyshev expansion of $f$, and the "Locally Self-Consistent
Multiple Scattering" algorithm calculates the argument's resolvent
or Green's function and then obtains $f$ via contour integration.
Because these approaches are all mathematically equivalent when
applied to analytic functions, they should all converge to
identical results, as long as one makes identical choices of which
matrix function to evaluate, of how to break up the function, of
which projection operator to use, and of a possible interpolation
scheme. Moreover, given an identical choice of matrix function,
variations in the other choices should obtain results that are
qualitatively the same.  In this work I use the Goedecker
algorithm, but I want to emphasize that the results obtained here
apply to the whole class of basis truncation algorithms.

The Goedecker algorithm is essentially a Chebyshev expansion of
the matrix function.  As long as all the eigenvalues of the
argument $H$ are between $1$ and $-1$, a matrix function may be
expanded in a series of Chebyshev polynomials of $H$: $f{(H)}
\cong { \sum_{s=0}^{S} {c_{s} T_{s}{(H)}}}$.  The coefficients
$c_{s}$ are independent of the basis size, and therefore can be
calculated numerically in the scalar case.  The Chebyshev
polynomials can be calculated in $O{(N)}$ time using the recursion
relation $T_{s+1} = {{({2 H T_{s}})}
 - T_{s-1}}$, $T_{1} = H$, $T_{0}$ = 1.  (Of course, one must also
 bound the highest and lowest eigenvalues of $H$ and then
 normalize.  In practice very simple heuristics are sufficient for
 estimating these bounds.)  If the matrix function $f{(H)}$ has a
 characteristic scale of variation $\alpha$, then the error induced by the Chebyshev
 expansion  is controlled by an exponential with argument of order
 $-\alpha S$.

\section{Measuring The Error} Previous efforts to test
numerically the accuracy of $O{(N)}$ algorithms have almost always been confined
to qualitative evaluations of whether their overall physical predictions (total energy, bond structure, etc.) are
reasonable, and tests of convergence.  The only exceptions that I am aware of are two papers examining the accuracy of the Green's function, which is computed as an intermediate step in the "Locally Self-Consistent Multiple Scattering" algorithm\cite{Smirnov01, Smirnov02}.   The research presented here sets itself apart by carefully checking basis truncation algorithms quantitatively in a systematic and detailed way.  I computed the density matrix function using both an $O{(N)}$ algorithm and an algorithm based on diagonalization, and then
compared the results.

This careful comparison required development of a metric for
comparing two matrices. First, note that the dot product used for
comparing two vectors can be easily generalized to matrices:
$${MDP{(A, B)}} \equiv {Tr{(AB)}} =
{\sum_{\vec{x},\vec{y}}{{\langle \vec{x} | A | \vec{y}
\rangle}{\langle \vec{y} | B | \vec{x} \rangle}}} $$ If $A = B$,
this matrix dot product is just the 	square of the Frobenius norm, one
of the traditional norms for matrices. Note also that matrix dot
product is invariant under change of basis. Moreover, it is simple
to show that ${-1} \leq {\frac{MDP{(A, B)}}{\sqrt{{MDP{(A, A)}}
{MDP{(B, B)}}}}} \leq {1}$, so one may define the angle $\theta$
between two matrices as the arcsin of this quantity. Bowler and
Gillan\cite{Bowler99} justified this, showing that the concept of
perpendicular and parallel matrices is valid and useful.

However the matrix dot product is not quite suited to my needs.
$O{(N)}$ algorithms have a preferred, local basis, and thus are
not well matched by a basis invariant measure.  Moreover, the
matrix functions which they compute are expected to agree best
with the exact matrix functions close to the diagonal, and to
agree not at all outside the truncation radius. Therefore, a more
sensitive metric is needed, one that distinguishes different
distances from the diagonal.  I define the Partial Matrix Dot
Product as:

$${MDP{(A, B, \vec{x})}} \equiv {\sum_{\vec{y}}{{\langle \vec{y} |
A | {\vec{y} + \vec{x}} \rangle}{\langle \vec{y} | B | {\vec{y} +
\vec{x}} \rangle}}} $$

The argument $\vec{x}$ of this dot product allows me to obtain
information about agreement at displacement $\vec{x}$ from the
diagonal.  It is still valid to call this a dot product, because
the magnitude of $\frac{MDP{(A, B, \vec{x})}}{\sqrt{{MDP{(A,
A,\vec{x})}} {MDP{(B, B, \vec{x})}}}}$ is bounded by one, and thus
one can compute a displacement-dependent angle $\theta{(\vec{x})}$
and relative magnitude $ m{(\vec{x})}$. The partial matrix dot
product has a simple sum rule relating it to the full matrix dot
product: $MDP{(A,B)} = {\sum_{\vec{x}} {MDP{(A, B, \vec{x})}}}$.

In my results I actually compute another dot product, an angular
average $MDP{(A, B, r)}$ over all $\vec{x}$ satisfying ${r} =
{|\vec{x}|}$.

In this work I used a projection operator $P$ for basis truncation which is independent of the left index
$\vec{x}$ and truncates all basis elements outside a sphere of
radius $R$ centered at $\vec{y}$.  This truncation strategy was chosen because it allows the matrix function to be calculated one column at a time instead of one element at a time.  However, this strategy does not respect transverse symmetry in the matrix function's argument, which should result in the matrix function also having transverse symmetry.  The asymmetric truncation also destroys the matrix dot product's reflection symmetry ${MDP{(A, B, \vec{x})}} = {MDP{(A, B, -\vec{x})}}$, which holds whenever $A$ and $B$ have transverse symmetry.  However, it is reassuring to note that the angular matrix dot product $MDP{(A, B, r)}$ is not sensitive this difficulty.  In any case, the symmetry breaking effects go to zero as the truncation radius $R$ is increased to infinity, and therefore are insignificant unless the $O{(N)}$ algorithm would fail irregardless of choice of truncation strategy.  

As far as I know, the first occurence of these matrix dot products and partial matrix dot products in the literature was in my paper on density matrix errors\cite{Sacksteder03}.  I believe that the partial matrix dot product defined here is the appropriate metric for evaluating the accuracy of $O{(N)}$ algorithms, and will be useful to other researchers.

 \section{The Density Matrix}
 In this paper I restrict my attention to a single matrix
 function, the density matrix.  This function is very important in
 quantum calculations of
 electronic structure in atoms and molecules, where its argument
 is the
 system's Hamiltonian, its diagonal
 elements describe the charge density, and its off-diagonal
 elements are used to compute forces on the atoms.  Eigenvalues of
 the Hamiltonian give the energies of their corresponding
 electronic states, and I
 will use the words eigenvalue and energy interchangeably
 throughout the rest of this thesis.

 The density matrix function $\rho{(\mu, T, H)}$
 is basically a projection operator
 which deletes eigenvectors having energy $E$ larger than the
 Fermi level $\mu$.
  Here I use the following form:
$$\rho {(\mu, T, H)} = {\frac{1}{2} {Erfc}{(\sqrt{2}{(\frac{H -
{\mu I}}{T})})}}$$
  For physical reasons, it is not quite a projection
  operator: it has a transition
 region around $\mu$ of width proportional to the temperature $T$
 where its eigenvalues interpolate between $0$ and $1$.
The error induced by a Chebyshev expansion is controlled by an
exponential with argument proportional to ${-T S} / { \Delta}$,
where $\Delta$ is the size of the energy band and $S$ is the
number of terms in the Chebyshev expansion\cite{Goedecker94}.

 The density matrix is well suited to $O{(N)}$ algorithms.  As $\mu$ becomes
large, it converges to the identity. Moreover, it is invariant
under unitary transformations acting on the set of eigenvectors
with energies below the Fermi level\cite{Scuseria99}. Even when
the Hamiltonian's eigenvectors are not a local basis set, often a
unitary transformation can be found which maps them to a basis
which is localized.  If such a transformation exists, the density
matrix is localized.

Several papers have examined density matrix locality in the
context of ordered systems; i.e. ones whose Hamiltonians possess
lattice translational invariance\cite{Baer97, Baer97b,
Goedecker98, IsmailBeigi99, Goedecker99b, Goedecker99, Zhang01,
Koch01, He01, Taraskin02}. (Lattice translational invariance can
be expressed quantitatively as $ {{\langle \vec{x}|} H { | \vec{y}
\rangle}} = {{\langle \vec{x} + \vec{\Delta}|} H { | \vec{y} +
\vec{\Delta} \rangle}}$ for all $\vec{\Delta}$ located on an
infinitely extended lattice.) It is well known that the
eigenvalues of such systems are arranged in bands separated by
energy gaps where there are no eigenvalues, and
 that the eigenvectors are extended through all
space.  Notwithstanding the nonlocality of the eigenvectors, there
are strong arguments for localization in all ordered systems. If
the system is metallic (meaning that the Fermi level lies in one
of the bands of eigenvalues) and the temperature is zero, then in
a three-dimensional system the density matrix is expected to fall
off asymptotically as ${R^{-2}}$, where $R$ is the spatial
distance from the diagonal. A non-zero temperature multiplies this
by an exponential decay.  If instead the system is an insulator(meaning that the Fermi level does not lie in one
of the bands of eigenvalues),
then the density matrix should decay exponentially even at $T =
0$.

Most systems of physical interest do not exhibit lattice
translational symmetry.  In particular, many exhibit
inhomogeneities at scales much smaller than that of the system
itself.  These are termed disordered systems.   In this thesis I
study the prototypical disordered system, the Anderson
model\cite{Anderson58}. It describes a basis laid out on a cubic
lattice, one basis element per lattice site, and a very simple
symmetric Hamiltonian matrix $H$ composed of two parts:
\begin{itemize}
\item A regular part: ${\langle \vec{x}|H| \vec{y} \rangle} = 1$
if $\vec{x}$ and $\vec{y}$ are nearest neighbors on the lattice.
This term is, up to a constant, just the second order
discretization of the Laplacian; its spectrum consists of a single
band of energies between $-2D$ and $2D$, where $D$ is the spatial
dimensionality of the lattice.

\item A disordered part: Diagonal elements $<\vec{x}|H|\vec{x}>$
have random values chosen according to some probability
distribution.  In this article I choose to use the Gaussian
distribution
$${P{(V)}} = {\frac{1}{\sqrt{2 \pi \sigma}} {\exp{(\frac{- V^{2}}{2 \sigma}
)}}}$$ I call $\sigma$ the disorder strength; this is related to
the disorder strength used in the literature by a factor of
$\sqrt{12}$\cite{Slevin01, Bulka87}.
\end{itemize}

At small disorder strengths, the Anderson model is dominated by
its regular part; in particular the eigenvectors are extended
throughout the whole system volume. However, there is a small but
important departure from the ordered behavior: at the band edges
one finds a few eigenvectors with volumes much smaller than the
system volume. In fact, there is an energy $E_{LOC}$ such that any
eigenvector with eigenvalue $E$ satisfying ${|E|}
> E_{LOC}$ is localized.  On average these eigenvectors
 decay exponentially with the
spatial distance from their maximum\cite{Kantelhardt02}. As the
disorder strength is increased, $E_{LOC}$ gets smaller and
smaller; i.e. more and more of the energy band becomes localized.
At a critical disorder strength the whole energy band becomes
localized.  This phenomenon is called the Anderson transition; for
the Gaussian probability distribution used in this paper it occurs
at the critical disorder ${\sigma}_{c} = {{6.149} \pm
{0.006}}$\cite{Slevin01}.

Note that these statements must all be understood as regarding the
ensemble of Hamiltonians determined by the probabilistic
distribution of the disorder: for instance, I am stating that
above the critical disorder the subset of Hamiltonians with
unlocalized eigenvectors is vanishingly small compared to the
total ensemble size. Moreover, these statements are valid for an
infinite lattice; the mapping to computations on a finite lattice
is not always absolutely clear.

Studies of the locality of disordered systems have traditionally
concentrated on computations of the Green's function, not the
density matrix. It is expected that the average of the Green's
function should decay exponentially as $
\exp{(\frac{-R}{\tilde{R}})}$, where $\tilde{R}$ is called the
coherence length\cite{Economou84}. The density matrix, as we will
see, is closely related to the Green's function, so one may hope
that its average will also decay exponentially.  However, there
are two reasons why this hope may be unjustified: First, we do not
need to know whether the average of all density matrices is
localized, but instead whether each individual density matrix is
localized. The difference between the two could be significant.
Second, in a system below the critical disorder there will be
unlocalized eigenvectors, and one might therefore expect the
${R^{-2}}$ behavior typical of a metal.

Many $O{(N)}$ computations have treated systems which are
disordered\cite{Goedecker94, Zhang01, Schubert03}.  However, the
$O{(N)}$ literature contains little theoretical material about the
applicability of $O{(N)}$ algorithms to disordered systems. The
originators of the "Locally Self-Consistent Green's Function"
algorithm, which does not truncate the basis but instead does a
sort of averaging outside of a radius $r$, suggested that $r$
should be related to the coherence length\cite{Abrikosov96}, and
also to the error induced by their averaging\cite{Abrikosov97}.
Zhang and Drabold computed the density matrix of amorphous Silicon
using exact diagonalization and found an exponential
decay\cite{Zhang01}. In the next sections, I will first argue
theoretically and then show numerically that $O{(N)}$ basis
truncation algorithms are applicable to disordered systems,
including ones far below the Anderson transition.

\section{Estimating the Error}
The quantity of interest is the relative error,
\begin{equation} \label{Definee}
{e{(r, R)}} \equiv {\frac{MDP{(\Delta{(R)}, \Delta{(R)},
r)}}{MDP{(f, f, r})}}
\end{equation}
where $R$ is the radius of the truncation volume and $\Delta{(R)}$
is the difference between the exact matrix function $f$ and the
approximate matrix function $\tilde{f}{(R)}$.

 A first guess can be
made from the intuition that the absolute error
\begin{equation}\label{DefineE}{E{(r, R)}} = {MDP{(\Delta{(R)},
\Delta{(R)}, r)}} \end{equation}
 is probably
bounded by its value at the boundary of the truncation region.
This allows a rough estimate of the relative error: $ {e{(r, R)}}
\lessapprox {\frac{E{(R, R)}}{MDP{(\tilde{f}, \tilde{f}, r})}}$,
suggesting that it can be made arbitrarily small if the matrix
function is localized. The numerator, however, is left undefined.
Perhaps it is reasonable to assume that on the boundary the
absolute error is equal to the approximate matrix function,
giving:

\begin{equation}
\label{RoughEstimate1} {e{(r,R)}} \lessapprox
{\frac{MDP{(\tilde{f}, \tilde{f}, R)}}{MDP{(\tilde{f}, \tilde{f},
r)}}}
\end{equation}

The following paragraphs develop further insight into the absolute
error by resolving $\Delta{(R)}$ into a multiple sum over dot
products between the argument's eigenstates $| \psi \rangle$ and the
position eigenstates $|\vec{x}\rangle$ . Knowledge of the
normalization and asymptotic behavior of these states sheds some
light on the magnitude of the matrix elements of the error: $
{\langle \vec{x} |} \Delta{(R)} {| {\vec{x} + \vec{r}} \rangle }$.

Basis truncation algorithms separate the basis into two projection
operators, $P_{A}$ for the part inside the localization cutoff
$R$, and $P_{B}$ for the part outside $R$. $P_{A}$ and $P_{B}$ are
then used to divide the argument $H$ into two parts: a part $H_{0}
= {{P_{A} H P_{A}} + {P_{B} H P_{B}}}$ which leaves $A$ and $B$
disconnected, and a boundary term connecting $A$ and $B$, $H_{1} =
{{P_{A} H P_{B}} + {P_{B} H P_{A}}}$.   The final result of a
basis truncation algorithm is just $P_{A} {f{(H_{0})}} P_{A}$.
Therefore, the error induced by a basis truncation algorithm is
entirely due to the boundary term $H_{1}$. In other words, ${P_{A}
{\Delta{(R)}} P_{A}} = {P_{A}{( {f{(H_{0} + H_{1})} - f{(H_{0})} }
)}P_{A}}$.

For matrix functions which are analytic on a region of the complex
plane which contains the poles of $H$ and $H_{0}$, an exact
equation for this boundary effect can be easily derived from the
Dyson equation. First define the Green's functions ${G{(E)}} =
{{(E - H)}^{-1}}$, ${G_{0}{(E)}} = {{(E - H_{0})}^{-1}}$.  Then
note that the matrix function can be obtained from the Green's
function through contour integration over the complex energy $E$:
${f{(H)}} = {\frac {1} {2 \pi \imath} \oint {G{(E)}f{(E)}}}$,
where the complex integral must contain the poles of $H$.  Next,
apply the Dyson equation $G = {G_{0} + {G H_{1} G_{0}}}$ twice to
obtain:
$$G = {G_{0} + {G_{0} H_{1} G_{0}} + {G_{0} H_{1} G H_{1}
G_{0}}}$$  This gives an exact relation between the correct
Green's function $G$ of the untruncated argument $ {H} = {H_{0} +
H_{1}}$ and the Green's function $G_{0}$ of the truncated argument
$H_{0}$. In order to obtain a similar relation for the matrix
function $f$, one must make the poles in this expression explicit
and then do a complex integration.  Define $| a \rangle $ and $| b
\rangle $ as two eigenvectors of $H_{0}$ which are both located
inside of the localization region, the set of $ | c \rangle $ as
the complete set of eigenvectors of $H$, and $E_{a}$, $E_{b}$, and
$E_{c}$ as their respective energies. Then:

\begin{eqnarray} \label{EqExpand}
{{\langle \vec{x} |} {\Delta{(R)}} {| {\vec{x} + \vec{r}}
\rangle}} & = &
 \int_{-\infty}^{\infty}\int_{-\infty}^{\infty}\int_{-\infty}^{\infty}{{dE_{a}}{dE_{b}}{dE_{c}}}
 \nonumber \\
 & &
 {{\langle \vec{x} |a \rangle} {{\langle a|}H_{1}{|c \rangle}}
  {{\langle c |} H_{1} {| b \rangle}} {\langle b | {\vec{x} + \vec{r} \rangle}}
{g{(E_{a},E_{b},E_{c})}}}
\nonumber \\
g{(E_{a}, E_{b}, E_{c})} & \equiv & {{n_{A}{(E_{a})}}{n_{A}{(E_{b})}}{n{(E_{c})}}}  
\nonumber \\ 
& \times & \oint{{dE}f{(E)}
\frac{1}{E - E_{a}} \frac{1}{E - E_{b}} \frac{1}{E - E_{c}}} 
\end{eqnarray}

$n{({E})}$ is the spectral density ${\sum_{c}{\delta{(E-E_{c})}}}$
and is often approximated as a continuous function.  Similarly,
$n_{A}{({E})}$ is the spectral density of the eigenstates of
$H_{0}$ which are located inside the truncation region. If the
matrix elements and matrix function are well-behaved, then this
integral is also well-behaved.  Consider the integral ${\gamma} =
{\int \int \int {{{dE_{a}}{dE_{b}}{dE_{c}}}g{(E_{a}, E_{b},
E_{c})}}}$.  When $f$ is the density matrix and one uses a simple
model with $n$ equal to a constant $\frac{N}{w}$ inside the energy
band $[{\frac{-\omega}{2}},{\frac{\omega}{2}}]$, this integral is
of order $\frac{N }{{\omega}^{2}} {(\frac{4 \pi}{3}R^{3})}^{2}$
when $\mu$ is inside the energy band and $0$ when it is outside
the band.

    Assuming that all eigenvectors of both $H$ and $H_{0}$
    are unlocalized, one can use Eq. \ref{EqExpand} to derive an
    upper bound on ${|{{\langle \vec{x} |} {\Delta{(R)}} {| {\vec{x} + \vec{r}}
\rangle}}|}$ of order $R^{4}$.  In the case of the density matrix
this is a gross overestimate.  Nonetheless Eq. \ref{EqExpand} can
teach three lessons:

\subsection{\label{LocalizedSystems} Localized systems} Suppose that all the
eigenvectors are bounded by $\exp{(-\frac{|{\vec{x} -
{\vec{x}}_{0}}|}{L})}$, where ${\vec{x}}_{0}$ is the point where the eigenvector has its maximum magnitude and $L$ is the minimum decay length of the system.
Then one can use Eq. \ref{EqExpand} to prove in the limit of large $R$ 
that ${|{{\langle \vec{x} |} {\Delta{(R)}} {| {\vec{x} + \vec{r}} \rangle}}|}$ is
bounded by a polynomial times $\exp{(-\frac{{2 R} - r}{L})}$, that the absolute error $E{(r,R)}$ is bounded by a
polynomial times $\exp{(-\frac{{4 R} - {2 r}}{L})}$, and that the
relative error can be made arbitrarily small. This suggests that
in localized systems the absolute error $E{(r,R)}$ depends
exponentially on $r$. If so, then Eq. \ref{RoughEstimate1} is a
gross overestimate.

\subsection{\label{UnlocalizedSystems} Unlocalized systems}
If the eigenvectors are unlocalized, then the magnitudes of
${\langle \vec{x} |a \rangle}$ and ${\langle b | {\vec{x} +
\vec{r}} \rangle}$ have no strong dependence on $\vec{x}$ and
$\vec{r}$. This suggests that ${|{{\langle \vec{x} |}
{\Delta{(R)}} {| {\vec{x} + \vec{r}} \rangle}}|}$ is roughly
independent of the position, and that $E{(r, R)}$ is roughly
independent of $r$, thus providing partial justification of Eq.
\ref{RoughEstimate1}.

\subsection{\label{FiniteCoherenceLength} Finite coherence length}
Eq. \ref{EqExpand} suggests that in systems with a finite
coherence length $\eta$ the absolute error is reduced, via
reduction of the matrix elements ${{\langle a|}H_{1}{|c \rangle}}$
and ${{\langle c|}H_{1}{|b \rangle}}$.  I assume a very crude
model of the incoherence where the
 eigenvectors are broken into domains with constant phase, each
  domain of size $\frac{4 \pi}{3}{\eta}^{3}$.  The main effect
   is to decrease any integral over an
  eigenvector by a factor of $\sqrt{N_{\eta}}$, where $N_{\eta}$
  is the number of different domains where the integrand is
  non-zero.  $H_{1}$ touches about $\frac{4 \pi R^{2}} {4 \pi {\eta}^{2}}$
  such domains.  Therefore if $R > \eta$,
  ${{\langle a|}H_{1}{|c \rangle}} \propto {\frac{{\eta}}{R}}$ and
  ${E{(r,R)}} \propto {\frac{{\eta}^{4}}{R^{4}}}$.  As I will show in section \ref{Matrix Elements}, analytic calculations of the second moment of the matrix element confirm the same scaling law.

\subsection{The Coherent Potential Approximation}
An alternative way of estimating the error is available via the Coherent Potential Approximation\cite{Soven67, Taylor67}. This approximation estimates the average of the Green's function, so one could use it to calculate the average Green's function in a large volume and also inside a truncation volume with radius $R$ and then estimate the error as being equal to the difference.  Having obtained the error in the average Green's function, one could calculate the average error in any analytic matrix function by doing a complex integration.  Of course this approach has the weakness that one is estimating the average error instead of the average magnitude of the error.  But I nonetheless began implementing a program which would calculate the Coherent Potential Approximation (CPA) numerically, using the IATA algorithm\cite{Chen73} which has been proved to always converge to a unique solution\cite{Ducastelle74}.  The only interesting result so far of my work has been a demonstration that the published proofs of convergence are wrong in the cases of finite or periodic systems.  In these systems the bare Green's function $G_{0}$ (which ignores the disordered potential) has discrete poles.  The IATA algorithm can (and, in my experience, always does) move the coherent potential variable toward values which correspond to those poles and make the bare Green's function blow up.  So the coherent potential is converging (very very slowly) to an incorrect solution, but the bare Green's function is diverging.  Also the solution is not unique - depending on the algorithm's initialization it may converge to any of $G_{0}$'s poles.  It is likely that standard algorithms for finding zeros of nonlinear functions could circumvent this problem, but I haven't managed to give this a try yet.  As I will show in section \ref{DensityMatrixResults} on numerical results, equation \ref{RoughEstimate1} gives a good estimate of the error, so obtaining an error estimate from the CPA is not so important.

\section{\label{DensityMatrixResults}Results}
I studied ensembles of Anderson hamiltonians at eleven disorder
values between ${\sigma} = {0.65}$ and ${\sigma} = {9.00}$,
including the critical disorder ${\sigma}_{c} = 6.15$.   In the
results presented here a truncation radius of $R = 5$ was used,
but the results are qualitatively similar to those obtained with
${R} = {1, 2, 3, 4}$.  A lattice size of ${16}^{3}$ was used, and
calculations with ${12}^{3}$ and ${20}^{3}$ lattices at the
critical disorder ${\sigma}_{c}$ indicate that finite size effects
are small. The largest such effect is an improvement of the basis
truncation algorithm's accuracy at smaller lattice volumes.  At
each disorder I calculated the density matrix at $13$ values of
the Fermi level $\mu$ ranging uniformly from $-12$ to $12$, which
covered the whole energy band at the lower disorders, and most of
it even at larger disorders.  Close to the edges of the energy
band the density matrix magnitude drops precipitously and the
other observables also change rapidly; the main results reported
here apply only to values of $\mu$ where the spectral density
remains high.

A low temperature ($T = {0.05}$) was chosen in order to minimize
any temperature effect.  A careful examination of the density
matrix's behavior at $\mu = 0$ showed that any temperature effect
was swamped by other effects. In particular, at low disorder the
density matrix's behavior is dominated by lattice effects. Because
of the low temperature a large number of Chebyshev terms was
needed; for each matrix I used a number of terms $S$ equal to $25$
times the total band width, which was enough to make the
truncation error quite small.

\begin{figure}
\includegraphics[width=5cm, height=8cm, angle=270]{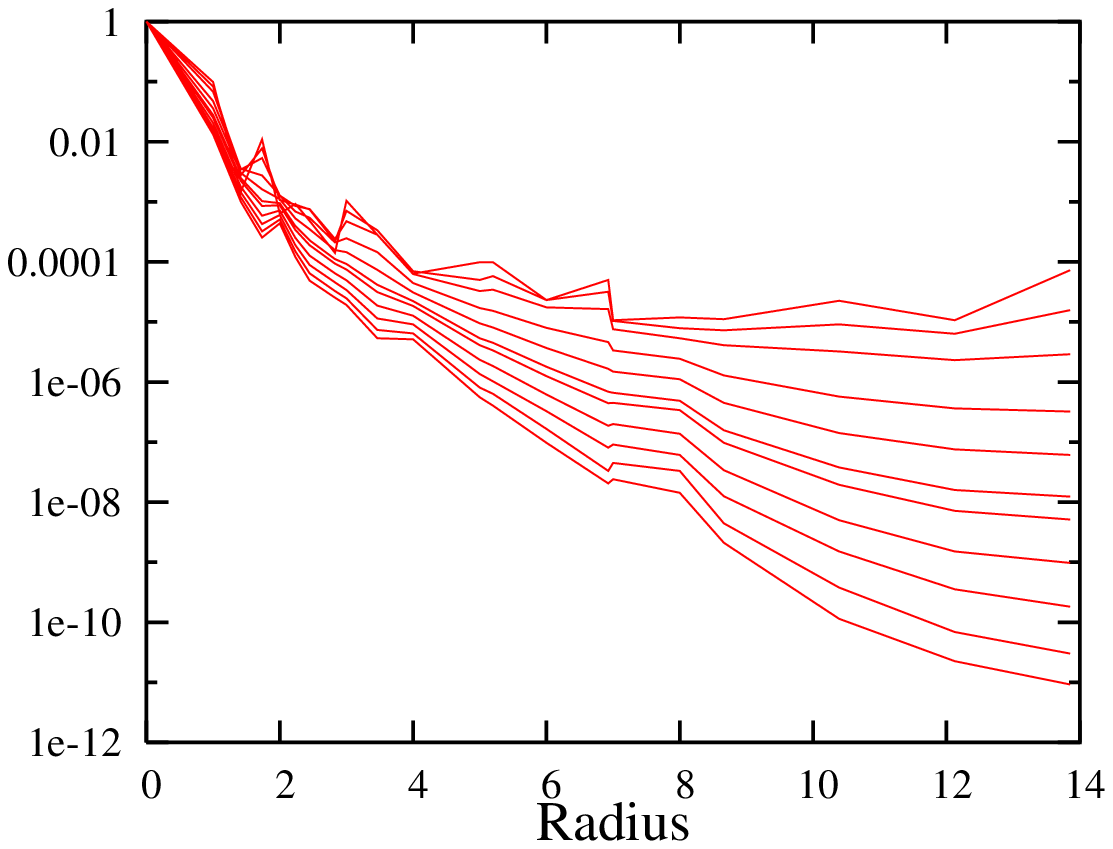}
\includegraphics[width=5cm, height=8cm, angle=270]{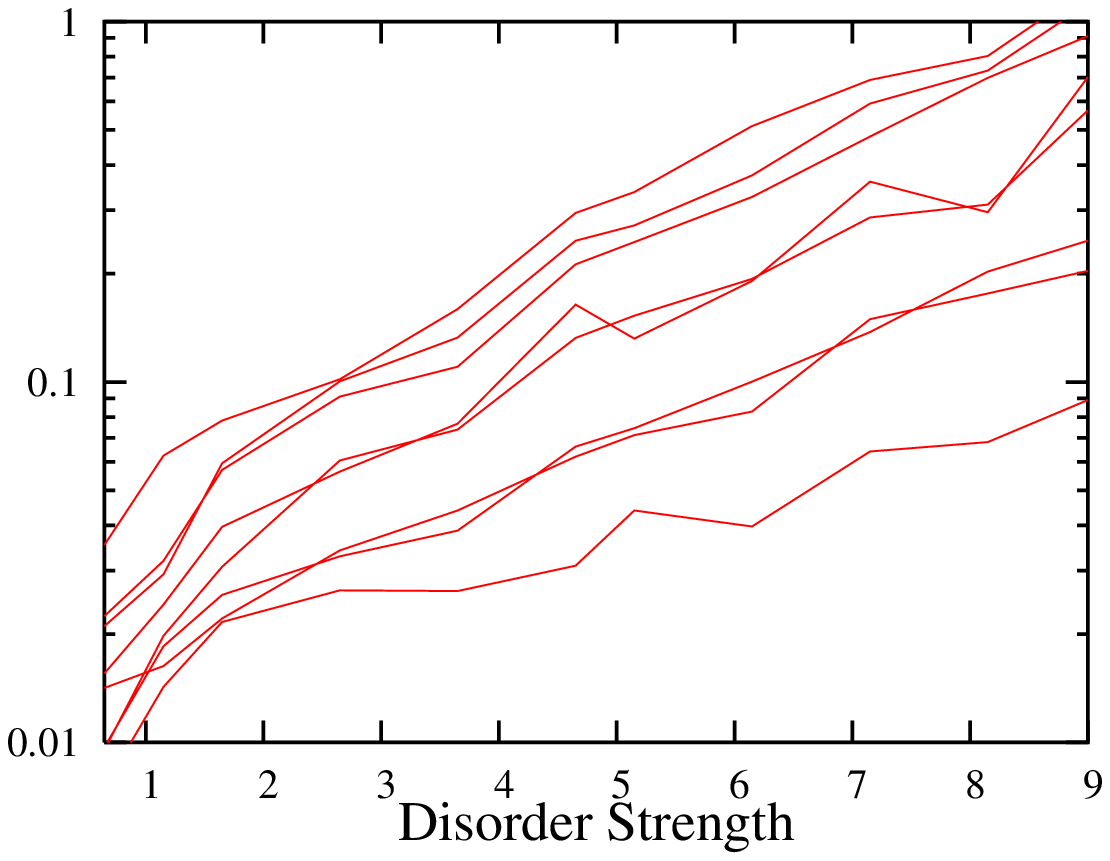}
\caption{\label{Figure2} The top graph shows the normalized
density matrix magnitude at ${\mu} = 0$.  Each line corresponds to
a different disorder strength between $0.65$ and $9.00$; lower
disorders are higher on the graph. The lower graph shows the ratio
of the square root of the second moment to the mean at $\mu = 0$.
The different lines correspond to different radii $r = n\sqrt{3}$,
with smaller radii lower on the graph.}
\end{figure}

The top graph of figure \ref{Figure2} shows the normalized
density matrix magnitude $ {\frac{MDP{(\rho, \rho, r)}}{MDP{(\rho,
\rho, 0)}} }$ at ${\mu} = 0$. For $r > 0$ and $\sigma \geq 1.65$ a
good fit to this quantity can be obtained by $ \gamma {r^{-4}
\exp{(\frac{-2r}{\tilde{R}})}}$, where the coherence length is
given by $\tilde{R} = {{({0.057 \sigma} - {0.089} - {0.064
{\sigma}^{-1}})}^{-1}}$ and $\gamma$ is a normalization constant. Note the almost inverse relation between
the coherence length $\tilde{R}$ and the disorder strength
$\sigma$.  Lattice effects cause a systematic uncertainty in the
first constant $0.057$ of roughly $30\% $. Similar fits can be
obtained at other Fermi levels within the band ${|\mu|} \leq {6}$;
the first constant has a minimum at $\mu = 0$ and a total
variation of about $30\%$.

Now I consider the statistical distribution of the density matrix
magnitude. The lower graph of figure \ref{Figure2} shows the ratio
of the square root of the second moment to the mean.  Note that
this ratio seems to grow roughly exponentially with the disorder
$\sigma$.  An examination of the kurtosis (the normalized fourth
moment of the statistical distribution) of the density matrix
magnitude shows that for $\sigma \geq {5.15}$ this quantity
becomes very large, starting at larger radii and larger Fermi
level $|\mu|$. These statistics suggest that at the Anderson
transition the statistical distribution of the density matrix
magnitude develops a long tail; it loses its self-averaging
property.

\begin{figure}
\includegraphics[width=5.5cm, height=8cm,angle=270]{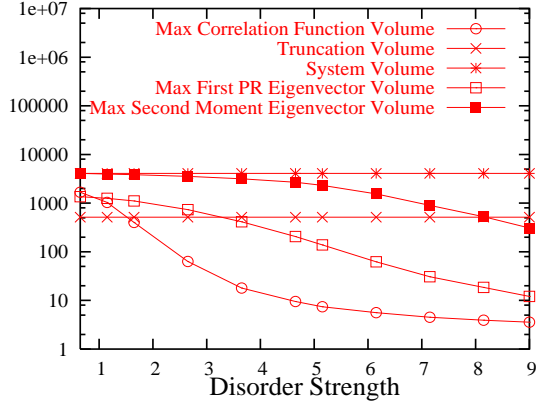}
\caption{\label{Figure3} Important volume scales.  Each volume
measure was averaged across the ensemble and across a small
interval of the energy spectrum.  Shown here are the maximum
values of these averages.}
\end{figure}

Figure \ref{Figure3} shows two measures of eigenvector volume. The
first is the inverse of the first participation ratio; i.e.
${({\sum_{\vec{x}}{{|{\langle \psi | \vec{x} \rangle}|}^2}})}^{2}
/ {\sum_{\vec{x}}{{|{\langle \psi | \vec{x} \rangle}|}^4}}$. This
quantity is a lattice friendly measure of volume because it has a
minimum value of one when  $ {\langle \psi | \vec{x} \rangle} $ is
a delta function and a maximum value of the system size when $
{\langle \psi | \vec{x} \rangle} $ is a constant. Figure
\ref{Figure3} shows that this volume becomes smaller than the
truncation volume in the range $\sigma = {3.00}$ to $\sigma =
{4.00}$.

My second volume measure is based on the second moment:
${({(2\pi)}^{D}{{det}{Q}})}^{\frac{-1}{2}} $ , where $Q$ is the
second moment (or quadrupole tensor) of ${{|{\langle \psi |
\vec{x} \rangle}|}^2}$.  Unlike the first volume measure, this
measure remains large even at large disorders, indicating that
each eigenfunction consists of several isolated peaks scattered
throughout the system volume.   This structure is caused by the
fact that states with similar energies will mix even if they are
connected by exponentially small matrix elements.  However, mixing
caused by such small matrix elements does not influence the
density matrix, because it essentially just induces a unitary
transformation of the mixed eigenvectors, and as we know the
density matrix is invariant under unitary transformations between
states that are all either less than $\mu$ or more than $\mu$.

 Figure \ref{Figure3} also shows the maximum value of the coherence volume;
 $V_{C} = {{max}_{E}{V{(E)}}}$, where $V{(E)}$ is the coherence
 volume.  I calculated this volume by first computing
the correlation function ${C{(\vec{x})}} =
{\int{{d\vec{k}}^{D}{{|{\langle \vec{k} | \psi \rangle
}|}^{2}{exp{(\imath {\vec{k} \cdot \vec{x}})}}}}} $, and then
applying my two volume measures to $ {C^{2}{(\vec{x})}}$.  Taking
the maximum value resolves an important ambiguity: at disorder
strengths $\sigma \leq {6.15}$ the coherence length shows two
peaks at energies ${|E|} = {6 - 8}$. These peaks become very
pronounced at $\sigma \leq 1.65$, where they are a factor of $10$
above the minimum.  Note that the coherence volume becomes small
much sooner than the eigenvector volume,
 in the range from $\sigma = {1.65}$ to $\sigma = {2.65}$.
For $\sigma \geq {1.15}$ and $V_{C} \gg 1$ it is roughly
proportional to ${{\sigma}^{-3}}$.  This fits well with the
density matrix's coherence length at large $\sigma$, but not at
small $\sigma$.

\begin{figure}
\includegraphics[width=5cm, height=8cm, angle=270]{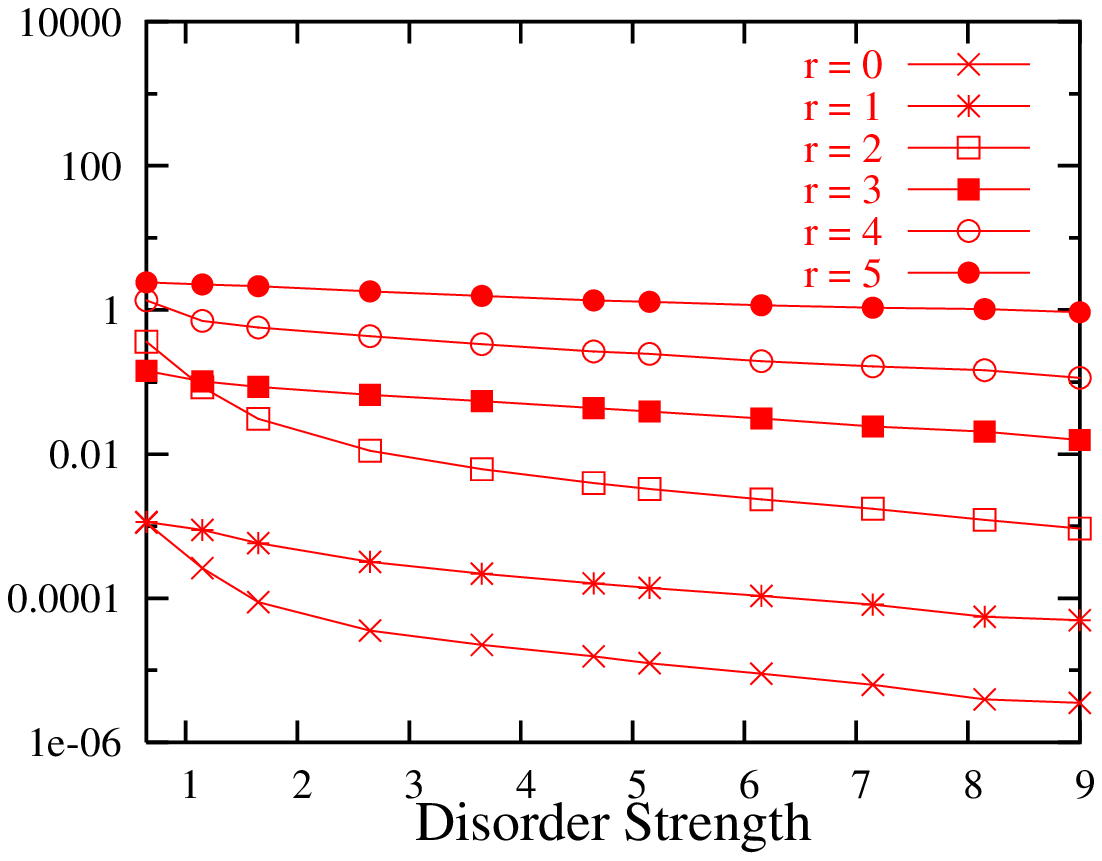}
\includegraphics[width=5cm, height=8cm, angle=270]{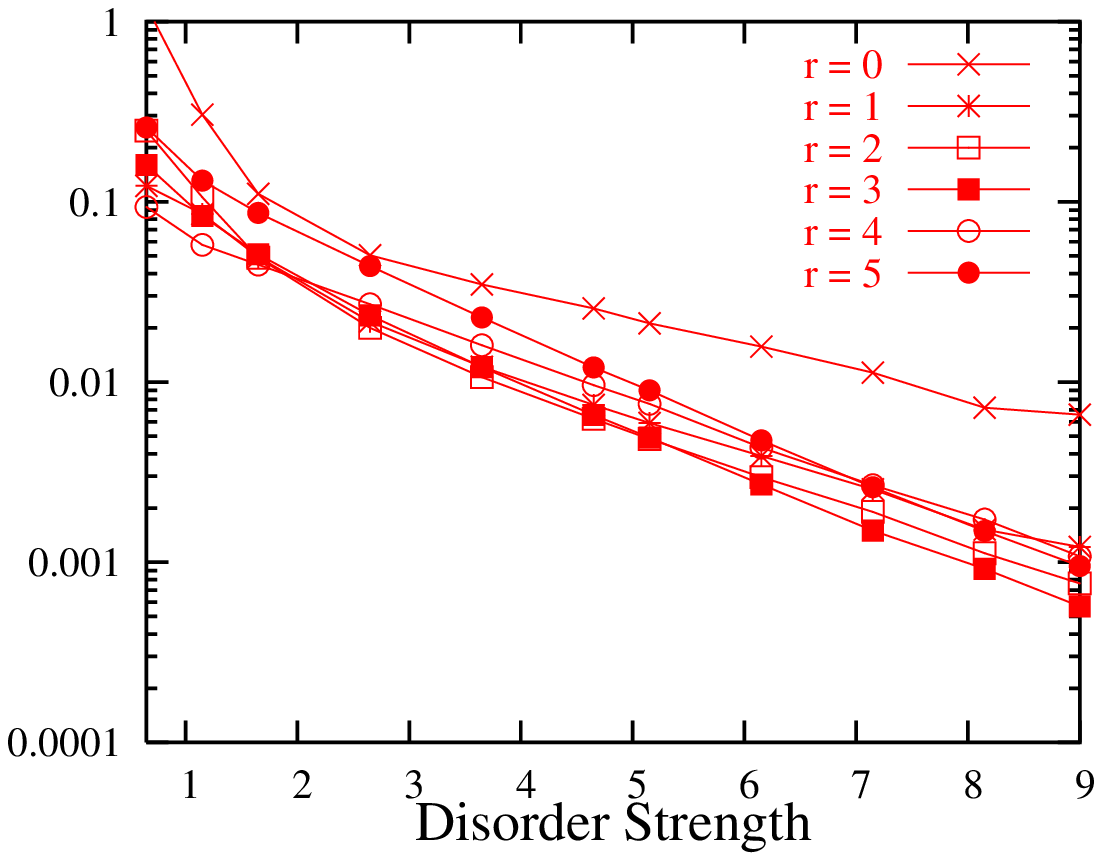}
\caption{\label{Figure4} The upper graph shows the relative error $e{(r,R)}$ (eq.
\ref{Definee}) as a function of disorder, while the lower graph shows the absolute error $E{(r,R)}$ (eq.\ref{DefineE}) as a function of disorder. ${\mu} = {0}$.}
\end{figure}

Figure \ref{Figure4} shows the relative error and absolute error
as a function of disorder. The $O{(N)}$ algorithm begins to work
well at $r = {0,1}$ at quite small disorders, and the $r = 2$
relative error falls to $1\%$ at about $\sigma \thickapprox {3}$.
Clearly the $O{(N)}$ algorithm's success is controlled by the
coherence volume, not by the eigenvector volumes.

The overlapping lines in the lower graph of figure 
\ref{Figure4} (the absolute error) confirm section \ref{UnlocalizedSystems}'s argument
that the absolute error $E{(r,R)}$ is roughly independent of $r$,
except at $r = 0$. The magnification at $r = 0$ is probably caused
by the density matrix's close relationship to the identity matrix.
It is more than compensated for by a corresponding magnification
of the density matrix at $r = 0$, so that the upper graph (the relative error)
shows that the value of the relative error $e{(r,R)}$ at $r = 0$
is smaller than its value at $r=1$.

Section \ref{FiniteCoherenceLength} suggested that a small
coherence length may cause a decrease in the absolute error
$E{(r,R)}$ of order $\frac{{\eta}^{4}}{R^{4}}$.  The $R=5$ line in
the relative error graph of figure \ref{Figure4} shows that the
decrease is actually even more pronounced: at the boundary $r =
R$, $E{(r,R)}$ is of the same order of magnitude as
$MDP{(f,f,r)}$, which is controlled by an exponential.  This is
just the ansatz used to obtain Eq. \ref{RoughEstimate1}; my
results fully support the validity of Eq. \ref{RoughEstimate1}
except - as discussed before - at $r = 0$, and at the edges of the
energy band. Therefore accurate estimates of the error of an
$O{(N)}$ calculation may be obtained from the results of the
$O{(N)}$ calculation itself.

Section \ref{LocalizedSystems} showed that for large $R$ the
absolute error $E{(r,R)}$ is bounded by a polynomial times
$\exp{(-\frac{{4 R} - {2 r}}{L})}$, where $L$ is the decay length.
This suggests an $r$ dependence which is not confirmed by the
absolute error graph of figure \ref{Figure4}, where it would cause a
splitting of the lines at $\sigma > {6.15}$. However, the previous
paragraph showed that $E{(r,R)}$ actually depends on $R$ as
$R^{-4} exp{(\frac{-2R}{\tilde{R}})}$, where $\tilde{R}$ is the
coherence length. Therefore as long as ${\tilde{R}} < {2 L}$, the
bound obtained in section \ref{LocalizedSystems} is automatically
verified at all $r$.  Moreover at small $R$ the unknown polynomial
in the bound may mask any $r$ dependence.

\section{Qualitative Reliability of These Results}
I have already discussed numerical due to the finite lattice size
and to the truncation of the Cheybyshev expansion.  I will discuss these errors in more depth in section \ref{ErrorBudget}, but my checks indicated that they have at most a
quantitative, not qualitative, effect on the results presented here. The main risks to the qualitative correctness of these results probably lie in two areas: finite ensemble size, and software reliability and
reproducibility.  I have taken steps to manage both issues:

\subsection{ Finite ensemble size.}  All results presented here
were obtained from ensembles of $33$ realizations, but repeating
some of the calculations with ensembles of $100$ realizations yielded the
same results. At the critical disorder the same quantities were
computed with three different lattice sizes ($100$ realizations at
both ${12}^{3}$ and ${16}^{3}$, and $10$ realizations at
${20}^{3}$), and the agreement is very good.  Graphing any
quantity across several disorders, one immediately notices that
there is little noise induced overlap of the two graphs.
Therefore, it seems likely that risks due to finite ensemble size
are under control.

\subsection{\label{ReliabilityRepro1} Software Reliability and Reproducibility.}  I have tried
very hard to reduce this risk, and discuss these efforts in some detail in chapter 
\ref{Reliability}.  The software includes an automated
test suite which tests all computational functions except the
highest level output (graph printing) routines. Moreover, I have
taken pains to enable other researchers to easily reproduce and
check my results, simply by installing my software and the
libraries it depends on, compiling it with the GNU gcc
compiler\cite{GNUGCC}, and starting it running. The software, with
all needed configuration files, is available under the GNU Public
License\cite{GNUGPL}; check www.sacksteder.com for further
details.

\section{\label{ErrorBudget}Quantitative Error Analysis}
A full quantitative analysis of $O{(N)}$ error would include an error budget: how much error came from the finite size effect, from the Chebyshev expansion, from the $O{(N^{3})}$ algorithm, and from statistical uncertainty caused by the finite sample size.   In the next sections I argue that the errors from the Chebyshev expansion and the $O{(N^{3})}$ algorithm are small, and I discuss the statistical uncertainty and the finite size effect.  

\subsection{The Finite Size Effect} 
In this chapter I calculated the density matrix for a ${16}^{3}$ system with both an $O{(N)}$ algorithm and an $O{(N^{3})}$ algorithm, and I presented a detailed calculation of the differences in results.  However, the ideal would be to go a step further and figure out how the $O{(N)}$ error depends on the system size.  I already mentioned that I computed several observables at the critical disorder with three different system sizes, and obtained good agreement.  However there was a systematic effect that the error improved as the system size decreased, which was caused by the fact that basis truncation methods become exact once the truncation volume is larger than the system volume.  Quantification of this effect, which I will call the finite size effect, would answer questions like "How much worse would the $O{(N)}$ errors be if the system were of size ${1000}^{3}$ instead of ${16}^{3}$?"   I present here a theoretical foundation for numerical estimates of the finite size effects.  While I have not yet been able to carry out the numerical work prescribed by this theory, it is already clear that if the $O{(N)}$ error is controlled by an exponential then the finite size effect is also controlled by an exponential.

The key intellectual step in this theory is to imagine that the disordered system is really infinitely big.  This puts my exact calculation of a matrix function $f$ in a ${16}^{3}$ system on the same level as the $O{(N)}$ calculation: both results were obtained by truncating the basis of the (imaginary) infinite system.  The quantity of interest is therefore the derivative of the matrix function with respect to the change in radius of the truncation volume.  I will use the symbol $\sigma$ to represent the disorder realization of the infinite system, and use the function ${\delta{(R, \sigma)}} = {-\frac{d}{dR}{f{(R, \sigma)}}}$ to represent the derivative of the matrix function $f$ with respect to the truncation radius $R$.  ${\delta{(R, \sigma)}}$ can be computed by simply calculating the matrix function at two nearby localization radii. In equation \ref{DefineE} I defined the absolute error ${E{(r, R)}} = {MDP{(\Delta{(R)}, \Delta{(R)}, r)}}$.  $\Delta{(R, \sigma)}$ is the difference between the exact matrix function and the approximate matrix function, and can be rewritten as  ${\Delta{(R, \sigma)}} = {{\int}_{R}^{\infty} {dR_{1}} \delta{(R_{1}, \sigma)}}$. Substituting this into equation \ref{DefineE}, I obtain the true error:
\begin{eqnarray}
{E{(r, R, \sigma)}} & = & {{\int}_{R}^{\infty} {dR_{1}} {dR_{2}} {MDP{(\delta{(R_{1}, \sigma)}, \delta{(R_{2}, \sigma)}, r)}}}
\nonumber \\
 & = & {\int}_{R}^{\infty} {dR_{1}} {dR_{2}} \sqrt{|MDP{(\delta{(R_{1}, \sigma)}, \delta{(R_{1}, \sigma)}, r)}|}
 \nonumber \\
 & \times & \sqrt{|MDP{(\delta{(R_{2}, \sigma)}, \delta{(R_{2}, \sigma)}, r)}|}
\label{FiniteSizeBound}
\end{eqnarray}
The finite size effect is caused by the fact that numerical studies do not compute the true error ${E{(r, R)}}$ but instead the difference ${E{(r, R)}} - {E{(r, R_{V})}}$, where $R_{V}$ is the system's radius.  Therefore the finite size effect can be calculated by computing the ensemble average of ${MDP{(\delta{(R_{1})}, \delta{(R_{2})}, r)}}$ over some range of radii, extrapolating these results out to $R = \infty$, and then doing the integration over $R_{1}$ and $R_{2}$.  

My numerical results for the density matrix $\rho$ showed that:
\begin{equation} {E{(r, R)}} \equiv {MDP{(\Delta{(R)}, \Delta{(R)}, r)}} \approx {MDP{(\rho, \rho, R)}}
\end{equation}.
This result was essential to the correctness of the error estimate given by equation \ref{RoughEstimate1}.  Moreover, I found that ${MDP{(\rho, \rho, r)}} \approx { \gamma {r^{-4} \exp{(\frac{-2r}{\tilde{R}})}}}$.  Therefore the magnitude of $\Delta{(R)}$ is proportional to ${ \gamma {r^{-2} \exp{(\frac{-r}{\tilde{R}})}}}$.  Similarly, the magnitude of ${\delta{(R)}} \equiv {\frac{d}{dR}{\Delta{(R)}}}$ is also controlled by the exponential $\exp{(\frac{-r}{\tilde{R}})}$.  Plugging this result into the equation \ref{FiniteSizeBound}'s bound immediately shows that the finite size effect is controlled by the exponential $\exp{(\frac{-2 R_{V}}{\tilde{R}})}$.  When the $O{(N)}$ error is exponentially small, the finite size effect is also exponentially small.

This analysis ignored one detail: in my calculations the ${16}^{3}$ system had repeating boundary conditions.  If the system were really part of a larger, infinite system - as I assumed for my analysis of the finite size effect - it would have fixed boundary conditions.  However, this difference of boundary conditions is just a change in the matrix elements at the boundary of the ${16}^{3}$ system.  My numerical results demonstrated that the density matrix's dependence on things far away is exponentially small; the effect of boundary conditions is also exponentially small.  In any case this difficulty can be avoided by choosing fixed boundary conditions for system.

\subsection{Statistical Uncertainty}
I computed the standard deviation and kurtosis (related to the fourth moment) of the matrix dot products and the absolute and relative errors, and spent a lot of time graphing them and thinkng about them.  The most interesting of these results is reported in figure \ref{Figure2}. That graph is the exception, however; in general each observable's standard deviation was small compared to the observable itself.   There is also an interesting pattern in the kurtosis of certain observables, which becomes inconsistent with a gaussian distribution as the disorder grows, starting at large radii and large fermi levels $|\mu|$.  In fact the statistical distribution of observables in disordered systems is not guaranteed to be gaussian, and may have a long tail.  Therefore a full quantitative analysis of the statistical uncertainty will require use of a jackknife, bootstrap, or similar method.
 
\subsection{Errors from the Chebyshev Expansion}
The error induced by truncating a Chebyshev expansion to the first $S$ terms is controlled by an exponential with argument proportional to ${-T S} / { \Delta}$,
where $\Delta$ is the size of the energy band\cite{Goedecker94}.  In my calculations I used $S = {25 \Delta}$ and $T = {0.05}$, giving this fraction a value of $-1.25$.  I believe that this value reduced Chebyshev errors to an insignificant level. I base this belief on a set of earlier calculations where I used a number of terms $S$ which was constant rather than proportional to the band width $\Delta$. At small disorders - and thus small band widths - there was no difference between results using $S = {500}$ terms and results obtained using $S = {1000}$ terms.  However, as the band width increased with disorder I began to see a difference, meaning that the Chebyshev truncation became the dominant source of error.  Moreover, the error began increasing as the disorder (and band width) were increased.  Therefore I redid all my calculations with the conservative heuristic $S = {25 \Delta}$.    

The Chebyshev expansion also introduces errors through finite precision arithmetic.  I am unaware any published work analyzing the effects of finite precision arithmetic on the stability and accuracy of Chebyshev expansions, so I don't know how to estimate these errors.

In section \ref{OrderN Algorithms}, I briefly mentioned that there are alternative basis truncation algorithms that can be used as alternatives to the Goedecker algorithm discussed here.  These alternatives avoid the Chebyshev expansion and therefore their error may be easier to analyze.  On the other hand, they introduce other sources of error, which may also be difficult to analyze.

In any case, probably the best practical way of estimating the Chebyshev expansion's error is simply to measure how the density matrix varies while increasing the number of terms.  As I already mentioned, the density matrix should converge exponentially, so one can fit an exponential to the error.  As long as the error due to truncating the expansion is large compared to the error due to finite precision arithmetic, an exponential fit should give a good estimate of the error.  However, finite precision arithmetic will cause the expansion to converge to a (slightly) wrong result, and the exponential fit will not be any help in measuring this effect.

Of course there are complications in doing the exponential fit.  One should really fit every 
matrix element of the density matrix, and then perform some sort of average of the fitting parameters.  Moreover, even though the error is guaranteed to decrease exponentially, the exponential decay may be multiplied by some oscillatory behavior, which can cause difficulties in the fitting.  At present my program generates data about the how the absolute error varies with the number of terms in the Chebyshev expansion, but I have not yet found the time to implement and test the exponential fit or the averaging.

\subsection{Errors in the "Exact" $O{(N^{3})}$ Algorithm}
I used the simple algorithm of diagonalizing the argument $H = {U D U^{\dagger}}$ and then evaluating the matrix function ${f{(H)}} = {U {f{(D)}}  U^{\dagger}}$.  It has been reported in the numerical analysis literature\cite{Golub83, Smith02} that for general arguments $H$ this algorithm is often not the best because errors in the eigenvalues can be multiplied by the square of the condition number of the transformation $U$.  However, the Anderson Hamiltonian has transverse symmetry, so $U$ is unitary, implying that the condition number is one.  Therefore this simple algorithm is fairly well suited to my needs.  

I used the LAPACK\cite{LAPACK} algorithm dsyevr to diagonalize the Hamiltonian, and checked its results by computing the eigenvector orthogonality relation $ {\langle {\psi}_{i} | {\psi}_{j} \rangle} = {{\delta}_{ij}}$ and the eigenvalue equation $ {\langle {\psi}_{i} | H | {\psi}_{j} \rangle} = {E_{i} {\delta}_{ij}}$.  One thing I noticed was that for $i \neq j$ the eigenvalue equation was violated by errors roughly proportional to ${(E_{i} - E_{j})}^{-1}$.  The proportionality constant was of course quite small, but if the eigenvalues were nearly degenerate then the violation could be important.  This effect motivated a move from single precision arithmetic to double precision arithmetic.  It is well known that in unlocalized disordered systems the eigenvalues repel each other and degeneracies are avoided, so that this effect is likely not a problem.

One non-trivial check available for any matrix function calculation is the trace identity ${{Tr}{(f{(H)})}} = {{Tr}{(f{(D)})}}$, which follows from the identity ${{Tr}{(AB)}} = {{Tr}{(BA)}}$.  I checked this identity for every density matrix I calculated and threw an error whenever the relative error was more than ${10}^{-9}$.  There was only one matrix that exceeded that limit, with a relative error of approximately $3 \times {10}^{-9}$.

In any case my numerical results clearly showed that the $O{(N)}$ results were converging to the $O{(N^{3})}$ results, which strongly indicates that errors within the $O{(N^{3})}$ algorithm did not dominate the final results.

\chapter{\label{FunctionAccuracy}Accuracy of $O{(N)}$ Algorithms for the Logarithm, Exponential, Inverse, Cauchy Distribution, and Gaussian Distribution}
\section{The Functions}
In the last chapter I examined the accuracy of $O{(N)}$ basis truncation algorithms for evaluating the density matrix in disordered systems.  In this chapter I apply the same analysis to five other matrix functions:
\subsection{The Natural Logarithm}
This function is non-analytic at the origin and is imaginary if the argument is negative.  If the logarithm's argument is real then the logarithm itself can be written as: ${\ln x} = {{\ln {|x|}} + {\imath \Theta{(-x)}}}$.  The reader will recognize that the theta function in the imaginary part is just the density matrix I have already examined in so much detail.  Therefore I only calculate the real part of the logarithm, $\ln{|x|}$.  Furthermore I regularize the singularity by using the following formula:  
\begin{eqnarray}
{f{(|x - \mu|)}} & \equiv & {{\ln {|x - \mu|}}, {|x - \mu|} > \tau}
\nonumber \\
& \equiv & {{\frac{1}{2}{[{(\frac{x - \mu}{\tau})}^{2}  - 1]}} + {\ln \tau}, {|x - \mu|} \leq \tau}
\end{eqnarray}
Just as with density matrix, the parameter $\mu$ simply displaces the matrix function along the real axis.  I will continue to call $\mu$ the fermi level for lack of a more descriptive name.  The temperature $\tau$ again determines the energy resolution of the matrix function.  Just as with the density matrix, I choose $\tau = {0.05}$ and vary the fermi level $\mu$ between $-6$ and $6$ in steps of $2$. 

Both this regularized logarithm and its derivative are continuous.  Two factors forced me to regularize: First, the unregularized matrix logarithm could be extremely sensitive to eigenvalues which lie near $\mu$.  In computations with finite matrices, such eigenvalues become exceedly rare and cannot be studied using numerical techniques.  Calculating the unregularized matrix function would not have given any information about the accuracy of $O{(N)}$ methods in calculating eigenvalues near $\mu$.  Second, the Goedecker algorithm used here is based on the Chebyshev expansion, which always regularizes singularities and discontinuities at a scale which is inversely related to the number of terms in the expansion.  I chose to impose a regularization of my own choice rather than rely on the Goedecker algorithm for the regularization.  This is a limitation of the Goedecker algorithm only; other basis truncation $O{(N)}$ algorithms do not impose any regularization and are able to calculate eigenvalues which are arbitrarily close to $\mu$.  

Note that if an $O{(N)}$ method can accurately calculate a regularized matrix function, then one can create a hybrid algorithm which accurately calculates the unregularized function.  This hybrid algorithm would use a Lanczos-based approach to calculate the difference between the regularized and unregularized matrix functions.  

\subsection{The Inverse}
The inverse function is the derivative of the real part of the logarithm.  Like the logarithm, it is nonanalytic at the origin.  Therefore I regularize the inverse function, setting it equal to the derivative of the regularized logarithm:
\begin{eqnarray}
{f{(x - \mu)}} & \equiv & {{\frac{1} {x - \mu}}, {|x - \mu|} > \tau}
\nonumber \\
& \equiv & {{(\frac{x - \mu}{{\tau}^{2}})}, {|x - \mu|} \leq \tau}
\end{eqnarray}
I choose $\tau = {0.05}$ and vary the fermi level between $-6$ and $6$ in steps of $2$.  This is the real part of the inverse in the limit where the argument has only a very small imaginary part.  The complex part of the inverse is just the Cauchy distribution, described next.

\subsection{The Gaussian and the Cauchy Distribution}
These two functions are interesting because they are regularizations of the Dirac delta function.  I define them as follows:
\begin{equation} {f{(|x - \mu|)}} \equiv {\exp{(-\frac{1}{2}{(\frac{x - \mu}{\tau})}^{2})}}
\end{equation}
\begin{equation} {f{(|x - \mu|)}} \equiv {{({({x - \mu})}^{2} + {\tau}^{2})}^{-1}}
\end{equation}

As with the other functions, I varied the fermi level $\mu$ between $-6$ and $6$ in steps of $2$.   However, I chose a bigger temperature $\tau = {0.3}$.  The reason for this choice is that in the case of the Gaussian and Cauchy Distribution $\tau$ defines both the largest energy scale and the smallest energy scale, while for the other functions $\tau$ defines only the smallest energy scale and the largest energy scale is the band width.  Therefore it seemed reasonable to choose a somewhat larger energy scale.

All of the functions discussed so far are either symmetric or antisymmetric under the parity transformation ${x - \mu} \rightarrow {\mu - x}$.  This implies that the statistical distributions which I measure should be symmetric under $\mu \rightarrow {- \mu}$.

\subsection{The Exponential}
I define the exponential:
\begin{equation} 
{f{(x)}} \equiv {e^{\alpha x}}
\end{equation}
There is only one parameter $\alpha$, a multiplier which controls the energy scale.
I varied $\alpha$ from $0.1$ to $3.1$ in steps of $0.25$, which corresponded to changing the scale of the exponential's variation between $10$ and $\sim 0.3$. 

\section{Results}
All of the results discussed here are obtained only for values of $\mu$ which lie inside the energy band.  On the edges of the energy band, and outside it, fitting becomes difficult and the relative error decreases very rapidly (except in the case of the Gaussian where the relative error converges to one.)  The graphs and numerical fits are all at $\mu = 0$; the results are qualitatively the same at other values of $\mu$ inside the energy band.
 
I computed all quantities at eleven different values of the disorder varying between $\sigma = {0.65}$ and $\sigma = {9.0}$.  At the lowest disorder value, $\sigma = {0.65}$, all quantities display a rapid variation with the position $r$.  I call this effect, which is even more pronounced at zero disorder, the lattice effect.  There is a transition as the disorder is increased to $\sigma = {1.15}$ and $\sigma = {1.65}$: these rapid variations smooth out and the $O{(N)}$ algorithms obtain good accuracy for the density matrix.  All observables which I calculate tend to maintain their qualitative features across all disorders above $\sigma = {1.65}$.   In particular, the localization transition, which occurs at $\sigma = {6.15}$, has little effect on the results.

I remind the reader that in the previous chapter I defined the Partial Matrix Dot Product, a new metric for measuring matrices in localized systems:
$${MDP{(A, B, \vec{x})}} \equiv {\sum_{\vec{y}}{{\langle \vec{y} |
A | {\vec{y} + \vec{x}} \rangle}{\langle \vec{y} | B | {\vec{y} +
\vec{x}} \rangle}}} $$
This metric allows one to compute both magnitudes of matrices and angles between matrices as a function of the displacement $\vec{x}$ from the matrix diagonal.  In my computations I actually calculate the angular average ${MDP{(A, B, r)}}$, which is also a dot product.  

I defined the error matrix $\Delta{(R)}$
as the difference between the exact matrix function $f$ and the
approximate matrix function $\tilde{f}{(R)}$.  An $O{(N)}$ algorithm's accuracy is measured in terms of the relative error: \begin{equation} \label{Definee1}
{e{(r, R)}} \equiv {\frac{MDP{(\Delta{(R)}, \Delta{(R)},
r)}}{MDP{(f, f, r})}}
\end{equation}
Therefore the matrix magnitude ${MDP{(f, f, r})}$ and the absolute error ${MDP{(\Delta{(R)}, \Delta{(R)}, r)}}$ are also of interest.  My numerical results showed that, in the case of the density matrix, the absolute error is roughly invariant throughout the $O{(N)}$ algorithm's truncation volume; i.e it doesn't depend on the coordinate $r$.  Therefore the question of $O{(N)}$ accuracy reduced down to examining how the matrix magnitude ${MDP{(f, f, r})}$ falls off with increasing displacement $r$ from the diagonal.  

The questions to be addressed for other matrix functions are:
\begin{itemize}
\item What is the matrix magnitude's dependence on $r$?
\item When does the relative error become small?
\item Is the absolute error roughly invariant throughout the truncation volume?
\item Is the relative error of order one at the boundary of the truncation volume?
\end{itemize}

If the answers to the third and fourth questions are yes, then the following formula is useful for estimating the relative error:
\begin{equation}
\label{} {e{(r,R)}} \lessapprox
{\frac{MDP{(\tilde{f}, \tilde{f}, R)}}{MDP{(\tilde{f}, \tilde{f},
r)}}}
\label{RoughEstimateRepeat}
\end{equation}

\subsection{The Matrix Magnitude}
\begin{figure}
\includegraphics[width=5cm, height=8cm, angle=270]{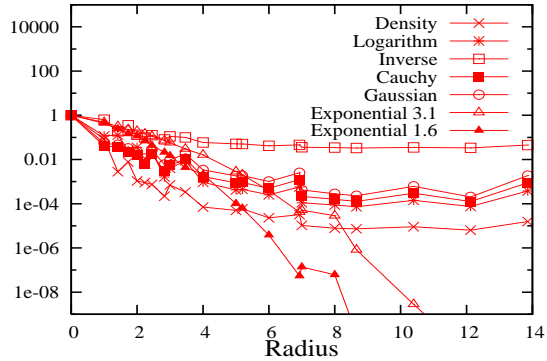}
\caption{\label{MatMag1} The normalized matrix magnitude at small disorder $\sigma = {1.15}$.}  \end{figure}
Graph \ref{MatMag1} shows the magnitudes of the various matrix functions at a low disorder $\sigma = {1.15}$.  This graph is typical of results at small disorders.  In particular, the $\sigma = {0.65}$ graph is the same, except that: 
\begin{itemize}
\item The lattice effect is much bigger, causing much bigger oscillations in the graphs.
\item The rise in the inverse function at large $r$, barely perceptible for $\sigma = {1.15}$, is much larger.
\item The Cauchy and logarithm results lie on top of each other.
\item The matrix magnitude is slightly bigger.
\end{itemize}
I choose to print the $\sigma = {1.15}$ results because the lattice effect is smaller and therefore the graph is prettier.  Qualitatively the same trends remain for all small	 disorders: 
\begin{itemize}
\item The exponentials do not very decay quickly at small $r$, but as $r$ increases they decrease precipitously.  Smaller values of the constant $\alpha$ decay more quickly.  The exponential's fast decrease is probably caused by the fact that the matrix exponential is dominated by eigenfunctions at the band edge, which are of course localized in the Anderson model.  Possibly the trend to fast decay at small $\alpha$ could be explained by the fact that at $\alpha = 0$ the matrix exponential becomes the identity.
\item The inverse function decays the least, with only a factor of about ten between $r = 0$ and $r = {14}$.
\item The density matrix, gaussian, logarithm, and Cauchy distribution all decay more quickly at small $r$ than at large $r$.  Among these, the density matrix has the smallest matrix magnitude at small disorders, while the gaussian, logarithm, and Cauchy distribution have similar matrix magnitudes.  As we will see throughout the numerical results, these four functions behave in generally the same way, except that the density matrix is better behaved at small disorders.
\end{itemize}

\begin{figure}
\includegraphics[width=5cm, height=8cm, angle=270]{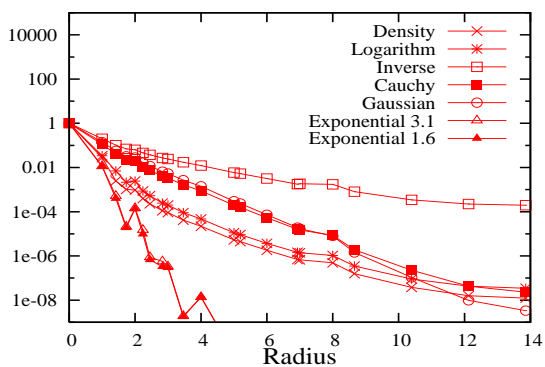}
\caption{\label{MatMag2} The normalized matrix magnitude at intermediate disorder $\sigma = {4.65}$.}
\end{figure}

The trends change a bit at larger disorders.  I present here the results  for $\sigma = {4.65}$, which are representative of all but the smallest disorders:
\begin{itemize}
\item All graphs are flattening out to become straight lines except at small $r$.  This is a sign of the coherence length, which I will shortly discuss in more detail.  
\item Again the inverse decays most slowly and the exponentials the most quickly.
\item The exponentials decrease extremely rapidly, losing nine orders of magnitude by $r = 5$.  This immense decrease is accentuated more and more as the disorder increases.  At the same time the statistical distribution of the exponential's observables becomes very strange.  On the graph shown here, the observed second moment of the matrix magnitude is of the same order as the matrix magnitude itself.  Moreover the kurtosis is huge, signifying that the statistical distribution is no longer gaussian and has a large tail. 
\item At small $r$ the magnitudes of the gaussian and the Cauchy distribution are larger than those of the logarithm and the density matrix.  However at large $r$ they catch up and the trend is reversed. The seeming correspondence between the gaussian and the Cauchy distribution is an artifact; as the disorder is increased the Gaussian becomes smaller and smaller compared to the Cauchy distribution.
\end{itemize}

In order to understand these results better, I fit the matrix magnitudes to a simple formula modeling a coherence length: ${\ln {MDP{(f, f, r})}} \propto {-{\eta \ln{r}} - {2 r {L}^{-1}}}$, where the coherence length $L$ is equal to $L^{-1} = {{\alpha \sigma } + {\beta} + {\gamma / \sigma}}$.  Note that this expression for the coherence length is just a perturbative expansion in powers of the inverse disorder $\frac{1}{\sigma}$ and thus can be expected to fail at small disorder.  In fact I was unable to obtain satisfactory fits for any function other than the inverse except by omitting small disorders ($\sigma < {2.0}$) from the fit.  I experimented with a couple of other parameterizations of the matrix magnitude and found that the fit was not significantly improved, which leads me to conjecture that the matrix magnitude at low disorders is determined by non trival physics. By obtaining the cutoff at low $\sigma$ I was able to obtain a chi squared per data point typically around  $0.30$, but ranging as low as $0.07$ (for the inverse at $\mu = 0$) to as high as $0.5$ (for the inverse at ${|\mu |} = 6$.)

The exponential function resisted any satisfactory fit.  The problem appeared to be rapid oscillations with $r$, which can be seen superimposed on the overall exponential decrease.  This suggests that the lattice effect is not damped in the exponential, but continues to play an important role at all disorders.

In contrast, the inverse could be fit satisfactorily even when including small disorders.  At $\mu = 0$ the parameter $\gamma$ which controls the $\frac {1}{\sigma}$ term comes out to a number very close to zero. I am not sure whether this really indicates that the inverse has relatively trivial small-disorder physics; in particular if I leave the lowest disorders out of the fit I get a somewhat better chi squared and $\gamma$ becomes non-zero.  I report here the fit with all disorders included because it is extraordinary that one can obtain this fit.

The following table of fitting parameters is meant to give only a qualitative picture of the results.  The quantitative aspects, for instance an estimate of the uncertainty in the fitting parameters, are plagued by systematic problems, including both the necessity of leaving small disorders out of the fit and also effects caused by the finite lattice size.

\[ \left( \begin{array}{rrrrr}
        & \eta & \alpha & \beta & \gamma \\
Inverse & 2 & 0.061 &  -0.19 &  0.0 \\ 
Log     & 4 & 0.062 & -0.090 & -0.24 \\
Density & 4 & 0.058 & -0.092 & -0.60 \\
Cauchy  & 2 & 0.057 & 0.21 & -0.19 \\
Gauss   & 0 & 0.062 & 0.50 & -0.35 \end{array} \right)\]   
 
 This table offers two pieces of qualitative information:
 \begin{itemize}
 \item The parameter $\alpha$ is almost the same for all of the fits.  The parameter controls large disorder limit of the coherence length, where  $L^{-1} \rightarrow {\alpha \sigma  }$.  So all of the five fitted functions are controlled by the same coherence length, but differ in their small-disorder physics.
 \item The power laws, indicated the parameter $\eta$, are also very interesting.  The literature contains predictions that in three dimensions the inverse (or Green's function) should have a $\frac{1}{r}$ behavior \cite{Schindlmayr00}, and that the density matrix should have a $\frac{1}{r^{2}}$ behavior\cite{Goedecker98}. These predictions are confirmed by the fits I report here.  However, I am not aware of similar theoretical predictions for the other matrix functions; developing such would be an interesting challenge for theorists, and perhaps a first step on the road to the more important problem of understanding the small-disorder physics.  
 \end{itemize}

 Note that I left $r = 0$ out of all the fits, and imposed an integer constraint on the power $\eta$.  When I let $\eta$ vary continuously, its best fits were usually fractional powers.  However the integer powers which I report here had chi squareds which were only ten percent worse.  Much worse chi squareds could be obtained by by choosing another integer power.  It is worth noting that the exponential fits, though unsatisfactory, showed a clear preference for $\eta = 4$.

 \subsection{The Absolute Error}
  \begin{figure}
\includegraphics[width=5cm, height=8cm, angle=270]{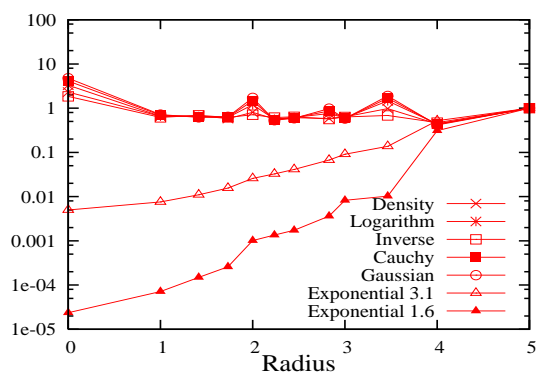}
\caption{\label{AbsErr1} The absolute error at small disorder $\sigma = {1.15}$, normalized by its value at $r = 5$.}  
\end{figure}

 \begin{figure}
\includegraphics[width=5cm, height=8cm, angle=270]{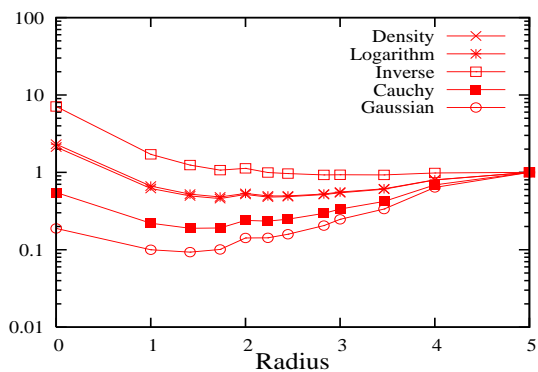}
\caption{\label{AbsErr2} The absolute error at intermediate disorder $\sigma = {4.65}$, normalized by its value at $r = 5$.}  
\end{figure}

As I mentioned earlier, the important question about the absolute error is: does it remain fairly constant within the truncation volume?  This condition is necessary for the validity of estimate \ref{RoughEstimateRepeat}.  Graph \ref{AbsErr1} displays the absolute error at small disorder $\sigma = {1.15}$, normalized by its value at $r = 5$.  In all cases except for the exponential, it is very close to constant.  Graph \ref{AbsErr2}, at $\sigma = {4.65}$, shows that as the disorder increases the absolute error begins to vary a bit, and in particular increases sharply at $r = 0$.  At this intermediate disorder the absolute error still varies by only an order of magnitude and can still be considered of order one.  However, as the disorder crosses the Anderson transition the increase at $r = 0$ gets bigger, so for the inverse it becomes a factor of $61$ at $\sigma = {9.00}$.  Moreover, in the case of the inverse function the magnification of the absolute error seems to start to extend to larger radii, as far as $r = 2$.  Unfortunately these results do not allow an extrapolation to large truncation volumes, so it is impossible to say whether the variations in the absolute error increase as the truncation volume is increased.
 
 Clearly the behavior of the absolute error depends on which matrix function is being considered. In the case of the exponential it decreases precipitously,  while for the other functions it is roughly of order one except at large disorders.  However one can clearly distinguish functions which tend to decrease away from the truncation volume's boundary (the gaussian and the Cauchy distribution), the inverse function which tends to increase, and functions which stay even closer to constant (the density matrix and the logarithm.)

I do not show the exponential in graph \ref{AbsErr2} because of rounding errors.  My software calculates the absolute error ${MDP{({f - \tilde{f}}, {f - \tilde{f}}, r)}}$ via the identity:
\begin{equation}
 {MDP{({f - \tilde{f}}, {f - \tilde{f}}, r)}} = {{MDP{(f, f, r)}} + {MDP{(\tilde{f},\tilde{f}, r)}} - {{2 \times }MDP{(f, \tilde{f}, r)}}} \end{equation}
   This formula works well for most cases and saved the work that would be required for implementing and testing subtraction and then rerunning all the previous calculations.  The exponential, however, varies over many many orders of magnitude, and therefore rounding errors ruined the absolute and relative errors except within certain ranges of the disorder, the radius, and the parameter $\alpha$. This was one example of the often very painful feature triage process that is required in writing and testing reliable software.

\subsection{The Relative Error}
\begin{figure}
\includegraphics[width=5cm, height=8cm, angle=270]{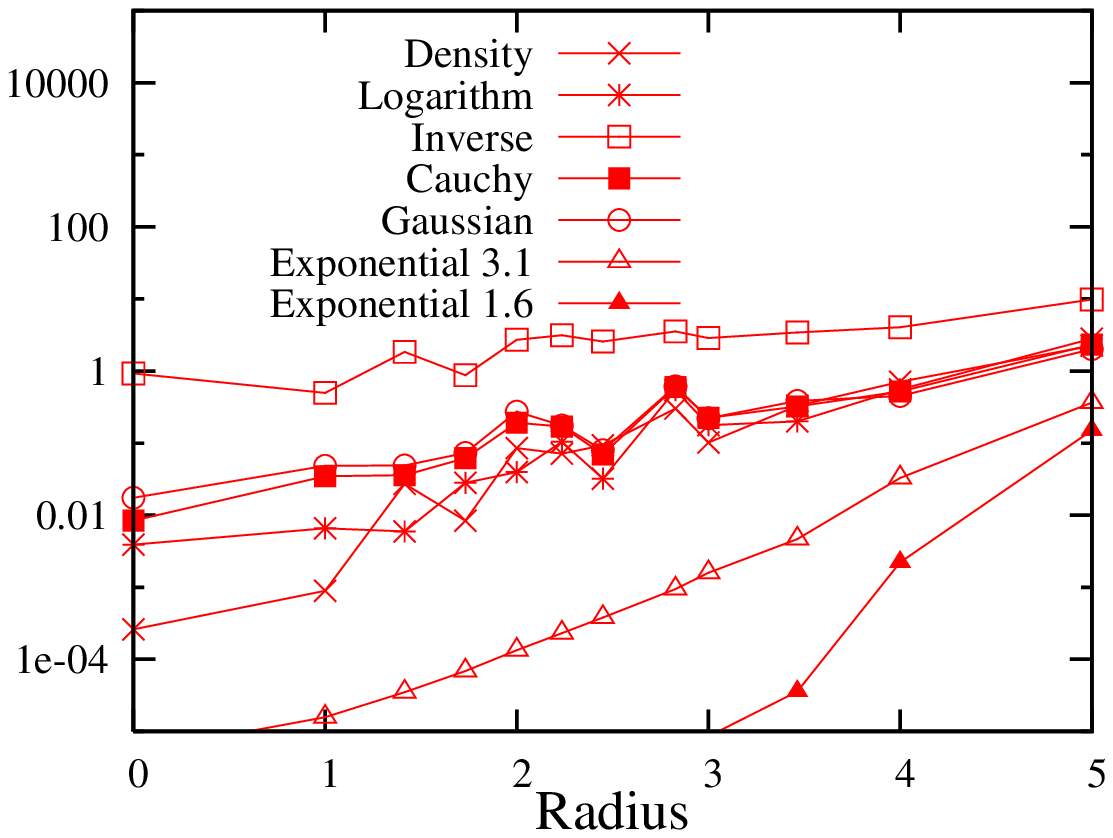}
\includegraphics[width=5cm, height=8cm, angle=270]{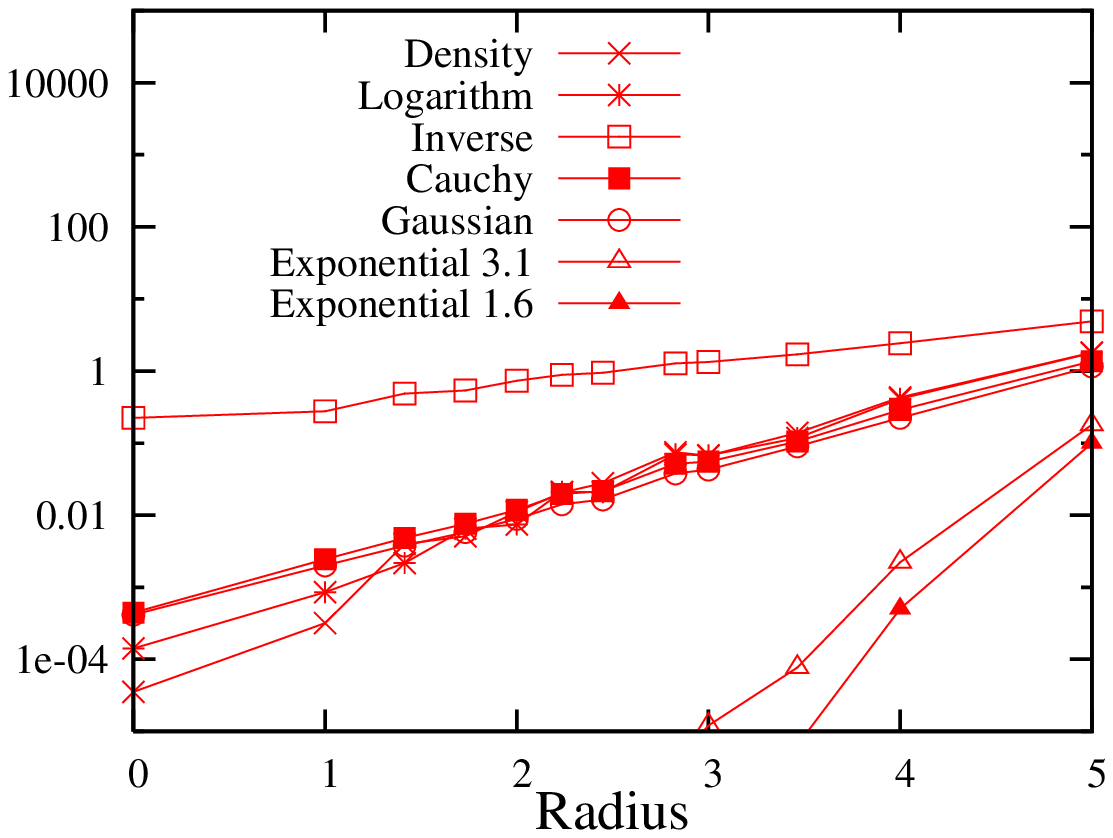}
\includegraphics[width=5cm, height=8cm, angle=270]{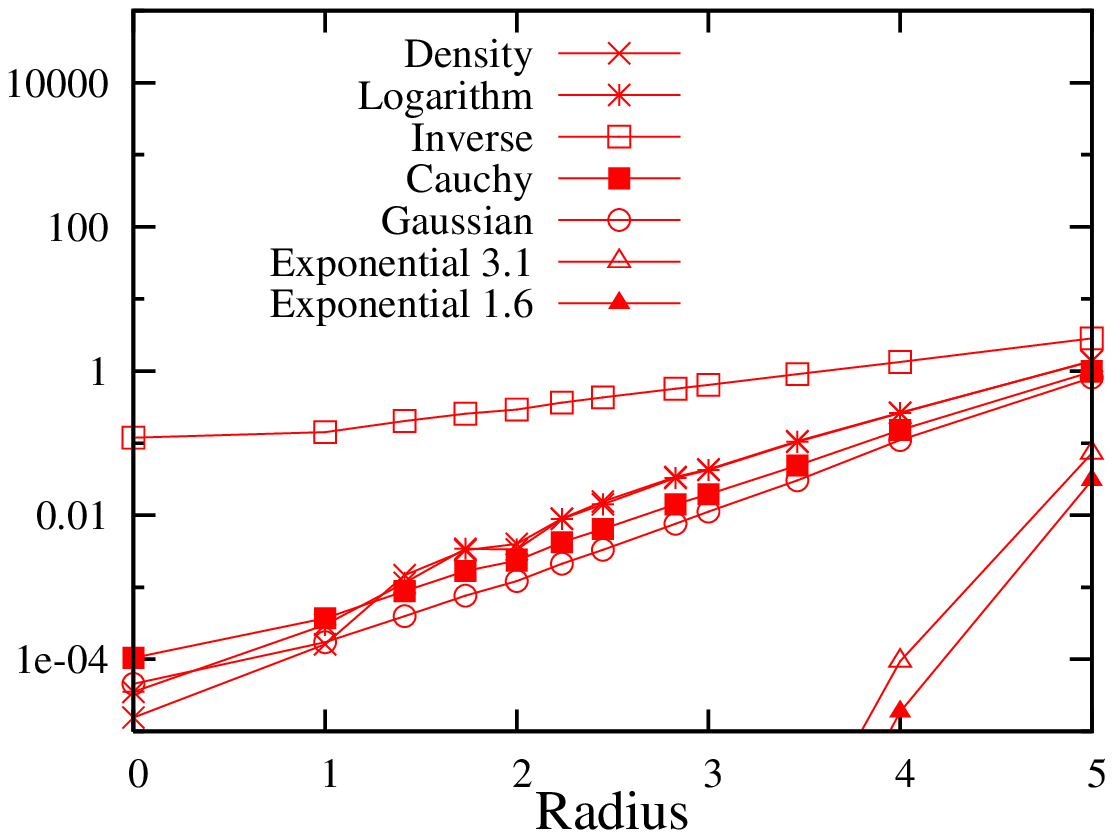}
\caption{\label{RelErr1} The relative error at small disorder $\sigma = {1.15}$ and intermediate disorders $\sigma = {2.65}$ amd $\sigma = {4.65}$.}  
\end{figure}

The question of greatest interest is: when does the relative error become small?   Figure \ref{RelErr1} shows the relative error at three disorders: $\sigma = {{1.15}, {2.65}, {4.65}}$.  I don't show higher disorders because the results don't change qualitatively.  Concentrating on the intermediate four matrix functions (the gaussian, logarithm, density matrix, and Cauchy distribution), one immediately notes that their relative errors are very similar, although a slight splitting appears as the disorder increases.  The exception to this rule is when both the radius and the disorder are small, in which case the density matrix is considerably smaller than the gaussian and the Cauchy distribution, and the logarithm lies somewhere in between.  When does the relative error of these four functions become small?  Clearly at $\sigma = {1.15}$ the $r = 0$ relative error is already of order ${10}^{-2}$ or better, while by $\sigma = {2.65}$ the error is ${10}^{-2}$ at radii up to about $2$.  Moreover, even at $\sigma = {1.15}$ the error shows an exponential dependence on $r$ caused by the coherence length, suggesting that it could be made arbitrarily small by increasing the radius of the truncation volume.  As the disorder increases, the slope of the error also increases, suggesting that smaller and smaller truncation volumes are necessary to obtain a given error. 

Turning to the exponential, its relative error is already very small at $\sigma = {1.15}$ and diminishes very quickly with the disorder.  This function is extremely well suited to $O{(N)}$ algorithms.

The inverse is the opposite case: the lowest value of its relative error is $0.05$, which is reached at ${\sigma = {9.0}}, {r = 0}$.  Its maximum variation with $r$ is a factor of $34$, again at $\sigma = {9.0}$.  This leaves little hope that the error could be decreased substantially by expanding the truncation volume.  Therefore, basis truncation $O{(N)}$ algorithms seem to be unsuitable for calculating the real part of the matrix inverse, at least when the inverse's argument is has only an infinitesimally small imaginary part.  

This conclusion must be qualified by three points:
\begin{itemize}
\item My results do not close the door on $O{(N)}$ algorithms which do not use basis truncation.
\item The imaginary part of the inverse is the Cauchy distribution, which is tractable for basis truncation algorithms.  The imaginary part of the inverse is the only thing that matters when evaluating a matrix function via the complex integration formula ${f{(X)}} = {{(2 \pi \imath)}^{-1} \oint {dE} {(E-H)}^{-1}f{(E)}}$.  Therefore, basis truncation algorithms can calculate the inverse as an intermediate step towards calculating other functions.  The real part of the intermediate results will be worthless, but that doesn't matter.
\item When the inverse's argument has a finite imaginary part, basis truncation algorithms may still be applicable. Smirnov and Johnson\cite{Smirnov01, Smirnov02} did a detailed study of the accuracy of a basis truncation algorithm when calculating the resolvent ${(E - H)}^{-1}$ in an ordered system.  They showed that when the energy $E$ has a finite imaginary part the error falls off exponentially with $r$.   
\end{itemize}

The last question of interest is whether the relative error is of order one at the boundary of the truncation volume.  For all the functions except the exponential, the answer is an unambiguous yes.  In fact, as the disorder increases the $r = 5$ values converge to one, so that at $\sigma = {9.0}$ the maximum disagreement from one is exhibited by the inverse with a value of $1.7$.  The exponential is, perhaps, another matter:   its $r = 5$ value gets progressively smaller, until at $\sigma = {9.0}$ the $\alpha = {1.6}$ exponential has a value of $3 \times {10}^{-3}$.  However maybe even this can be considered to be order one in light of the exponential's extremely fast variation with $r$.

Finally I discuss the validity of the error estimate in equation \ref{RoughEstimateRepeat}, which rests on both the order one question discussed in the previous paragraph and on the uniformity of the absolute error throughout the truncation volume.  All the functions examined here (with the possible exception of the exponential) justify the first assumption.  However, as discussed earlier, the spatial uniformity is more dubious, especially for the inverse and for large disorders.  The gaussian, logarithm, density matrix, and Cauchy distribution are more or less invariant within the truncation volume except at very high disorder.  On the other hand, the inverse function is elevated at small radii, and this is important in making the inverse's relative error so large.  Lastly, the matrix exponential varies by many orders of magnitude within the truncation volume.  Equation \ref{RoughEstimateRepeat} could be used nonetheless as an upper bound for the exponential's relative error.

\subsection{Software Reliability and Reproducibility}
The software used to obtain these results is similar to that used to obtain the results in chapter \ref{DensityMatrixAccuracy}, and should be able to reproduce exactly that chapter's results, although I haven't checked.  Only a small percentage of the total number of lines of code was changed.  This software includes an automated
test suite which tests all computational functions except the
highest level output (printing and graphing) routines. Moreover, I have
taken pains to enable other researchers to easily reproduce and
check my results, simply by installing my software and the
libraries it depends on, compiling it with the GNU gcc
compiler\cite{GNUGCC}, and starting it running. The software, with
the configuration files needed to reproduce all the numerical results and graphs presented in this paper, will be made available under the GNU Public
License\cite{GNUGPL}; check www.sacksteder.com for further
details.

For the reader's benefit it is important to discuss the biggest new risk, associated with the exponential function.  I have already mentioned that some of the (unpublished) results for that function are incorrect; in particular sometimes the partial matrix magnitude of the absolute error, which is positive definite, is computed to be negative.  Moreover, there is a consistent trend for the standard deviation of observables associated with the exponential to be of the same order as the observables themselves.  This second effect, while disturbing, may accurately reflect the true physics.  I believe that any unreliability in the computations is caused by rounding errors related to the exponential's huge changes in magnitude.  I have verified that the algorithm in use is in fact vulnerable to rounding errors when operating on numbers with widely varying magnitudes, have checked that the negative magnitudes do not occur when the parameters are such that the exponential varies less, and have avoided publishing any data about the absolute error except in the cases where the exponential's variation is most restricted.  I also note that exactly the same code was used to produce analogous results for other matrix functions which do not exhibit such large changes in magnitude, and that the results for these functions are free from any negative magnitudes. There is still some risk that the negative magnitudes are due to some other problem or bug, and that the results are more thoroughly wrong than I imagine.  I believe that the chances of this being true are sufficiently small to allow publication in good conscience of the results about the exponential's absolute error.

\chapter{\label{CoherenceLength}A Phenomenology of Eigenfunctions in Disordered Systems}

In disordered systems, eigenfunctions and other physical quantities
generally display a finite coherence length, and possibly also a localization length\cite{Soven67, Economou83, Anderson58}.   The coherence length often has a value much different than the localization length, and both lengths can be physically important. Several theoretical approaches have been developed to calculate observables in these systems, including the coherent potential approximation\cite{Soven67, Economou83}, the replica sigma model\cite{Wegner79, Lerner03}, and the supersymmetric sigma model \cite{Efetov97}. 

When studying a disordered system with hamiltonian $H$, a practical problem of real importance is how to efficiently evaluate matrix functions $f{(H)}$.  For instance, quantum calculations of condensed matter systems are typically limited by the computational cost of calculating the density matrix function $\rho$.  Recently a class of approximate but speedy algorithms has	 been introduced for evaluating matrix functions in an amount of time which is proportional to the basis size\cite{Goedecker99}.  One subclass of these $O{(N)}$ algorithms is the basis truncation class, which truncates all the basis elements outside of a truncation volume and then does the function evaluation within the truncation volume.   

In chapter \ref{DensityMatrixAccuracy} I tried to estimate the size of the absolute error $\Delta$ incurred by basis truncation algorithms, and derived equation \ref{EqExpand}, which gives an exact expression for the error in terms the eigenfunctions and eigenvalues of the truncated and untruncated systems.  I repeat it here:  
\begin{eqnarray}
{{\langle \vec{x} |} {\Delta{(R)}} {| {\vec{x} + \vec{r}}
\rangle}} & = &
 \int_{-\infty}^{\infty}\int_{-\infty}^{\infty}\int_{-\infty}^{\infty}{{dE_{a}}{dE_{b}}{dE_{c}}}
 \nonumber \\
 & &
 {{\langle \vec{x} |a \rangle} {{\langle a|}H_{1}{|c \rangle}}
  {{\langle c |} H_{1} {| b \rangle}} {\langle b | {\vec{x} + \vec{r} \rangle}}
{g{(E_{a},E_{b},E_{c})}}}
\nonumber \\
g{(E_{a}, E_{b}, E_{c})} & \equiv & {{n_{A}{(E_{a})}}{n_{A}{(E_{b})}}{n{(E_{c})}}}  
\nonumber \\ 
& \times & \oint{{dE}f{(E)}
\frac{1}{E - E_{a}} \frac{1}{E - E_{b}} \frac{1}{E - E_{c}}} 
\end{eqnarray}

$H_{1}$ is the part of the hamiltonian which connects the truncation volume with the rest of the system, $| a \rangle$ and $ | b \rangle$ are eigenstates of the truncated system, $| c \rangle$ is an eigenstate of the untruncated system, $n$ and $n_{A}$ are the level densities of the untruncated and truncated systems, and $g$ is a function describing the eigenvalues.

This exact formula is useless without phenomenological estimates for the following quantities:
\begin{itemize}
\item The level densities $n$ and $n_{A}$.  These quantities can be estimated from fits to numerical studies, theoretical estimates using the coherent potential approximation, or convenient models.
\item  The correlations between the eigenvalues $E_{a}, E_{b}, E_{c}$.  Eigenvalue correlations have been studied in detail in the unlocalized regime, and precise predictions are available when the conductance $g$ is large.  It is well known that in localized systems the eigenvalue correlation is nill unless two states are close together.  However, the eigenvalue correlations between an untruncated system and its truncated version have not been studied so much, except on the level of perturbation theory.  Perturbation theory is of course invalid when there is strong mixing of states.
\item  The matrix elements ${\langle \vec{x} |a \rangle}$,  ${{\langle a|}H_{1}{|c \rangle}}$, ${{\langle c |} H_{1} {| b \rangle}}$, and ${\langle b | {\vec{x} + \vec{r} \rangle}}$.  One also needs their correlations with each other and with the eigenvalues. 
\end{itemize}
The phenomenology of these quantities must apply to systems where the coherence length is smaller than the truncation volume, because this is the regime where $O{(N)}$ basis truncation algorithms work best\cite{Sacksteder03}.  It should apply both to systems which are unlocalized and systems which are localized.

Of the three items listed above, the matrix elements may be the most challenging, because one expects a complete change as the disorder becomes large.  There is a well developed theory of eigenfunction amplitudes and correlations in the limit of infinite conductance, and some work has been done to extend these results to the regime of large but not infinite conductance.  However, I am not aware of any phenomenology for estimating matrix elements in either the localized regime or the regime which is unlocalized but still has a small conductance.  In this chapter I propose a phenomenological model of wavefunctions and matrix elements which incorporates only information about the coherence length and the decay length, and
leaves out all the details of the disorder.  This model facilitates some calculations which would otherwise be impracticable, and can also provide qualitative physical insight
about effects of length scales on physical quantities.

This phenomenological model is designed to allow calculation of expectation values of observables.  I start by constructing an ensemble of random functions designed to have a specific coherence length and
decay length.  With this random ensemble in hand, I compose the
desired observable, and then average to obtain the expectation value. The result of the average
is naturally expressed in the momentum basis; evaluation of results
in the position basis requires evaluation of integrals which are
mathematically similar to loop integrals in quantum field theory.

I begin this chapter by explaining the random ensemble and the
averaging procedure. Next comes an explanation of how to compute the participation ratios and a demonstration that this model predicts that the eigenfunctions are multifractal. Then there is a check of the model's description of localization and a simple example of calculating the value of an observable.  I end with brief discussions about the possibility of modeling correlations between eigenfunctions and about the relationship between this phenomenological model and the supersymmetric sigma model.

\section{The Model}
I start by incorporating into my model the correlation function, from which
the correlation length can be derived: 
\begin{equation}
{C{(\vec{x})}} =
{\int{{d\vec{k}}^{D}{{{|{\psi{(\vec{k})} }|}^{2}{exp{(\imath
{\vec{k} \cdot \vec{x}})}}}}}  = {\int{{d\vec{y}}^{D}{{{
\psi{(\vec{y})} }} {{{\psi}^{*}{({\vec{x} + \vec{y}})} }} }}}}
\label{CoherenceFunction}
\end{equation}

This function completely determines the magnitudes of the amplitudes ${\psi{(\vec{k})}}$. An eigenfunction centered at a
position $\vec{o}$ must include a phase factor $\exp{(i \vec{k}
\cdot \vec{o})}$; I consider $\vec{o}$ as a random variable and
average over it.  In addition, some other phase factor
$\phi{(\vec{k})}$ may be present.  Thus I have:
\begin{equation} {\psi{(\vec{k})}} = {{\beta{(\vec{k})}} \exp{(-i \vec{k} \cdot
\vec{o})} \exp{(i \phi{(\vec{k})})}}
 \label{WavefunctionParameterization}
 \end{equation}
 $\beta{(\vec{k})}$ is the normalized square root of the correlation function ${\beta{(\vec{k})}} =  { {{(\frac{1}{{2\pi}})}^{\frac{d}{4}}} {\sqrt{|{ C{(\vec{k})}}|}}}$.  I require that the wave functions be normalized by setting $1 = {C{(\vec{x} = 0)}} = {\int{{d\vec{k}}^{D} {{\beta}^{2}{(\vec{k})}} }}$.

I next insert information about the localization of the spatial wave function $\phi{(\vec{x})}$ by choosing a rule for averaging the phase over the disorder:
\begin{equation}
{\langle {\exp{(i \phi{(\vec{k})})} \exp{(-i \phi{(\vec{k'})})}}
\rangle} = {f{(\vec{k} - \vec{k'})}}
 \label{PhaseParameterization}
\end{equation}

I will call ${f{(\vec{k} - \vec{k'})}}$ the phase correlation function. Obviously $f{(\vec{0})} = 1$.   I decompose products of even
numbers of phase factors into pairs of phase factors (a la Wick's
theorem), and set averages of odd numbers of phase factors to
zero.  The above rules apply only to phase factors of different
components $\psi{(\vec{k})}$ of the same eigenfunction $\psi$; for the moment I
assume that phase factors of two different eigenfunctions are
uncorrelated. Each eigenfunction will also have its own  translational phase $\exp{(i \vec{k} \cdot \vec{o})}$ which must be averaged over.

These rules complete the definition of the phenomenological model presented in this chapter.  They are equivalent to a generating function: 
\begin{eqnarray}
{f{(J, J^{*})}} & = &{\frac{1}{V}\int {{{d\vec{o}}^{D}} {e}^{K}}}
\nonumber \\
 {\textit{K}} & = & {\int{{d\vec{k}}^{D} {d\vec{\acute{k}}}^{D} {J{(\vec{k})}} {J^{*}{(\vec{\acute{k}})}} {\beta{(\vec{k})}} {{\beta}{(\vec{\acute{k}})}}}}
 \nonumber \\
 & \times &
  {f{(\vec{k} - \vec{\acute{k}})}} {{exp}{(-\imath\vec{o} \cdot {(\vec{k} - \vec{\acute{k}})})}}
\label{Kernel}
\end{eqnarray}

As usual, correlation functions are obtained by taking derivatives with respect to the sources $J$, $J^{*}$, and then setting the sources equal to zero.  $V$ is the system volume, and both $J$ and $\beta$ carry units of $\sqrt{V}$.

Berry conjectured that eigenfunctions in a quantized chaotic system are sums of plane waves, each with a random phase\cite{Berry77}.  This corresponds to choosing the phase correlation function $f$ to be a delta function, in which case the kernel reduces to:
\begin{equation}
{\textit{K}} = {\frac{1}{V} \int{{d\vec{k}}^{D} {|J{(\vec{k})}|}^{2}  {{\beta}^{2}{(\vec{k})}}  }}
\end{equation}

In this case, the phenomenological model presented here corresponds exactly with Srednicki's mathematical formulation of Berry's conjecture\cite{Srednicki96}, and can be used to calculate any observable.  As shown by Prigodin\cite{Prigodin95}, Prigodin et al.\cite{Prigodin95a}, and Srednicki\cite{Srednicki96}, calculations using the supersymmetric sigma model result in the same values of observables if the system being computed is not spatially extended or, equivalently, has an infinite conductance $g$.   Several authors have used the supersymmetric sigma model to compute corrections to the values obtained via Berry's conjecture in powers of the inverse conductance $\frac{1}{g}$ \cite{Fyodorov95, Blantner97, Prigodin98, Falko96}.  Their results are necessarily applicable only to the regime of large $g$.  I will discuss the relation between their results and this phenomenological model at opportune moments.
 
As an exercise, I compute the average two point correlation function,
\begin{equation}
{\langle {{\psi{(\vec{k})}}{{\psi}^{*}{(\vec{k'})}}} \rangle} =
{{(\frac{1}{{2\pi}})}^{\frac{D}{2}}
\sqrt{|{{C{(\vec{k})}}{C{(\vec{k'})}}}|} {\langle
{\exp{(\imath{(\vec{k} - \vec{k'})\cdot \vec{o}})}
\exp{(\imath{(\phi{(\vec{k})} - \phi{(\vec{k'})} )})}\rangle}}}
\end{equation}

The average over $\vec{o}$ results in a delta function divided by
the system volume $V$, enforcing momentum conservation.  Thus one
obtains ${\langle {{\psi{(\vec{k})}}{{\psi}^{*}{(\vec{k'})}}}
\rangle} = {V^{-1} {(2 \pi)}^{\frac{D}{2}}\, {|C{(\vec{k})}|}\,
{\delta^{D}{(\vec{k} - \vec{k'})}}}$.  Conversion of this result
to position space requires two integrals over momentum; however
momentum conservation gets rid of one of the integrals.  Inspection of equation \ref{CoherenceFunction} shows that the Fourier transform $C{(\vec{k})}$  of the coherence function is real;
\begin{equation}
{\langle {{\psi{(\vec{x})}}{{\psi}^{*}{(\vec{x'})}}}
\rangle} = \frac{C{(\vec{x} - \vec{x'})}}{V}
\label{WavefunctionCorrelation}
\end{equation}  

Equation \ref{WavefunctionCorrelation} is a consequence of the assumption that each individual wave function in the ensemble shares the same correlation function $C{(\vec{x})}$. This assumption is probably the weakest point of the model presented in this paper.  In general one would expect it to be wrong; that each individual eigenfunction would have a distinct shape and correlation function.  

\section{The Participation Ratios and Multifractality}
Next I calculate the averages of the participation ratios:
\begin{eqnarray}
{ \langle P^{n} \rangle } & \equiv & {\langle \int{{{d\vec{x}}^{D}} {({{\psi{(\vec{x})}} {{\psi}^{*}(\vec{x})}})}^{n}}\rangle}
\nonumber \\ 
& = & {{n!} \int{{d{\vec{o}}}^{D} {h^{n}{(\vec{o})}}}}
\nonumber \\
{h{(\vec{o})}} & \equiv & {\int{ {d\vec{k}}^{D} {d\vec{\acute{k}}}^{D} \beta{(\vec{k})} \beta{(\vec{\acute{k}})} f{(\vec{k} - \vec{\acute{k}})} {exp}{(- \imath \vec{o} \cdot {(\vec{k} - \vec{\acute{k}})})}}}
\label{ParticipationRatiosEq}
\end{eqnarray}

In the case of Berry's conjecture, $h = V^{-1}$ and $P^{n} = {{n!}\,V^{1-n}}$.  This result can be rewritten in terms of the probability distribution $P{({|\psi |}^{2})}$ of eigenfunction magnitudes: ${P{({|\psi |}^{2})}} = {V \, \exp{(-V {|\psi |}^{2})}}$.  Setting $V = 1$ reduces this to a famous result from random matrix theory; the result with $V \neq 1$ can be derived using supersymmetry in the limit $g = \infty$.

 The participation ratios $P^{n}$ are closely related to the singularity strength of fractals and multifractals; in fact fractal can be characterized by how its $P^{n}$ scale with the system volume $V$\cite{Benzi84, Frisch85, Halsey86, Halsey86a}.  A function with fractal dimension $D^{*}$ which lives in a world with dimension $D$ will have participation ratios whose volume dependence is given by $P^{n} \propto V^{{(1-n)}D^{*}/D}$.  This relation can be used to define the fractal dimension.  In contrast, the multifractal has a fractal dimension for each participation ratio; $D^{*}$ is a function of $n$. 
 
 Theoretical arguments and numerical calculations indicate that wave functions near the localization transition are multifractals\cite{Castellani86, Cain02}.  The model presented in this paper clearly allows wave functions at criticality to be modeled by simply choosing functions $\beta$ and $f$ which result in a multifractal $h$.  For instance, if $f$ is taken equal to $1$, implying no localization, then $\beta$ would have to be chosen to be a multifractal.  Note that this model only allows calculation of average participation ratios; however I believe that if the average ratios indicate multi-fractality then the individual wave functions should also be multi-fractal.

It would be nice to find a choice of $\beta$ and $f$ which reproduces supersymmetric predictions for the regime of large conductance $g$.  Fyodorov and Mirlin\cite{Fyodorov95} obtained the participation ratios $P^{n} = {{{n!}\,V^{1-n}} {(1 + {\frac{\Pi}{2}n{(n-1)}})}}$, which corresponds to the probability distribution ${P{(y \equiv {V {|\psi |}^{2}})}} = {{\exp{(-y)}}{(1  + {{\Pi}(1 - {2y} + {y^{2} / 2})})}}$.  The diffuson loop $\Pi$ is a geometry dependent constant determined by the spectrum of the system's diffusion operator ${\nabla}^{2}$; in two dimensions it is related to the volume $V$ via the equation $\Pi = {\frac{1}{2 \pi g} {\ln\frac{V}{V_{0}}}}$.   This result assumes that ${V {|\psi |}^{2}}$ is small and therefore is valid only for participation ratios $P^{n}$ which have small $n$. Working in two dimensions, Falko and Efetov\cite{Falko96} were able to go a step further by using an instanton technique; i.e. they chose a saddle point which was not translationally invariant.  Using this improved prescription, they were able to calculate the probability of large amplitudes\cite{Mirlin99}: 
 \begin{eqnarray}
 {P{(y \equiv {V {|\psi |}^{2}})}} & = & {\exp{(-\frac{1}{2 {\Pi}} {{\ln}^{D}{({\Pi}y)}})}}, {{\frac{1}{\Pi} \lesssim y}}
 \nonumber \\
 & = & {\exp{(-y + {{\Pi}y^{2}/2})}}, {{\frac{1}{\sqrt{\Pi}}} \lesssim y \lesssim {\frac{1}{\Pi}} }
 \end{eqnarray}

The most intriguing aspect of these results is their prediction that the eigenfunctions are weakly multifractal.  The diffuson loop $\Pi$ gives Fyodorov and Mirlin's results in an extra volume dependence which translates to multifractality.  One can see this from the fact\cite{Mirlin99} that at large $g$ their participation ratio is equal to: 
\begin{eqnarray}
 {1 + {\frac{\Pi}{2}n{(n-1)}}} = {1 + { \frac{n{(n-1)}}{4 \pi g} \ln\frac{V}{V_{0}}}} \approx {{\exp}{({\frac{n{(n-1)}}{4 \pi g}\ln\frac{V}{V_{0}}})}} = {{(\frac{V}{V_{0}})}^{n{(n-1)} / { 4 \pi g}}}
\end{eqnarray}
Therefore the participation ratios derived by Fyodorov and Mirlin scale with the volume as $V^{{(n - 1)}{(1 + {n / { 4 \pi g}})}}$. The nonlinear exponent signifies multifractality.  Falko and Efetov's results are also multifractal.

Interestingly enough, the model presented here can easily reproduce the structure of Fyodorov and Mirlin's results and therefore describe a multifractal system, with the following choice of the phase correlation function $f$: 
\begin{equation}
{f{(\vec{k})}} \propto {{V^{-1}{\delta}^{D}{(\vec{k})}} + {a_{1} \nabla{\delta}^{D}{(\vec{k})}} + {a_{2} {\nabla}^{2}{\delta}^{D}{(\vec{k})}}} 
\end{equation}
The $a$'s are just small (perturbative) constants, with $a_{2} \propto a_{1}^{2}$.  Then, to  second order in $\vec{o}$,  the function $h$ defined in equation \ref{ParticipationRatiosEq} is given by:
\begin{equation}
 {h{(\vec{o})}} = {1 + {{\vec{a}}_{3} \cdot \vec{o}}  + {\frac{1}{2} \vec{o} \cdot {\ddot{a}}_{4} \cdot \vec{o}} }
\end{equation}
  ${\vec{a}}_{3}$ is the first derivative, while ${\ddot{a}}_{4}$ is the second derivative, which is a $D \times D$ tensor.  Again keeping only lowest order terms, one obtains the following value for the $n$-th power of $h$:
   \begin{equation}
   1 + {n {\vec{a}}_{3} \cdot \vec{o}}  + {\frac{n}{2} \vec{o} \cdot {\ddot{a}}_{4} \cdot \vec{o}}  + {\frac{n {(n-1)}}{2} {({\vec{a}}_{3} \cdot \vec{o})}^{2}} + ...
   \end{equation}
     Equation \ref{ParticipationRatiosEq} requires integration over $\vec{o}$. During this integration, the second term integrates to zero and the second derivative in the third term is replaced by its trace ${Tr}{({\ddot{a}}_{4})}$.  If the trace is zero then the lowest order term is of order $O{(n{(n-1)})}$ and equation \ref{ParticipationRatiosEq} gives $P^{n} = {{{n!}\,V^{1-n}} {(1 + {a_{5} \frac{n{(n-1)}}{2}})}}$, which is just Fyodorov and Mirlin's multifractal result.  It is very interesting that such a simple model can reproduce the results of a very complicated supersymmetric calculation.
  
The odd thing though is that this perturbative calculation predicts multifractality for any perturbation to Berry's conjecture, in any dimension.   Usually multifractality is expected only close to the Anderson transition.  Fyodorov and Mirlin's prediction of multifractality was expected only because they derived their result in two dimensions.  The critical dimension for the Anderson transition is two dimensions and therefore every eigenfunction is expected to be multifractal.  In other dimensions multifractality is not generally expected.  Perhaps the reason for the result here is that it is based on Berry's conjecture, which assumes that the system is fully chaotic.

\section{The Localization Length}

The phenomenological model presented here contains two parameters: $\beta{(\vec{k)}}$ which is related to the correlation function $C{(\vec{x})}$, and the phase correlation function $f{(\vec{k})}$.  I have proved that $\beta$ determines the model's correlation function, and clearly if $\beta$ decays with a momentum scale $R^{-1}$ then the wave functions have correlation length $R$.  On the other hand, what determines the localization length $L$ of localized wave functions?  It turns out that $L$ is equal to the maximum of two length scales: the correlation length $R$, and the length $\rho$ which governs the phase correlation function $f$.  To prove this assertion, I calculate
the four point correlator ${F{(\vec{\Delta})}} \equiv {\langle {{|\psi{(\vec{x})}|}^{2}
{|\psi{(\vec{x}+\vec{\Delta})}|}^{2}} \rangle} $.  Its long distance behavior will allow me determine the model's localization length.

Wick's theorem breaks $F$ into the sum of three terms, and some algebraic manipulation results in the following identity:
\begin{equation}
{F{(\vec{\Delta})}} = 
{\frac{1}{V}\int{{(\frac{d\vec{k}}{2 \pi})}^{D}
 {{f}^{2}{(\vec{k})}} 
 [{{exp}{(\imath \vec{k} \cdot {\vec{\Delta}})} 
 {I{(\vec{0}, \vec{k})}} } +
  {2I{(\vec{\Delta}, \vec{k})}}]}}
\label{FourPoint1}
\end{equation}
The function $I$ is real and is defined as: 
\begin{equation}
I{(\vec{\delta}, \vec{k})} \equiv {[\int {{d\vec{\acute{k}}}^{D} {{exp}{(\imath \vec{\delta} \cdot \vec{\acute{k}})}} {\beta{(\vec{\acute{k}} - {\frac{1}{2} \vec{k}})}} {\beta{(\vec{\acute{k}} + {\frac{1}{2} \vec{k}})}}}]^{2}}
\label{FourPoint2}
\end{equation}
 $I$ completely encapsulates the influence of the spatial correlation function $C{(\vec{x})}$ on the four-point function $F$, and ${I{(\vec{\delta}, \vec{0})}} = {C^{2}{(\vec{\delta})}}$.  The second term in equation \ref{FourPoint1} describes the two Wick terms which connect $\vec{x}$ and $\vec{x} + \vec{\Delta}$, and clearly its long distance behavior is always governed by the correlation length $R$.  In contrast, the decay of the first term in equation \ref{FourPoint1} is influenced by the phase correlation $f$ as well.  To be more concrete, I consider two different choices of $\beta$ and $f$: gaussian decay, and exponential decay.
 
 \subsection{\label{Gaussian Decay} Gaussian Decay}
 I choose ${\beta{(\vec{k})}} = { {(\frac{R^{2}}{2 \pi})}^{D / 4} {{exp}{(-k^{2} R^{2} / 4)}} }$ and ${f{(\vec{k})}} = {{exp}{(- k^{2} {\rho}^{2} / 4)}}$.  Integrating over the gaussians just produces more gaussians;  the final result is:
 \begin{equation}
 {F{(\vec{\Delta})}} = { V^{-1} {(2 \pi {\lambda}^{2})}^{-D / 2}  [ {{exp}{(\frac{- {\Delta}^{2}}{2 {\lambda}^{2}})}} + {2 {{exp}{(- {\Delta}^{2} / R^{2})}}}] }
 \label{FourPointGaussian} 
 \end{equation}
 The length scale $\lambda$ is defined by ${{\lambda}^{2}} = {{\rho}^{2} + {\frac{1}{2}{R}^{2}}}$. Note that the localization length $L$ is never smaller than the correlation length $R$, even if the phase decay length $\rho$ is much smaller.  In the opposite situation where $\rho \gg R$, the four point function resolves into two terms, the first decaying slowly with scale $\rho$ and the second more quickly with scale $R$.  As the phase decay length $\rho$ goes to infinity, the first term becomes a constant, signalling that there is no localization.  Thus I have $L = {{max}{(\rho, R)}}$.
 
 \subsection{\label{Exponential Decay} Exponential Decay}
 I work in three dimensions, assume that the phase correlation function $f{(\vec{k})}$ can be described in terms of its poles, and choose a coherence function $C{(\vec{x})}$ proportional to ${exp}{(-x / R)}$.  The corresponding $\beta$ is:  
 \begin{equation}
 {\beta{(\vec{k})}} = {\frac{2 \sqrt{2}}{\pi} R^{3 / 2} {(1 + {R^{2} k^{2}})}^{-2}}
 \label{ExponentialFunction}
 \end{equation}
 By introducing a Feynman parameter I reduce the three integrals in the definition of $I{(\vec{\delta}, \vec{k})}$ down to a single integral:  
 \begin{equation}
 {\sqrt{I{(\vec{\delta}, \vec{k})}}} = {{6}  \int_{0}^{1} {dx} {(1 - x) x}\, {{exp}{({-\delta / R} + {\imath (x - \frac{1}{2}) {\vec{\delta} \cdot \vec{k}})}}} {(\mu / R)}^{5} {\lbrace 1 + {(\delta / \mu)} + \frac{1}{3}{(\delta / \mu)}^{2} \rbrace}} 
 \label{FourPointExponential}
 \end{equation}
The new length scale $\mu$ is ${{\mu}^{2}} \equiv {{({ R^{-2}} + {k^{2} x {(1-x)}})}^{-1}} \leq {R^{2}}$.  Turning to equation \ref{FourPoint1} for the four point correlator, it is obvious that the second term is dominated by an exponential which decays at least as fast as ${exp}{(-2 \Delta / R)}$.  The first term is a little more tricky: one must do the integral over $\vec{k}$ before the integrals over the Feynman parameters.  After angular integration, the $\vec{k}$ integral looks like $ \int_{-\infty}^{\infty} {{dk} {{exp}{(\imath k \Delta)}}{f^{2}{(k)}} {(\mu / R)}^{5} {(\acute{\mu} / R)}^{5}}$.  Now assume that the phase correlation function $f{(\vec{k})}$ has the same structure as in equation \ref{ExponentialFunction}, but with its poles at $\pm\imath {\rho}^{-1}$.  Then the integrand in the $k$ integral contains poles  at $k = {\pm\imath{\rho}^{-1}}$, at $k = {\pm\imath{(R \sqrt{x {(1-x)}})}^{-1}}$, and at $k = {\pm\imath{(R \sqrt{\acute{x} {(1-\acute{x})}})}^{-1}}$.  It also has a branch cut caused by the non-analyticity of the square roots, but this can be hidden in the lower half plane.   The poles are multiplied by an exponential ${{exp}{(\imath k \Delta)}}$, which causes decays lengths of $\rho$, ${R \sqrt{x {(1-x)}}} $, and ${R \sqrt{\acute{x} {(1-\acute{x})}}} $.  Thus I obtain the same qualitative behavior as with gaussian functions; in particular the localization length $L$ is given by $L = {{max}{(\rho, R)}}$.

\subsection{Correspondence with Supersymmetric Results}
Blantner and Mirlin\cite{Blantner97} used the supersymmetric sigma model to derive the following result:
\begin{equation}
{V^{2} {F{(\vec{\Delta})}}} = {1 + {{C^{2}{(\vec{x})}}{(1 + \Pi)}} + {\Pi{(\vec{\Delta})}}}
\end{equation}
$\Pi{(\vec{x})}$ is the diffusion propagator; i.e. the inverse of ${\nabla}^{2}$, omitting the zero-momentum mode.  $\Pi \equiv {\Pi{(\vec{0})}}$ is the diffuson loop that I have already discussed.  Prigodin and Altshuler\cite{Prigodin98} used a simplified model (a Liouville model)
to derive the joint probability distribution of wavefunction magnitudes at two different points $P{({{|\psi{(\vec{x})}|}^{2}}, {{|\psi{(\vec{y})}|}^{2}})}$.  This probability distribution is equivalent to knowing the moments ${\langle {{|\psi{(\vec{x})}|}^{2n}
{|\psi{(\vec{x}+\vec{\Delta})}|}^{2m}} \rangle}$ for all $n, m$. 

As I already mentioned, the model presented here is a generalization of Berry's model, and therefore reproduces these results in the limit of infinite conductivity, where the diffusion propagator is zero and the phase correlation function $f$ is a delta function.  Clearly one would hope to reproduce Blantner and Mirlin's result using ${f{(\vec{k})}} = {{V^{-1}{\delta}^{D}{(\vec{k})}} + {\alpha {\tilde{f}{(\vec{k})}}}}$, where $\alpha$ is a small parameter.  However I haven't yet checked whether this is actually possible.    

\section{\label{Matrix Elements}Matrix Elements}
The original motivation of this work was to estimate certain matrix elements in a disordered system, given only information about the coherence length and localization length.    As a simple example, I consider the case where the matrix element is a surface of area $A$ described by the equation ${\textit{s}{(\vec{x})}} = 0$.  Then the matrix element between two wave functions ${\psi}_{1}$ and ${\psi}_{2}$ is ${\textit{M}} = {\int {d\vec{x}}^{D} {\delta{(\textit{s}{(\vec{x})})}} {{\psi}_{1}{(\vec{x})}} {{\psi}_{2}^{*}{(\vec{x})}} }  $.  Since ${\psi}_{1}$ and ${\psi}_{2}$ must be averaged separately, the average matrix element is trivially zero.  However, the normalized average of the squared matrix element does not vanish:
\begin{equation}
{{(V^{2} A^{-2})}{\langle {|\textit{M} |}^{2} \rangle}} = {{(V^{2} A^{-2})} {\langle {\int {d\vec{x}}^{D} {d\vec{y}}^{D} } {\delta{(\textit{s}{(\vec{x})})}} {\delta{(\textit{s}{(\vec{y})})}} {{\psi}_{1}{(\vec{x})}} {{\psi}_{2}^{*}{(\vec{x})}} {{\psi}_{1}^{*}{(\vec{y})}} {{\psi}_{2}{(\vec{y})}}  \rangle}}
\label{MatrixElement1}
\end{equation}
 This quickly converts to:
 \begin{equation}
{ A^{-2} {\int {d\vec{x}}^{D} {d\vec{y}}^{D} } {\delta{(\textit{s}{(\vec{x})})}} {\delta{(\textit{s}{(\vec{y})})}} {|{C{(\vec{x}- \vec{y})}}|}^{2}   }
\label{MatrixElement2}
\end{equation}
If the surface $\textit{s}$ has a curvature which is small compared to the correlation length $R$, then ${(V^{2} A^{-2})} {\langle {|\textit{M} |}^{2} \rangle}$ will be proportional to $4 \pi R^{2} A^{-1}$.  This is the same scaling law that one obtains heuristically by arguing that the matrix element may be modeled as a sum over incoherent volumes of radius $R$.  It is quite remarkable that the phase coherence has no effect whatsoever on the second moment of this matrix element.

\section{Final Remarks}
\subsection{Correlations between Different Eigenfunctions}
So far I have ignored correlations between eigenfunctions and correlations between eigenfunctions and their energies.  Supersymmetric results are again available in the limit of large conductance\cite{Blantner97}:
 \begin{equation}
 {{V^{2} {\langle {{|\psi{(\vec{x}, E)}|}^{2} {|\psi{({\vec{x}+\vec{\Delta}}, {E + \omega})}|}^{2}} \rangle}} = {1 + {{C^{2}{(\vec{x})}}{\Pi}}}}, {\omega \lesssim E_{c}}
 \end{equation}
 The same paper gives a more complicated formula for the case of the energy separation $\omega$ being large compared to the Thouless energy $E_{c}$.  These formulas predict that the correlation between eigenfunctions which is quite weak; this prediction is a consequence of the assumption that the disorder is small.  As the disorder becomes large, one should expect eigenfunctions to be strongly correlated around individual peaks and valleys of the disordered potential.  In the localized regime, the orthogonality of eigenfunctions implies that eigenfunctions that are close together will be strongly correlated.  Eigenfunctions that are far apart will be entirely uncorrelated.
 
 The phenomenological model presented here can be extended to include these correlations by allowing the phase of different eigenfunctions to be correlated.  One would introduce a phase correlation function between two different eigenfunctions $i$ and $j$:  
 \begin{equation} {f_{ij}{(\vec{k} - \vec{k'})}} \equiv{\langle {\exp{(i \phi{(\vec{k})})} \exp{(-i \phi{(\vec{k'})})}}} \neq 0
\end{equation}
In the localized regime the functions $f_{ij}$ would have to be zero when the eigenfunctions are far apart; i.e. when $|{\vec{o}}_{i} - {\vec{o}}_{j}|$ is large compared to the localization length.

\subsection{Relation to Supersymmetric Results}
At several points in this chapter I have cited supersymmetric results and discussed whether the phenomenological model presented here can reproduce those results.  However the validity of this phenomenological model cannot be judged by its ability to reproduce supersymmetric results.  This model's purpose is specifically to understand the physics of systems where the coherence length is not that large, even systems that are localized.  The diffusive supersymmetric sigma model which was used to produce the cited results is not applicable to such systems. The main strength of the phenomenological model presented here is that it offers the hope of describing those systems. Its success or failure in reproducing supersymmetric results is more of a curiosity.

\chapter{\label{NoGradedMatrices}Supersymmetric Results Without Graded Matrices}
\section{Introduction}

In this chapter I carefully derive a new sigma model which is suitable for calculations in random matrix theory and in mesoscopic physics.  The new sigma model is a very important contribution to mesoscopic theory because it greatly simplifies the current state of the art in mesoscopic theory (the supersymmetric sigma model) while retaining the same reliability in nonperturbative calculations.  As an exercise, I use the new sigma model to derive some basic results in random matrix theory.  

In addition to containing novel results, this chapter can be used as a tutorial. It gives an unusually thorough derivation of a sigma model, and offers the most detailed and thorough explanation in the literature of how to use a field theory to do reliable non-perturbative calculations of the gaussian unitary ensemble of random matrices.  I do not presume that the reader already understands random matrix theory, mesoscopic physics, or the supersymmetric sigma model, but instead derive everything from first principles.  I do presume a knowledge of linear algebra, calculus, and field theory.

\subsection{Random Matrix Theory}
Random matrix theory is the study of matrices containing random numbers.  One starts by defining the set of possible matrices.  Then one defines the probability that each matrix will occur, a different probability for each matrix.  These two pieces of information, taken together, define an ensemble.  For instance, consider the set of $N \times N$ hermitian matrices.  Define the probability of a particular matrix $H$ occuring to be proportional to the gaussian:
\begin{equation}
 P{(H)} = {{\exp}{(-\frac{1}{N}{{Tr}{(H^{2})}})}}
 \label{GUEDefinition}
 \end{equation}  The choice of Hermitian matrices, in combination with equation \ref{GUEDefinition}, defines the gaussian unitary ensemble.  Technically speaking, the size $N$ is part of the ensemble definition, but usually this is not specified because one expects that calculations won't depend much on $N$.  Other ensembles can be defined easily; for instance the gaussian orthogonal ensemble has the same probability measure as the gaussian unitary ensemble but differs by requiring that the matrices be real and symmetric.  This reduces the number of degrees of freedom in the matrices $H$ and changes the results of most calculations. In this chapter I will be treating hermitian matrices, but the same techniques explained here can be applied to other ensembles.

Given a particular matrix $H$, one can compute all sorts of things about it: its eigenvalues and eigenvectors, as well as any number of matrix functions $f{(H)}$. I will call these quantities observables, denote them with the symbol $\mathfrak{O}$, and denote their possible values with the symbol $\mathfrak{o}$. Given an ensemble of possible matrices, a particular observable $\mathfrak{O}$ may have any number of values $\mathfrak{o}$, and its actual value will depend in a complicated way on which member of the ensemble has been (randomly) chosen.  
It can be very interesting to calculate the probability that observable $\mathfrak{O}$ will have a certain value $\mathfrak{o}$.   The average value of $\mathfrak{O}$ can also be interesting. Random matrix theory concerns itself with probabilities and average values of observables.  For example, one of the most important observables is the two point correlator, which describes whether eigenvalues like to cluster close to each other or instead stay far apart. This quantity is one of random matrix theory's most famous results, and will be calculated in meticulous detail later in this chapter.  The significance of this calculation is not that the result is novel - the two point correlator has been calculated in other ways - but instead that the means by which it is obtained is novel and unusually simple.

Random matrix theory is useful for describing physical problems that are extremely complicated.  It offers a way of calculating results while ignoring the details of the problem; the unspecified details are represented by the random numbers in the random matrices.  In other words, random matrix theory is is a way of mathematically representing one's knowledge that complicated things are going on while at the same time not specifying what those complicated things are.  It has proven to give accurate predictions of certain quantities in nuclear physics and chaos\cite{Guhr98}.

\subsection{Mesoscopic Physics}
Random matrix theory has become one of the most important theoretical techniques in the field of mesoscopic physics.  Mesoscopic physics is a branch of condensed matter physics which is devoted to studying systems at a length scale from a few atomic spacings to a few hundreds of atomic spacings.  Typically it is most concerned with the electrons, and treats the atoms as a medium through which the electrons move.  In terms of quantum mechanics, this means that the atoms are represented as a more or less unchanging potential, while the electrons are represented as dynamic variables moving in that potential.  

Now usually the atoms are not under precise human control: one does not know or control precisely how many of which types of atoms there are, or exactly where they are.  There may be $90\% \pm 2\%$ silicon atoms, mixed with $10\%$ other atoms, and one may not know anything about which atom is where.  This situation is called "disorder."  Despite not knowing this important information, one needs to make predictions about the electronic behavior.  One way of handling the problem of disorder is a statistical approach: one enumerates the set of all possible ways the atoms could be arranged, including not only all the possible positions of the atoms but also the fact that there could be different numbers of each type of atom.   One then defines the probability of each atomic arrangement.  Then to calculate electronic behavior one calculates the average behavior, averaged over all possible atomic arrangements.  

Quantum mechanics uses matrices to describe electronic motion.  There is a matrix, called the Hamiltonian, whose numbers represent the potential that the electons are moving in.  One makes predictions about how the electrons will move by computing various observables $\mathfrak{O}$ of the Hamiltonian, just as I described earlier in relation to random matrices.  In disordered systems one doesn't know some of the numbers in the Hamiltonian, so instead one defines an ensemble of possible Hamiltonians and then averages the observables associated with electronic motion over this ensemble.  In other words, the quantum mechanical theory of electronic motion in disordered media is a type of random matrix theory.  In systems that are extended in one or more dimensions, the random matrices - random Hamiltonia - have a very special structure reflecting how electrons move from place to place.   Only in systems that look like a dot - a single point in space - do the random Hamiltonia have the gaussian random matrix structure which I showed earlier.  The term "random matrix theory" is generally taken to refer only to the simpler case, often called a zero-dimensional system.  Nevertheless, properly speaking the quantum mechanical theory of extended systems is also random matrix theory.

The prototypical quantum mechanical model of a disordered system was the Anderson model\cite{Anderson58}. It uses a basis laid out on a cubic
lattice, one basis element per lattice site. (I mean "cubic lattice" in a sense which applies to any number $D$ of dimensions; i.e. a lattice which is a uniformly spaced along each orthonormal direction in coordinate space.  In two dimensions it would be a square lattice, et cetera.)
The Anderson model specifies an ensemble of random Hamiltonian matrices which are symmetric and composed of two parts:
\begin{itemize}
\item A regular part: ${\langle \vec{x}|H| \vec{y} \rangle} = 1$
if $\vec{x}$ and $\vec{y}$ are nearest neighbors on the lattice.
This term is, up to a constant, just the second order
discretization of the Laplacian; its spectrum consists of a single
band of energies between $-2D$ and $2D$, where $D$ is the spatial
dimensionality of the lattice.
\item A disordered part: Diagonal elements $<\vec{x}|H|\vec{x}>$
have random values chosen according to some probability
distribution.  One must select a probability distribution; a popular choice is the gaussian distribution:
\begin{equation}
{P{(V)}} = {\frac{1}{\sqrt{2 \pi \sigma}} {\exp{(\frac{- V^{2}}{2 \sigma})}}}
\end{equation}
\end{itemize}
These rules completely specify the ensemble of random matrices which is called the Anderson model.  This is the third random matrix ensemble which I have described so far.  

\subsection{The Supersymmetric Sigma Model}
Although random matrix ensembles are very easy to specify, it can be very challenging to compute averages or probabilities of observables in these ensembles.  A number of involved mathematical techniques have been developed for such calculations, including several field theories of the sigma model type. The two most popular sigma models are Efetov's supersymmetric sigma model\cite{Efetov97} and the replica sigma model\cite{Lerner03}.  The supersymmetric sigma model's strong point is its reliability in non-perturbative calculations, which contrasts with fact that the replica sigma model can not be guaranteed to be correct in non-perturbative calculations. The supersymmetric model is called supersymmetric because its degrees of freedom include both Grassman variables and normal bosonic variables.  It does not involve the graded Poincare symmetry which is the starting point of the supersymmetric theory of elementary particles\cite{Srivastava86, Wess92}. 

The key non-perturbative calculations which are the supersymmetric model's claim to fame are all zero-dimensional; they are essentially just calculations of gaussian random matrix ensembles.  One dimensional systems are unusually tractable and can be solved exactly via several techniques, including the supersymmetric technique.  In contrast, systems with two or more dimensions are too complicated and one has to make a perturbative approximation.  This new approximation is called the diffusive approximation, and assumes that the system's conductance is large; i.e. that the disorder is small and electrons are conducted easily.  In many supersymmetric calculations, one employs the diffusive approximation to show that $D$-dimensional highly conducting systems are mathematically equivalent to certain specially chosen zero-dimensional random matrix ensembles.  With this equivalence in place, one is able to calculate observables.  Thus supersymmetric calculations of extended systems (with two or more dimensions) are an interesting mix of the perturbative diffusive approximation and a non-perturbative zero dimensional calculation.
 
One of the biggest difficulties with the supersymmetric sigma model is the complexity of dealing with both Grassman variables and non-Grassman (bosonic) variables.  The variables are combined into graded matrices, i.e. matrices where half of the entries are Grassman variables and the rest are bosonic.  Mathematical manipulation of graded matrices quickly becomes overwhelming, not because of any deep theoretical difficulty, but instead because of the huge number of details that one must keep track of.  

\subsection{A New Sigma Model}
Recently Fyodorov derived a zero-dimensional model without Grassman variables which can be used to reliably do non-perturbative calculations of random matrix ensembles\cite{Fyodorov02}.  The new model is very attractive because it avoids much of the complexity of the supersymmetric model.  This chapter (the one you are reading) presents a novel result: it generalizes Fyodorov's model to disordered systems in $D$ dimensions.  Both for the sake of rigorousness and to assist the newcomer, this chapter derives then new model step by step in a tutorial sort of way.

Both the new sigma model and the supersymmetric sigma model are derived from exactly the same starting theory, so both should make identical predictions about the values and probability distributions of observables.  However, one should always check the equivalence.  Using the new sigma model and working on random matrices in zero dimensions, this chapter contains detailed calculations of the two most basic observables in random matrix theory: the density of states and the two point correlator.  Shortly I will explain what these are, but for now let me just mention that the two point correlator is a non-perturbative result, which shows that the new sigma model equals the supersymmetric sigma model's strength with non-perturbative calculations.  As I mentioned earlier, the extension of the supersymmetric model to extended systems is purely perturbative; after the new model's success with zero dimensional calculations, one has very good reason to expect success with extended systems as well.  I do have plans to demonstrate the equivalence of the extended versions of the two sigma models by going on and calculating the two point correlator in extended systems, but this calculation will have to wait until after this thesis is complete.

\subsection{Outline Of This Chapter}
Before starting the mathematics, here is a rough outline of the steps I will go through.  In section \ref{OriginalProblem} I present the random matrix ensemble of concern and the observables that I'll be calculating.  Then I set up the mathematical formalism, which is a generating function whose derivatives are various observables.  Following the standard supersymmetric approach, I show that the generating function is equal to an integral over both bosonic variables and Grassman variables, and then - before doing the bosonic and fermionic integrals - I integrate over the disorder.  Up to this point I am following exactly the same steps as the derivation of the supersymmetric sigma model, though I consider a somewhat larger class of problems than is usual.  Returning to the usual equations is just a matter of setting certain parameters to have certain values.

After the integration over disorder, one is left with a field theory with both bosonic and Grassman degrees of freedom.  The next step in the supersymmetric method is to show that this field theory is mathematically equivalent to another theory where the degrees of freedom are matrices, one matrix for every point in space.  The new matrices are graded matrices; they contain equal numbers of Grassman variables and bosonic variables.  If the matrices are $N \times N$ matrices with $N^{2}$ degrees of freedom, then $N^{2}/2$ of those degrees of freedom will be Grassman variables, and the other $N^{2}/2$ will be bosonic variables. The reason for this conversion is that - in zero dimensional models - the number of degrees of freedom in the new matrices is much less than the number of degrees of freedom in the previous theory. In other words, the supersymmetric method integrates out almost all of the degrees of freedom and is left with these graded matrices.  This step of integrating out those degrees of freedom is called an Hubbard-Stratonovich transformation.  It is an exact integration without approximations, but the graded matrices makes it very complicated.  In fact, everything about the graded matrices is quite complicated, and the details of supersymmetric calculations are very taxing even when there are no conceptual difficulties.

At this point (section \ref{ConversionToMatrix}) I part ways from the supersymmetric method, though only in the details, not in the philosophy.  Like the supersymmetric method, I do change to a new set of variables which (in the case of zero dimensional models) is much smaller than before.  However the new variables are not supersymmetric matrices.  Whereas the supersymmetric method would arrive at one $N \times N$ graded matrix, I arrive at two $\frac{N}{2} \times \frac{N}{2}$ matrices, each of them having $N^{2}/4$ bosonic variables.  Thus my two matrices, taken together, have the same number of bosonic variables as the graded matrices derived by the supersymmetric method.  But they contain no Grassman variables!  This is the principle advantage of the new sigma model: it avoids all the complications of Grassman variables and graded matrices.  

The conversion to these new matrices happens in two steps.  The first is an Hubbard-Stratonovich tranformation which converts the Grassman variables in the original model into a single $\frac{N}{2} \times \frac{N}{2}$ matrix, which I call the fermionic matrix even though it contains no Grassman variables.  This step is exact, and one arrives at a theory in terms of the fermionic matrix and the original bosonic variables.  

At this point I would like to convert the bosonic variables into a second $\frac{N}{2} \times \frac{N}{2}$ matrix, which I will call the bosonic matrix.  It is possible to do this conversion exactly only in zero-dimensional systems.  In extended systems one cannot perform the conversion without at the same time regularizing the field theory, which means requiring that the bosonic matrix may not vary appreciably between points which are close together.   With the regularization in place, the conversion to bosonic variables becomes an exercise in evaluating an effective Lagrangian. Thus I arrive at the new theory in terms of a fermionic matrix and a bosonic matrix, both composed entirely of bosonic variables.  This is still conceptually the same as the supersymmetric approach; the only differences are that in the supersymmetric approach one ends up with only one matrix, and it is a graded matrix.

At this point one still has to do a complicated matrix integral, with one matrix per point in space.  My next step (in section \ref{NewSigmaModel}) is again conceptually the same as in the supersymmetric approach: I make a saddle point approximation.  Since the degrees of freedom are matrices, the saddle point equations are matrix equations, so I spend some time explaining how to take matrix derivatives.  The saddle point approximation requires me to evaluate the second derivative (the Hessian) of the action.  This is equivalent to a one loop integral, so I spend some time going through a detailed evaluation of the one loop integral.  

It is impossible to calculate the saddle point equations or the second derivative of the action without regularizing the fermionic matrix, which means requiring that the fermionic matrix may not vary appreciably between points which are close together.  The mathematical mechanism for introducing the regularization is called the diffusive approximation.  Because I regularize the bosonic and fermionic matrices separately, I can use the regularizations to introduce a disparity between the two matrices.  However, I can also choose to match the regularizations so that the end result shows a remarkable symmetry between the bosonic and fermionic matrices.  In this way I would match the results of the supersymmetric method, where there is only one matrix and only one regularization, so it is natural (but not required) to regularize all of the matrix elements in an identical fashion.  I would like to point out that the derivation of the new sigma model is no less rigorous than the derivation of the supersymmetric sigma model: in both cases one must regularize the fields, in both cases the details of the regularization procedure can have a large impact on the final result, and in both cases these details are represented in the final result via the introduction of new parameters whose values can not be rigorously derived from the original theory.

After one has regularized the fields and found the saddle point, one finds that the saddle point has a continuous symmetry when certain parameters are set to zero.  Therefore, there are Goldstone modes which have either zero mass or small mass, depending on whether the parameters are zero.  In contrast to the Goldstone modes, there are other degrees of freedom which are always very massive, and the saddle point approximation constrains these degrees of freedom to have their equilibrium values.  From a matrix point of view, this just amounts to requiring that the eigenvalues of the matrices have fixed values.  In this way one arrives a sigma model - "sigma model" just means a field theory with a constraint on the fields.   The supersymmetric method and the new method discussed here do qualitatively the same thing - both find the same sort of saddle point, and derive a sigma model.  However, the details are significantly different: in particular, the mathematics of factoring a graded matrix into massive modes and Goldstone modes is very challenging.  

After deriving the sigma model, one still has to do an integration over the Goldstone modes. Both the supersymmetric model and the new model presented here are able to do the Goldstone integration exactly when treating zero-dimensional systems; this is why these models are reliable for non-perturbative results.   In sections \ref{DOSZero} and \ref{TwoPointZero} I calculate two observables for a zero-dimensional system, the gaussian unitary ensemble.  Although the new model developed here is simpler than the supersymmetric sigma model, there are still a fair number of details to attend to, particularly in the calculation of the two point correlator.  In zero dimensions it turns out to be a bit easier to do the saddle point calculation after integrating out the Goldstone modes, which is the reverse of the order followed by the supersymmetric sigma model.  I hope to redo the calculation in the more normal order, but that will be done after this thesis.  I also have plans to calculate an observable in an extended system, but this too is being postponed.
  
\section{\label{OriginalProblem}The Original Problem}
\subsection{The Ensemble}
The Hamiltonians in the ensemble which I calculate in this chapter live in a lattice with $V$ different lattice points, or sites.  The number of dimensions and other structural details of the lattice are modeled by the kinetic energy matrix (operator) $K$, which I will not specify at this time except to say that it is not a random variable.  In a zero-dimensional system $V = 1$ and $K = 0$.  There are $N$ different basis elements at each site, so the total basis size is is $N \times V$. I use the lower case letters $n, v$ to denote indices in the basis.  The $v$ index would normally be written as $\vec{x}$ or $\vec{p}$, but my notation saves space.  The kinetic energy matrix $K$ is diagonal in the index $n$ and independent of $n$; I write this statement mathematically as $K = {{K{(v_{1} v_{2})}} {\delta{(n_{1} n_{2})}}}$.  In my notation the arguments in parenthesis specify a matrix or vector's indices, and there is an implied sum whenever two matrices or vectors share an index.  Throughout this chapter I will use the words "operator" and "matrix" interchangeably.

The random matrices themselves are a sum $H + K$ of the kinetic operator $K$ with a random potential $H$.  $H$ is diagonal in the spatial index; i.e. it is a "local" potential.  However, it does vary from site to site.  Mathematically, this is written as $H = {{H{(n_{1} n_{2} v_{1})}} {\delta{(v_{1} v_{2})}}}$.  

 I require that $H$ be hermitian.  Mathematically the hermitian case is the simplest, but the model presented here should be easy to generalize to other ensembles.  I choose the probability distribution:
\begin{equation}
{dP{(H)}} = {{dH} \times {(N / 2 \pi {\xi}^{2})}^{NV/2} {2}^{N{(N-1)V}/2} { { e^{-\frac{N}{2 {\xi}^{2}} {{Tr}(H^{2})}} }} {{\det}^{F}(H+K)}}
\label{MyProbabilityDistribution}
\end{equation}
In this and all other uses of a matrix measure I use the following normalization convention: ${dH} \equiv {{({\prod}_{l} {d{H}_{ll}})}{({\prod}_{l<m}{d{H}_{lm}^{R}}{d{H}_{lm}^{I})}}}$. 
$\xi$ has units of energy and sets the scale of the disorder. 

$F$ is the power of the determinant, and should be an even number to make the probability distribution $dP$ be positive definite.  Physically a model with non-zero $F$ corresponds to letting the disordered potential interact with $F$ species of fermions; integrating out the fermions gives these determinants.  I added these fermions to the ensemble in order to create a toy model of QCD, where one does indeed integrate out fermions and does obtain a probability distribution which is weighted by a determinant.  

Late in this chapter I will set $F = 0$ and thus obtain a standard sort of disorder.  This standard disorder is spatially uncorrelated; i.e. its value at one point is not correlated with its value at any other point.   It is also Gaussian, which implies that its higher moments can be expressed as products of its second moment a la Wick's theorem.   It is easy to calculate the second moment of the $F = 0$ potential using formula \ref{MyProbabilityDistribution}; the result is:
\begin{equation} {\overline{{H{(n_{1} n_{2} v_{1} v_2})} {H{(n_{3} n_{4} v_{3} v_4})} }} = {\frac{{\xi}^{2}}{N} {\delta{(n_{1} n_{4})}} {\delta{(n_{2} n_{3})}} {\delta{(v_{1} v_{2} v_{3} v_{4})}} }
\label{TheGaussianPotentialSecondMoment}
\end{equation}  

When $F=0$ and the number $N$ of degrees of freedom per site is equal to one, a suitable choice of kinetic energy operator $K$ turns this ensemble into the Anderson model, which is the usual starting point of the supersymmetric sigma model.   On the other hand, when $F = 0$, $V = 1$, and, of course, $K = 0$, this ensemble turns into the gaussian unitary ensemble described earlier. 

This completes the ensemble's specification.

\subsection{The Observables.  Their Relation to Averages of Green's Functions.}
In this chapter I calculate two observables: the density of states and the two point correlator.  The simplest of the two is the density of states $\rho{(E)}$. In a zero dimensional system ($V = 1$), the density of states is defined as the number of eigenvalues per unit energy.  It is a function of energy $E$ because one will find more eigenvalues in some parts of the spectrum than in others.  Given a Hamiltonian $h = {H + K}$, the zero dimensional density of states has the mathematical definition $ {\rho{(E)}} \equiv {{Tr}{(\delta{(E - h)})}} $.

In a spatially extended system ($V \neq 1$) one must generalize these formulas a little bit.  Define the local trace ${Tr}_{v}$ as a trace over all the basis elements which share the same spatial coordinate $v$.  Thus the usual trace is a sum of local traces; ${Tr} = {{\sum}_{v} {Tr}_{v}}$.  With this formalism in place, I can define the local density of states:  $ \rho{(E, v)} \equiv {{Tr}_{v}{(\delta{(E - h)})}} $.  The global density of states is just a sum over the local density of states; $\rho{(E)} = {{\sum}_{v} \rho{(E, v)}}$.  

For the rest of this section I will discuss only the global density of states.  In unlocalized systems, this quantity is sufficient for calculating observables based on the eigenvalues.  If, however, spatial information is desired, one needs to use the local density of states.  All of the formulas in this section are given in terms of the global density of states.  However all of them, with the exception of the global unitary ensemble results in equations \ref{DOSZeroCorrectResult} and \ref{TwoPointZeroCorrectResult}, can be easily converted to the local density of states by simply replacing the usual trace ${Tr}$ with the local trace ${Tr}_{v}$.

I will be calculating the average of the density of states, averaging over the ensemble of random Hamiltonians $h = {H + K}$:
   \begin{equation}
 \overline{\rho{(E)}} \equiv {{Tr}{(\overline{\delta{(E - h)}})}}
 \label{DOSDefinition}
 \end{equation}
 The average density of states in the gaussian unitary ensemble is well known;
  \begin{equation}
 \overline{\rho{(E)}} = {\frac{N}{\pi \xi} \sqrt{1 - \frac{E^{2}}{4{\xi}^{2}}} }
 \label{DOSZeroCorrectResult}
 \end{equation}
 Section \ref{DOSZero} gives a detailed derivation of this result.
 
 I will also calculate the two point correlator, which is defined as:
 \begin{equation}
 {R_{2}{(E_{1}, E_{2})}} \equiv {\frac{\overline{{{Tr}({\delta{(E_{1} - h)}})} \, {{Tr}({\delta{(E_{2} - h)}})} }}{\overline{\rho{(E_{1})}} \quad  \overline{\rho{(E_{2})}}}}
 \label{TwoPointDefinition}
 \end{equation}
 Note immediately that the numerator is the average of the multiple of the two delta functions, not a multiple of their separate averages.  $R_{2}$ is the precisely the former divided by the latter.  Its qualitative meaning is best understood by analyzing its dependence on the energy difference $\omega \equiv {E_{1} - E_{2}}$.  If eigenvalues are most often separated by an energy ${\omega}_{1}$, then $R_{2}$ will have a peak at $\omega =  {\omega}_{1}$.   If, on the other hand, eigenvalues are never separated by an energy ${\omega}_{2}$, then $R_{2}{({\omega}_{2})}$ will be exactly zero.  In other words, the two point correlator describes how eigenvalues prefer to be spaced relative to each other.  In the gaussian random matrix ensemble, when $\omega $ is small the two point correlator is given by the expression:
 \begin{equation}
  {R_{2}{(E_{1}, E_{2})}} = {{\delta{(\omega \overline{\rho})}} + {1} - {(\frac{\sin{(\pi \omega \overline{\rho})}}{\pi \omega \overline{\rho}})}^{2}}
 \nonumber 
 \label{TwoPointZeroCorrectResult}
 \end{equation}
 This result, minus the delta function, will be derived in section \ref{TwoPointZero}.  $\overline{\rho}$ is the density of states at $\overline{E} \equiv {E_{1} + E_{2}}$. The missing delta function is not physically significant; it just represents the fact that  $\rho{(E)}$ is perfectly correlated with $\rho{(\acute{E})}$ when $E = \acute{E}$.
 
 The field theory approach to random matrix problems starts by reformulating observables in terms of Green's functions.  The Green's function of a matrix $h$ is defined by:
 \begin{equation}
 {G{(E)}} \equiv {(E - h)}^{-1}
 \end{equation}
 This matrix function has poles at energies $E$ equal to the eigenvalues of $h$.  The poles encode in a mathematically convenient way a precise description of $h$'s eigenvalues and eigenvectors, which is one of the biggest reasons why Green's functions are so popular with physicists. In order to set Green's functions on a firm mathematical footing, one must add an infinitesimally small imaginary part $\imath \epsilon$ to the energy.  This moves the poles slightly off the real axis.  If the imaginary part is negative, the poles move into the positive imaginary half plane.  The corresponding Green's function ${G_{A}{(E)}} \equiv {(E - {\imath \epsilon }- h)}^{-1}$ is called the advanced Green's function.  If instead the imaginary part is positive, then the poles move into the negative imaginary half plane and one has the retarded Green's function ${G_{R}{(E)}} \equiv {(E + {\imath \epsilon }- h)}^{-1}$.  Clearly the advanced and retarded Green's functions are related by complex conjugation: ${G_{A}{(E)}} = {G_{R}^{*}{(E)}}$.

 The density of states is related to the Green's functions via the following equation:
\begin{equation}  
{\rho{(E)}} = {\frac{1}{2 \pi} \lim_{\epsilon \rightarrow 0} {{Im}{({Tr}{(G_{A}{(E)})})}}}
\label{DOSGreens}
\end{equation}
One proves equation \ref{DOSGreens} by changing to the basis which diagonalizes $h$ and then applying the identity ${\delta{(E - \acute{E})}}  = {\frac{1}{2 \pi} \lim_{\epsilon \rightarrow 0} {{Im}{(E - {\imath \epsilon} - \acute{E})}^{-1}}}$.  This parameterization of the $\delta$ function can be verified by checking that for every function $f{(E)}$ the following equation holds: 
\begin{equation}
{\int_{-\infty}^{\infty} { f{(E)} {\frac{1}{2 \pi} \lim_{\epsilon \rightarrow 0} {{Im}({(E - {\imath \epsilon} - \acute{E})}^{-1})}}}} = {f{(\acute{E})}}
\end{equation}
The check also involves noticing that the integrand is nonzero only at $E = \acute{E}$.

Following equation \ref{DOSGreens}, I will calculate the average level density by calculating the average of the trace of the advanced Green's function, taking its imaginary part, and then dividing by $2 \pi$. A similar strategy can be used to calculate the two point correlator, which can be rewritten in terms of products of two Green's functions:
\begin{eqnarray}  
{{\rho{(E)}}{\rho{(\acute{E})}}} & = & {\frac{1}{4 {\pi}^{2}} \lim_{\epsilon \rightarrow 0} {{Im}{({Tr}{(G_{A}{(E)})})}} {{Im}{({Tr}{(G_{A}{(\acute{E})})})}}}
\nonumber \\
& = & {-\frac{1}{16 {\pi}^{2}} \lim_{\epsilon \rightarrow 0} [ {({{Tr}{(G_{A}{(E)}})} - {{Tr}{(G_{R}{(E)}})})} \times {{({{Tr}{(G_{A}{(\acute{E})})}}  - {{Tr}{(G_{R}{(\acute{E})})}})}} ]}
\nonumber \\
& = & -\frac{1}{16 {\pi}^{2}} \lim_{\epsilon \rightarrow 0}  [{{{Tr}{(G_{A}{(E)}})}{{Tr}{(G_{A}{(\acute{E})})}}} 
+ {{{Tr}{(G_{R}{(E)}})}{{Tr}{(G_{R}{(\acute{E})})}}}
\nonumber \\ 
&  & \qquad\qquad - {{{Tr}{(G_{A}{(E)}})}{{Tr}{(G_{R}{(\acute{E})})}}} 
- {{{Tr}{(G_{R}{(E)}})}{{Tr}{(G_{A}{(\acute{E})})}}}]
\nonumber \\
& = & \frac{1}{8 {\pi}^{2}} \lim_{\epsilon \rightarrow 0} [ {{Re}({{{Tr}{(G_{A}{(E)}})}{{Tr}{(G_{R}{(\acute{E})})}}} )}
\nonumber \\
&  & \qquad\qquad - {{Re}({{{Tr}{(G_{A}{(E)}})}{{Tr}{(G_{A}{(\acute{E})})}}})} ]
\label{TwoLevelZeroGreensA}
\end{eqnarray}

I define the advanced-advanced two point correlator \linebreak  $R_{AA} \equiv \lim_{\epsilon \rightarrow 0} \overline{{{Tr}{(G_{A}{(E)}})}{{Tr}{(G_{A}{(\acute{E})})}}}$ and the advanced-retarded two point correlator $R_{AR} \equiv \lim_{\epsilon \rightarrow 0} \overline{{{Tr}{(G_{A}{(E)}})}{{Tr}{(G_{R}{(\acute{E})})}}}$.  I will calculate $R_{AR}$ in section \ref{ARCorr} and $R_{AA}$ in section \ref{AACorr}.  With these results in hand, the following formula gives the two point correlator:
\begin{equation}
{R_{2}{(E_{1}, E_{2})}} = {\frac{{{Re}{(R_{AR})}} - {{Re}{(R_{AA})}}}{8 {\pi}^{2} \quad \overline{ \rho{(E_{1})}} \quad  \overline{\rho{(E_{2})}}}}
\label{TwoLevelZeroGreensB}
\end{equation}

\subsection{The Generating Function}
I will be calculating averages of an advanced Green's function, of the multiple of two advanced Green's functions, and of the multiple of an advanced and a retarded Green's function.   When calculating other observables, one could end up averaging the multiple of any number of advanced and retarded Green's functions.   One powerful way of calculating multiples of Green's functions is to calculate instead a generating function whose derivatives are the desired multiples of Green's functions.  The nicest advantages of this approach are that it allows you to avoid repeating certain steps for every observable one calculates, and also that many sorts of mistakes are easier to detect.  In this section I introduce a generating function which is suitable for calculating the observables that I'm interested in.  This generating function is the mathematical starting point of my calculations; in fact most of this chapter will be spent on various manipulations and approximations of the generating function, and only at the end will I take derivatives and obtain the observables. 

First let me present a generating function which is sufficient for generating a single Green's function:
\begin{equation}
{Z{(E, \overline{E}, J)}} = {\frac{{{det}{(E - H - K)}}}{{det}{(\overline{E} - H - K - J)}}}
\label{SimpleGeneratingFunction}
\end{equation}
$J$ is a source matrix ${J} = {{J}{(n_{1} n_{2} v_{1} v_{2})}  }$.  Note that when the source $J$ is set to zero and the two energies are set to be equal, the generating function ${Z{({E = \overline{E}}, {J = 0})}}$ is equal to one.  Using the identity relating the determinant of a matrix to its logarithm ${{det}{(A)}} = {{e}^{{Tr}{(\ln{(A)})}}}$, one can easily prove that: 
\begin{equation}
{\frac{dZ}{dJ{(n_{1} n_{2} v_{1} v_{2})}}} = {Z \times { {{\langle n_{2} v_{2} |} {(\overline{E} - H - K - J)}^{-1} {| n_{1} v_{1} \rangle} }}}                            \end{equation}
(If you are not familiar with how to take this sort of derivative, skip ahead for a moment to section \ref{MatrixDerivative} on matrix derivatives, and then come back.)  Therefore, one can calculate the Green's function by first taking the derivative with respect $J$ and then setting $J = 0$ and $E = \overline{E}$: 
\begin{equation}
{[\frac{dZ}{dJ{(n_{1} n_{2} v_{1} v_{2})}}]_{{J = 0}, {E = \overline{E}}}} = { {{\langle n_{2} v_{2} |} { {(E - H - K)}^{-1} } {| n_{1} v_{1} \rangle}}} = {{\langle n_{2} v_{2} |}{G{(E)}} {| n_{1} v_{1} \rangle}}
\end{equation}

As I saw in the last section, I typically want the local trace of the Green's function, which can be obtained by making the source $J$  diagonal in indices $n$ and $v$ and independent of the index $n$; ${J} = {{J}{(v)} {\delta{(n_{1} n_{2})}}  {\delta{(v_{1} v_{2})}} }$.  With this choice of source, one has:
\begin{equation}
{[\frac{dZ}{{dJ{(v)}}}]_{{J = 0}, {E = \overline{E}}}} =  {{ {Tr}_{v}{( G{(E)} )}}}
\end{equation}
This is exactly the result that is needed.

How can this approach be generalized to obtain the product of two traces of Green's functions?  Taking a second derivative of the generating function in equation \ref{SimpleGeneratingFunction} results in:
\begin{eqnarray}
{[\frac{d^{2}Z}{{dJ{(v_{1})}}{dJ{(v_{2})}}}]_{{J = 0}, {E = \overline{E}}}} & = & {{ {Tr}_{v_{1}}{( G{(E)} )}} { {Tr}_{v_{2}}{( G{(E)} )}}} 
\nonumber \\
& + & {{ {{\langle n_{2} v_{2} |}{G{(E)}} {| n_{1} v_{1} \rangle}} {{\langle n_{1} v_{1} |}{G{(E)}} {| n_{2} v_{2} \rangle}}}}
\label{BadSecondDerivative}
\end{eqnarray}
Note that the indices $v_{1}$ and $v_{2}$ are not summed over although they are repeated, while $n_{1}$ and $n_{2}$ are summed over. The first term in equation \ref{BadSecondDerivative} is almost what we want, except that it would be better to have two different Green's functions with two different energies.  The second term is not wanted at all.

A better approach is to use the following generating function:
\begin{equation}
{Z{(E_{1}, J_{1}, E_{2}, J_{2})}} = {\frac{{{det}{(E_{1} - H - K)}} \; {{det}{(E_{2} - H - K)}}}{ {{det}{({\overline{E}}_{1} - H - K - J_{1})}} \; {{det}{({\overline{E}}_{2} - H - K - J_{2})}} }}
\label{SecondGeneratingFunction}
\end{equation}
With this generating function, one obtains the desired result:
\begin{equation}
{[\frac{d^{2}Z}{{dJ_{1}{(v_{1})}}{dJ_{2}{(v_{2})}}}]_{{J_{1} = J_{2} = 0}, {E_{1} = {\overline{E}}_{1}}, {E_{2} = {\overline{E}}_{2}}}} =  {{ {Tr}_{v_{1}}{( G{(E_{1})} )}} \times { {Tr}_{v_{2}}{( G{(E_{2})} )}}}
\end{equation}
Proceeding along the same lines, it is obvious that in order to produce the multiple of $X$ Green's functions one must use a generating function with $X$ determinants in the numerator and $X$ determinants in the denominator.

At this point I would like to remind the reader of the probability distribution $dP{(H)}$ of the random matrix ensemble that I'm interested in calculating, which was given in equation \ref{MyProbabilityDistribution}.  It included a factor of ${\det}^{F}{(H + K)}$, which represented the addition of $F$ species of fermions to the usual gaussian ensemble.  I will be calculating the average $\overline{Z} \equiv {\int {dP{(H)}} Z}$ of the generating function. I would like to point out that the extra determinants in $dP{(H)}$ can be moved into the generating function $Z$.  This doesn't complicate the generating function much, since it was already composed of determinants.  However it simplifies $dP{(H)}$ a lot, reducing it to a gaussian function. 

To repeat, I am saying that averaging a generating function with $X$ determinants in the numerator and $X$ determinants in the denominator in an ensemble with $F$ extra fermions is equivalent to averaging a generating function with $X + F$ determinants in the numerator and $X$ determinants in the denominator in a gaussian ensemble.  Therefore I am interested in calculating a generating function with $I^{f} = {X + F}$ determinants in the numerator and $I^{b} = X$ determinants in the denominator.

Instead of keeping track of all those determinants separately, it is convenient to combine them into one determinant in the numerator and one in the denominator.  Here is an example of the basic concept:  Suppose that I have two matrices $A$ and $B$ which each inhabit a basis with $M$ basis elements, and I want to combine the determinants $\det{(A)}$ and $\det{(B)}$ into one determinant.  I therefore invent a new basis with $2M$ basis elements, and then assert that $M$ of the new basis elements correspond exactly with $A$'s old basis and that the other $M$ new basis elements correspond exactly with $B$'s old basis elements.  In the new basis $AB = BA = 0$ and therefore ${{\det{(A)}} \times {\det{(B)}}} = {\det{(A+B)}}$.  This is sleight of hand: I have gotten rid of the complication of having two determinants at the expense of using a basis which is twice as large. 

I now apply this trick to the case of $I^{f}$ determinants in the numerator and $I^{b}$ determinants in the denominator, with each determinant inhabiting a basis with $N \times V$ basis elements.  I create two new bases. For the numerator I create a basis with $N \times V \times I^{f}$ basis elements; there are $N \times I^{f}$ basis elements living at each site, and $V$ sites.  I use the indices $n v i$ to denote the numerator's basis elements.  For the denominator I create a basis with $N \times V \times I^{b}$ basis elements, and denote them with the indices $n v j$. I choose to use the index $j$ for the denominator and the index $i$ for the numerator in a deliberate effort to contrast the two bases.

I now adjust my matrices to match the new bases.  All the new matrices must be diagonal in the indices $i$ and $j$.  The kinetic energy $K$ and the random potential $H$ must be independent of $i$ and $j$; the kinetic matrix in the numerator is $K = {{K{(v_{1} v_{2})}} {\delta{(n_{1} n_{2})}} {\delta{(i_{1} i_{1})}}}$ and the potential in the numerator is $K = {{K{(n_{1} n_{2} v)}} {\delta{(v_{1} v_{2})}} {\delta{(i_{1} i_{1})}}}$. The kinetic matrix and the potential in the denominator have exactly the same form, but with $j$ substituted for $i$.  The source $J$ occurs only in the denominator, and has the structure ${J} = {{{J}{(v i)}} {\delta{(n_{1} n_{2})}} {\delta{(v_{1} v_{2})}} {\delta{(i_{1} i_{2})}}}$

There were $I^{f}$ determinants in the numerator, each with with a different energy.  (Of course, $F$ of the energies will be set to zero.)  I denote the $I^{f}$ energies in the numerator with the notation
${\lbrace E^{f}_{i} \rbrace} = {\lbrace E^{f}_{1}, ... , E^{f}_{I^{f}} \rbrace}$.
Likewise there are $I^{b}$ energies associated with the denominator, which I denote with ${\lbrace E^{b}_{j} \rbrace}  = {\lbrace E^{b}_{1}, ... , E^{b}_{I^{b}} \rbrace} $.  These energies then define two energy matrices:  ${{{\hat{E}}^{f}}} \equiv {E^{f}_{i} \delta{(n_{1} n_{2})} \delta{(v_{1} v_{2})} \delta{(i_{1} i_{2})}}$, and ${{{\hat{E}}^{b}}} \equiv {E^{b}_{j} {\delta{(n_{1} n_{2})}} {\delta{(v_{1} v_{2})}} {\delta{(j_{1} j_{2})}} }$.  
 
Now that I have defined all the matrices, I can write the generating function:
\begin{equation}
{Z{({{\hat{E}}^{f}}, {{\hat{E}}^{b}}, J)}} = {\frac{{{det}{({{\hat{E}}^{f}} - H - K)}}}{{det}{({{\hat{E}}^{b}} - H - K - J)}}}
\end{equation}

This is the most general generating function which I will consider, and I will spend the rest of this chapter evaluating its average and derivatives.  It can generate multiples of up to $I^{b}$ different Green's functions.  After taking the derivatives, one sets $I^{b}$ of the $I^{f}$ energies $E^{f}_{i}$ in the numerator to be equal to the energies $E^{b}_{j}$ in the denominator.  The remaining energies (if any) in the numerator are set to zero.
                                             
\subsection{The Supersymmetric Formula for the Generating Function}
There are several ways of converting the generating function $Z$ into a field theory.  Here I concentrate on the supersymmetric method, which converts the determinant in the numerator into an integral over Grassman variables, and the determinant in the denominator into an integral over bosonic variables. I now briefly state the two identities used to perform this conversion.  If $A$ is an $M \times M$ matrix and $\psi, \overline{\psi}$ are vectors of $M$ Grassman variables, then:
\begin{equation}
{ {{det}{(A)}}} = {{(-2\imath)}^{M}{\int{{{d\overline{\psi}}{d{\psi}}} e^{\frac{\imath}{2} \psi A \overline{\psi}}}}}
\label{FermionDeterminant}
\end{equation}
I am using the convention ${{d\overline{\psi}}{d{\psi}}} \equiv {\prod_{m}^{M} {\overline{\psi}}_{m} {\psi}_{m} }.$ One can find a similar result with vectors $S$ composed of $M$ complex bosonic variables ($S = {S^{R} + \imath S^{M}}$):
\begin{equation}
{{ {{det}^{-1}{(A)}}}} = {{(2 \pi \imath)}^{M}{\int{{{dS^{R}}{dS^{I}}}} e^{\frac{\imath}{2} S A {S^{*}}}}}
\label{BosonDeterminant}
\end{equation}
However the imaginary part of $A$ must be positive definite in order for the bosonic integral to converge.  This seemingly easy point is actually quite deep and the cause of many of the interesting mathematical intricacies in the supersymmetric model.

Using these identities, I rewrite the generating function in supersymmetric form:
\begin{eqnarray}
{Z} & = & {(2 \pi)}^{-V N I^{b}}  {(\imath / 2)}^{-V N I^{f}} {(\imath)}^{- N V {Tr}{(L)}} 
\nonumber \\
& \times & {\int {d\overline{\psi}}{d{\psi}} {{dS^{R}}{dS^{I}}}  e^{\frac{\imath}{2} S L {({{\hat{E}}^{b}} - H - K - J)} {S^{*}}} e^{\frac{\imath}{2} \psi {({{\hat{E}}^{f}} - H - K)} \overline{\psi}} }
\label{SUSYGeneratingFunction}
\end{eqnarray}
 $\psi$ and $\overline{\psi}$ are vectors composed of $N \times V \times I^{f} $ Grassman variables, while $S$ is a vector composed of $N \times V \times I^{b}$ complex bosonic variables. Examination of the exponents reveals that $\psi$ and $S$ must have dimensions of $\sqrt{[ Energy ]}$.  The differentials cause the generating function $Z$ to have dimensions of ${[Energy]}^{N V {(I^{b} - I^{f})}}$, but this can be ignored without ill effect.  
 
 $L$ is the diagonal matrix with all diagonal elements determined by the sign of the imaginary parts of the bosonic energies; 
 \begin{equation}
 {L{(n_{1} n_{2} v_{1} v_{2} i_{1} i_{2})}} \equiv {{{sign}{(Im{({E}^{b}_{i})})}} {\delta{(n_{1} n_{2})}} {\delta{(v_{1} v_{2})}} {\delta{(j_{1} j_{2})}}}
 \label{LDefinition}
 \end{equation}
  One has to introduce this matrix $L$ in order to insure the convergence of the bosonic integrals.  I am assuming, as is usual, that the operators $H, K$, and $J$ are real.

 One could also introduce a sign matrix similar to $L$ in the fermionic integrals, and their guaranteed convergence allows one to choose any combination of signs one likes.  Verbaarschot et al\cite{Verbaarschot85} explored this freedom in the context of the supersymmetric sigma model, and discovered that in that context one must choose the signs to all be the same, and thus obtain a compact representation for the fermionic variables.  In the method I am presenting here, one can choose any signs one likes, but pretty soon (just before equation \ref{FermionlessGamma}) this fermionic sign matrix factors out entirely, and one is again forced to use a compact representation.

\subsection{Averaging Over the Disorder}
I want to calculate the average of products of Green's functions, so the next step is to average the generating function $Z$ over realizations of the random potential, averaging according to the probability distribution specified in equation \ref{MyProbabilityDistribution}.  This would be difficult if I had not moved the fermion factor ${{\det}^{F}{(H + K)}}$ into the generating function.  But with the move accomplished, I just have to average over a gaussian distribution, which is easy enough.  One expands formula \ref{SUSYGeneratingFunction} as a power series in the random potential $H$, counts pairings of $H$, and then substitutes the second moment specified in equation \ref{TheGaussianPotentialSecondMoment} for each pairing.  The term to be averaged is: 
\begin{equation}
{\overline{e^{\frac{-i}{2} {({SLHS^{*}} + {\psi H \overline{\psi}})}}}} = {\sum_{l=0}^{\infty}{\frac{1}{l!} {(\frac{-\imath}{2})}^{l} \overline{{({SLHS^{*}} + {\psi H \overline{\psi}})}^{l}}}}
\end{equation}
Remembering that odd moments of $H$ average to $0$ and that even moments average to the number of pairs $N_{p}$ times the second moment to some power, one obtains:
\begin{equation}{\sum_{l=0}^{\infty}{\frac{N_{p}}{(2l)!} {{(\frac{-\imath}{2})}^{2l} { \brace{ \overline{{({SLHS^{*}} + {\psi H \overline{\psi}})}^{2}} }}^{l}}}}
\end{equation}

${N_{p}{(2a)}}$ is the number of distinct ways of pairing $2a$ objects.  It can be calculated as follows:  
\begin{eqnarray}
{N_{p}{(2a)}} & = & {{[{(\frac{d}{dJ})}^{2a} e^{J^{2}/2} ]}_{J=0}}
\nonumber \\
& = & {{(2 \pi)}^{-1/2} {[{(\frac{d}{dJ})}^{2a} {\int {d\alpha} e^{{-\frac{1}{2}{\alpha}^{2}} + {\alpha J}}} ]}_{J=0}}
\nonumber \\
& = & {{(2 \pi)}^{-1/2} {\int {d\alpha} {\alpha}^{2a} e^{-\frac{1}{2}{\alpha}^{2}}}}
= {(2a - 1)!}
\end{eqnarray}
(Thanks to John Keating for helping me with this.)

Resumming the series, one finds:
\begin{equation}
e^{-\frac{1}{8} \overline{{({SLHS^{*}} + {\psi H \overline{\psi}})}^{2}}}
\end{equation}

Turning now to evaluating the second moment, I have to evaluate three averages.  The first is:
 \begin{eqnarray}
\overline{{({SLHS^{*}})}^{2}} & = & {\frac{{\xi}^{2}}{N}{S{(n_{1} v_{1} j_{1})}} {L{(j_{1})}} {S^{*}{(n_{2} v_{1} j_{1})}} {S{(n_{2} v_{1} j_{2})}} {L{(j_{2})}} {S^{*}{(n_{1} v_{1} j_{2})}}}
\nonumber \\
& = & {\frac{{\xi}^{2}}{N} {{Tr}{(\hat{S}L\hat{S}L)}}}
\end{eqnarray}
I define the new quantity $\hat{S}$ as:
 \begin{equation}
 {\hat{S}{(v_{1} v_{2} j_{1} j_{2})}} \equiv {{S^{*}{(n_{1} v_{1} j_{1})}} {S{(n_{1} v_{1} j_{2})}} {\delta{(v_{1} v_{2})}}}
 \label{SHatDefinition}
 \end{equation}
 Remember that summations are implied wherever an index is repeated.  Note that $\hat{S}$ has no dependence on $n$ and that the trace does not sum over $n$.

The second average is:
\begin{equation}
\overline{{({\psi H \overline{\psi}})}^{2}}
= {\frac{{\xi}^{2}}{N}{\psi{(n_{1} v_{1} i_{1})}} {\overline{\psi}{(n_{2} v_{1} i_{1})}} {\psi{(n_{2} v_{1} i_{2})}} {\overline{\psi}{(n_{1} v_{1} i_{2})}}}
={- \frac{{\xi}^{2}}{N} {{Tr}{(\hat{\psi}\hat{\psi})}}}
\end{equation}
where $\hat{\psi}$ is defined by ${\hat{\psi}{(v_{1} v_{2} i_{1} i_{2})}} \equiv {{\overline{\psi}{(n_{1} v_{1} i_{1})}} {\psi{(n_{1} v_{1} i_{2})}} {\delta{(v_{1} v_{2})}}}$.  The minus sign will be important in subsequent calculations and was caused by the anticommutation of fermions.

The third average is:
\begin{equation}
\overline{2 {SLHS^{*}}  {\psi H \overline{\psi}}} 
= {\frac{2 {\xi}^{2}}{N} {\psi{(n_{1} v_{1} i_{1})}} {\overline{\psi}{(n_{2} v_{1} i_{1})}} {S{(n_{2} v_{1} j_{1})}} {S^{*}{(n_{1} v_{1} j_{1})}}}
\equiv {\frac{2 {\xi}^{2}}{N} X}
\end{equation}

Combining these results, I find that the averaged generating function is:
\begin{equation}
{\bar{Z}} = {\gamma {\int {d\overline{\psi}}{d{\psi}} {{dS^{R}}{dS^{I}}}{\exp}{(\mathcal{L})}}}
\end{equation}
where the newly introduced Lagrangian $\mathcal{L}$  and prefactor $\gamma$ are:
\begin{eqnarray}
{\mathcal{L}} & \equiv & {\frac{\imath}{2} S L {({{\hat{E}}^{b}} - K - J)} {S^{*}}} + {\frac{\imath}{2} \psi {({{\hat{E}}^{f}} - K)} \overline{\psi}}  
\nonumber \\
& - & {\frac{{\xi}^{2}}{8N}{(-{{Tr}{(\hat{\psi}\hat{\psi})} } + {{Tr}{(\hat{S}L\hat{S}L)}} + 2X )}}
\nonumber \\
\gamma & \equiv & {{(2 \pi)}^{-V N I^{b}}  {(\imath  / 2)}^{-V N I^{f}} {(\imath)}^{- N V {Tr}{(L)}} } 
\end{eqnarray}

This Lagrangian and prefactor are the starting point of the supersymmetric method.  

\section{\label{ConversionToMatrix} Conversion of the Theory to Matrix Variables}
The main content of the supersymmetric method consists in first converting from the theory in terms of vector variables $S$ and $\psi$ to a theory  in terms of graded matrices, and then performing various approximations in order to arrive at an integral which can be evaluated.  Traditionally the process begins by transforming $S$ and $\psi$ to graded matrices.  I will now begin to diverge from the traditional path, and transform $S$ and $\psi$ separately, obtaining a theory in terms of matrices which contain only bosonic variables.  However, I will still be following the overall strategy of converting to matrix variables and then performing the same approximations that are used by the supersymmetric method.

\subsection{Hubbard-Stratonovich Conversion of the Fermionic Variables}
First, I do an exact transformation of the fermionic vector $\psi$ into a bosonic $I^{f} \times I^{f}$ matrix $Q^{f}$.  This step is called an Hubbard-Stratonovich transformation because it involves using a gaussian integral to rewrite a quartic term as a quadratic term coupled to a new field variable.  Note that if $Q$ is an $I^{f} \times I^{f}$ matrix, 
\begin{equation}
{(N / 2 \pi )}^{{I^{f}}^{2} / 2} 2^{I^{f} {(I^{f} -1)} / 2} {\int {{dQ} e^{-\frac{N}{2} {{Tr}({Q}^{2})}} }}
= 1 
\label{GaussianMatrixNormalization}
\end{equation}

Using this identity and completing the square, one obtains:
\begin{eqnarray}
{{\exp}{(\frac{{\xi}^{2}}{8N} {{Tr}{(\hat{\psi}\hat{\psi})}})}} & = & {(N / 2 \pi)}^{{I^{f}}^{2} / 2} 2^{I^{f} {(I^{f} -1)} / 2} 
\nonumber \\
& \times & {\int {{dQ} {{\exp}{(-\frac{ N}{2} {{Tr}({Q^{f}}^{2}) + {{\frac{\xi }{2}}{{Tr}{(Q^{f}\hat{\psi})}}}})}} }} 
\end{eqnarray}

I apply this transformation at each point in the spatial lattice, with a different $Q^{f}$ at each point.  Thus I alter the equations for the generating function:
\begin{eqnarray}
{\bar{Z}} & = & {\gamma {\int {dQ^{f}} {d\overline{\psi}}{d{\psi}} {{dS^{R}}{dS^{I}}}{\exp}{(\mathcal{L})}}}
\nonumber \\
{\mathcal{L}} & = & {\frac{\imath}{2} S L {({{\hat{E}}^{b}} - K - J)} {S^{*}}} + {\frac{\imath}{2} \psi {({{\hat{E}}^{f}} - K)} \overline{\psi}} 
\nonumber \\
& - & {\frac{{\xi}^{2}}{8N}{({{Tr}{(\hat{S}L\hat{S}L)}} + 2X )}} -\frac{ N}{2} {{Tr}({Q^{f}}^{2}) + {{\frac{\xi }{2}}{{Tr}{(Q^{f}\hat{\psi})}}}}
\nonumber \\
\gamma & = & {{(2 \pi)}^{-V N I^{b}}  {(\imath  / 2)}^{-V N I^{f}} {(\imath)}^{- N V {Tr}{(L)}}  {(N / 2 \pi)}^{{I^{f}}^{2}V / 2} 2^{I^{f} {(I^{f} -1)} V / 2} }
\end{eqnarray}

The new Lagrangian is linear in $\psi$ and $\overline{\psi}$, so I can use equation \ref{FermionDeterminant} to integrate out these variables.  I get:
\begin{eqnarray}
{\bar{Z}} & = &{\gamma {\int {dQ^{f}} {{dS^{R}}{dS^{I}}}{\exp}{(\mathcal{L})}}}
\nonumber \\
{\mathcal{L}} & = & {\frac{\imath}{2} S L {({{\hat{E}}^{b}} - K - J)} {S^{*}}} - {\frac{{\xi}^{2}}{8N}{{{Tr}{(\hat{S}L\hat{S}L)}}}} 
\nonumber \\
& - & \frac{ N}{2} {{Tr}({Q^{f}}^{2}) }
+ {{Tr}{(\ln A)}}
\nonumber \\
\gamma & = & {{(2 \pi)}^{-V N I^{b}}  {(\imath)}^{- N V {Tr}{(L)}}  {(N / 2 \pi )}^{{I^{f}}^{2}V / 2} 2^{I^{f} {(I^{f} -1)} V / 2} }
\nonumber \\
{A{(n_{1} n_{2} v_{1} v_{2} i_{1} i_{2})}} & \equiv & 
{+ \imath \frac{{\xi}^{2}}{2N} {S{(n_{1} v_{1} j_{1})}} {L{(j_{1})}} {S^{*}{(n_{2} v_{1} j_{1})}} {\delta{(v_{1} v_{2})}} {\delta{(i_{1} i_{2})}}} \nonumber \\
&+& {{\imath \xi } {Q^{f}{(v_{1} i_{1} i_{2})}} {\delta{(n_{1} n_{2})}} {\delta{(v_{1} v_{2})}}}
\nonumber \\
& + & {{({{\hat{E}}^{f}{(i_{1})}{\delta{(v_{1}v_{2})}}} - {K{(v_{1} v_{2})}}) {\delta{(n_{1} n_{2})}} {\delta{(i_{1} i_{2})}} }}
\label{FermionlessGamma}
\end{eqnarray}

\subsection{Effective Lagrangian for the Bosonic Variables}
I just integrated out the fermionic vectors $\psi$ and $\overline{\psi}$ and obtained the matrix $Q^{f}$, which I will call the fermionic matrix even though it is composed of bosonic variables.  I would like to do the same sort of thing for the bosonic variables $S$ and obtain a matrix $Q^{b}$.  However the Hubbard-Stratonovich trick I used before won't work anymore, because the logarithm in the Lagrangian contains all even powers of $S$.  Therefore I must simply introduce an $I^{b} \times I^{b}$ bosonic matrix $Q^{b}$ at each site and then integrate out $S$.  Introducing the matrix is an exact step:

\begin{equation}
{\bar{Z}} = {\gamma {\int {dQ^{f}}  {dQ^{b}} {{dS^{R}}{dS^{I}}} {\delta{(Q^{b} - \hat{S})}} {\exp}{(\mathcal{L})}}}
\end{equation}
$\hat{S}$ was defined in equation \ref{SHatDefinition}. I use the convention:
\begin{equation}{\delta{(Q^{b} - \hat{S})}} \equiv {{\prod_{j}{\delta{(Q^{b}_{jj} - {\hat{S}}_{jj})}}} {\prod_{jk}{\delta{(Q^{b,I}_{jk} - {\hat{S}}_{jk}^{I})}}{\delta{(Q^{b,R}_{jk} - {\hat{S}}_{jk}^{R})}}} }\end{equation}  

In order to complete the conversion to matrix degrees of freedom, I must get rid of the $S$ variables by calculating the effective Lagrangian ${\mathcal{L}}_{eff}$:
\begin{eqnarray}
{{e}^{{\mathcal{L}}_{eff}}}
& \equiv & {\int {{dS^{R}}{dS^{I}}} {\delta{(Q^{b} - \hat{S})}} {{det}{(A)}} 
e^{\frac{\imath}{2} S L {({{\hat{E}}^{b}} - K - J)} {S^{*}}}  }
\nonumber \\
{\bar{Z}} & = & {\gamma {\int {dQ^{f}}  {dQ^{b}} {\exp}{(\mathcal{L})}}}
\nonumber \\
{\mathcal{L}}  & = & {{{\mathcal{L}}_{eff}} + {\frac{{\xi}^{2}}{8N}{{{Tr}{(Q^{b}LQ^{b}L)}}}} -\frac{ N}{2} {{Tr}({Q^{f}}^{2}) }}
\nonumber \\
\gamma & = & {{(2 \pi)}^{-V N I^{b}}  {(\imath)}^{- N V {Tr}{(L)}}  {(N / 2 \pi )}^{{I^{f}}^{2}V / 2} 2^{I^{f} {(I^{f} -1)} V / 2} }
\nonumber \\
{A{(n_{1} n_{2} v_{1} v_{2} i_{1} i_{2})}} & = & 
{+ \imath \frac{{\xi}^{2}}{2N} {S{(n_{1} v_{1} j_{1})}} {L{(j_{1})}} {S^{*}{(n_{2} v_{1} j_{1})}} {\delta{(v_{1} v_{2})}} {\delta{(i_{1} i_{2})}}} \nonumber \\
&+& {{\imath \xi } {Q^{f}{(v_{1} i_{1} i_{2})}} {\delta{(n_{1} n_{2})}} {\delta{(v_{1} v_{2})}}}
\nonumber \\
& + & {{({{\hat{E}}^{f}{(i_{1})}{\delta{(v_{1}v_{2})}}} - {K{(v_{1} v_{2})}}) {\delta{(n_{1} n_{2})}} {\delta{(i_{1} i_{2})}} }}
\label{EffectiveLagrangianDefinition}
\end{eqnarray}

Fyodorov showed how to calculate this effective Lagrangian exactly in a zero dimensional ($V = 1$) system\cite{Fyodorov02}.  Calculating ${{\mathcal{L}}_{eff}}$ non-perturbatively in systems with more than one site is much more difficult, and has not been done yet.  In a system which is a continuum instead of a lattice, it is not clear whether the integral is even well defined.  The source of difficulty, both on the lattice and on the continuum, is that the $S$ variables at neighboring sites are allowed to vary wildly with respect to each other, and the Lagrangian's value at any given point depends on every one of the wildly varying $S$ variables in the system.
 
In order to evaluate the continuum effective Lagrangian, one must regularize $S$, requiring that it not vary significantly between nearby points.  In the next section I will introduce a regularization of $S$.  Having applied this regularization, I will do a non perturbative calculation of the effective Lagrangian in a diffusive limit where the fields are permitted to vary only over very long distance scales.  I will then use effective Lagrangian arguments to determine the leading corrections to the diffusive Lagrangian.  This procedure will give me both the non perturbative information necessary to determine the saddle point and also the perturbative information necessary to determine the effective Lagrangian of the final sigma model.

\subsection{Diffusive Limit of the Effective Lagrangian}
I will evaluate the diffusive limit of the effective Lagrangian given in equation \ref{EffectiveLagrangianDefinition} by first using a diffusive approximation to move the determinant outside of the integral, then regularizing the theory, and then evaluating the remaining integral in a diffusive approximation.

\subsubsection{The Determinant}
First let's address the determinant of $A$.  This is hard to evaluate because all of the indices are coupled: the $S L S^{*}$ term couples $n$ to $v$, while the $Q^{f}$ term couples $i$ to $v$ and the kinetic term $K$ ensures that different points $v_{1}$ and $v_{2}$ are coupled.   However, if we assume that $S$ does not depend on the spatial index $v$, then the index $n$ decouples from everything else, and one can begin simplifying the determinant.   We choose a unitary transformation $U$ which diagonalizes $A_{0} = {U \Lambda {U}^{\dagger}}$, the part of $A$ with a spatial dependence:
\begin{eqnarray}
{A_{0}}  & \equiv &  
 {{\imath \xi } {Q^{f}{(v_{1} i_{1} i_{2})}} {\delta{(n_{1} n_{2})}} {\delta{(v_{1} v_{2})}}}
\nonumber \\
& + & {{({{{\hat{E}}^{f}{(i_{1})}} {\delta{(v_{1} v_{2})}}}- {{K{(v_{1} v_{2})}} }) {\delta{(n_{1} n_{2})}} {\delta{(i_{1} i_{2})}} }}
\label{SInvariantApprox}
\end{eqnarray}
Then the determinant decomposes into $I^{b}V$ small determinants, one for each eigenvalue $\lambda$ of $A_{0}$:
\begin{equation}
{{det}{(A)}} = {\prod_{\lambda } {{det}{({\lambda} + { \imath \frac{{\xi}^{2}}{2N} {S{(n_{1} j_{1})}} {L{(j_{1})}} {S^{*}{(n_{2} j_{1})}}  })}}} 
\end{equation}
Note that these small determinants are each in very small bases, each basis having only $N$ basis elements.  Now we're in business;
\begin{equation}
{{det}{(A)}} = {\prod_{\lambda} {({\lambda})}^{N}{{\exp}{({Tr}{({{\ln}{(1 + { \imath \frac{{\xi}^{2}}{2N \lambda} {S{(n_{1} j_{1})}} {L{(j_{1})}} {S^{*}{(n_{2} j_{1})}}  })}})})}}} 
\end{equation}
Expanding the logarithm as a power series, it becomes obvious that:
\begin{eqnarray}
{{det}{(A)}} & = & {\prod_{\lambda} {({\lambda})}^{N}{{\exp}{({Tr}{({{\ln}{(1 + { \imath \frac{{\xi}^{2}}{2N \lambda} {S^{*}{(n_{1} j_{1})}} {S{(n_{1} j_{2})}} {L{(j_{2})}}   })}})})}}}
\nonumber \\
& = & {\prod_{\lambda} {({\lambda})}^{N}{{\exp}{({Tr}{({{\ln}{(1 + { \imath \frac{{\xi}^{2}}{2N \lambda} \hat{S} L     })}})})}}}
\end{eqnarray}
The trace has now changed from being over the $n$ index to being over the $j$ index; all the $S$ dependence is now hidden in the matrix $\hat{S}$, which is what I was aiming for. Now to re-insert the dependence on $v$ and $i$:
\begin{equation}
{{det}{(A)}} = { {({{\det}{({A_{0}})}})}^{N - I^{b}}
{{\det}{({A_{0}} + { \imath \frac{{\xi}^{2}}{2N } \hat{S} L     })}}
}
\label{FactorizedDeterminant}
\end{equation} 
The first determinant is in the $ni$ basis, while the second is in the $nij$ basis.  

This simplification relied on the approximation that $S$ does not depend on the spatial index $v$, an approximation which I will now discuss in more detail.  First, some new notation: I divide the $S L S^{*}$ term into a translationally invariant part $\overline{S L S^{*}}$ and a fluctuating part $\delta S$, and define a Green's function $g \equiv {({\hat{E}}^{b} - K + {\imath \xi Q^{f}} + {\imath \frac{{\xi}^{2}}{2M} \overline{S L S^{*}}})}^{-1}$.  In this notation, one has:
\begin{equation}{\det{(A)}} = {\det{(g^{-1} + {\delta S})}} = {{\det{(g^{-1})}}{\det{(1 + {g {\delta S}})}}}
\label{ApproxInDetail}
\end{equation}
 In mathematical terms, the approximation consisted of two steps:
 \begin{enumerate} 
\item \label{ApproxStep1} Just before equation \ref{SInvariantApprox}, I threw away the last determinant in equation \ref{ApproxInDetail}. This step is valid when the Green's function $g$ varies at length scales which are much longer than the characteristic scale of variation of $S$.  This is called the diffusive limit, because it means that the Green's function allows particles to diffuse over distances that are much bigger than the length scale of the disorder represented by $S$.  The diffusive limit is a sort of mean field approximation.  
\item \label{ApproxStep2} I have put $\hat{S}$ in equation \ref{FactorizedDeterminant}, even though the correct term to insert would be the average $\overline{\hat{S}}$.  This is a diffusive approximation similar to the previous one, and one can hope that the two will partially cancel each other out.
\end{enumerate}
Note that these approximations are not needed in the $V = 1$ zero dimensional system, and the factorizations are exact.

Applying equation \ref{FactorizedDeterminant}, I simplify the effective Lagrangian:
\begin{eqnarray}
{{e}^{{\mathcal{L}}_{eff}}}
&= & {{({{\det}{(A_{0})}})}^{N - I^{b}} {{\det}{(A_{0} + { \imath \frac{{\xi}^{2}}{2N } Q^{b} L     })}}
 }
 \nonumber \\
& \times & { {\int {{dS^{R}}{dS^{I}}} {\delta{(Q^{b} - \hat{S})}}  
e^{\frac{\imath}{2} S L {({{\hat{E}}^{b}} - K - J)} {S^{*}}}}}
\label{EffAct1}
\end{eqnarray}

Note that all of the effective Lagrangian's dependence on $Q^{f}$ is now outside of the integral.  This is very good news, because it means the I have obtained the exact $Q^{f}$ dependence, up to corrections to the diffusive approximation.  Moreover, because the approximations which I have made so far are exact when $S$ and $\hat{S}$ are translationally invariant, I expect (but can't prove) that any terms correcting the effective Lagrangian's dependence on $Q^{f}$ will include at least one spatial derivative of $Q^{b}$.

The remaining integral in \ref{EffAct1} can be evaluated exactly in the large $N$ limit $I^{b} V \leq N$; this is an old way of taking the diffusive limit\cite{Wegner79, Wegner79a, Schafer80, Pruisken82, Jungling80}, and was recently used by Fyodorov to derive a sigma model\cite{Fyodorov04}.  In order to maintain a close correspondence with the supersymmetric sigma model, I will avoid taking the large $N$ limit, and instead examine what further approximations are required in order to evaluate the effective Lagrangian when $N$ is small. I will obtain a result which is similar to the one obtained by the large $N$ limit, which gives some additional confirmation that my regularization and approximations are correct.

\subsubsection{The Continuum Regularization}
The integral in equation \ref{EffAct1} can also be evaluated exactly for any value of $N$ when the system is zero dimensional; $V = 1$.  One can also imagine evaluating it numerically on a lattice.  However, as it stands it is ill-defined on the continuum, because it permits $Q^{b}$ and $S$ to be totally different at neighboring points.  In order to calculate the continuum effective action, one must regularize $Q^{b}$ and $S$, forcing them to be constant at some small length scale.  I now do this regularization explicitly by breaking the system's volume $V $ into $V \kappa$ blocks with ${\kappa}^{-1}$ sites per block, and requiring that $S$ and $Q^{b}$ are constant within each block.   In a zero dimensional system $\kappa = 1$.

\subsubsection{The Diffusive Approximation}
If the kinetic energy operator $K$ is zero, then the integral in equation \ref{EffAct1} factorizes into $V \kappa$ integrals, one for each block of constant $S$ and $Q^{b}$. Each of these integrals is effectively a zero dimensional integral and can be evaluated exactly.  I now make an approximation and do this factorization even though $K \neq 0$.  This is a diffusive approximation, because it is is valid if at the block volume scale ${\kappa}^{-1}$ the kinetic energy $K$ is small compared to the other operators.  

While this approximation gives a correct treatement of the effective Lagrangian's zero momentum (translationally invariant) behavior, it does lose some important information about the effective Lagrangian's low momentum behavior.  This is not a problem because later I will restore these low momentum terms to the Lagrangian by making the usual arguments from effective field theory.

I now calculate the remaining integral in equation \ref{EffAct1} for a single block of ${\kappa}^{-1}$ points.  This derivation, up through equation \ref{EffAct3}, is a very slight generalization of the proof due to Fyodorov\cite{Fyodorov02}. Doing the sums over the spatial index $v$, I obtain:
\begin{equation}
{ {\int {{dS^{R}}{dS^{I}}} {\delta{(Q^{b} - \hat{S})}}  
e^{\frac{\imath}{2 \kappa} S L {({{\hat{E}}^{b}} - K - J)} {S^{*}}}}}
\label{BlockIntegral1}
\end{equation}

The delta function in equation \ref{BlockIntegral1} can be be Fourier transformed using the identity ${\delta{(G)}} = {{(2 \pi)}^{-L^{2}} 2^{L {(L-1)}} \int{{dF}e^{\imath{{Tr}{(FG)}}}}}$ to obtain:
\begin{eqnarray}
&  & {(2 \pi)}^{-{I^{b}}^{2}} 2^{-{I^{b} } + {I^{b} I^{b} }} 
 \int{{dF} {{dS^{R}}{dS^{I}}} e^{\frac{\imath}{2}{{Tr}{({(Q^{b} - \hat{S})}F)}}} e^{\frac{\imath}{2 \kappa} S L {({{\hat{E}}^{b}} - K - J)} {S^{*}}}  }
\label{EffAct2}
\end{eqnarray}

Finally using equation \ref{BosonDeterminant} to integrate over $S$, I get:
\begin{eqnarray}
{\imath}^{N {{Tr}{(L)}}} {(2 \pi)}^{I^{b}N -{I^{b}}^{2}} &  &  2^{{-I^{b} } + {I^{b} I^{b} }} (i)^{{{Tr}{(L)}} N} 
 {{\eta}^{N, I^{b}}{(L {({{\hat{E}}^{b}} - K - J)}/ \kappa)}}, 
\nonumber \\
{{{\eta}^{N, I^{b}}{(\mu)}}} & \equiv & {{\int{{dF}  e^{\frac{\imath}{2}{{Tr}{({(Q^{b} - \hat{S})}F)}}}  {(\det {(F - \mu)} )}^{-{N}} }}} 
\label{EtaDefinition}
\end{eqnarray}

The second superindex on the new function $\eta$ specifies the rank of the matrices $F$ and $\mu$.  $\eta$ is easy to evaluate, and can be reduced to a pretty form if $I^{b} \leq N$.  I will evaluate it by establishing a recursion relation between ${\eta}^{N, I^{b}}$ and ${\eta}^{N-1, I^{b} - 1}$.    The first step is to transform $F$ to a basis where $Q^{b}$ is diagonal.  I adopt the notation $Q^{b} \equiv { W Q^{D} {W}^{\dagger}}$, $\hat{\mu} \equiv {W \mu {W}^{\dagger}}$.  The measure $dF$ is, of course, invariant under this transformation.  I obtain:
\begin{eqnarray}
{{\eta}^{N, V}{(\mu)}} 
& = & {\int{{dF}  e^{\frac{\imath}{2}{{Tr}{(Q^{D}F)}}}  {\det {({WFW^{\dagger}} - \mu)} )}^{-{N}} }} 
\nonumber \\
& = & {\int{{dF}  e^{\frac{\imath}{2}{{Tr}{(Q^{D}F)}}}  {\det {(F - \hat{\mu})} )}^{-N} }}  
\end{eqnarray}

Preparing for the recursion, I decompose the $I \times I$ matrices $F$, $\hat{\mu}$, and $Q^{D}$ into parts: \[ \left( \begin{array}{lr}
f_{11} & f_{I-1} \\
f_{I-1}^{\dagger} & F_{I-1} \end{array} \right)\] 
The ${(I - 1)} \times {(I - 1)}$ matrix $F_{I- 1}$ is just $F_{I}$ with the topmost row and leftmost column removed.  There is a useful relation between the determinant of $F_{I}$ and the determinant of $F_{I-1}$:  
\begin{equation}
{\det{(F_{I})}} = { {\det{(F_{I-1})}} {(f_{11} - {{f^{\dagger}}_{I-1}} {(F_{I-1})}^{-1} f_{I -1})}}  
\end{equation}
Therefore I can rewrite $\eta$:
\begin{eqnarray}
{{\eta}^{N, I}{(\mu)}} 
& = & {\int{{dF_{I - 1}  e^{\frac{\imath}{2}{{Tr}{({Q^{D}}_{I - 1}F_{I - 1})}}}  {\det {(F_{I - 1} - {\hat{\mu}}_{I - 1})} }^{-N} }} } 
\nonumber \\
& \times & \int{\prod_{l>1}{df^{R}_{1l}}{df^{I}_{1l}}{df_{11}}}  e^{\frac{\imath}{2} {Q^{D}}_{11}f_{11}} 
(f_{11} - {\hat{\mu}}_{11} 
\nonumber \\
 & - & {({f^{\dagger}}_{I-1} - {{\hat{\mu}}^{\dagger}}_{I-1})}{(F_{I-1} - {\hat{\mu}}_{I-1})}^{-1}{(f_{I -1} - {\hat{\mu}}_{I -1})} ) 
\end{eqnarray}
The integral over $f_{11}$ is a simple contour integral.  Recall that $\mu \equiv {L {({{\hat{E}}^{b}} - K - J)}}$ and therefore its imaginary part is positive definite.  Therefore the pole is in the positive imaginary half plane.  If ${Q^{D}}_{11}$ is positive, the contour integral must extend into the positive imaginary half plane and therefore includes the pole. If, on the other hand, ${Q^{D}}_{11}$ is negative, then the contour integral extends into the negative imaginary half plane so that the pole is not included and the integral is identically zero. We have:
\begin{eqnarray}
{{\eta}^{N, I}{(\mu)}} 
& = & \frac{2 \pi \imath}{(I-1)!} {(\frac{\imath {Q^{D}}_{11}}{2})}^{N-1} {\theta{({Q^{D}}_{11})}} e^{\frac{\imath}{2} {Q^{D}}_{11}{\hat{\mu}}_{11}} 
\nonumber \\
& \times &
{\int{{dF_{I - 1}  e^{\frac{\imath}{2}{{Tr}{({Q^{D}}_{I - 1}F_{I - 1})}}}  {\det {(F_{I - 1} - {\hat{\mu}}_{I - 1})} }^{-N} }} } 
\nonumber \\
& \times & \int{\prod_{l>1}{df^{R}_{1l}}{df^{I}_{1l}}}   
{exp}(\frac{\imath}{2} {Q^{D}}_{11}({({f^{\dagger}}_{I-1} - {{\hat{\mu}}^{\dagger}}_{I-1})} 
\nonumber \\
& \times & {(F_{I-1} - {\hat{\mu}}_{I-1})}^{-1}{(f_{I -1} - {\hat{\mu}}_{I -1})} )) 
\end{eqnarray}

After a shift in the remaining integration variables $f_{I-1}$, we apply formula \ref{BosonDeterminant} for bosonic determinants  to obtain:
\begin{eqnarray}
{{\eta}^{N, I}{(\mu)}} 
& = & {(\frac{2 \pi \imath}{{Q^{D}}_{11}})}^{I-1} \frac{2 \pi \imath}{(I-1)!} {(\frac{\imath {Q^{D}}_{11}}{2})}^{N-1} {\theta{({Q^{D}}_{11})}} e^{\frac{\imath}{2} {Q^{D}}_{11}{\hat{\mu}}_{11}}
\nonumber \\
& \times & 
{\int{{dF_{I - 1}  e^{\frac{\imath}{2}{{Tr}{({Q^{D}}_{I - 1}F_{I - 1})}}}  {\det {(F_{I - 1} - {\hat{\mu}}_{I - 1})} }^{-N+1} }} } 
\nonumber \\
& = & {{\eta}^{N-1, I-1}{(\mu)}} \frac{\theta{({Q^{D}}_{11})}}{(I-1)!} e^{\frac{\imath}{2} {Q^{D}}_{11}{\hat{\mu}}_{11}} {(\frac{\imath {Q^{D}}_{11}}{2})}^{N-I}  2 {\pi}^{I} {i}^{2I - 1}  
\end{eqnarray}
The formula for ${{\eta}^{N, 1}{(\mu)}} $ gives the same result if one defines ${{\eta}^{N, 0}{(\mu)}} \equiv 1$.  Using the identities ${\sum_{i} {{Q^{D}}_{ii}{\hat{\mu}}_{ii}}} = {{Tr}{(Q^{b}\mu)}}$ and ${\prod_{i} {({Q^{D}}_{ii})}^{N-I}} = {{(\det Q^{b})}^{N-I}}$, one obtains the final expression for $\eta$, which is valid if $N \geq I$:
\begin{equation}
{{\eta}^{N, I}{(\mu)}} = {{\Theta{(Q^{b})} e^{\frac{\imath}{2}{{Tr}{(Q^{b}\mu)}}} {{(\det Q^{b})}^{N-I}} {{(\prod_{l=N-I}^{N-1}{l!})}^{-1}} {\imath}^{NI} {\pi}^{I{(I+1)}/2} 2^{{-NI} + {I{(I+3)}/2}}}}
\end{equation}

Substituting this result into equation \ref{EtaDefinition}, I obtain the final value of expression \ref{BlockIntegral1}, the integral of a single block:
\begin{eqnarray}
&  & {\Theta{(Q^{b})} e^{\frac{\imath}{2 \kappa}{{Tr}{(Q^{b}L {({{\hat{E}}^{b}} - K - J)})}}} {{(\det Q^{b})}^{N-I^{b}}} }
\nonumber \\
& \times & {\imath}^{N {{Tr}{(L)}}} {(2 \pi)}^{I^{b}N -{I^{b}}^{2}}  2^{{-I^{b} } + {I^{b} I^{b} }} (i)^{{{Tr}{(L)}} N}  { {{(\prod_{l=N-I^{b}}^{N-1}{l!})}^{-1}} {\imath}^{NI^{b}} {\pi}^{I^{b}{(I^{b}+1)}/2} 2^{{-NI^{b}} + {I^{b}{(I^{b}+3)}/2}}}
\nonumber \\
\label{EffAct3}
\end{eqnarray}

The matrices in result \ref{EffAct3} are all single site matrices, without spatial indices.  The effective Lagrangian contains an instance of this expression for every one of the $V \kappa$ blocks.  Multiplying them all together and substituting them into equation \ref{EffAct1} for the effective action, I obtain the following expression for the diffusive limit of the effective action:
\begin{eqnarray}
{{e}^{{\mathcal{L}}_{eff}}} &=& {{{({{\det}{({A_{0}})}})}^{N - I^{b}} {{\det}{({A_{0}} + { \imath \frac{{\xi}^{2}}{2N } Q^{b} L     })}}
 } {{\Theta{(Q^{b})}} e^{\frac{\imath}{2}{{Tr}{(Q^{b}{(L {({{\hat{E}}^{b}} - K - J)})})}}} {{(\det Q^{b})}^{\kappa{(N-I)}}}}}
 \nonumber \\  & \times & {{{(\prod_{l=N-I}^{N-1}{l!})}^{-V}} {\imath}^{{NI^{b}V} + {{{Tr}{(L)}}NV}} {\pi}^{{-I^{b}{(I^{b}-1)}V/2} + {I^{b}NV} }  2^{I^{b}{(I^{b} + 1)}V/2}  }
 \label{EffAct4}
\end{eqnarray}
In equation \ref{EffAct4} the matrices have regained their spatial indices and the traces once again include sums over the volume.  I have omitted factors of $\kappa$ in the exponents of the multiplicative constants; there should be a $\kappa$ to match each explicit factor of $V$.  However in the continuum these multiplicative constants are ignored, and on the $V = 1$ lattice $\kappa = 1$, so this omission doesn't matter except on extended lattices.

Combining equation \ref{EffAct4} with equation \ref{EffectiveLagrangianDefinition}, I obtain the diffusive Lagrangian:
\begin{eqnarray}
{\bar{Z}} & = &  {\gamma {\int_{Q^{b} > 0} {dQ^{f}}  {dQ^{b}} {\exp}{(\mathcal{L})}}}
\nonumber \\
{\mathcal{L}}   & = & { -{\frac{{\xi}^{2}}{8N}{{{Tr}{(Q^{b}LQ^{b}L)}}}} -{\frac{ N}{2} {{Tr}({Q^{f}}^{2}) }} + {{{{(N - I^{b})}{{{Tr}\ln}{({A_{0}})}}}}}} 
\nonumber \\ 
& + & {{ {{{{{Tr}\ln}{({A_{0}} + { \imath \frac{{\xi}^{2}}{2N } Q^{b} L })}}} { + {\frac{\imath}{2}{{Tr}{(Q^{b}{(L {({{\hat{E}}^{b}} - K - J)})})}}} + {\kappa{(N-I^{b})}{({Tr}\ln{ {Q^{b}}})}}}}}}
\nonumber \\
{A_{0}} & = & {{\imath \xi  Q^{f}} + {\hat{E}}^{f} - K}
\nonumber \\
\gamma & = &  N^{I^{f} I^{f} V / 2}  {{{(\prod_{l=N-I}^{N-1}{l!})}}^{-V}}  2^{{I^b{(I^{b} + 1)V}/2} - {I^{f} V / 2} - {N I^{b} V}}
\nonumber \\
& \times &  {\imath}^{{NI^{b} V} } {\pi}^{{-I^{b}{(I^{b}-1)}V/2} - {I^{f} I^{f} V / 2}}   
\label{CloseToFinalQbQfLagrangian}
\end{eqnarray}

I switch the units of $Q^{b}$ so that it will be dimensionless, just like $Q^{f}$ already is: $ Q^{b} \rightarrow {(\frac{2 Q^{b} N}{\xi}}$.  Since the integral contains a factor of ${{\det}^{N - I^{b}}{(Q^{b})}} {dQ^{b}}$, this tranformation multiplies it by a factor of ${(\frac{2N}{\xi})}^{NI^{b}V}$.  Moreover, I multiply $\det{(Q)}$ by ${\det{(L)}}={\imath}^{{{Tr}{(L)}}-I^{b}}$, and compensate by multiplying the constant $\gamma$ also by $det{(L)}$.

\begin{eqnarray}
{\bar{Z}} & = &  {\gamma {\int_{Q^{b} > 0} {dQ^{f}}  {dQ^{b}} {\exp}{(\mathcal{L})}}}
\nonumber \\
\gamma & = & { {\xi}^{-I^{b} N V} N^{{{N I^{b} V } } + {I^{f} I^{f} V / 2}} {\imath}^{{ {(I^{b} - N)}{{Tr}{(L)}V} - {I^{b}I^{b}V} + {2 I^{b} NV}} } } \nonumber \\
& \times & { {{{(\prod_{l=N-I^{b}}^{N-1}{l!})}}^{-V}}  2^{{I^b{(I^{b} + 1)V}/2} - {I^{f} V / 2}}   {\pi}^{{-I^{b}{(I^{b}-1)}V/2} - {I^{f} I^{f} V / 2}}   } 
\nonumber \\
{\mathcal{L}}   & = & {  -{\frac{ N}{2} {{Tr}({Q^{f}}^{2}) }} + {{{{(N - I^{b})}{{{Tr}\ln}{({{\imath \xi  Q^{f}} + {\hat{E}}^{f} - K}})}}}}} 
\nonumber \\ 
& - & {{ {  {\frac{N}{2}{{{Tr}{(Q^{b}LQ^{b}L)}}}} { + {\imath N{{Tr}{(Q^{b}{L {({{\hat{E}}^{b}} - K - J) / \xi}})}}} + {\kappa{(N-I^{b})}{({Tr}\ln{( {Q^{b} L})})}}}}}} 
\nonumber \\
& + & {{{{Tr}\ln}{({{\hat{E}}^{f} - K} + { \imath {\xi} {(Q^{f} + {Q^{b} L})} })}}}
\label{FinalQbQfLagrangian}
\end{eqnarray}

The two terms in the top line are in the $vi$ basis, the three terms in the second line are in the $vj$ basis, and the last term is an amphibian, living in the $vij$ basis.  

Equation \ref{FinalQbQfLagrangian} completes the conversion to matrix coordinates and gives the final expression for the Lagrangian in the diffusive limit.  This Lagrangian will be the starting point for sections \ref{DOSZero} and \ref{TwoPointZero}, which calculate observables in the gaussian unitary ensemble.  

The effective Lagrangian has two continuous global symmetries when the source $J = 0$ is equal to zero, the energy matrices ${\hat{E}}^{b}$ and ${\hat{E}}^{f}$ are proportional to the identity, and $Q^{b}$ and $Q^{f}$ are translationally invariant.  There is a symmetry under $Q^{f} \rightarrow {U Q^{f} U^{\dagger}}$, where $U$ is a unitary transformation which is diagonal in the spatial coordinate and also translationally invariant.  There is also a symmetry under ${Q^{b}} \rightarrow {T Q^{b} T^{-1}}$, where $T$ satisfies the constraint ${T L} = {L T}$, is diagonal in the spatial coordinate, and is also translationally invariant.  Both of these symmetries were in the original exact expression for the effective Lagrangian, given in equation \ref{EffectiveLagrangianDefinition}, where the $Q^{b}$ symmetry is accompanied by transformations of $S$ and $S^{*}$. These symmetries are very important for the physics of our model; in section \ref{NewSigmaModel} I will do a saddle point approximation and find Goldstone modes around the saddle point corresponding to these symmetries.
  
\subsection{Corrections to the Diffusive Limit}
Shortly I will make a saddle point approximation and find a saddle point which is translationally invariant.  Because equation \ref{FinalQbQfLagrangian} is exact in the zero dimensional ($V = 1$) limit and because the steps of its derivation were rigorous within the bounds of the chosen regularization and the diffusive approximation,  its result should be exactly correct when $Q^{b}$ and $Q^{f}$ are translationally invariant.   However, when I made the diffusive approximation, the Lagrangian lost its correct dependence on fluctuations in $Q^{b}$.  It may have also lost similar information about $Q^{f}$, but I believe that $Q^{f}$ escaped with much lighter damages. Because the final sigma model will require a correct dependence on fluctuations in $Q^{b}$ and $Q^{f}$, I now analyze what are the perturbative corrections to the diffusive Lagrangian.

\subsubsection{Perturbation Theory}
One option for evaluating the perturbative corrections is to return to the original definition of the effective Lagrangian, given equation \ref{EffectiveLagrangianDefinition}.  If one knows the Lagrangian's correct saddle point, then perturbative corrections to the effective Lagrangian may be evaluated via a Taylor expansion in $Q^{b}$ and $Q^{f}$:
\begin{eqnarray}
{{e}^{{\mathcal{L}}_{eff}}}
& \equiv & {\int {{dS^{R}}{dS^{I}}} {\delta{(Q^{b} - \hat{S})}} {{det}{(A)}} 
e^{\frac{\imath}{2} S L {({{\hat{E}}^{b}} - K - J)} {S^{*}}}  }
\nonumber \\
& = & \sum_{({{\tilde{Q}}^{b})}^{n}, {({\tilde{Q}}^{f})}^m} \frac{({{\tilde{Q}}^{b})}^{n} {({\tilde{Q}}^{f})}^{m}}{C{(n,m)}} \int {{dS^{R}}{dS^{I}}} 
e^{\frac{\imath}{2} S L {({{\hat{E}}^{b}} - K - J)} {S^{*}}} 
\nonumber \\
& \times &
{[{{(\frac{d}{dQ^{b}})}^{n} {\delta{(Q^{b} - \hat{S})}}}]}_{Q^{b} = Q_{b}^{0}} {[{(\frac{d}{dQ^{f}})}^{m} {{det}{(A)}}]}_{Q^{f} = Q_{0}^{f}}  
\end{eqnarray}
 
$Q_{0}^{b}$ and $Q_{0}^{f}$ are the values of $Q^{b}$ and $Q^{f}$ at the saddle point, while ${\tilde{Q}}^{b}$ and ${\tilde{Q}}^{f}$ are the deviations from the saddle point.  The powers in $n$ and $m$ of ${\tilde{Q}}^{b}$ and ${\tilde{Q}}^{f}$ must be understood formally, since ${\tilde{Q}}^{b}$ is a matrix with ${I^{b}}^{2} V$ elements, and ${\tilde{Q}}^{f}$ is a matrix with ${I^{f}}^{2} V$ elements.  The above Taylor expansion is an expansion ${({I^{f}}^{2} + {I^{b}}^{2})} V$ distinct variables.  $C{(n, m)}$ is meant to represent the multiple of all the Taylor expansion coefficients.

This integral can be further simplified by turning the derivative in $Q^{b}$ into a derivative in $S$.  This trick can be easily demonstrated using the following example:
\begin{eqnarray}
& & {\int{{dx}{{g{(a)}}{{\frac{d}{dx}{\delta{(x-{f{(a)}})}}}}}}}
\nonumber \\
& = & {-\int{{dx}{{g{(a)}}{\frac{da}{df}\frac{d}{da}{\delta{(x-{f{(a)}})}}}}}}
\nonumber \\
& = & {\int{{dx}{\delta{(x-{f{(a)}})}}{\frac{da}{df}\frac{d}{da}{{g{(a)}}}}}}
\label{DerivativeShiftingTrick}
\end{eqnarray}
The last step was just an integration by parts and relied on an assumption - conventional in field theory - that total derivatives can be ignored. This derivative-shifting trick is easily generalized to multiple integrals.  As long as the dimensionality of the delta function is equal to or greater than the dimensionality of the integral, meaning that the delta function kills all the integrations, equation \ref{DerivativeShiftingTrick} changes in only two respects: the derivative $\frac {da} {df}$ turns into a matrix, and $\frac {d} {da}$ becomes a gradient.  If the delta function does not kill all the integrations, then things are a little more complicated: some elements of the matrix $\frac {da} {df}$ diverge because the delta function constraint describes a surface.  However these same matrix elements are the ones multiplying portions of the gradient which run parallel to the constraint surface. In this case one must be a little more careful and first determine the space $\hat{n}$ of directions normal to the delta function constraint.  One then takes the gradient  $\frac{d}{da}$ only along directions in $\hat{n}$, and multiplies that restricted gradient by the matrix $\frac {da} {df}$.   Hiding these complications in the formalism $\frac{d}{d\hat{S}}$, I obtain the simplified effective Lagrangian:

\begin{eqnarray}
{{e}^{{\mathcal{L}}_{eff}}}
& = & \sum_{({{\tilde{Q}}^{b})}^{n}, {({\tilde{Q}}^{f})}^m} \frac{({{\tilde{Q}}^{b})}^{n} {({\tilde{Q}}^{f})}^{m}}{C{(n,m)}} \int {{dS^{R}}{dS^{I}}}  {\delta{(Q_{0}^{b} -\hat{S})}} 
\nonumber \\
& \times & {[{{(\frac{d}{d\hat{S}})}^{n} {(\frac{d}{d{\hat{Q}}^{f}})}^{m} {{det}{(A)}} e^{\frac{\imath}{2} S L {({{\hat{E}}^{b}} - K - J)} {S^{*}}}  }]}_{Q^{f} = Q_{0}^{f}, Q^{b} = Q_{0}^{b} }  
\end{eqnarray}

This is a nonlinear sigma model with constraint $\hat{S} = Q_{0}^{b}$.  The effective Lagrangian ${{\mathcal{L}}_{eff}}$ is equal to the sum of all of this sigma model's connected diagrams, and probably can be evaluated perturbatively using conventional field theory techniques. 

Of course this approach is no good for evaluating non-perturbative terms in the Lagrangian, including the terms which determine the saddle point.  Furthermore, the perturbative calculations will require us to regularize the theory, just as we already did when deriving the diffusive Lagrangian.  The strengths of the corrections which we calculate will be determined by the details of the regularization, not by the original Lagrangian specified in equation \ref{EffectiveLagrangianDefinition}.

\subsubsection{Effective Field Theory}
Instead of deriving the form of the perturbative corrections from the original Lagrangian, I will determine them by using arguments from effective field theory.

I assume that the effective Lagrangian is local.  I also assume that it will have the same symmetries that the original exact Lagrangian, given in equation \ref{EffectiveLagrangianDefinition}, has when ${\hat{E}}^{f}$ and ${\hat{E}}^{b}$ are proportional to the identity, the source $J$ is zero, and both $Q^{b}$ and $Q^{f}$ are translationally invariant.  In particular, $Q^{b}$ always comes paired with exactly one $L$ and there are no constant operators (other than the identity) which I can insert into the effective Lagrangian.  Corrections to these assumptions will be suppressed by powers of $k^{2}$, ${\hat{E}}^{f}$'s deviation from the identity, and ${\hat{E}}^{b}$'s deviation from the identity.
  
  I am only interested in the low momentum corrections, so I expand the Lagrangian in powers of the gradient, and keep only terms which are linear or quadratic; i.e. terms with no gradients, one gradients, two gradients, or a ${\nabla}^{2}$.  I assume that the theory is rotationally invariant, so terms with one $\vec{\nabla}$ are necessarily zero.  Moreover the Lagrangian  already includes the correct result for translationally invariant fields.  So the only possible corrections have either two gradients or a ${\nabla}^{2}$.

I anticipate the saddle point approximation in section \ref{NewSigmaModel}, and assume that the fields $Q^{b}$ and $Q^{f}$ are small.  Therefore I consider only the terms ${\nabla}^{2}Q^{f}$, ${\nabla}^{2}Q^{b}L$, $Q^{f}{\nabla}^{2}Q^{f}$, ${\vec{\nabla} Q^{f}} \cdot {\vec{\nabla}Q^{f}}$, $Q^{b}L{\nabla}^{2}Q^{b}L$, ${\vec{\nabla} Q^{b}L} \cdot {\vec{\nabla}Q^{b}L}$,
$Q^{f}{\nabla}^{2}Q^{b}L$, $Q^{f}{\nabla}^{2}Q^{b}L$, and ${\vec{\nabla} Q^{f}} \cdot {\vec{\nabla}Q^{b}L}$.  As is conventional in perturbative field theory, I assume that terms which are total derivatives can be ignored, which eliminates ${\nabla}^{2}Q^{f}$ and ${\nabla}^{2}Q^{b}L$.  Moreover partial integration reveals that several of the other terms are equal up to a total derivative.  After weeding these out, the remaining possible corrections are ${\vec{\nabla} Q^{f}} \cdot {\vec{\nabla}Q^{f}}$, ${\vec{\nabla} Q^{b}L} \cdot {\vec{\nabla}Q^{b}L}$, $Q^{f}{\nabla}^{2}Q^{b}L$, and ${\vec{\nabla} Q^{f}} \cdot {\vec{\nabla}Q^{b}L}$.

The Lagrangian contains a trace over all indices. The ${\vec{\nabla} Q^{f}} \cdot {\vec{\nabla}Q^{b}L}$ term, for instance, should be written as $\sum_{ijv}{\vec{\nabla} {Q^{f}{(vi)}}} \cdot {\vec{\nabla} {Q^{b}{(vj)}L_{j}}}$.  In this term the sum over $j$ acts only on $Q^{b}L$, so the trace can be moved inside the derivative, resulting in: $\sum_{v}{\vec{\nabla} {(\sum_{i}Q^{f}{(vi)})}} \cdot {\vec{\nabla} {(\sum_{j}Q^{b}{(vj)}L_{j})}}$.  I again anticipate the saddle point approximation, which will constrain the traces of $Q^{f}$ and $Q^{b}L$ to be constant.  Therefore both the $Q^{f}{\nabla}^{2}Q^{b}L$ term and the ${\vec{\nabla} Q^{f}} \cdot {\vec{\nabla}Q^{b}L}$ term are surpressed by the saddle point approximation.

The only two remaining terms are the kinetic terms ${\vec{\nabla} Q^{f}} \cdot {\vec{\nabla}Q^{f}}$ and ${\vec{\nabla} Q^{b}L} \cdot {\vec{\nabla}Q^{b}L}$.  As I mentioned earlier, $Q^{f}$ seemed to escape lightly from the diffusive approximations.  Moreover, we will see in section \ref{SigmaModelLagrangian2} that the diffusive lagrangian already contains a kinetic term for $Q^{f}$ which looks just right.  Therefore I doubt that the ${\vec{\nabla} Q^{f}} \cdot {\vec{\nabla}Q^{f}}$ correction is non zero.  However, section \ref{SigmaModelLagrangian2} will also show that the diffusive approximation totally annihilated $Q^{b}$'s kinetic term, making it absolutely necessary to restore it.

Adding these perturbative corrections to the diffusive Lagrangian, I obtain the result:
\begin{eqnarray}
{\bar{Z}} & = &  {\gamma {\int_{Q^{b} > 0} {dQ^{f}}  {dQ^{b}} {\exp}{(\mathcal{L})}}}
\nonumber \\
\gamma & = & { {\xi}^{-I^{b} N V} N^{{{N I^{b} V } } + {I^{f} I^{f} V / 2}} {\imath}^{{ {(I^{b} - N)}{{Tr}{(L)}V} - {I^{b}I^{b}V} + {2 I^{b} N V}} } } \nonumber \\
& \times & { {{{(\prod_{l=N-I}^{N-1}{l!})}}^{-V}}  2^{{I^b{(I^{b} + 1)V}/2} - {I^{f} V / 2}}   {\pi}^{{-I^{b}{(I^{b}-1)}V/2} - {I^{f} I^{f} V / 2}}   } 
\nonumber \\
{\mathcal{L}}   & = & {  -{\frac{ N}{2} {{Tr}({Q^{f}}^{2}) }} + {{{{(N - I^{b})}{{{Tr}\ln}{({{\imath \xi  Q^{f}} + {\hat{E}}^{f} - K})}}}}}} 
\nonumber \\ 
& - & {{ {  {\frac{N}{2}{{{Tr}{(Q^{b}LQ^{b}L)}}}} { + {\imath N{{Tr}{(Q^{b}{L {({{\hat{E}}^{b}} - K - J) / \xi}})}}} + {\kappa{(N-I^{b})}{({Tr}\ln{( {Q^{b} L})})}}}}}} 
\nonumber \\
& + & {{{{Tr}\ln}{({{\hat{E}}^{f} - K} + { \imath {\xi} {(Q^{f} + {Q^{b} L})} })}}} 
+ {\frac{1}{2}\frac{\nu {D}^{f}}{2}{Tr}{({\vec{\nabla} Q^{f}} \cdot {\vec{\nabla}Q^{f}})}} 
+ {\frac{1}{2}\frac{\nu {D}^{b}}{2}{Tr}{({{\vec{\nabla} Q^{b}L} \cdot {\vec{\nabla}Q^{b}L}})}}
\nonumber \\ & &
\label{CorrectedQbQfLagrangian}
\end{eqnarray}

\section{\label{ContinuousSystems}Reparameterization for Continuous Systems}
Three changes to equation \ref{CorrectedQbQfLagrangian} are required before obtaining the continuum model: 
\begin{enumerate}
 \item \label{ContNum1} One must throw away the prefactor $\gamma$ because on the continuum it diverges and contains no physical meaning.  
 \item \label{ContNum2} One must change the constants in the Lagrangian.  I define an intensive constant - the level density $\nu \equiv {N  {\textit{v}}^{-1} {\xi}^{-1}}$  - which will replace $N$. ($\textit{v}$ is the volume of a single point.)  Also I define $\alpha \equiv {I^{b} / N}$.  Since $\nu \xi$ has units of inverse volume, I now allow the traces over the spatial index to acquire units of volume and become integrals $d^{D}\vec{x}$.  However, I will continue using the trace notation as before.  The trace now includes an integral over the volume index $v$ and sums over the fermion index $i$ and the boson index $j$.  
\item I will have to regularize the spatial variation of the fields, prohibiting variations at small length scales.  I have already imposed this regularization on the bosonic matrix $Q^{b}$ while calculating the diffusive limit of the effective Lagrangian.  The regularization of the fermionic matrix $Q^{f}$ will occur later, when I derive the effective Lagrangian of the sigma model.  
\end{enumerate}

If I carry out changes \ref{ContNum1} and \ref{ContNum2} I obtain the following equation, which can be applied to both discrete lattices and also the continuum:
\begin{eqnarray}
\overline{Z} & = & {\int_{Q^{b} > 0} {{dQ^{b}}{dQ^{f}}} e^{\mathcal{L}}}
\nonumber \\
\mathcal{L}  & =  & {  -{\frac{\nu \xi}{2} {{Tr}({Q^{f}}^{2}) }} + {{{{\nu \xi(1 - \alpha)}{{{Tr}\ln}{({\hat{E}}^{f} - K + {\imath \xi  Q^{f}})}}}}}} 
\nonumber \\ 
& - & {{ {  {\frac{\nu \xi }{2}{{{Tr}{(Q^{b}LQ^{b}L)}}}} { + {\imath \nu{{Tr}{(Q^{b}{L {({{\hat{E}}^{b}} - K - J)}})}}} + {\nu \xi \kappa {(1-\alpha})}}{({Tr}\ln{( {Q^{b} L})})}}}} \nonumber \\
& + & {{{\frac{\nu \xi \alpha } {I^{b}}}{{Tr}\ln}{(\hat{E}}^{f} - K  + { \imath {\xi} {(Q^{f} + {Q^{b} L})} })}}
\nonumber \\
& + & {\frac{1}{2}\frac{\nu {D}^{f}}{2}{Tr}{({\vec{\nabla} Q^{f}} \cdot {\vec{\nabla}Q^{f}})}} 
+ {\frac{1}{2}\frac{\nu {D}^{b}}{2}{Tr}{({{\vec{\nabla} Q^{b}L} \cdot {\vec{\nabla}Q^{b}L}})}}
\label{ContinuumQbQfLagrangian}
\end{eqnarray}
Recall that the traces on the first line are over the volume and fermionic indices $v$ and $i$, the traces on the second line are over the volume and bosonic indices $v$ and $j$, and the last trace is over all three indices. 

\section{\label{NewSigmaModel}The Sigma Model}
My strategy for evaluating the continuum generating function (equation \ref{ContinuumQbQfLagrangian}) will be to use the saddle point approximation, which will create constraint equations and thus give me a sigma model.  Because the degrees of freedom are matrices, the saddle point equations are matrix equations, and are derived by taking matrix derivatives.  Therefore I now briefly review how to take matrix derivatives.

\subsection{\label{MatrixDerivative} Matrix Derivatives}
On a conceptual level, taking matrix derivatives is just a matter of taking derivatives with respect to individual matrix elements of the matrices.  I use the notation $\delta$ to represent two things: the total variation $\delta F$ of a function $F$, and also the matrix infinitesimal $\delta$ which can be assumed to have all but one of its matrix elements equal to zero.  
\begin{itemize}
\item \textbf{The First Derivative of a Polynomial:} One can verify that ${\delta{({Tr}{(X^{m})})}} = {m{Tr}{(\delta X^{m-1})}}$ by writing $X = X + \delta$, expanding the trace to first order in $\delta$ giving ${{Tr}{(X^{m})}}={{{Tr}{(X^{m})}} + {\sum_{j=0}^{m-1}{Tr}{(X^{j}\delta X^{m-j-1})}}}$, reorganizing the trace using the cyclic identity ${{Tr}{(AB)}} = {{Tr}{(BA)}}$ to obtain ${{Tr}{(X^{m})}}={{{Tr}{(X^{m})}} + {m{Tr}{(\delta X^{m-1})}}}$, and then taking the derivative with respect to the infinitesimal.  \item \textbf{The Second Derivative of a Polynomial:} The matrix second derivative can be obtained by taking the derivative of the first derivative; in this case the cyclic indentity can not be used again and one obtains ${{\delta}_{1}{\delta}_{2}{{Tr}{(X^{m})}}}={m{\sum_{j=0}^{m-2}{Tr}{({\delta}_{1}X^{j}{\delta}_{2} X^{m-j-2})}}}$.  
\item \textbf{The First Derivative of the Inverse:} Turning to $\delta{{Tr}{( A X^{-1})}}$, one writes ${{Tr}{(A{(X + \delta)}^{-1})}} = {{Tr}{(A X^{-1}{(1 + X^{-1}\delta)}^{-1})}}$, does a Taylor series expansion ${(1 + {X^{-1}\delta})}^{-1} \simeq {1 - {\delta X^{-1}}}$, and obtains $\delta{{Tr}{( A X^{-1})}} = {-{Tr}{( A X^{-1} \delta X^{-1})}}$.  
\item \textbf{The First Derivative of the Logarithm:}
${\delta{Tr{(\ln{(X + \delta)})}}} = {\delta{Tr{(\ln{(1 + X^{-1}\delta)})}}} = {{Tr{(X^{-1}\delta)}}}$. 
\end{itemize}
In general there is a close analogy between the formulas for matrix derivatives and those for scalar derivatives.

\subsection{\label{ContinuumSaddlePoint}The Saddle Point}
I want to find the saddle point determined by the equations:
\begin{eqnarray}
 {{[\frac{d\mathcal{L}}{d{(Q^{b}L)}}]}_{\delta = Q_{0}^{b}} =  0} , \qquad { {[\frac{d\mathcal{L}}{dQ^{f}}]}_{\delta = Q_{0}^{f}} =  0} 
 \label{MatrixSaddlePointEquations}
 \end{eqnarray}
  
  $Q_{0}^{b}$ and $Q_{0}^{f}$ are the solutions of the saddle point equation \ref{MatrixSaddlePointEquations}.  In order to simplify my equations, I define the Green's functions:
\begin{eqnarray}
{ G^{f}{(X)}} & \equiv & {({\hat{E}}^{f} - K + {\imath \xi X})}^{-1}
\nonumber \\
{ G^{b}{(X)}} & \equiv & {({\hat{E}}^{b} - K + {\imath \xi X})}^{-1}
\label{GreensFunctionDefinition}
\end{eqnarray}

I also neglect the source $J$ because it is infinitesimally small and should not determine the saddle point solutions.  Referring to the Lagrangian in equation \ref{ContinuumQbQfLagrangian}, one finds that the derivatives in equation \ref{MatrixSaddlePointEquations} are:
\begin{eqnarray}
{\frac{d\mathcal{L}}{dQ^{f}}} & = & {-{\nu \xi {Tr}{(\delta Q_{0}^{f})}} + {\imath {\nu} {\xi}^{2} {(1 - \alpha)} {Tr}{(\delta {G^{f}{(Q_{0}^{f})}})}}} + 
{\frac{\nu {D}^{f}}{2}{Tr}{({\vec{\nabla} \delta} \cdot {\vec{\nabla}Q_{0}^{f}})}}
\nonumber \\
& + &{\frac{\imath {\nu} {\xi}^{2} \alpha} {I^{b}} {Tr}{(\delta {G^{f}{(Q_{0}^{f} + {Q_{0}^{b}L})}}})}
\nonumber \\
{\frac{d\mathcal{L}}{dQ^{b}L}} & = & {{ {  {- {\nu \xi }{{{Tr}{(\delta Q_{0}^{b}L)}}}} { + {\imath \nu{{Tr}{(\delta{ {({{\hat{E}}^{b}} - K)}})}}} + {\nu \xi \kappa {(1-\alpha})}}{{Tr}({\delta({Q_{0}^{b} L})}^{-1})}}}} 
\nonumber \\
& + & {\frac{\nu {D}^{b}}{2}{Tr}{({{\vec{\nabla} \delta} \cdot {\vec{\nabla}Q_{0}^{b}L}})}} + {\frac{\imath {\nu} {\xi}^{2} \alpha} {I^{b}} {Tr}{(\delta {G^{f}{(Q_{0}^{f} + {Q_{0}^{b}L})}}})}
\label{LagrangianFirstDerivative}
\end{eqnarray}

Remember that most of the traces are over only two indices, while the last term in each equation is traced over all three.  I would also like to point out that the quantity $\xi Q_{0}^{f}$ in the Green's function ${G^{f}{(Q_{0}^{f})}} = {({\hat{E}}^{f} - K + {\imath \xi Q_{0}^{f}})}^{-1}$ is exactly the self energy of the Green's function.   (This is because the self energy of a Green's function is defined as being $-\imath$ times the quantity which is added to $E - K$.)  The physical origins of a Green's function's self-energy lie in scattering off of disorder.  In fact the self energy is equal to the inverse of the scattering time $\tau$ when using units where $\hbar = 1$.  $Q_{0}^{f}$ is a sort of dimension-free self energy and its inverse is proportional to the scattering time. \linebreak

Another very important point: these saddle point equations say nothing about matrix elements that are not diagonal in the position basis.  This is because $Q^{f}$ and $Q^{b}$ are local operators, diagonal in the position basis, so their infinitesimals are also diagonal in the position basis.

Remembering that the Lagrangian has a continuous global symmetry when the source $J$ is equal to zero and the energy matrices ${\hat{E}}^{f}$ and ${\hat{E}}^{b}$ are proportional to the identity, for the rest of this section I will enforce this proportionality by choosing ${\hat{E}}^{f} = {1 \otimes {\overline{E}}^{f}}$, where ${\overline{E}}^{f} \equiv {{I^{f}}^{-1}{\sum}_{i}E_{i}}$.  Similarly I choose ${\hat{E}}^{b} = {1 \otimes {\overline{E}}^{b}}$, where ${\overline{E}}^{b} \equiv {{I^{b}}^{-1}{\sum}_{j}E_{j}}$. 

After I have made these choices for ${\hat{E}}^{f}$ and ${\hat{E}}^{b}$ and have set $J = 0$, the saddle point equation \ref{MatrixSaddlePointEquations} does not depend on any aspect of $Q_{0}^{b}$ and $Q_{0}^{f}$ except their eigenvalues. I introduce the notation $q_{i}$ to signify the eigenvalues of $Q_{0}^{f}$, $p_{j}$ to signify the eigenvalues of $Q_{0}^{b}$, and $L_{j}$ to signify the diagonal elements of $L$.  In order to simplify the saddle point equations, I temporarily require that $Q_{0}^{b}$ and $Q_{0}^{f}$ be diagonal.  This does not change the eigenvalues and therefore does not change the content of the saddle point equations.  
 
 Because $Q_{0}^{f}$ is diagonal, the fermionic Green's function $G^{f}$ defined in equation \ref{GreensFunctionDefinition} is diagonal in the index $i$.  Therefore I define the $i$'th component of $G^{f}$ to be:
\begin{equation}
 { G_{i}^{f}{(X)}}  \equiv  {(E_{i}^{f} - K + {\imath \xi X})}^{-1}  
 \label{GreensFunctionDefinitionComponentwise}
\end{equation}
 
The saddle point equations become:
\begin{eqnarray}
 {q_{i}} & = & {\imath  {\xi} {(1 - \alpha)} { {\langle \vec{x} |} {G_{i}^{f}{(q_{i})}} {| \vec{x} \rangle}}} + {\frac{\imath {\xi} \alpha}  {I^{b}} \sum_{j}{ {\langle \vec{x} |} {G_{i}^{f}{(q_{i} + {p_{j}L_{j}})}}  {| \vec{x} \rangle}}}
 \label{DiagonalSaddlePointFermionic1}
 \nonumber \\
{p_{j} L_{j}}  & =  &  {\imath {\xi}^{-1} {\overline{E}}^{b} } - {\imath {\xi}^{-1} {\langle \vec{x} |}K{ | \vec{x} \rangle}}  + { {\kappa (1-\alpha}){{(p_{j}L_{j})}^{-1}}} 
+ {{\frac{\imath {\xi} \alpha } {I^{b}}}{\sum_{i} {\langle \vec{x} |} G_{i}^{f}{({{p_{i}}} + { p_{j} L_{j} })} { | \vec{x} \rangle} }}
\nonumber \\ & &
\label{DiagonalSaddlePointBosonic1}
\end{eqnarray}

Equation \ref{DiagonalSaddlePointBosonic1} includes terms containing the diagonal element of the fermionic Green's function $\langle \vec{x} | G_{i}^{f} | \vec{x} \rangle$.  One can estimate the real part of this matrix element, and according to Mirlin \cite{Mirlin99} it is small compared to other quantities if we regularize the continuum behavior of $Q^{f}$ and force it to be smooth at small distance scales.  Moreover, Mirlin claims the freedom to shift the definition of the energy ($\overline{E} \rightarrow {\overline{E} + \epsilon}$) and thus set the real part of $\langle \vec{x} | G_{i}^{f} | \vec{x} \rangle$ to exactly zero.  Therefore I will only calculate the imaginary part.  The first few steps are:

\begin{eqnarray}
{{\langle \vec{y} |}{G_{i}{(X)}}{| \vec{y} \rangle}} & \equiv & {{\langle \vec{y} |}{({\overline{E}}^{f} - K  + {i \xi X})}^{-1}{| \vec{y} \rangle}} 
\nonumber \\
& = & {\int \frac{d^{D}\vec{p}}{{(2 \pi)}^{D}} {({\overline{E}}^{f} - {{\langle \vec{p} |}K{| \vec{p} \rangle}}  + {i \xi X})}^{-1}} 
\nonumber \\
& = & { \int_{0}^{\infty} {dE} {({\overline{E}}^{f} - E  + {i \xi X})}^{-1} {\int \frac{d^{D}\vec{p}}{{(2 \pi)}^{D}}  {\delta{(E - {{\langle \vec{p} |}K{| \vec{p} \rangle}} )}}}} 
\nonumber \\ & &
\end{eqnarray}

I define a new quantity, the average spacing of energy levels, usually called the level spacing: 
\begin{equation}
{\Delta}^{-1}{(E)} \equiv {\int \frac{d^{D}\vec{p}}{{(2 \pi)}^{D}}  {\delta{(E - {{\langle \vec{p} |}K{| \vec{p} \rangle}} )}}}
\label{LevelSpacingDefinition}
\end{equation} 
The level spacing has units of inverse energy.  
Then I have:
\begin{equation}
{{Im}{{\langle \vec{y} |}{G_{i}{(X)}}{| \vec{y} \rangle}}} = { \int_{0}^{\infty} {dE} {{\Delta}^{-1}{(E)}} { {Im}{({({\overline{E}}^{f} - E  + {i \xi X})}^{-1})}}  }
\end{equation}
I have assumed that the kinetic energy ${{\langle \vec{p} |}K{| \vec{p} \rangle}}$ is never negative, which seems quite reasonable.  

I now assume that ${|\xi X|}  =  {\tau}^{-1} \ll {\overline{E}}^{f}$,  and that therefore ${ {Im}{({\overline{E}}^{f} - E  + {i \xi X})}^{-1}} $ is zero for all negative energies.  $\xi X$ is the self energy and $\tau$ is the scattering time; this approximation asserts that the particle's energy is large compared to its scattering-induced self energy, which is a way of saying that the particle is only weakly affected by scattering.  In other words, this approximation is a way of  taking the diffusive limit and thus regularizing the short distance behavior of the continuum theory.  This diffusive approximation probably should not be construed as a statement about the energy ${\overline{E}}^{f}$, which I already have claimed (following Mirlin) can be set to zero.
 
Next I play a trick: I define a level density at negative kinetic energies $ {{\Delta}^{-1}{(-E)}} = {{\Delta}^{-1}{(E)}}$, and extend the integral down to negative infinity:
\begin{equation}
{{Im}{{\langle \vec{y} |}{G_{i}{(X)}}{| \vec{y} \rangle}}} = { \frac{1}{2} \int_{-\infty}^{\infty} {dE} {{\Delta}^{-1}{(E)}} { {Im}{({({\overline{E}}^{f} - E  + {i \xi X})}^{-1})}}  }
\end{equation}

The diagonal matrix element of the Green's function is now equivalent to a contour integral over $E$.  This was the whole point of tranforming the integral over $\vec{p}$  into an integral over $E$, making the diffusive approximation, and then extending the integral down to $-\infty$.  This approach is standard in the literature, and will be reused later in this chapter. 

Doing the contour integral, I obtain:
\begin{equation}
{{\langle \vec{y} |}{G_{i}{(X)}}{| \vec{y} \rangle}} = {-\imath \pi {{\Delta}^{-1}{({\overline{E}}^{f})}} {{sign}{(X)}}}.
\end{equation}

It is conventional to neglect the dependence of the level spacing $\Delta$ on the energy $E$.  The fermionic saddle point equations become:

\begin{eqnarray}
 {q_{i}} & = & {\pi {\xi} {(1 - \alpha)} {\Delta}^{-1}}   {  {{sign}{(q_{i})}} } + {\frac{\pi {\xi} \alpha {\Delta}^{-1}}  {I^{b}} \sum_{j}{  {{sign}{(q_{i} + {p_{j}L_{j}})}} }}
 \label{DiagonalSaddlePointFermionic2}
 \nonumber \\
{p_{j} L_{j}}  & =  &  {\imath {\xi}^{-1} {\overline{E}}^{b} } - {\imath {\xi}^{-1} {\langle \vec{x} |}K{ | \vec{x} \rangle}}  + { {\kappa (1-\alpha}){{(p_{j}L_{j})}^{-1}}} 
+ {\frac{\pi {\xi} \alpha }  {I^{b} \Delta} \sum_{i}{  {{sign}{(q_{i} + {p_{j}L_{j}})}} }}
\nonumber \\ & &
\label{DiagonalSaddlePointBosonic2}
\end{eqnarray}

Without any good justification, I now neglect ${p_{j}L_{j}}$ in the sum $q_{i} + {p_{j}L_{j}}$.  (As I will explain later, it will turn out that the only possible effects on the final sigma model would be changes in the values of the saddle points and corrections to certain constants in the sigma model Lagrangian.)  This simplifies the saddle point equations:
\begin{eqnarray}
 {q_{i}} & = & {{\pi {\xi} {\Delta}^{-1}}   {  {{sign}{(q_{i})}} }} 
 \label{DiagonalSaddlePointFermionic3}
 \nonumber \\
{p_{j} L_{j}}  & =  &  {\imath {\xi}^{-1} {\overline{E}}^{b} } - {\imath {\xi}^{-1} {\langle \vec{x} |}K{ | \vec{x} \rangle}}  + { {\kappa {(1-\alpha)}}{{(p_{j}L_{j})}^{-1}}} 
+ {\frac{\pi {\xi} \alpha }  {I^{b} \Delta} \sum_{i}{  {{sign}{(q_{i})}} }}
\nonumber \\ & &
\label{DiagonalSaddlePointBosonic3}
\end{eqnarray}

The solutions of the fermionic saddle point equations are clearly:
\begin{equation}
{{q_{i}}  =  {s_{i} \pi {\xi} {\Delta}^{-1}}}, \: s_{i} = \pm 1
\label{FermionicSaddlePointSolution}
\end{equation}

The bosonic saddle point equation is a quadratic equation \linebreak $0 = {{a {(p_{j}L_{j})}^{2}} - {\imath b p_{j} L_{j}} + c}$, where the coefficients are $a = 1$, $b = { {\xi}^{-1} {\overline{E}}^{b} } - {{\xi}^{-1} {\langle \vec{x} |}K{ | \vec{x} \rangle}} - {\imath \frac{\pi {\xi} \alpha }  {I^{b} \Delta} \sum_{i}{  {{sign}{(q_{i})}} }}$, and $c = {- \kappa {(1-\alpha)}}$.  The solutions have the form:
\begin{equation}
{{p_{j}L_{j}} = {{\imath b} + {s_{j} \sqrt{{\kappa{(1 - \alpha)}} - b^{2}}}}}, {s_{j} = {\pm 1}}
\label{FermionicSaddlePointSolution1}
\end{equation}

If the imaginary part of $b$ can be neglected and $b^{2} < {\kappa{(1 - \alpha)}}$, then the solution can be rewritten in the form:
\begin{equation}
{{p_{j}L_{j}} = {{s_{j} {\sqrt{\kappa{(1 - \alpha)}}}} {{exp}{(i s_{j} \phi)}}}}, {{\sin{(\phi)}} \equiv {\frac{b} {\sqrt{\kappa{(1 - \alpha)}}}}}, {s_{j} = {\pm 1}}
\label{FermionicSaddlePointSolution2}
\end{equation}
Interestingly enough, the energy dependence of the solution is entirely inside the phase $\phi$.  In zero dimensions the saddle point solution is analogous to the one found here, but with $\kappa = 1$.  I surmise that if I had not neglected the real part of the Green's function when solving the fermionic saddle point equation, I would have found an energy dependent phase in the solution.  However I got rid of the fermionic phase by making a suitable shift in the energy ${\overline{E}}^{f}$.  I repeat the step here and set ${b = 1}, {\phi = 0}$.  The saddle point solutions now read:
\begin{equation}
{{q_{i}}  =  {s_{i} \pi {\xi} {\Delta}^{-1}}}, \qquad {{p_{j}L_{j}} = {{s_{j} {\sqrt{\kappa{(1 - \alpha)}}}}}}
\end{equation}

Remember that the eigenvalues of $Q^{b}L$ were constrained to be positive; therefore $s_{j} = L_{j}$.

Both $\kappa$ and $\Delta$ entered into the equations during the step of regularizing the continuum theory.  I regularized the bosonic matrix $Q^{b}$ by breaking the lattice into blocks of size $\kappa$, requiring $Q^{b}$ to be constant in each block, and requiring that the blocks were big enough that the diffusive Lagrangian could be factorized into a multiple of its values at each block. I regularized $Q^{f}$ by converting a momentum integral for  ${{\langle \vec{y} |}{G_{i}{(X)}}{| \vec{y} \rangle}}$ into a contour integral, and in the process introduced $\Delta$.    Therefore I redefine $\kappa$ in terms of a bosonic level spacing: $\sqrt{\kappa} \equiv {\pi \xi {\Delta}_{b}^{-1}}$, and I rename the fermionic level spacing to be ${\Delta}_{f}$.

In this notation, the final saddle point solutions are the set of all Hermitian matrices $Q_{0}^{f}$ and $Q_{0}^{b}$ that have the following eigenvalues:
\begin{equation}
{{q_{i}}  =  {s_{i} \pi {\xi} {\Delta}_{f}^{-1}}}, \qquad {{p_{j}} = {{ {{\pi \xi {\Delta}_{b}^{-1} \sqrt{1 - \alpha}}}}}}, \: {s_{i} = {\pm 1}} 
\label{FinalExtendedSaddlePointSolutions}
\end{equation}

One would hope that ${\Delta}_{b} = {\Delta}_{f}$.  If so, my earlier neglect of ${p_{j}L_{j}}$ in the sum $q_{i} + {p_{j}L_{j}}$ did not affect the saddle point solutions at all. 

Values of $Q^{b}$ and $Q^{f}$ which do not satisfy equation \ref{FinalExtendedSaddlePointSolutions} at every point in space are massive and are therefore strongly disfavored by the Lagrangian.  This results in a constraint that $Q^{b}$ and $Q^{f}$ satisfy the saddle point equations at every point in space.  The presence of a constraint implies that this model is a sigma model.

\subsection{\label{SigmaModelLagrangian1}The Sigma Model Lagrangian, Part I}

In the last section I derived the saddle point solutions, which define a sigma model with the constraint that the eigenvalues of $Q^{f}$ and $Q^{b}$ are fixed.  The saddle point constraint actually defines a continuous manifold at each point in space, since unitary transformations of $Q^{f}$ are not constrained by the saddle point equations, and similar degrees of freedom exist within $Q^{b}$.  When the source $J$ is zero and the energy operators ${\hat{E}}^{f}$ and ${\hat{E}}^{b}$ are both proportional to the identity, the model has degrees of freedom with exactly zero mass which correspond to varying $Q^{b}$ and $Q^{f}$ while both preserving their translational invariance and also obeying the saddle point constraint.  I will call these degrees of freedom zero modes.  There are also many other degrees of freedom which do not preserve the saddle point constraint, or exhibit a variation in space, or both.  These are all massive.

Now I begin a process of deriving the sigma model's action.  This amounts to making a perturbative expansion of the Lagrangian given in equation \ref{CorrectedQbQfLagrangian} around the saddle point manifold.  Clearly one must expand in $Q^{b}$ and $Q^{f}$.  However, as Kamenev and Mezard\cite{Kamenev99a} pointed out in their paper explaining how to do certain non-perturbative calculations with the replica technique, the zero modes of $Q^{f}$ and $Q^{b}$ can not be treated perturbatively because they are free to vary over the whole saddle point manifold.  Therefore I perform the decompositions $Q^{b} = {Q_{0}^{b} + {\tilde{Q}}^{b}}$ and $Q^{b} = {Q_{0}^{b} + {\tilde{Q}}^{b}}$.  $Q_{0}^{b}$ and $Q_{0}^{f}$ are the zero modes; the translationally invariant degrees of freedom which obey the saddle point constraint.  They will be treated non-perturbatively.  In contrast, ${\tilde{Q}}^{b}$ and ${\tilde{Q}}^{f}$ represent all the other degrees of freedom; i.e. the ones which disobey the saddle point constraint, or are not translationally invariant, or both.  These will be treated perturbatively.  

Because I chose a saddle point with the source $J $ equal to zero and both of the energy matrices ${\hat{E}}^{f}$ and ${\hat{E}}^{b}$ proportional to the identity, I also have to expand in these quantities.  I will call the actual deviation of ${\hat{E}}^{f}$ from the identity ${\hat{\omega}}^{f} \equiv {{\hat{E}}^{f} - {1 \otimes {\overline{E}}^{f}}} $, and similarly ${\hat{\omega}}^{b} \equiv {{\hat{E}}^{b} -  {1 \otimes {\overline{E}}^{b}}} $.

I decompose the Lagrangian into the part ${\mathcal{L}}_{0}$ depending on the zero modes $Q_{0}^{b}$ and $Q_{0}^{f}$ and the part depending on the other degrees of freedom. To second order in ${\tilde{Q}}^{b}$, ${\tilde{Q}}^{f}$, ${\hat{E}}^{f}$, ${\hat{E}}^{b}$, and $J$, the sigma model Lagrangian ${\mathcal{L}}_{\sigma}$ is:
\begin{eqnarray}
{\mathcal{L}}_{\sigma} & = & {\mathcal{L}}_{0} + {\imath \nu {{Tr}{({\tilde{Q}}^{b}L{({\hat{\omega}}^{b} - J)})}}} + {[{\frac{d^{2}\mathcal{L}}{{{d{\hat{E}}^{f}}{dQ^{f}}}}}]}_{{{\delta}_{E}={{\hat{\omega}}^{f}}}, {{\delta}_{f}={\tilde{Q}}^{f}}}  
+{[{\frac{d\mathcal{L}}{{{d{\hat{E}}^{f}}}}}]}_{{{\delta}_{E}={{\hat{\omega}}^{f}}}}
\nonumber \\
& + &  {[\frac{d\mathcal{L}}{d{(Q^{b}L)}}]}_{\delta = {\tilde{Q}}^{b}} +  {[\frac{d\mathcal{L}}{dQ^{f}}]}_{\delta ={\tilde{Q}}^{f}}
\nonumber \\
& + & {\frac{1}{2}  {[{\frac{d^{2}\mathcal{L}}{{d{(Q_{2}^{b}L)}}{d{(Q_{1}^{b}L)}}}}]}_{{{\delta}_{1}={{\tilde{Q}}_{1}^{b}L}}, {{\delta}_{2}={\tilde{Q}}_{2}^{b}L}}}  + {\frac{1}{2} {[{\frac{d^{2}\mathcal{L}}{{dQ_{2}^{f}{dQ_{1}^{f}}}}}]}_{{{\delta}_{1}={{\tilde{Q}}_{1}^{f}}}, {{\delta}_{2}={\tilde{Q}}_{2}^{f}}}  }
\nonumber \\
& + & {[{\frac{d^{2}\mathcal{L}}{{{d(Q^{b}L)}{dQ^{f}}}}}]}_{{{\delta}_{b}={{\tilde{Q}}^{b}L}}, {{\delta}_{f}={\tilde{Q}}^{f}}}
\label{SigmaLagrangian1} 
\end{eqnarray}

The Lagrangian ${\mathcal{L}}_{0}$ of the zero modes is given by:
\begin{eqnarray}
{\mathcal{L}}_{0}   & \equiv & {  -{\frac{ \nu \xi}{2} {{Tr}({Q_{0}^{f}}^{2}) }} + {{{\nu \xi{(1 - \alpha)}{{{Tr}\ln}{({\hat{E}}^{f} + {{\imath \xi  Q_{0}^{f}} })}}}}}} 
\nonumber \\ 
& - & {{ {  {\frac{\nu \xi}{2}{{{Tr}{(Q_{0}^{b}LQ_{0}^{b}L)}}}} { + {\imath \nu \xi {{Tr}{(Q_{0}^{b}{L {({{\hat{\omega}}^{b}} - J) / \xi}})}}} + {{\nu \xi {(1 - \alpha)}}{(\frac{\pi \xi}{{\Delta}_{b}})}^{2} {({Tr}\ln{( {Q_{0}^{b} L})})}}}}}} 
\nonumber \\
& + & {\frac{\nu \xi \alpha}{I^{b}}{{{Tr}\ln}{({{\hat{E}}^{f} + { \imath {\xi} {(Q_{0}^{f} + {Q_{0}^{b} L})} })}}}} 
\nonumber \\ & &
\label{ZeroModeLagrangian}
\end{eqnarray}

  The first derivatives $\frac{d\mathcal{L}}{d{(Q^{b}L)}}$ and $\frac{d\mathcal{L}}{dQ^{f}}$ required by equation \ref{SigmaLagrangian1} are given by equation \ref{LagrangianFirstDerivative}.  The second derivatives required by equation \ref{SigmaLagrangian1} are:
\begin{eqnarray}
{\frac{d^{2}\mathcal{L}}{{{d{\hat{E}}^{f}}{dQ^{f}}}}} 
& = & {- \imath {\nu} {\xi}^{2} {(1 - \alpha)} {Tr}{({\delta}_{f} {G^{f}{(Q_{0}^{f})}} {\delta}_{E} {G^{f}{(Q_{0}^{f})}})}}
\nonumber \\
& - & { \frac{\imath {\nu} {\xi}^{2} \alpha}{I^{b}} {\sum}_{j}  {Tr}{({\delta}_{f} {G^{f}{(Q_{0}^{f} + {q_{j}L_{j}})}} {\delta}_{E} {G^{f}{(Q_{0}^{f} + {q_{j}L_{j}})}})}}
\nonumber \\
{\frac{d^{2}\mathcal{L}}{{dQ_{2}^{f}{dQ_{1}^{f}}}}} 
& = & {-{\nu \xi {Tr}{({\delta}_{1} {\delta}_{2} )}}} + 
{\frac{\nu {D}^{f}}{2}{Tr}{({\vec{\nabla} {\delta}_{1}} \cdot {\vec{\nabla}{\delta}_{2}})}} 
\nonumber \\
& + & { {\nu} {\xi}^{3} {(1 - \alpha)} {Tr}{({\delta}_{1} {G^{f}{(Q_{0}^{f})}} {\delta}_{2} {G^{f}{(Q_{0}^{f})}})}}
\nonumber \\
& + & { \frac{{\nu} {\xi}^{3} \alpha}{I^{b}} {\sum}_{j}  {Tr}{({\delta}_{1} {G^{f}{(Q_{0}^{f} + {q_{j}L_{j}})}} {\delta}_{2} {G^{f}{(Q_{0}^{f} + {q_{j}L_{j}})}})}}
\nonumber \\
{\frac{d^{2}\mathcal{L}}{{{d(Q^{b}L)}{dQ^{f}}}}} & = & { \frac{{\nu} {\xi}^{3} \alpha}{I^{b}} {\sum}_{i} {Tr}{({{\delta}_{f}{(i\, i)}}\;{G_{i}^{f}{(Q_{0}^{f} + {Q_{0}^{b}L})}} {\delta}_{b} {G_{i}^{f}{(Q_{0}^{f} + {Q_{0}^{b}L})}})}} 
\nonumber \\
 {\frac{d^{2}\mathcal{L}}{{d{(Q_{2}^{b}L)}}{d{(Q_{1}^{b}L)}}}}  & =  & {{ {  {- \nu \xi {{{Tr}{({\delta}_{1} {\delta}_{2})}}}} {  - {{\pi}^{2} \nu {\xi}^{3} {\Delta}_{b}^{-2} {(1-\alpha)}}}{{Tr}({\delta}_{1}{(Q_{0}^{b}L)}^{-1}{\delta}_{2}{(Q_{0}^{b}L)}^{-1})}}}} 
 \nonumber \\
& + & {\frac{\nu {D}^{b}}{2}{Tr}{({\vec{\nabla} {\delta}_{1}} \cdot {\vec{\nabla} {\delta}_{2}})}}
 \nonumber \\
 & + & {{\frac{\nu {\xi}^{3} \alpha } {I^{b}}} {\sum}_{i} {Tr}{({\delta}_{1}  {G_{i}^{f}{({{q_{i}}} + { Q_{0}^{b} L })}} {\delta}_{2} {G_{i}^{f}{({{q_{i}}} + { Q_{0}^{b} L })}} )}}
 \label{TheHessian}
\end{eqnarray}

I have been able to get rid of the triple traces over the $v i j$ indices, reducing them to traces over either $v i$ or $v j$.  The notation ${{\delta}_{f}{(i\, i)}}$ means the diagonal elements of ${\delta}_{f}$.

Throughout the analysis of the sigma model Lagrangian ${\mathcal{L}}_{\sigma}$ I will use the two bases where $Q_{0}^{b}$ and $Q_{0}^{f}$ are diagonal to analyze all the matrix multiples and traces that I run into.  This will simplify a lot of calculations.  However it will also cause ${\mathcal{L}}_{\sigma}$ to depend on $Q_{0}^{b}$ and $Q_{0}^{f}$, because as these matrices vary over the saddle point manifold the bases in which they are diagonal also changes.

The derivative $\frac{d\mathcal{L}}{d{(Q^{b}L)}}$ is exactly zero because of the saddle point equation.  The $\frac{d\mathcal{L}}{dQ^{f}}$ is also drastically simplified by the saddle point, and reduces to a term proportional to ${\nabla}^{2}{\tilde{Q}}^{f}$.  This is a total derivative and therefore can be ignored. The ${{\frac{d\mathcal{L}}{{{d{\hat{E}}^{f}}}}}}$ term in equation \ref{SigmaLagrangian1} has no effect on ${\tilde{Q}}^{f}$ and ${\tilde{Q}}^{b}$, and therefore I have taken the liberty of moving this term from ${\mathcal{L}}_{\sigma}$ to the zero mode Lagrangian ${\mathcal{L}}_{0}$.  If I had not moved this term, then the zero mode Lagrangian would have contained ${\overline{E}}^{f}$'s instead of ${\hat{E}}^{f}$'s.

There are four terms in equation \ref{TheHessian} that are proportional to $\alpha$, all of which contain two instances of the propagator $G^{f}$.  In the last two terms, the two propagators share the same index $i$.  I will show in sections \ref{OneLoop} and \ref{SigmaModelLagrangian2} that the multiple of two propagators with a shared index evaluates to zero in the diffusive limit.  Therefore these two terms can be ignored.  On the other hand, the first two terms that are proportional to $\alpha$ have exact analogues that are proportional to $1 - \alpha$.  The only difference is that they have 
the argument $Q_{0}^{f} + {q_{j}L_{j}}$ instead of just $Q_{0}^{f}$.  If one neglects $q_{j}L_{j}$, equation \ref{TheHessian} loses all its dependence on $\alpha$.    So I will simplify my life by setting $\alpha = 0$, which is equivalent to neglecting $q_{j}L_{j}$.   This is consistent with my choce to neglect $q_{j}L_{j}$ when finding the fermionic saddle point.  As it turns out, the only possible ill effects upon the final sigma model Lagrangian will be changes of order $\alpha$ in the fermionic diffusion constant,the dependence on ${\hat{\omega}}^{f}$, and on the saddle point eigenvalues of $Q_{0}^{f}$.

I now combine equations \ref{SigmaLagrangian1} and \ref{TheHessian}.  I also use the fact that $L {(Q_{0}^{b}L)}^{-1}$ is proportional to the identity in the basis where $Q_{0}^{b}$ is diagonal.   Thus I obtain:

\begin{eqnarray}
{\mathcal{L}}_{\sigma} & = & {\mathcal{L}}_{0} + {\imath \nu {{Tr}{({\tilde{Q}}^{b}L{({\hat{\omega}}^{b} - J)})}}} - { \imath {\nu} {\xi}^{2}  {Tr}{({\tilde{Q}}^{f} \: {G^{f}{(Q_{0}^{f})}}  \: {\hat{\omega}}^{f} \: {G^{f}{(Q_{0}^{f})}})}}
\nonumber \\
& + & {-{\nu \xi {Tr}{({\tilde{Q}} \: {\tilde{Q}})}}} + 
{\frac{\nu {D}^{f}}{2}{Tr}{({\vec{\nabla} {\tilde{Q}}^{f} } \cdot {\vec{\nabla}{\tilde{Q}}^{f}})}} 
\nonumber \\
& + & { {\nu} {\xi}^{3}  {Tr}{({\tilde{Q}}^{f} \: {G^{f}{(Q_{0}^{f})}} \: {\tilde{Q}}^{f} \: {G^{f}{(Q_{0}^{f})}})}}
\nonumber \\
  & -  & {{ {  { \nu \xi {{{Tr}{({\tilde{Q}}^{b}L {\tilde{Q}}^{b}L)}}}} {  -  \nu {\xi}  }}{{Tr}({{\tilde{Q}}^{b}} \:{{\tilde{Q}}^{b}})}}} 
 + {\frac{\nu {D}^{b}}{2}{Tr}{({\vec{\nabla} {\tilde{Q}}^{b}L \cdot {\vec{\nabla} {\tilde{Q}}^{b}L})}}}
 \label{SigmaLagrangian2}
\end{eqnarray}

Equation \ref{SigmaLagrangian2} contains two terms of the form ${{Tr}{(A^{1} \: {G^{f}{(Q_{0}^{f})}}  \: A^{2} \: {G^{f}{(Q_{0}^{f})}})}}$, and I need to find the low momentum behavior of these terms.  I start by decomposing the trace in its $i$ index.  Because I have chosen to use the bases where $Q_{0}^{f}$ and $Q_{0}^{b}$ are diagonal, $G^{f}$ is also diagonal.  Therefore:
\begin{eqnarray}
& & {{Tr}{(A^{1} \: {G^{f}{(Q_{0}^{f})}}  \: A^{2} \: {G^{f}{(Q_{0}^{f})}})}}
\nonumber \\
& = & {\sum_{i_{1}, i_{2}} {A^{1}{(i_{1} i_{2})}} \: {G_{i_{2}}^{f}{(Q_{0}^{f})}} \: {A^{2}{(i_{2} i_{1})}} \: {G_{i_{1}}^{f}{(Q_{0}^{f})}}}
\end{eqnarray}

Next I expand the $A$'s into their momentum components, using the identities $A \equiv {\int \frac{d^{D}\vec{t}}{{(2 \pi)}^{D}} {A{(\vec{t}\:)}}} $,  ${\langle \vec{s} | {A{(\vec{t}\:)}} | \vec{u} \rangle} = {{\delta}^{D}{(\vec{s} + \vec{t} - \vec{u})}}$, and $1 = {\int \frac{d^{D}\vec{t}}{{(2 \pi)}^{D}} {{| \vec{t} \rangle}{\langle \vec{t} |}}}$.  Then the trace turns into:

\begin{eqnarray}
& & {{Tr}{(A^{1} \: {G^{f}{(Q_{0}^{f})}}  \: A^{2} \: {G^{f}{(Q_{0}^{f})}})}}
\nonumber \\
& = & \int {\frac{d^{D}\vec{t}}{{(2 \pi)}^{D}} \frac{d^{D}\vec{u}}{{(2 \pi)}^{D}} \frac{d^{D}\vec{v}}{{(2 \pi)}^{D}} \frac{d^{D}\vec{w}}{{(2 \pi)}^{D}}} \sum_{i_{1}, i_{2}} {A^{1}{(i_{1} i_{2}, \vec{t}\:)}} {\langle {\vec{v} + \vec{t}} \:|{G_{i_{2}}^{f}} | \vec{w} \rangle} {A^{2}{(i_{2} i_{1}, \vec{u})}} {\langle {\vec{w} + \vec{u}} | G_{i_{1}}^{f} | \vec{v} \rangle}
\nonumber \\
& = & 
\int {\frac{d^{D}\vec{t}}{{(2 \pi)}^{D}}   \frac{d^{D}\vec{w}}{{(2 \pi)}^{D}}} \sum_{i_{1}, i_{2}} {A^{1}{(i_{1} i_{2}, \vec{t}\:)}} \; {\langle {\vec{w}} |{G_{i_{2}}^{f}} | \vec{w} \rangle} \; {A^{2}{(i_{2} i_{1}, -\vec{t}\:)}} \; {\langle {\vec{w} - \vec{t}} \: | G_{i_{1}}^{f} | {\vec{w} - \vec{t}\:} \rangle}
\nonumber \\
& = & 
\int {\frac{d^{D}\vec{t}}{{(2 \pi)}^{D}}  \sum_{i_{1}, i_{2}} {A^{1}{(i_{1} i_{2}, \vec{t}\:)}} \; {A^{2}{(i_{2} i_{1}, -\vec{t}\:)}}}  \; {\int \frac{d^{D}\vec{w}}{{(2 \pi)}^{D}}  {\langle {\vec{w}} |{G_{i_{2}}^{f}} | \vec{w} \rangle} \; {\langle {\vec{w} - \vec{t}} \: | G_{i_{1}}^{f} | {\vec{w} - \vec{t}\:} \rangle} }
\nonumber \\ & &
\label{OneLoop1}
\end{eqnarray}

The integral over $\vec{w}$ describes two Green's functions forming a loop which carries a total momentum equal to $\vec{t}$.  $\vec{t}$ is the momentum of the $A$ matrices, which correspond to ${\tilde{Q}}^{f}$ and ${\hat{\omega}}^{f}$.   A careful calculation of this integral involves some complexities, and will be the subject of the next section.

\subsection{\label{OneLoop}Calculating the One Loop Integral}
In this section I calculate the low momentum behavior of the one loop interaction of two fields, one field with momentum $\vec{k}$ and the other with momentum $-\vec{k}$.  ($\vec{k}$ corresponds to $\frac{1}{2}\vec{t}$ in the last section.) The fields are connected by two Green's functions with the form given by equation \ref{GreensFunctionDefinitionComponentwise}.  The loop integral is:
\begin{eqnarray}
M & \equiv & {\int \frac{d^{D}\vec{p}}{{(2 \pi)}^{D}} \frac{1}{E_{1} - {E{(\vec{p} + \vec{k})}} + {\imath \xi I_{1}} } \frac{1}{E_{2} - {E{(\vec{p} - \vec{k})}} + {\imath \xi I_{2}} }}
\nonumber \\
& = & 
\int {d \epsilon} \frac{d^{D}\vec{p}}{{(2 \pi)}^{D}} {\delta{(\epsilon - {E{(\vec{p})}})}} \frac{1}{E_{1} - \epsilon - {({E{(\vec{p} + \vec{k})}} - \epsilon )}+ {\imath \xi I_{1}} }
\nonumber \\
&  & \qquad \times \frac{1}{E_{2} - \epsilon - {({E{(\vec{p} - \vec{k})}} - \epsilon )} + {\imath \xi I_{2}} }
\label{OneLoop2}
\end{eqnarray}

The integral over energy $\epsilon$ has zero as its lower limit and infinity as its upper limit, but if I assume that the imaginary factors $\xi I_{i}$ are much smaller than the energies $E_{i}$, then the lower limit can be extended down to minus infinity, so that I have a contour integral. This is just the diffusive approximation which I introduced in section \ref{ContinuumSaddlePoint}.  One immediate result of the diffusive approximation is that if ${{sign}{(I_{1})}} = {{sign}{(I_{2})}} $, then the loop integral evaluates to exactly zero.  Another consequence is that when $\vec{k} = 0$ the loop integral $M$ can be computed exactly; it comes out to:
\begin{equation} M{(\vec{k} = 0)} = {\frac{2 \imath \pi {{sign}{(I_{1})}} {\Theta{(I_{1}I_{2})}}}{\Delta {(\omega + {\imath \xi {(I_{1} - I_{2})})}}}}
\label{LoopAtZeroMomentum}
\end{equation}  This expression is valid for any value of the energy difference $\omega \equiv {E_{1} - E_{2}}$. $\Delta$ is the level spacing which I defined in equation \ref{LevelSpacingDefinition}.

Note two properties of the loop integral:
\begin{itemize}
\item If $\omega = 0$, the loop integral has an interesting symmetry: ${M{(E,  I_{1},  I_{2}, \vec{k})}} = {M{(E, \ I_{2},  I_{1}, -\vec{k})}}$.  In systems with translational symmetry, $M$ will depend on $k^{2}$ instead of $\vec{k}$; therefore in these systems ${M{(E,  I_{1},  I_{2}, k^{2})}} = {M{(E,  I_{2}, I_{1}, k^{2})}}$.  
\item In systems which are invariant under the parity transformation $\vec{p} \rightarrow {-\vec{p}}$, the loop integral is real when $\omega = 0$ and $I_{1} = {-I_{2}}$.
\end{itemize}

I now develop a perturbative expansion of $M$ in the parameters $k^{2}$ and $\omega$, expanding around the point $\omega = 0, \vec{k} = 0$.  This approach is justified for computing low momentum contributions to the loop integral.  In particular, consider the term which is first order in $k^{2}$ and zeroth order in $\omega$.  The most general form for this term that is possible is ${\Theta{(I_{1}I_{2})}} k^{2} d{(E, I_{1},  I_{2})}$, where $d$ is some function that I have not yet determined.  I will show in equation \ref{MContourResult} that each term in the loop integral contains at least one power of ${(I_{1} - I_{2})}^{-1}$.  Since ${d{(E, I_{1},  I_{2}, k^{2})}} = {d{(E,  I_{2},  I_{1}, k^{2})}}$, any analytic function $d$ must have an even number of powers of $\xi (I_{1} - I_{2})$.  These points lead me to write the first order correction in $k^{2}$ as:
\begin{equation}
{\frac{2 {\pi}^{2}{\Theta{(I_{1}I_{2})}}}{{\xi} {{\Delta}^{2}{(E)}}  {(I_{1} - I_{2})}^{2}} k^{2} d{(E,  I_{1}, I_{2})}}
\label{GeneralDParameterization}
\end{equation}
There has been no loss of generality except that required by the points I just raised.  The extra factor of $2 {\pi}^{2} \xi / {\Delta}^{2}$ redefines $d$ in order to give it the correct dimensions for a diffusion constant.  The second property that I listed earlier implies that $d$ is real at $I_{1} = {- I_{2}}$, which makes it natural to not include an explicit $i$ in equation \ref{GeneralDParameterization}.

I can determine the other low order terms in the Taylor series expansion from equation \ref{LoopAtZeroMomentum}, which contains all the terms in the Taylor series which have $k^{2}$ to the zeroth power.  Doing this analysis, I find that to first order in $k^{2}$ and $\omega$, but neglecting the term which is proportional to $\omega k^{2}$, the loop integral is:
\begin{eqnarray}
{M} & = & {{\frac{2 \pi {\Theta{(I_{1}I_{2})}}}{\Delta {\xi} {|I_{1} - I_{2}|}}} + {\frac{2 {\pi} {\Theta{(I_{1}I_{2})}}}{{\Delta} {\xi}^{2} {(I_{1} - I_{2})}^{2}} {({\frac{\pi \xi {d{(E,  I_{1},  I_{2})}}}{\Delta} k^{2} } + {\imath \,{{sign}{(I_{1})}}{(E_{1} - E_{2})} }) } }}
\nonumber \\ & &
\label{LoopInt1}
\end{eqnarray}

In summary, general arguments and the $\vec{k} = 0$ integral have allowed me to evaluate the loop integral $M$ at the lowest perturbative orders.  The only thing I did not calculate was the exact expression for the diffusion function $d$.

Now I will show how to do a thorough calculation of the loop integral given in equation \ref{OneLoop2}, with no assumptions about the size of $\omega$ other than that it is small compared to $I_{i}$.  Examining equation \ref{OneLoop2}, one finds that the interesting parts of the integrand - the poles - are very small unless either  $ E_{1} \approx {E{(\vec{p} + \vec{k})}} \approx {E{(\vec{p})}} = \epsilon$ or $E_{2} \approx {E{(\vec{p} - \vec{k})}} \approx {E{(\vec{p})}} = \epsilon$.  Therefore ${E{(\vec{p} \pm \vec{k})}} - \epsilon$ is bounded by either $\omega$ or the change which $\vec{k}$ makes in the kinetic energy, whichever is bigger.   I assume that these quantities are small compared to $I_{i}$; therefore ${E_{i} - \epsilon + {\imath \xi I_{i}} }  \gg {({E{(\vec{p} \pm \vec{k})}} - \epsilon )} $. Doing a Taylor expansion in the small parameter and using the notation $G_{i} \equiv {(E_{i} - \epsilon + {\imath \xi I_{i}})}^{-1}$, I obtain:
\begin{eqnarray}
M & = & \int {d \epsilon} \frac{d^{D}\vec{p}}{{(2 \pi)}^{D}} {\delta{(\epsilon - {E{(\vec{p})}})}} 
\nonumber \\
& \times &
{(G_{1} + {G_{1} {({E{(\vec{p} + \vec{k})}} - \epsilon )} G_{1}} + {G_{1} {({E{(\vec{p} + \vec{k})}} - \epsilon )} G_{1} {({E{(\vec{p} + \vec{k})}} - \epsilon )} G_{1}})}^{-1}
\nonumber \\
& \times & 
{(G_{2} + {G_{2} {({E{(\vec{p} - \vec{k})}} - \epsilon )} G_{2}} + {G_{2} {({E{(\vec{p} - \vec{k})}} - \epsilon )} G_{2} {({E{(\vec{p} - \vec{k})}} - \epsilon )} G_{2}})}^{-1} 
\nonumber \\ & &
\end{eqnarray}

I again assume that $\vec{k}$ makes only small changes in the kinetic energy of $\vec{p} + \vec{k}$.  Expanding the kinetic energy function perturbatively in powers of $\vec{k}$ gives:
\begin{equation}
{E{(\vec{p} - \vec{k})}} \simeq {\epsilon + {{\vec{\nabla}E} \cdot \vec{k}} + {\frac{1}{2} \vec{k} \cdot {\ddot{\nabla}E} \cdot \vec{k} } }
\end{equation}
The notation $\ddot{\nabla}$ signifies the tensor second derivative $\sum_{a,b} {| \hat{a} \rangle} \frac{d^{2}}{{dx_{a}} {dx_{b}}} {\langle \hat{b} |}$, while the notation $\vec{\nabla}$ just makes the fact that the gradient is a vector even more explicit.

Keeping only terms to second order in $\vec{k}$, I obtain:
\begin{eqnarray}
M  & = & {\int {d \epsilon} \frac{d^{D}\vec{p}}{{(2 \pi)}^{D}} {\delta{(\epsilon - {E{(\vec{p})}})}} }   \times  { {[{G_{1}G_{2}} + {\frac{1}{2}{({G_{1}^{2}G_{2}} + {G_{1}G_{2}^{2}})\vec{k} \cdot {\ddot{\nabla}E} \cdot \vec{k}}}}}
\nonumber \\
& + & {{ {{({G_{1}^{3}G_{2}} + {G_{1}G_{2}^{3}} - {G_{1}^{2} G_{2}^{2}})}{({\vec{\nabla}E}\cdot \vec{k})}^{2} } + {{({G_{1}^{2}G_{2}} - {G_{1}G_{2}^{2}})} {{\vec{\nabla}E}\cdot \vec{k}}}]}} 
\end{eqnarray}

The integral's $\vec{p}$ dependence is now contained entirely in the derivatives of $E$, so I am able to average these derivatives over the constraint $\epsilon = {E{(\vec{p})}}$. I define  the average as:
\begin{eqnarray}
{\langle {\mathcal{O}{(E)}} \rangle} \equiv {{\Delta{(E)}}  {\int  \frac{d^{D}\vec{p}}{{(2 \pi)}^{D}} {\delta{(\epsilon - {E{(\vec{p})}})}} } {\mathcal{O}{(\vec{p})}}} 
\end{eqnarray}
Using this definition, I can finally do the integral over $\vec{p}$:
\begin{eqnarray}
M  =  {\int {d \epsilon} {{\Delta^{-1}{(\epsilon)}}}} & \times & {{  { {[{G_{1}G_{2}} + {\frac{1}{2}{({G_{1}^{2}G_{2}} + {G_{1}G_{2}^{2}})\vec{k} \cdot {\langle {\ddot{\nabla}E} \rangle} \cdot \vec{k}}}}}}}
\nonumber \\
& + & {{({G_{1}^{3}G_{2}} + {G_{1}G_{2}^{3}} - {G_{1}^{2} G_{2}^{2}})}{(\vec{k} \cdot {\langle {\vec{\nabla}E} {\vec{\nabla}E} \rangle}\cdot \vec{k})} } 
\nonumber \\
& + & {{({G_{1}^{2}G_{2}} - {G_{1}G_{2}^{2}})} {{\langle {\vec{\nabla}E} \rangle }\cdot \vec{k}}}] 
\nonumber \\ 
& \equiv &
M_{0} + {\frac{1}{2}\vec{k} \cdot \ddot{M_{1}} \cdot \vec{k}} + {\vec{k} \cdot \ddot{M_{2}} \cdot \vec{k}} + {\vec{M_{3}} \cdot \vec{k}}
\end{eqnarray}

In the last line I have simply broken up the one loop integral into its four parts.  \linebreak 

Before performing the integral over $\epsilon$, let me briefly estimate the  effect of taking a derivative with respect to $\epsilon$. 
I will use a simple model, choosing $E = {\frac{E_{0}}{2} k^{2}}$.  With this model 
$ {\frac{d}{d\epsilon}{({{\Delta}^{-1}{(\epsilon)}{\langle {\ddot{\nabla}E} \rangle}})}} = {\frac{d^{2}}{d{\epsilon}^{2}}{({{\Delta}^{-1}{(\epsilon)}{\langle {\vec{\nabla}E} {\vec{\nabla}E}\rangle}})}} = {\frac{E_{0} {(D-2)}}{2 E}}$.  This indicates that taking a derivative corresponds (at least when estimating orders of magnitude) to dividing by the kinetic energy.
 
I now turn to the task of performing the contour integral over $\epsilon$.  Where one Green's function occurs to the $n$-th power and the other to the $m$-th power, the contour integral will result in an expression like this: $\frac{d^{n-1}}{d{\epsilon}^{n-1}}{[{{\Delta^{-1}{(\epsilon)}}} {\langle {\mathcal{O}{(\epsilon)}} \rangle} G^{m}{(\epsilon)}]}$.  If the derivative acts on $G^{m}$, the resulting term will have an extra power of ${(E_{1} - E_{2} + \imath \xi {(I_{1} - I_{2})})}^{-1}$.  On the other hand, if the derivative acts on the combination ${{\Delta^{-1}{(\epsilon)}}} {\langle {\mathcal{O}{(\epsilon)}} \rangle}$, the effect will be more or less a division by a constant proportional to the kinetic energy.    I will simplify my life by assuming that the kinetic energy is much smaller than the energy difference $ {(E_{1} - E_{2} + \imath \xi {(I_{1} - I_{2})})}$, and therefore neglect derivatives of the Green's functions.  (I could evaluate $M$ without making this approximation, but the resulting expressions are considerably more complicated.) Throwing out all terms with extra powers of $ {(E_{1} - E_{2} + \imath \xi {(I_{1} - I_{2})})}$ in the denominator, I find that the contour integrals are:
\begin{eqnarray}
M_{0} & = & {2 \pi \imath {(E_{1} - E_{2} + \imath \xi {(I_{1} - I_{2})})}^{-1} {{\Delta}^{-1}{({E_{1} + \imath \xi I_{1}})}}}
\nonumber \\
\ddot{M_{1}} & = & {- 2 \pi \imath {(E_{1} - E_{2} + \imath \xi {(I_{1} - I_{2})})}^{-1} {\frac{d}{d\epsilon}{({{\Delta}^{-1}{(\epsilon)}{\langle {\ddot{\nabla}E} \rangle}})}}{\mid}_{\epsilon = {E_{1} + \imath \xi I_{1}}}}
\nonumber \\
\ddot{M_{2}} & = & {\pi \imath {(E_{1} - E_{2} + \imath \xi {(I_{1} - I_{2})})}^{-1} {\frac{d^{2}}{d{\epsilon}^{2}}{({{\Delta}^{-1}{(\epsilon)}{\langle {\vec{\nabla}E} {\vec{\nabla}E}\rangle}})}}{\mid}_{\epsilon = {E_{1} + \imath \xi I_{1}}}}
\nonumber \\
\vec{M_{3}} & = & {- 2 \pi \imath {(E_{1} - E_{2} + \imath \xi {(I_{1} - I_{2})})}^{-1} {\frac{d}{d{\epsilon}}{({{\Delta}^{-1}{(\epsilon)}{\langle {\vec{\nabla}E} \rangle}})}}{\mid}_{\epsilon = {E_{1} + \imath \xi I_{1}}}}
\label{MContourResult}
\end{eqnarray}

These are the results for ${{{sign}{(I_{1})}} = 1}, {{{sign}{(I_{2})}} = -1} $.  When the signs of the imaginary parts $I_{1}$ and $I_{2}$ are reversed, the results remain the same except for two things: each of the four equations is multiplied by $-1$, and also $\Delta$ and the derivatives are evaluated at ${E_{2} + \imath \xi I_{2}}$ instead of ${E_{1} + \imath \xi I_{1}}$.  This second difference should be substantial because the typical scale of change of $\Delta$ is proportional to the kinetic energy.  However, if we assume yet again that $\omega$ is small compared to $\xi I_{i}$, $M$'s symmetry under interchange of $I_{1}$ and $I_{2}$ ensures that changing the sign of the $I$'s has no effect on equations \ref{MContourResult} except to change their sign.  For simplicity I will assume that $\omega$ is small compared to $\xi I_{i}$, but one could avoid this assumption if one wished to treat the case of large $\omega$.  

I assume that the kinetic energy is spherically symmetric; $ \vec{M_{3}}$ is zero and $\ddot{M_{1}}$ and $\ddot{M_{2}}$ are both proportional to the identity.  Therefore I can write:
\begin{eqnarray}
{M} & = & {\frac{2 \imath \pi {{sign}{(I_{1})}} {\Theta{(I_{1}I_{2})}}}{\Delta {(E_{1} - E_{2} + \imath \xi {(I_{1} - I_{2})})}} {(1 + \frac{\pi D k^{2}}{\Delta {|I_{1} - I_{2} |}})}}
\nonumber \\
D & \equiv & {\frac{\Delta}{2 \pi}{|I_{1} - I_{2}|} \Delta {(-{\frac{d}{d\epsilon}{({{\Delta}^{-1}{(\epsilon)}{\langle {\ddot{\nabla}E} \rangle}})}} + {\frac{\Delta}{\pi \xi}\frac{d^{2}}{d{\epsilon}^{2}}{({{\Delta}^{-1}{(\epsilon)}{\langle {\vec{\nabla}E} {\vec{\nabla}E}\rangle}})}}) } }
\nonumber \\ & & 
\label{MoreExactSingleLoop}
\end{eqnarray}

I once again use the approximation that $ {|\omega|} \ll {\xi {|I_{1} - I_{2}|}}$ to obtain the formula: 
\begin{eqnarray}
{M} & = & {{\frac{2 \pi {\Theta{(I_{1}I_{2})}}}{\Delta \xi {|I_{1} - I_{2}|}}} + {\frac{2 \pi {\Theta{(I_{1}I_{2})}}}{\Delta {\xi}^{2} {(I_{1} - I_{2})}^{2}} {({\frac{\pi \xi D}{\Delta} k^{2}} + {\imath \,{{sign}{(I_{1})}}{(E_{1} - E_{2})} }) } }}
\nonumber \\ & & 
\label{LessExactSingleLoop}
\end{eqnarray}
This is the same result as in equation \ref{LoopInt1}, except that now I have an explicit expression for the diffusion function $d{(E, \xi I_{1}, \xi I_{2})}$, and now $d$ is manifestly real.  In the literature it is conventional to use a diffusion constant (not function) $D$ which is real, is constant under interchange of $I_{1}$ and $I_{2}$, and does not depend on the energy $E$.  The above derivation shows that these assumptions can be justified only in very specific limits, and that in general $D$ is correctly understood as a function, not a constant.  The diffusion constant $D$ found in the literature is best considered as a phenomenological constant appearing in the effective Lagrangian and nothing more; the expression derived here should not be taken seriously except to demonstrate the overall physics.  In fact the main value of this derivation was not its final result, but instead its explanation of how one would derive the loop integral $M$ for large $\omega$.

A few words about the validity of this integral.  Because the integral includes the factor ${\Theta{(I_{1}I_{2})}}$ and the saddle point solutions imply that $I_{1} = {\pm I_{2}}$,  the quantity $I_{1} - I_{2}$ is proportional to $I_{1}$.  In the course of evaluating the one loop integral I have made the following approximations and assumptions:
\begin{enumerate}
\item Spherical symmetry.
\item The diffusive approximation: ${\xi I_{i}} \ll E_{i}$.
\item \label{SmallKAssumption} $\vec{k}$ makes only small changes in the kinetic energy, allowing a perturbative expansion in $\vec{k}$.  The change in kinetic energy can be estimated as $D k^{2}$, which has a typical value given by the Thouless energy $E_{c}$.  
\item $E_{c} \ll {\xi I_{i}} \ll E_{i}$.
\item $\omega \ll {\xi I_{i}} \ll E_{i}$
\end{enumerate} 
 
\subsection{\label{SigmaModelLagrangian2}The Sigma Model Lagrangian, Part II}
I now return to the task of evaluating the effective action of the Goldstone bosons.  I start by substituting the saddle point solutions, given by equation \ref{FinalExtendedSaddlePointSolutions}, into equation \ref{LessExactSingleLoop} for the one loop integral.  Because I chose to evaluate the saddle point at ${\hat{E}}^{f} = {1 \otimes {\overline{E}}^{f}}$; the energy difference occuring in equation \ref{LessExactSingleLoop} is exactly zero.  The result of this substition is:

\begin{eqnarray}
{M} & = & {{{ {\Theta{({s(i_{1})} {s(i_{2})})}} \: {\xi}^{-2} }}  {(1 + \frac{  {Dk^{2} }}{ 2 {\xi}}  )} }
\end{eqnarray}

I remind the reader that the $s$ variables are just the signs of the fermionic saddle points.  

Substituting this result into equation \ref{OneLoop1}, I obtain:
\begin{eqnarray}
& & {{Tr}{(A^{1} \: {G^{f}{(Q_{0}^{f})}}  \: A^{2} \: {G^{f}{(Q_{0}^{f})}})}}
\nonumber \\
& = & {{\xi}^{-2} \int {\frac{d^{D}\vec{t}}{{(2 \pi)}^{D}}  \sum_{i_{1}, i_{2}} {A^{1}{(i_{1} i_{2}, \vec{t}\:)}} \; {A^{2}{(i_{2} i_{1}, -\vec{t}\:)}}}  \; {{{ {\Theta{({s(i_{1})} {s(i_{2})})}} }}  {(1 + \frac{{Dt^{2} }}{ 2 {\xi}}  )} }}
\nonumber \\ & &
\label{OneLoop3}
\end{eqnarray}
I have taken the liberty of moving the factor of one half in the equation $\vec{k} = {\frac{1}{2} \vec{t}}$ into the diffusion constant $D$.

Because the $s$ variables have unit magnitude, $ {\Theta{({s(i_{1})} {s(i_{2})})}} = {\frac{1}{2}{(1 - {s{(i_{1})} s{(i_{2})}} )}} $.  I define the sign matrix $S$ of the saddle points as the diagonal matrix whose entries are composed of the $s$ variables:
\begin{equation} 
{S{(i_{1} i_{2})}} \equiv {s{(i_{1})} {\delta {(i_{1} i_{2})}} }
\label{SignMatrixDefinition}
\end{equation}
The new sign matrix allows me to rewrite equation \ref{OneLoop3} in matrix notation:
\begin{eqnarray}
& &{{Tr}{(A^{1} \: {G^{f}{(Q_{0}^{f})}}  \: A^{2} \: {G^{f}{(Q_{0}^{f})}})}}
\nonumber \\
& = & \frac{1}{2 {\xi}^{2}} \int \frac{d^{D}\vec{t}}{{(2 \pi)}^{D}}  {(1 + \frac{ {Dt^{2} }}{ \pi {\xi}}  )}  {[ {{Tr}{({A^{1}{(\vec{t}\:)}} \; {A^{2}{( -\vec{t}\:)}} )}} - {{Tr}{({A^{1}{(\vec{t}\:)}} \;S \; {A^{2}{( -\vec{t}\:)}} \; S )}}]}
\nonumber \\
& & 
\label{OneLoop4}
\end{eqnarray}
The traces in equation \ref{OneLoop4} are over the $i$ index only.  I exploit the fact that $A^{1}$ has a momentum of $\vec{k}$ and $A^{2}$ has a momentum of $-\vec{k}$ to turn the $t^{2}$ into the $\vec{\nabla} \cdot \vec{\nabla}$ operator.  Thus I am able to return to original trace over both the volume index $v$ and the $i$ index:
\begin{eqnarray}
& & {{Tr}{(A^{1} \: {G^{f}{(Q_{0}^{f})}}  \: A^{2} \: {G^{f}{(Q_{0}^{f})}})}}
\nonumber \\
& = & {{\frac{1}{2 {\xi}^{2}}{{Tr}{({{A^{1}A^{2}} - {A^{1}SA^{2}S}})}}} + {\frac{ D}{ 4 {\xi}^{3}}{Tr}{({{{\vec{\nabla}}A^{1}} \cdot {{\vec{\nabla}}A^{2}}} - {{{\vec{\nabla}}A^{1}} \: S \cdot {{\vec{\nabla}}A^{2}} \: S} )}}}
\nonumber \\ & &
\label{OneLoop5}
\end{eqnarray}

I now plug this result into expression \ref{SigmaLagrangian2} for the Lagrangian of the sigma model:
\begin{eqnarray}
{\mathcal{L}}_{\sigma} & = & {\mathcal{L}}_{0} + {\imath \nu {{Tr}{({\tilde{Q}}^{b}L{({\hat{\omega}}^{b} - J)})}}} - { \frac{\imath \nu}{2} {Tr}{({{{\tilde{Q}}^{f} {\hat{\omega}}^{f}} - {{\tilde{Q}}^{f} S \: {\hat{\omega}}^{f} S}})}}
\nonumber \\
& + & {\frac{\nu {D}^{b}}{2}{Tr}{({\vec{\nabla} {\tilde{Q}}^{b}L \cdot {\vec{\nabla} {\tilde{Q}}^{b}L})}}} 
\nonumber \\
& + & {  \frac{\nu D}{4}  {Tr}{({{\vec{\nabla} {\tilde{Q}}^{f} } \cdot {\vec{\nabla}{\tilde{Q}}^{f}}} - {{\vec{\nabla} {\tilde{Q}}^{f} } S \cdot {\vec{\nabla}{\tilde{Q}}^{f}}} S)}} 
+ {\frac{\nu {D}^{f}}{2}{Tr}{({{\vec{\nabla} {\tilde{Q}}^{f} } \cdot {\vec{\nabla}{\tilde{Q}}^{f}}})}}
\nonumber \\
  & -  & {{ {  { \nu \xi {{{Tr}{({{\tilde{Q}}^{b} \: {\tilde{Q}}^{b}} + {{\tilde{Q}}^{b}L {\tilde{Q}}^{b}L})}}}} }} }
 - {\frac{\nu \xi}{2} {Tr}{({{\tilde{Q}}^{f} \: {\tilde{Q}}^{f}} + {{\tilde{Q}}^{f}S{\tilde{Q}}^{f}S})}} 
 \label{SigmaLagrangian3}
\end{eqnarray}

The terms on the last line of equation \ref{SigmaLagrangian3} are mass terms.  It is instructive to analyze them in terms of the individual matrix elements of ${\tilde{Q}}^{b}$ and ${\tilde{Q}}^{f}$.  The ${\tilde{Q}}^{f}$ mass term reads:
\begin{equation}
- {\frac{\nu \xi}{2} \sum_{v i_{1} i_{2}} {(1 + {{s{(i_{1})}} {s{(i_{2})}}})} \: {|{\tilde{Q}}^{f}{(v i_{1} i_{2})}|}^{2}   }  
\end{equation}
I have used the fact that ${\tilde{Q}}^{f}$ is Hermitian.  The mass term for ${\tilde{Q}}^{b}$ is entirely analogous, with ${s{(i)}}$ replaced by ${L{(j)}}$. When ${{s{(i_{1})}} \neq {s{(i_{2})}}}$, the term in parenthesis evaluates to zero.  The corresponding matrix elements ${{\tilde{Q}}^{f}{(v i_{1} i_{2})}} \; \forall {{s{(i_{1})}} \neq {s{(i_{2})}}}$  and ${{\tilde{Q}}^{b}{(v j_{1} j_{2})}} \; \forall {{L{(j_{1})}} \neq {L{(j_{2})}}}$ have very small masses caused only by their kinetic energy.  These are precisely the degrees of freedom which do not change the eigenvalues of $Q^{f} = {Q_{0}^{f} + {\tilde{Q}}^{f}}$ and $Q^{b}L = {Q_{0}^{b}L + {\tilde{Q}}^{b}L}$.

The other matrix elements ${{\tilde{Q}}^{f}{(v i_{1} i_{2})}} \; \forall {{s{(i_{1})}} = {s{(i_{2})}}}$  and ${{\tilde{Q}}^{b}{(v j_{1} j_{2})}} \; \forall {{L{(j_{1})}} = {L{(j_{2})}}}$ are the very massive degrees of freedom.  These are precisely the degrees of freedom which change the eigenvalues of $Q^{f} = {Q_{0}^{f} + {\tilde{Q}}^{f}}$ and $Q^{b}L = {Q_{0}^{b}L + {\tilde{Q}}^{b}L}$.  If we assume that energy operators ${\hat{\omega}}^{f}$  and ${\hat{\omega}}^{b}$ and the Thouless energies ${D^{f} k^{2}}$ and ${D^{b} k^{2}}$ are all small compared to $\xi$, then we can immediately integrate out these very massive degrees of freedom, and fix the eigenvalues to have exactly the values prescribed by the saddle point solutions.  Thus we arrive at the sigma model.    

I will avoid integrating out the traslationally invariant degrees of freedom ${{\tilde{Q}}^{f}{(\vec{p} = \vec{0})}}$ and ${{\tilde{Q}}^{b}{(\vec{p} = \vec{0})}}$.  I make this choice because some observables (including the two point correlator) evaluate to zero in the zeroth order saddle point approximation.  To evaluate the higher order corrections to the saddle point approximation, one must postpone this integration until the last moment.  Instead I will explicitly separate the $\vec{p} = \vec{0}$ degrees of freedom from the $\vec{p} \neq \vec{0}$ degrees of freedom with the notation change ${{\tilde{Q}}^{f} \rightarrow {{{\tilde{Q}}^{f}{(\vec{p} = \vec{0})}} + {{\tilde{Q}}^{f}}}}, \: {{\tilde{Q}}^{b} \rightarrow {{{\tilde{Q}}^{b}{(\vec{p} = \vec{0})}} + {{\tilde{Q}}^{b}}}}$.  Henceforth ${\tilde{Q}}$ refers only to the degrees of freedom which are translationally invariant, while ${{\tilde{Q}}{(\vec{p} = \vec{0})}}$ refers to the degrees of freedom which are translationally invariant but do not obey the saddle point constraint.

The fermionic matrix ${\tilde{Q}}^{f}$ contains ${{(\frac{V}{4}{({Tr}{(S)})}^{2}} + {\frac{1}{4}{I^{f}}^{2}}}$ degrees of freedom which violate the constraint on the eigenvalues, and after integration each of these degrees of freedom multiplies the generating function $\overline{Z}$ by a factor proportional to $\frac{1}{\sqrt{\nu \xi}}$.  In contrast, one can estimate that the other degrees of freedom each multiply the generating function by a bigger factor: either $\frac{1}{\sqrt{\nu \omega}}$ or $\frac{1}{\sqrt{\nu E_{c}}}$, depending on whether $\omega$ or $E_{c}$ is larger.  Therefore the generating function depends on the trace of the saddle point sign matrix roughly as:
\begin{equation}
\overline{Z} \propto {\exp{({-{\frac{V}{8}{({Tr}{(S)})}^{2}} \ln{(\xi / E_{c})}})}}
\end{equation}
The net result is that saddle points with the minimum value of ${({Tr}{(S)})}^{2}$ are favored by exponentially large factors.  The number of negative saddle point eigenvalues should be as close as possible to the number of positive saddle point eigenvalues.

The saddle point sign matrix $S$ occurs in two other terms in equation \ref{SigmaLagrangian3}.  After applying the saddle point constraint, these terms simplify.  The resulting sigma model Lagrangian is:
\begin{eqnarray}
{\mathcal{L}}_{\sigma} & = & {\mathcal{L}}_{0} + {\imath \nu {{Tr}{({\tilde{Q}}^{b}L{({\hat{\omega}}^{b} - J)})}}} - { {\imath \nu} {Tr}{({{{\tilde{Q}}^{f} {\hat{\omega}}^{f}}})}}
\nonumber \\
& + & {\frac{\nu {D}^{b}}{2}{Tr}{({\vec{\nabla} {\tilde{Q}}^{b}L \cdot {\vec{\nabla} {\tilde{Q}}^{b}L})}}} 
+ {\frac{\nu {(D + D^{f})}}{2}{Tr}{({{\vec{\nabla} {\tilde{Q}}^{f} } \cdot {\vec{\nabla}{\tilde{Q}}^{f}}})}}
 \label{SigmaLagrangian4}
\end{eqnarray}

Recall that $D^{b}$ and $D^{f}$ were Wilson coefficients of the perturbative corrections to the effective Lagrangian.  In contrast, $D$ came the original Lagrangian, but via the approximations entailed in computing the low momentum behavior of the one loop integral. All three are phenomenological constants.  I take the liberty of setting ${\tilde{Q}}^{f}$'s total diffusion constant to be equal to ${\tilde{Q}}^{b}$'s diffusion constant.  I am now ready to write the final sigma model.

\pagebreak

\subsection{\label{FinalSigmaModel}The Final Sigma Model}
The final sigma model can be described briefly in only five points:
\begin{enumerate}  
\item The generating function is given by: 
\begin{eqnarray}
\overline{Z} & = &  {\int} {dQ^{b}{(\vec{p} = \vec{0})} } \;\; {dQ^{b}{(\vec{p} = \vec{0})} }  \;\; e^{{\mathcal{L}}{(\vec{p} = \vec{0})}}
\nonumber \\ 
& \times & {\int {{d{\tilde{Q}}^{b}}{d{\tilde{Q}}^{f}}} e^{{\tilde{\mathcal{L}}}}}
\nonumber \\
{{\mathcal{L}}{(\vec{p} = \vec{0})}}   & = & {  -{\frac{ \nu \xi}{2} {{Tr}({(Q^{f}{(\vec{p} = \vec{0}))}^{2}) }} + {{{\nu \xi {(1 - \alpha)}{{{Tr}\ln}{({{\hat{E}}^{f} + {\imath \xi  {Q^{f}{(\vec{p} = \vec{0})} }}})}}}}}}} 
\nonumber \\ 
& - & {\frac{\nu \xi}{2}{{{Tr}{({Q^{b}{(\vec{p} = \vec{0})} } \: L \: {Q^{b}{(\vec{p} = \vec{0})} } \:L)}}}}  + {\imath \nu {{Tr}{({Q^{b}{(\vec{p} = \vec{0})} } \:{L \: {({{\hat{\omega}}^{b}} - J) }})}}} 
\nonumber \\ 
&+& {\nu \xi {(1 - \alpha)} {(\frac{\pi \xi}{{\Delta}_{b}})}^{2}{{Tr}\ln{( {{Q^{b}{(\vec{p} = \vec{0})} } \: L})}}} 
\nonumber \\
& + & {\frac {\nu \xi \alpha}{I^{b}} {{{Tr}\ln}{({{\hat{E}}^{f} + { {\imath {\xi} {Q^{f}{(\vec{p} = \vec{0})} }} + {\imath \xi {Q^{b}{(\vec{p} = \vec{0})} } \: L})} }}}} 
\nonumber \\
\tilde{{\mathcal{L}}} & = & {\imath \nu {{Tr}{({\tilde{Q}}^{b}L{({\hat{\omega}}^{b} - J)})}}} - { {\imath \nu} {Tr}{({{{\tilde{Q}}^{f} {\hat{\omega}}^{f}}})}}
\nonumber \\
& + & {\frac{\nu D}{2}{Tr}{({\vec{\nabla} {\tilde{Q}}^{b}L \cdot {\vec{\nabla} {\tilde{Q}}^{b}L})}}} 
+ {\frac{\nu D}{2}{Tr}{({{\vec{\nabla} {\tilde{Q}}^{f} } \cdot {\vec{\nabla}{\tilde{Q}}^{f}}})}}
 \label{SigmaLagrangian5}
 \end{eqnarray}
 \item The matrix ${Q^{b}{(\vec{p} = \vec{0})} }$ contains all of $Q^{b}$'s translationally invariant degrees of freedom.  It is composed of two parts: the saddle point solution $Q_{0}^{b}$, and the small perturbations ${{\tilde{Q}}^{b}{(\vec{p} = \vec{0})} }$.  $Q_{0}^{b}$ can vary freely, with the constraint that its eigenvalues be equal to $ p_{i} = {\pi \xi {\Delta}^{-1}}$.
 ${{\tilde{Q}}^{b}{(\vec{p} = \vec{0})} }$ is a small perturbation to $Q_{0}^{b}$, and consists of the translationally invariant degrees of freedom which change the eigenvalues of $Q^{b}$.   
\item The matrix ${Q^{f}{(\vec{p} = \vec{0})} }$ contains all of $Q^{f}$'s translationally invariant degrees of freedom.  It is composed of two parts: the saddle point solution $Q_{0}^{f}$, and the small perturbations ${{\tilde{Q}}^{f}{(\vec{p} = \vec{0})} }$.  $Q_{0}^{f}$ can vary freely, with the constraint that its eigenvalues be equal to $ q_{i} = {s_{i} \pi {\xi} {\Delta}^{-1}}, \: {s_{i} = {\pm 1}} $. These eigenvalues $q_{i}$ are constrained to minimize ${({Tr}{({\sum}_{i}s_{i})})}^{2}$.
 ${{\tilde{Q}}^{f}{(\vec{p} = \vec{0})} }$ is a small perturbation to $Q_{0}^{f}$, and consists of the translationally invariant degrees of freedom which change the eigenvalues of $Q^{f}$.    
 \item The matrix  ${\tilde{Q}}^{b}$ is composed of all the degrees of freedom which are not translationally invariant, and is constrained to not change the eigenvalues of $ {Q_{0}^{b} + {\tilde{Q}}^{b}}$.  This constraint is the mechanism by which ${\tilde{Q}}^{b}$ interacts with the other degrees of freedom. 
  \item The matrix  ${\tilde{Q}}^{f}$ is composed of all  the degrees of freedom which are not translationally invariant, and is constrained to not change the eigenvalues of $ {Q_{0}^{f} + {\tilde{Q}}^{f}}$.  This constraint is the mechanism by which ${\tilde{Q}}^{f}$ interacts with the other degrees of freedom. 
 \end{enumerate} 

The form of Lagrangian $\tilde{{\mathcal{L}}}$ is very similar to that of the supersymmetric sigma model\cite{Mirlin00a}.  When $I^{b} = I^{f} = 2$ and $J = 0$, this reads: 
\begin{eqnarray}
{\mathcal{L}}_{SUSY} & = & {\frac{\imath \pi \nu \omega}{2 } {{Str}{(\Lambda Q)}}} 
+ {\frac{\nu \pi D}{4}{Str}{({\vec{\nabla} Q \cdot {\vec{\nabla} Q})}}} 
 \label{SigmaLagrangian6}
 \end{eqnarray}
However this similarity hides a huge difference in complexity: the $Q$'s in the supersymmetric Lagrangian are graded matrices, and are quite challenging to manage.  Just changing $I^{b}$ and $I^{f}$ is very challenging, which is why the literature has confined itself to $I^{b} = I^{f} = 1$ and $I^{b} = I^{f} = 2$.  This is why the sigma model presented here is so important.

A few words about the validity of this model, which is best understood in terms of energy scales.  One of the important energ scales is the inverse scattering time ${\tau}^{-1} = {\xi I_{i}} = {\pi {\xi}^{2} {\Delta}^{-1}}$.  The assumptions used to derive this model were:
\begin{enumerate}
\item Spherical symmetry.
\item The dominant saddle points are translationally invariant.
\item The final field theory is local.
\item The diffusive approximation: $\tau$ is much less than the particles' energy.
\item $\omega \ll \tau$
\item $E_{c} \ll \tau$ 
\item $1 \ll {\tau / \xi}$, which implies that $\Delta \ll \xi$. This last inequality was necessary to justify the perturbative expansion of the Lagrangian; one can see this by estimating the next order corrections and seeing that they are divided by roughly ${\tau / \xi}$.
\item $E_{c} \ll \xi$ and $\omega \ll \xi$.  These assumptions were necessary to justify integrating out the very massive modes.  I would still have obtained a quite valid model without them.
\end{enumerate} 
 
This completes my derivation of the new sigma model.

\section{\label{DOSZero}The Density of States in a Zero Dimensional System}
In this section I calculate the density of states $\rho{(E)}$ of a non-extended system ($V = 1$).  Equation \ref{DOSGreens} indicates that I need to compute the average of the trace of the advanced Green's function. I will obtain this by first computing the average generating function $\overline{Z}$ with $I^{b} = I^{f} = 1$, and then taking the first derivative.  Equation \ref{LDefinition} implies that $L = -1$.   When $V = 1$, $I^{b} = I^{f} = 1$, and $L = -1$, equation \ref{FinalQbQfLagrangian} simplifies to:

\begin{eqnarray}
{\bar{Z}} & = &  {\gamma {\int_{q^{b} > 0} {dq^{f}}  {dq^{b}} {\exp}{(\mathcal{L})}}}
\nonumber \\
\gamma & \equiv & { {\xi}^{ -N } N^{N + 1/2} {\imath}^{{3N}+2}  \frac{1}{(N-1)!}  2^{1/2}   {\pi}^{-1/2}   }    
\nonumber \\
{\mathcal{L}}   & \equiv & {  -{\frac{ N}{2} {q^{f}}^{2}  + {{(N - 1)}{\ln}{({{\imath \xi q^{f}} + E^{f}})}} } } 
\nonumber \\ 
& - &  {  \frac{N}{2}{q^{b}}^{2} - {\imath Nq^{b}{ ({{E}^{b}} - J) / \xi}} + {{(N-1)}}\ln{q^{b}}} \nonumber \\
& + & {{{\ln}{({{\imath \xi q^{f}} + E^{f}} - { \imath {\xi} q^{b} })}}}
\nonumber \\
{A_{0}} & = & {{\imath \xi  q^{f}} + E^{f}}
\end{eqnarray} 

$q^{b}$ and $q^{f}$ are just real numbers.  When dealing with the non-extended system, it is always helpful to immediately make a shift $q^{f} \rightarrow {q^{f} + {\imath E^{f} / {\xi}}}$ to obtain:

\begin{eqnarray}
\overline{Z} & = & { -{ N^{N + 1/2}  \frac{1}{(N-1)!}  2^{1/2}   {\pi}^{-1/2}   \exp{(N {E^{f}}^{2} / {{\xi}^{2}})}}}    
\nonumber \\
& \times & {\int_{q^{b} > 0} {dq^{f}} {dq^{b}} {(q^{f} - q^{b})}}
\nonumber \\
& \times & {{\exp}{(  -{\frac{ N}{2} {q^{f}}^{2} - {\imath Nq^{f}{{{E}^{f}/ \xi}}} + {{(N - 1)}{\ln}{q^{f}}} } } )} 
\nonumber \\ 
& \times &{{\exp}{(  {  -\frac{N}{2}{q^{b}}^{2} - {\imath Nq^{b}{ ({{E}^{b}} - J) / \xi}} + {{(N-1)}}\ln{q^{b}}} )}}
\label{SimplifiedZeroDensity1}
\end{eqnarray} 

Remember that I will calculate the Green's function by taking a derivative with respect to $J$ and then setting $E^{b} = E^{f}$ and $J = 0$.  When $E^{b} = E^{f}$ and $J = 0$,  the integrand of equation \ref{SimplifiedZeroDensity1} is antisymmetric under interchanges of $q^{b}$ and $q^{f}$.  The only thing that is not antisymmetric is the limit of integration $q^{b} > 0$.  I will shortly be making the saddle point approximation, in which the limit of integration will be neglected; in this approximation $\overline{Z}{(E^{b} = E^{f}, J=0)}$ is exactly zero.  Therefore I need to either retain information about $J$ in the saddle point approximation, or else take the derivative with respect to $J$ now.  I opt for the latter option, which just multiplies the integrand by a factor of $\imath N q^{b} / \xi$.

I will evaluate the integral \ref{SimplifiedZeroDensity1} by making the saddle point approximation in both variables.  But first I briefly review the saddle point approximation.  

\subsection{\label{SaddlePointReview}The Saddle Point Approximation: A Review}
In order to be general, I assume a Lagrangian ${\mathcal{L}} = {-{N q^{2}/2} + {\imath t N q E / \xi} + {N{(1 - \alpha)} {\ln q}}}$, with $t = {\pm 1}$.  The saddle point equation for this Lagrangian is $0 = {\frac{d\mathcal{L}}{dq}}$, which implies that $0 = {-N{(q - {\imath t E / \xi} - {{(1-\alpha)}/q)}}}$ at the saddle point.  This saddle point equation has two solutions: $q{(s)} \equiv {st \sqrt{1-\alpha} \times {{\exp}{(\imath s \phi)}}}$, $s = {\pm 1}$, ${\sin{\phi}} \equiv {\frac{E}{2 \xi \sqrt{1 - \alpha}}}$.  

At the saddle point, the Lagrangian has second derivative ${\frac{d^{2}\mathcal{L}}{dq^{2}}} = {-2N {{\exp}{(-\imath s \phi)}} {\cos{(\phi)}}}$.  The value of the Lagrangian at the saddle point is:
\begin{eqnarray} {\mathcal{L}}_{0} & = & {{-\frac{ N{(1 - \alpha)}}{2}} {{\exp}{(\imath 2 s \phi)}}} + {\imath s N \sqrt{1 - \alpha} {E / \xi} {{\exp}{(\imath s \phi)}}} 
\nonumber \\
& + & {N{(1 - \alpha)}{\ln}{(st\sqrt{1-\alpha})}} + {N{(1 - \alpha)}\imath  s \phi}
\label{SaddlePointAction}
\end{eqnarray}
To the leading order in $N$, the saddle point approximation is given by: 
\begin{eqnarray}
{\int {dq} {f{(q)}} {\exp{({\mathcal{L}}(q))}}} & \approx &
{\sum_{s}   {f{(q{(s)})}} {\exp{{\mathcal{L}}_{0}(q{(s)})}}  {\int {dq} \exp{(\frac{q^{2}}{2}  {[\frac{d^{2}\mathcal{L}}{dq^{2}}]}_{q = {q{(s)}}} })} }
\nonumber \\ 
& = &{\sum_{s} \sqrt{\frac{\pi}{N {\cos{(\phi)}}}} {{\exp}{(\imath s \phi /2 )}} {f{(q{(s)})}} {\exp{{\mathcal{L}}_{0}(q{(s)})}}   }
\nonumber \\
&&
\label{SaddlePointSum}
\end{eqnarray}
If $f$ is zero at the saddle point, or perturbative corrections are desired, one instead uses the formula:
\begin{eqnarray}
{\int {dq} {f{(q)}} {\exp{({\mathcal{L}}(q))}}}
& = & \sum_{s} {\exp{({\mathcal{L}}_{0} (s))}} \int {dq} {f{(q)}}  {\exp{(-q^{2} N {{\exp}{(-\imath s \phi)}} {\cos{(\phi)}}})} 
\nonumber \\
& \times & \qquad {\exp{(-N{(1 - \alpha)} \sum_{j = 3}^{\infty} {j}^{-1} {(-q / {q{(s)}})}^{j})}} 
\label{SaddlePointIntegral}
\end{eqnarray}
One evaluates this integral by first expanding the last exponential perturbatively in powers of $q$; after the integration, each power of $q$ is weighted by a power of $\sqrt{\frac{1}{N}}$, which justifies the perturbative expansion.  However, the expansion can be expected to diverge at high orders.  Moreover, there are certain effects which are totally lost in the expansion; for instance, when evaluating $\overline{Z}$, the integration constraint $q^{b} > 0$ is not respected.  In the saddle point approximation, this constraint's only effect is to eliminate half of the bosonic saddle points. 
 
\subsection{Applying the Saddle Point Approximation}
 I am only interested in the leading order, so I take $\alpha = 0$.  I take the derivative of equation \ref{SimplifiedZeroDensity1} with respect to $J$, set $E^{b} = E^{f}$ and $J = 0$, and apply equation \ref{SaddlePointSum}'s version of the saddle point approximation.  Without the constraint $q^{b} > 0$ there would be $2 \times 2$ saddle points, two from the bosonic integral and two from the fermionic integral.  The constraint eliminates one of the bosonic saddle points, leaving only two ($1 \times 2$) saddle points.  Here is the result:
\begin{eqnarray}
{{Tr}{(\overline{G_{A}{(E)}})}} = {\frac{d\overline{Z}}{dJ}} & = & { 
\imath N^{N + {1/2}} 
s
{\xi}^{-1}
\frac{1}{(N-1)!}  
2^{1/2}   
{\pi}^{1/2}   
\exp{(N E^{2} / {2 {\xi}^{2}})} 
}  
\nonumber \\
& \times & 
{{\sum}_{s}
{\frac{-s \times {{\exp}{(\imath s {\phi})}} - 
{{\exp}{(-\imath {\phi})}}}
{\sqrt{\cos{({\phi})}{\cos{({\phi})}}}}  } \times {{\exp}{(-\imath {\phi})}} }
\nonumber \\
& \times &
{\exp}( {{-\frac{ N}{2}} {{\exp}{(\imath 2 s {\phi})}}} + {\imath s N {E / \xi} {{\exp}{(\imath s {\phi})}} } 
\nonumber \\
& & \qquad \qquad + {{(N - 1)}{\ln}{(-s)}} + {{(N-{1/2})}\imath  s {\phi}})
\nonumber \\
& \times & 
{\exp}( {{-\frac{ N}{2}} {{\exp}{(-\imath 2 {\phi})}}} - {\imath N{{E/\xi}} {{\exp}{(-\imath {\phi})}} }  - {{(N-{1/2})}\imath {\phi}} ) 
\nonumber \\ & &
\label{DensitySaddlePointSum}
\end{eqnarray}

Equation \ref{DensitySaddlePointSum} gives the $s = {-1}$ saddle point a weight of zero.  This is a consequence of the integrand's original antisymmetry in $q^{b}$ and $q^{f}$, and of the fact that the fermionic saddle point $s={-1}$ is symmetric with the bosonic saddle point, which has a (bosonic) index $s_{b} = {-1}$.  One can evaluate higher order corrections to the $s = {-1}$ saddle point by using the integral formulation of the saddle point, equation \ref{SaddlePointIntegral}. However, these corrections can be neglected to leading order in $N$.  Simplifying equation \ref{DensitySaddlePointSum}, I find:

\begin{eqnarray}
{{Tr}{(\overline{G_{A}{(E)}})}} & = & { 
{(-1)}^{N-1} N^{N + {1/2}}  
\frac{1}{(N-1)!}  
2^{3/2}  
{\pi}^{1/2}  
\exp{(N E^{2} / {2 {\xi}^{2}})}  }
\nonumber \\
& \times & 
  {  \frac{\imath}{\xi} {{\exp}{(- \imath \phi)}}}
{{\exp}{( {{-N} {{\cos}{(2 {\phi})}}} - {2 N {E / \xi} {{\sin}{({\phi})}} }  ) }}
\end{eqnarray}

The last exponential can be reduced to $\exp{(-N{(1 + {E^{2} / {2 {\xi}^{2}}})})}$ by using the definition of $\phi$, ${\sin{(\phi)}} = {E / {2 \xi}}$, and various trigonometric identities.  Stirling's formula gives ${(N-1)!} = {\sqrt{2 \pi} \exp{(-N + {{(N - {1/2})}{{ln}{(N)}}})}}$.  Applying these identities and equation \ref{DOSGreens}, I obtain:
 
\begin{eqnarray}
{\rho{(E)}} = {\frac{1}{2 \pi} {Im}{({Tr}{(\overline{G_{A}{(E)}})})}} & = & {{Im}  {( {(-1)}^{N-1} \frac{ \imath N}{\pi \xi} {( { \sqrt{1 - \frac{E^{2}}{4{\xi}^{2}} } - \frac{\imath E}{2\xi}})})} } 
\nonumber \\ & &
\end{eqnarray}

The correct result would be:
\begin{equation}
{\rho{(E)}}  = {\frac{N}{\pi \xi} \sqrt{1 - \frac{E^{2}}{4{\xi}^{2}}} }
\end{equation}
So my result is correct up to a multiplicative constant of $ {(-1)}^{N-1} $.  This phase is probably caused by some personal error.   I haven't had time to figure out figure this out yet. 

\section{\label{TwoPointZero}The Two Point Correlator in a Zero Dimensional System}
In this section I calculate the two point correlator $R_{2}{(E_{1}, E_{2})}$ of a non-extended system ($V = 1$). I will calculate this observable using equation \ref{TwoLevelZeroGreensB}, which I repeat here:
\begin{equation}
\nonumber 
{R_{2}{(E_{1}, E_{2})}} = {\frac{{{Re}{(R_{AR})}} - {{Re}{(R_{AA})}}}{8 {\pi}^{2} \quad \overline{ \rho{(E_{1})}} \quad  \overline{\rho{(E_{2})}}}}
\end{equation}
First I will calculate the advanced-retarded correlator \linebreak $R_{AR} \equiv \lim_{\epsilon \rightarrow 0} \overline{{{Tr}{(G_{A}{(E)}})}{{Tr}{(G_{R}{(\acute{E})})}}}$, and then the advanced-advanced correlator $R_{AA} \equiv \lim_{\epsilon \rightarrow 0} \overline{{{Tr}{(G_{A}{(E)}})}{{Tr}{(G_{A}{(\acute{E})})}}}$ .

\subsection{\label{ARCorr}The Advanced-Retarded Correlator}
 I will obtain the advanced-retarded correlator by computing the average generating function $\overline{Z}$ and then taking the second derivative, as shown in the following equation:
 \begin{equation}
R_{AR} = {\overline{{{Tr}{G_{R}{(E_{1})}}} {{Tr}{G_{A}{(E_{2})}}} }} 
= {\frac{d^{2}}{{dJ}_{1} {dJ}_{2}} \overline{Z{({\hat{E}}^{f}, {\hat{E}}^{b})}}}{\vert}_{{{\hat{E}}^{f} =  {\hat{E}}^{b}}, {\hat{J} = 0}}
\end{equation}  

Because I am calculating the average of a multiple of two Green's functions, I must choose $I^{b} = I^{f} = 2$; i.e. now the degrees of freedom will be $2 \times 2$ matrices.  

Because one Green's function is advanced and the other retarded, the imaginary parts of their energies have opposite sign; equation \ref{LDefinition} implies that:
\begin{equation}
L_{11} = 1, L_{22} = -1
\label{L2DefAR}
\end{equation}
The sign of $L$ could be reversed, but this doesn't  matter.  Remember that $L$ is diagonal.  
 
 When one chooses $V = 1$, $I^{b} = I^{f} = 2$, $L$ as in equation \ref{L2DefAR}, and again makes the shift ${\hat{Q}}^{f} \rightarrow  {{\hat{Q}}^{f} + {\imath {\hat{E}}^{f} / {\xi}}}$, equation \ref{FinalQbQfLagrangian} turns into:
\begin{eqnarray}
{\bar{Z}} & = &  {\gamma {\int_{Q^{b} > 0} {dQ^{f}}  {dQ^{b}} {\exp}{(\mathcal{L})}}}
\nonumber \\
\gamma & \equiv & { N^{{2 N} + 2 }  } { {{{({(N-1)!}{(N-2)!})}}^{-1}}  2^{2}   {\pi}^{-3} {(-1)}^{N}  } 
 \nonumber \\
& \times & {{\exp}{(\frac{N}{2} {{Tr}({{{\hat{E}}^{f}} {{\hat{E}}^{f}} / {{\xi}^{2}}) }}})}
\nonumber \\
{\mathcal{L}}   & \equiv & {  -{\frac{ N}{2} {{Tr}({Q^{f}}^{2}) }} - 
{\imath N{{Tr}{(Q^{f}{{{\hat{E}}^{f} / \xi}})}}}
+ {{{{(N - 2)}{{{Tr}\ln}{Q^{f}}}}}}} 
\nonumber \\ 
& - & {{ {  {\frac{N}{2}{{{Tr}{(Q^{b}LQ^{b}L)}}}} { + {\imath N{{Tr}{(Q^{b}{L {({{\hat{E}}^{b}} - J) / \xi}})}}} + {{(N-2)}}{({Tr}\ln{( {Q^{b} L})})}}}}} \nonumber \\
& + & {{{{Tr}\ln}{(Q^{f} + {Q^{b}L})}}}
\label{StartingAdvancedRetarded}
\end{eqnarray}

\subsubsection{The Angular Integrals}
Now I'm confronted with two matrix integrals.  When integrating over a matrix $M$, it is frequently helpful to decompose the matrix into $M = {U D U^{\dagger}}$, where $D$ is the diagonal matrix formed from $M$'s eigenvalues, and $U$ is the unitary transformation required to diagonalize $M$.  Having made this conceptual step, one then changes variables from $dM$ to ${dD}{dU}$.  This is the approach I will take with both $Q^{b}$ and $Q^{f}$.  All but two of the terms in the Lagrangian depend only on eigenvalues. The two exceptions are: ${\imath N {Tr}{(Q^{b}{L {({{\hat{E}}^{b}} - J) / \xi}})}}$ and ${-\imath N {Tr}{(Q^{f}{{{\hat{E}}^{f} / \xi}})}}$.    

The matrix $Q^{b}$ is special because both of its eigenvalues are required to be positive semidefinite.  Because $Q^{b}$ is always accompanied by the matrix $L$, I will actually decompose $Q^{b} L$ into eigenvalues and a transformation.  Fyodorov\cite{Fyodorov02} discussed the properties of $Q^{b}L$ with a lot of clarity, and also explained how to decompose any $2N \times 2N$ matrix $Q^{b} L$ where $N$ of  $L$'s eigenvalues are $+1$ and the other $N$ are $-1$.  Probably this proof can be generalized easily to arbitary combinations of $+1$ and $-1$.  However, here I won't bother with the details of the general solution, but instead just consider a $2 \times 2$ matrix.  

As Fyodorov showed, $Q^{b}L$ is non-Hermitian, so it can not be decomposed into ${U D U^{\dagger}}$. However, its eigenvalues are still real, one positive and the other negative, so one can decompose ${Q^{b}L} = {T P T^{-1}}$, where $P$ is diagonal and $T$ is a special "pseudo-unitary" matrix.  $Q^{b}L$ has four degrees of freedom, while $P$ has two degrees of freedom; $T$ should have two degrees of freedom in order to conserve degrees of freedom .  Because one wants to respect the special eigenvalue structure of $Q^{b}L$, one chooses a special parameterization of $T$:

\[ \left( \begin{array}{lr}
{\cosh{(\psi)}} & {{\sinh{(\psi)}} {\exp{(-\imath \theta)}}} \\
{{\sinh{(\psi)}} {\exp{(\imath \theta)}}} & {\cosh{(\psi)}} \end{array} \right)\]   
Its inverse $T^{-1}$ is similar:
\[ \left( \begin{array}{lr}
{\cosh{(\psi)}} & {-{\sinh{(\psi)}} {\exp{(-\imath \theta)}}} \\
{-{\sinh{(\psi)}} {\exp{(\imath \theta)}}} & {\cosh{(\psi)}} \end{array} \right)\]   

$\theta$ varies from $0$ to $2 \pi$, while $\psi$ varies from $0$ to $\infty$.  With this parameterization of $T$, $Q^{b}L$ is parameterized as:
\[ \left( \begin{array}{lr}
{\frac{1}{2}{{(p_{1} + p_{2}) + {\frac{1}{2}{(p_{1} - p_{2})} {\cosh}{(2 \psi)}}}}} & {-\frac{1}{2} {(p_{1} - p_{2})} {\sinh{(2\psi)}}  {\exp{(-\imath \theta)}}} \\
\frac{1}{2} {{(p_{1} - p_{2})} {\sinh{(2 \psi)}} {\exp{(\imath \theta)}}} & {\frac{1}{2}{{(p_{1} + p_{2}) - {\frac{1}{2}{(p_{1} - p_{2})} {\cosh}{(2 \psi)}}}}}
\end{array} \right)\]   

Of course when you change variables you must also introduce a Jacobian; i.e. a function which keeps track of the relation between the old variables and the new ones.  The most straightforward way to do this is simply to take derivatives of the four old variables with respect to the four new variables, form these sixteen derivatives into a $4 \times 4$ matrix, and then calculate the determinant of that matrix, which is the desired Jacobian.  This is straightforward; after a page of algebra I obtain $J = {\frac{1}{2} {(p_{1} - p_{2})}^{2} {\sinh{(2 \psi)}} }$.  The ${(p_{1} - p_{2})}^{2}$ dependence on the eigenvalues is quite typical of matrix decompositions.  

I now turn to the term ${{Tr}{(Q^{b}{L {({{\hat{E}}^{b}} - J) / \xi}})}}$, which is parametrized as:
\begin{equation}
{\frac{1}{2 \xi}{({{\hat{E}}_{1}^{b}} + {{\hat{E}}_{2}^{b}} - J_{1} - J_{2}){(p_{1} + p_{2})}}}
+ {\frac{1}{2 \xi}{({{\hat{E}}_{1}^{b}} - {{\hat{E}}_{2}^{b}} - J_{1} + J_{2}){{(p_{1} - p_{2})} {\cosh{(2\psi)}}}}}
\end{equation}

One must integrate over the angular variables $\psi$ and $\theta$.  The integral is:
\begin{eqnarray}
& \frac{1}{2} & {(p_{1} - p_{2})}^{2} {{exp}{(\frac{\imath N}{2 \xi}( {{\hat{E}}_{1}^{b}} + {{\hat{E}}_{2}^{b}} - J_{1} - J_{2}) {(p_{1} + p_{2})} )}}
\nonumber \\
& \times &
\int {{d\psi}{d\theta}} {\sinh{(2 \psi)}} {{exp}{(\frac{\imath N}{2 \xi}{({{\hat{E}}_{1}^{b}} - {{\hat{E}}_{2}^{b}} - J_{1} + J_{2}){{(p_{1} - p_{2})} {\cosh{(2\psi)}}}} )}}
\nonumber \\ & &
\end{eqnarray}

It is convenient to change variables to $x = {\cosh{(2 \psi)}}$; $x$ varies from one to infinity.  Note that the $\psi$ integral converges because the imaginary parts of ${\hat{E}}^{b}$ have signs $(+1, -1)$.  The final result is:
\begin{equation}
\frac{\imath \pi \xi {(p_{1} - p_{2})}} {N({{\hat{E}}_{1}^{b}} - {{\hat{E}}_{2}^{b}} - J_{1} + J_{2})} {{\exp} {(i N {{Tr}{(P {({\hat{E}}^{b} - J)})}})}}
\label{HyperbolicIntegrationResult}
\end{equation}

I now turn to the angular integration over $Q^{f}$ with integrand ${{\exp}{(-\imath N {{Tr}{(Q^{f}{{{\hat{E}}^{f} / \xi}})}})}}$.  Since $Q^{f}$ is hermitian, it can be decomposed into a diagonal part and a unitary part, as mentioned above.  There is a well-known formula for doing this integral, called the Harish-Chandra-Itzykson-Zuber formula because of three people (Harish-Chandra, Itzykson, and Zuber) who derived it.  I will simply quote it as listed in Mehta's book\cite{Mehta91}:
\begin{eqnarray}
& & {\int {dU}\; {{exp}{(-\frac{1}{2t}{(A - {U b U^{-1}})})}^{2}}} = {\frac{c \; \det N}{{\Delta{(a)}}{\Delta{(b)}}}},
\nonumber \\
& & {\langle i | N | j \rangle} \equiv {{exp}{(-\frac{1}{2t}{(a_{i} - b_{j})})}^{2}}
\label{ItzyksonZuber}
\end{eqnarray}
$U$ is a unitary matrix; the integral is over all unitary matrices, and occurs when you factorize a matrix $B$ into $B = U b U^{-1}$ and then integrate over $U$ before integrating over the eigenvalues in the diagonal matrix $b$.  $A$ is another Hermitian matrix, $a_{i}$ and $b_{i}$ are the eigenvalues of $A$ and $b$,  and $\Delta$ is the Vandermonde determinant 
${\Delta{(x)}} \equiv {{\prod}_{i=1}^{N} {\prod}_{j=1}^{i-1} {(x_{i} - x_{j})}}$.   $t$ is a constant of one's choice, and $c$ is a normalization constant which can be easily determined by making sure that the normalization equation \ref{GaussianMatrixNormalization} holds.  For $2 \times 2$ matrices, $c = {\frac{\pi t}{2}}$.

Recall that the term we are integrating came from a sum of squares:
\begin{equation}
{{\exp}{(-{N/2} {{Tr}({({Q^{f} + {\imath {\hat{E}}^{f} / {\xi}}})}^{2} ) })}}
\end{equation}

I skip the details and give the final result.   ${{\exp}{(\pm \imath N {{Tr}{(Q^{f}{{{\hat{E}}^{f} / \xi}})}})}}$ turns into:
\begin{equation}
\pm \frac{ \imath \pi \xi {(q_{1} - q_{2})}} {N({{{E}}_{1}^{f}} - {{{E}}_{2}^{f}} )} {{\exp}{(\pm \frac{\imath N}{\xi} {{Tr}{({q_{1}^{f}E_{1}^{f}} + {q_{2}^{f}E_{2}^{f}})}})}}
\label{UnitaryIntegrationResult}
\end{equation}
To derive this result I had to use the problem's symmetry under switching the values of $q_{1}$ and $q_{2}$.  Note that there is no difference between this result derived with unitary matrices and formula \ref{HyperbolicIntegrationResult}, which was derived using the special "pseudo-unitary" matrices. 

Substituting equations \ref{HyperbolicIntegrationResult} and \ref{UnitaryIntegrationResult} into equation \ref{StartingAdvancedRetarded}, I obtain the following expression for the generating function in terms of integrals over the eigenvalues of $Q^{f}$ and of $Q^{b}L$.  I use the symbols $q_{1}$ and $q_{2}$ to represent the eigenvalues of $Q^{f}$, and the symbols $p_{1}$ and $p_{2}$ to represent the eigenvalues of $Q^{b}L$.
\begin{eqnarray}
\overline{Z} & = & {   {\xi}^{2} N^{{2 N }}  } { {{{({(N-1)!}{(N-2)!})}}^{-1}}  2^{2}  {(-1)}^{N} {\pi}^{-1}   } 
 \nonumber \\
& \times & { {{\exp}{(\frac{N}{2} {{Tr}({{{\hat{E}}^{f}} {{\hat{E}}^{f}} / {{\xi}^{2}}) }}})}
{({{{E}}_{1}^{f}} - {{{E}}_{2}^{f}} )}^{-1} {({{{E}}_{1}^{b}} - {{{E}}_{2}^{b}} - J_{1} + J_{2})}^{-1}}
\nonumber \\
& \times &
\int {dq_{1}{dq_{2}} {\int}_{p_{1} > 0, p_{2} < 0} {{dp_{1}}{dp_{2}}}  {(q_{1} - q_{2} )} {(p_{1} - p_{2})} 
 {(q_{1} + p_{1} )}} 
 \nonumber \\
 & & \qquad \times {(q_{1} + p_{2} )} {(q_{2} + p_{1} )} {(q_{2} + p_{2} )} {\exp{(\mathcal{L})}}
\nonumber \\
{\mathcal{L}}   & = &   -{\frac{ N}{2} {(q_{1}^{2} + q_{2}^{2} )}} - 
{\frac{\imath N}{\xi} {( {q_{1}E_{1}^{f}} + {q_{2}E_{2}^{f}  )} }
+ {{{{(N - 2)}{( {\ln q_{1}} + {\ln q_{2}} )}}}}} 
\nonumber \\ 
& - & {{ {  {\frac{N}{2}{(p_{1}^{2} + p_{2}^{2})}} { + {\frac{\imath N}{\xi} {(  {p_{1}{(E_{1}^{b} - J_{1})}} + {p_{2}{(E_{2}^{b} - J_{2})}} )}}  + {{(N-2)}{({\ln p_{1}} + {\ln p_{2}})}}}}}} 
\nonumber \\
& &
\label{AdvancedRetardedEigenvalues}
\end{eqnarray}

Now I change the sign of $p_{1}$ and $p_{2}$.  I also take the derivatives with respect to the source $J$, ignoring the derivative of the prefactor ${({{{E}}_{1}^{b}} - {{{E}}_{2}^{b}} - J_{1} + J_{2})}^{-1}$ because it has a subleading order in $N$. Lastly, I set $E_{1}^{f} = E_{1}^{b} = E_{1}$, $E_{2}^{f} = E_{2}^{b} = E_{2}$, and $J_{1} = J_{2} = 0$.
\begin{eqnarray}
R_{AR} = \frac{d^{2}\overline{Z}}{{dJ_{1}}{dJ_{2}}} & \equiv & { {(-1)}^{N} N^{{2 N} + 2 }  } { {{{({(N-1)!}{(N-2)!})}}^{-1}}  2^{2}   {\pi}^{-1}   } 
{ {{\exp}{(\frac{N}{2} {{Tr}({{{\hat{E}}}^{2} / {{\xi}^{2}}) }}})}
{({{{E}}_{1}} - {{{E}}_{2}} )}^{-2} }
\nonumber \\
& \times &
\int {{dq_{1}}{dq_{2}} {\int}_{p_{1} < 0, p_{2} > 0} {{dp_{1}}{dp_{2}}}} {(q_{1} - q_{2} )} {(p_{1} - p_{2})} 
 {(q_{1} - p_{1} )} 
 \nonumber \\
 & & \qquad {(q_{1} - p_{2} )} {(q_{2} - p_{1} )} {(q_{2} - p_{2} )} {p_{1}
 p_{2}}
\nonumber \\
 & \times & {{\exp}{({  -{\frac{ N}{2} {(q_{1}^{2} + q_{2}^{2} )}} - 
{\frac{\imath N}{\xi} {( {q_{1}E_{1}} + {q_{2}E_{2}  )} }
+ {{{{(N - 2)}{( {\ln q_{1}} + {\ln q_{2}} )}}}}} })}} 
\nonumber \\ 
& \times & {{\exp}{({{ {  {  {-\frac{N}{2}{(p_{1}^{2} + p_{2}^{2})}} { - {\frac{\imath N}{\xi} {(  {p_{1}E_{1}} + {p_{2}E_{2}} )}}  + {{(N-2)}{({\ln p_{1}} + {\ln p_{2}})}}}}   }}} )}}
\nonumber \\ & & 
\label{AREig1}
\end{eqnarray}

I will drop the factor of ${(-1)}^{N}$ because I believe it is in error.

\subsubsection{The Saddle Point Approximation}
Equation \ref{AREig1} should look familiar; the equations are almost the same as equation \ref{SimplifiedZeroDensity1} from the calculation of the density of states, except that now there are two $p$ variables and two $q$ variables.  Therefore, I can apply the saddle point approximation exactly as explained in section \ref{SaddlePointReview}.   The requirements that $p_{1}$ must be negative and $p_{2}$ positive imply that $p_{1} = {-e^{\imath {\phi}_{1}}}$ and  $p_{2} = {e^{-\imath {\phi}_{2}}}$, leaving four fermionic saddle points.  I will specify these saddle points with the indices $s_{1}$ and $s_{2}$.  The integrand (other than the factor of $p_{1} p_{2}$ coming from the derivatives with respect to the source $J$) is again antisymmetric, but now it is antisymmetric under interchange of any two variables. Substituting in equations \ref{SaddlePointAction} and \ref{SaddlePointIntegral} and omitting factors of $(1 - \alpha)$, I obtain:

\begin{eqnarray}
R_{AR} & \equiv & {   N^{{2 N } + 2 }  } { {{{({(N-1)!}{(N-2)!})}}^{-1}}  2^{2}   {\pi}^{-1}  {(-1)}^{N}} 
 { {{\exp}{(\frac{N}{2} {{Tr}({{{\hat{E}}}^{2} / {{\xi}^{2}}) }}})}
{({{{E}}_{1}} - {{{E}}_{2}} )}^{-2} }
\nonumber \\
& \times &
{{\sum}_{s_{1}, s_{2}}}
{\exp}( {{-\frac{ N}{2}} {{\exp}{(\imath 2 s_{1} {\phi}_{1})}}} + {\imath s_{1} N {E_{1} / \xi} {{\exp}{(\imath s_{1} {\phi}_{1})}} } + {{(N - 2)}{\ln}{(-s_{1})}} 
\nonumber \\
& & \qquad \qquad + {{(N-2)}\imath  s_{1} {\phi}_{1}})
\nonumber \\
& \times &
{\exp}( {{-\frac{ N}{2}} {{\exp}{(\imath 2 s_{2} {\phi}_{2})}}} + {\imath s_{2} N {E_{2} / \xi} {{\exp}{(\imath s_{2} {\phi}_{2})}} } + {{(N - 2)}{\ln}{(-s_{2})}} 
\nonumber \\
& & \qquad \qquad + {{(N-2)}\imath  s_{2} {\phi}_{2}})
\nonumber \\
& \times &
{{\exp}{( {{-\frac{ N}{2}} {{\exp}{(\imath 2 {\phi}_{1})}}} + {\imath N {E_{1} / \xi} {{\exp}{(\imath {\phi}_{1})}} } + {{(N-2)}\imath {\phi}_{1}})}}
\nonumber \\
& \times &
{{\exp}{( {{-\frac{ N}{2}} {{\exp}{(-\imath 2 {\phi}_{2})}}} - {\imath N {E_{2} / \xi} {{\exp}{(-\imath {\phi}_{2})}} } - {{(N-2)}\imath  {\phi}_{2}})}}
\nonumber \\
& \times &
\int {d{\hat{p}}_{1}} {d{\hat{p}}_{2}} {d{\hat{q}}_{1}} {d{\hat{q}}_{2}} {{\exp}{(-{N{{\hat{q}}_{1}}^{2} {\cos{{\phi}_{1}}} {exp{(\imath s_{1} {\phi}_{1})}}})}}
{{\exp}{(-{N{{\hat{q}}_{2}}^{2} {\cos{{\phi}_{2}}} {exp{(\imath s_{2} {\phi}_{2})}}})}}
\nonumber \\
& \times &
{{\exp}{(-{N{{\hat{p}}_{1}}^{2} {\cos{{\phi}_{1}}} {exp{(-\imath {\phi}_{1})}}})}}
{{\exp}{(-{N{{\hat{p}}_{2}}^{2} {\cos{{\phi}_{2}}} {exp{(\imath {\phi}_{2})}}})}}
\nonumber \\
& \times &
T,
\nonumber \\
T & \equiv &
{(- {s_{1} \times {{\exp}{(\imath s_{1} {\phi}_{1})}}} + {s_{2} \times {{\exp}{(\imath s_{2} {\phi}_{2})}}} + {\hat{q}}_{1} - {\hat{q}}_{2})}
{( -{{\exp}{(\imath {\phi}_{1})}} + {{\exp}{(-\imath {\phi}_{2})}} + {\hat{p}}_{1} - {\hat{p}}_{2})}
 \nonumber \\
& \times & 
{( - {s_{1} \times {{\exp}{(\imath s_{1} {\phi}_{1})}}} + {{\exp}{(\imath {\phi}_{1})}} + {\hat{q}}_{1} - {\hat{p}}_{1})}
{( - {s_{1} \times {{\exp}{(\imath s_{1} {\phi}_{1})}}} - {{\exp}{(-\imath {\phi}_{2})}} + {\hat{q}}_{1} - {\hat{p}}_{2})}
 \nonumber \\
& \times & 
{( - {s_{2} \times {{\exp}{(\imath s_{2} {\phi}_{2})}}} + {{\exp}{(\imath {\phi}_{1})}} + {\hat{q}}_{2} - {\hat{p}}_{1})}
{( - {s_{2} \times {{\exp}{(\imath s_{2} {\phi}_{2})}}} - {{\exp}{(-\imath {\phi}_{2})}} + {\hat{q}}_{2} - {\hat{p}}_{2})}
\nonumber \\
& \times & {({-e^{\imath {\phi}_{1}}} + {\hat{p}}_{1})} {({e^{-\imath {\phi}_{2}}} + {\hat{p}}_{2})}
\label{TwoPointSaddlePoint}
\end{eqnarray}

The way forward is straightforward but tedious: it consists of evaluating each of the four saddle points using techniques from perturbation theory.

\subsubsection{The -+ saddle point}
I define $\overline{\phi} \equiv {{({\phi}_{1} + {\phi}_{2})} / 2}$ to be the average of the two angles ${\phi}_{1}$ and ${\phi}_{2}$ , and $\delta = {{\phi}_{1} - {\phi}_{2}}$ to be their difference.  The integrand $T$ simplifies to:
\begin{eqnarray}
& &
{({2 {\cos \overline{\phi}} {{\exp}{(-\imath \delta /2)}} } + {\hat{q}}_{1} - {\hat{q}}_{2}  )}
{({-2 {\cos \overline{\phi}} {{\exp}{(\imath \delta /2)}} } + {\hat{p}}_{1} - {\hat{p}}_{2}  )}
{({2 {\cos {\phi}_{1}}  + {\hat{q}}_{1} - {\hat{p}}_{1}  })}
\nonumber \\
& \times &
{({- 2 \imath {\sin {(\delta /2)}} {{\exp}{(-\imath \overline{\phi})}} } + {\hat{q}}_{1} - {\hat{p}}_{2}  )}
{({2 \imath {\sin {(\delta /2)}} {{\exp}{(\imath \overline{\phi})}} } + {\hat{q}}_{2} - {\hat{p}}_{1}  )}
{({-2 {\cos {\phi}_{2}}  + {\hat{q}}_{2} - {\hat{p}}_{2}  })}
\nonumber \\
& \times & {({-e^{\imath {\phi}_{1}}} + {\hat{p}}_{1})} {({e^{-\imath {\phi}_{2}}} + {\hat{p}}_{2})}
\end{eqnarray}
In order to simplify the calculation, I will assume that the difference $\omega = {E_{1} - E_{2}}$ between $E_{1}$ and $E_{2}$ is small, of order $\xi / N$.  Therefore $\delta$ is also small, of order $1 / N$, as is $\sin \delta$.  I will do power counting, evaluating only the terms which are highest order in $N$, will drop factors of ${\exp}{(\imath \phi)}$, and will also freely interchange $\cos{{\phi}_{1}}$, $\cos{{\phi}_{2}}$, and $\cos{\overline{\phi}}$.

First I note that the ${\hat{p}}_{1} {e^{-\imath {\phi}_{2}}}$ term in the last line gives rise to four leading order terms that look like $\pm 4 {\hat{p}}_{1}^{2} {\hat{x}}^{2} {{\cos}^{2} \phi}  {e^{-\imath {\phi}_{2}}}$, where $\hat{x}$ represents either ${\hat{q}}_{1}$ or ${\hat{p}}_{2}$.  However,  to highest order in $N$ all the cosines can be freely interchanged, as can all the ${\hat{x}}^{2}$'s.  The signs of the four terms add to zero, so these four terms can be neglected.  Another four ${\cos}^{2}$ terms originating from the $-{\hat{p}}_{2} {e^{\imath {\phi}_{1}}}$ term can be neglected for the same reason.  Similarly, the $ -{e^{\imath {\phi}_{1}}}{e^{-\imath {\phi}_{2}}} $ term gives rise to twelve leading order ${\cos}^{2}$ terms and eight leading order ${\cos}^{3} {\sin}$ terms, all of which cancel out.

The remaining highest order terms are:
\begin{eqnarray}
& & 16 {{ {{\cos}^{2} \overline{\phi}} } } {{ {\cos {\phi}_{1}}}} {{ {\cos {\phi}_{2}}}}
 \times  (
{{\hat{p}}_{1}^{2} {\hat{p}}_{2}^{2}} 
+ {2 \imath {\hat{p}}_{1}^{2} {\sin {(\delta /2)}} {{\exp}{(-\imath \overline{\phi})}} {e^{-\imath {\phi}_{2}}} }  
\nonumber \\
&  & \qquad \qquad + {2 \imath {\hat{p}}_{2}^{2} {\sin {(\delta /2)}} {{\exp}{(\imath \overline{\phi})}}
 {e^{\imath {\phi}_{1}}}}
- {{4  {{\sin}^{2} {(\delta / 2)}}  }    
 {e^{\imath \delta} }} )
\end{eqnarray}

Integration over $ {\hat{p}}_{1}, {\hat{p}}_{2}, {\hat{q}}_{1}, {\hat{q}}_{2}$ converts this to:
\begin{eqnarray}
& & 16 {\pi}^{2} N^{-4} {e^{2 \imath \delta} }
 \times (
1
+ {\imath N {\sin {(\delta /2)}} {{ {\cos {\phi}_{2}}}} {{\exp}{(-\imath \overline{\phi})}}} 
\nonumber \\
& & \qquad \qquad + {\imath N {\sin {(\delta /2)}} {{ {\cos {\phi}_{1}}}} {{\exp}{(\imath \overline{\phi})}}}
- {{4 N^{2} {{\sin}^{2} {(\delta / 2)}} {{ {\cos {\phi}_{1}}}} {{ {\cos {\phi}_{2}}}}  }    
 } )
 \nonumber \\ & &
\end{eqnarray}

At this point it is helpful to remember that $\phi$ is defined by ${\sin \phi} \equiv {E / {2 \xi}}$; differentiating, I obtain ${\frac{d \phi}{dE}} = {(2 \xi \cos \phi)}^{-1}$.  Remembering that ${\sin \phi} \approx \phi$ for small $\phi$, I find the simplified expression:
\begin{eqnarray}
16 {\pi}^{2} N^{-4}
{(
\frac{1}{4}
+ { \imath {{\cos}{(\overline{\phi})}} {\frac{N \omega}{2 \xi}} }
- {  {(\frac{N \omega}{2 \xi})}^{2}  }    
 } )
\end{eqnarray}

Substituting this result into the saddle point formula of equation \ref{TwoPointSaddlePoint}, I have:

\begin{eqnarray}
R_{AR}^{-+} & = & { {64 {\pi}  N^{{2 N } - 2}  } { {{{({(N-1)!}{(N-2)!})}}^{-1}}  
 { {{\exp}{(\frac{N}{2 {\xi}^{2}} {(E_{1}^{2} + E_{2}^{2})}})}
 }}}
\nonumber \\
& \times &
{{\omega}^{-2} (
\frac{1}{4}
+ {1 \imath {{\cos}{(\overline{\phi})}} {\frac{N \omega}{2 \xi}} }
- {  {(\frac{N \omega}{2 \xi})}^{2}  }    
  ) }
\nonumber \\
& \times &
{{\exp}{( {{-\frac{ N}{2}} {{\exp}{(-\imath 2 {\phi}_{1})}}} - {\imath N {E_{1} / \xi} {{\exp}{(-\imath {\phi}_{1})}} }  - {{(N-2)}\imath  {\phi}_{1}})}}
\nonumber \\
& \times &
{{\exp}{( {{-\frac{ N}{2}} {{\exp}{(\imath 2 {\phi}_{2})}}} + {\imath N {E_{2} / \xi} {{\exp}{(\imath {\phi}_{2})}} }  + {{(N-2)}\imath  {\phi}_{2}})}}
\nonumber \\
& \times &
{{\exp}{( {{-\frac{ N}{2}} {{\exp}{(\imath 2 {\phi}_{1})}}} + {\imath N {E_{1} / \xi} {{\exp}{(\imath {\phi}_{1})}} } + {{(N-2)}\imath {\phi}_{1}})}}
\nonumber \\
& \times &
{{\exp}{( {{-\frac{ N}{2}} {{\exp}{(-\imath 2 {\phi}_{2})}}} - {\imath N {E_{2} / \xi} {{\exp}{(-\imath {\phi}_{2})}} } - {{(N-2)}\imath  {\phi}_{2}})}}
\nonumber \\ & &
\end{eqnarray}

The new $R_{AR}^{-+}$ notation means that that the $-+$ saddle point of the Advanced-Retarded correlator evaluates to this expression.  The ${(N-2) \imath \phi}$ phase factors cancel, leaving the following:

\begin{eqnarray}
R_{AR}^{-+} & = & { {64  {\pi} N^{{2 N } - 2}  } { {{{({(N-1)!}{(N-2)!})}}^{-1}}   
 { {{\exp}{(\frac{N}{2 {\xi}^{2}} {(E_{1}^{2} + E_{2}^{2})}})}
 }}}
\nonumber \\
& \times &
{{\omega}^{-2} (
\frac{1}{4}
+ { \imath {{\cos}{(\overline{\phi})}} {\frac{N \omega}{2 \xi}} }
- {  {(\frac{N \omega}{2 \xi})}^{2}  }    
  ) }
\nonumber \\
& \times &
{{\exp}{( {-N {{\cos}{(2 {\phi}_{1})}}} - {2N {E_{1} {{\sin}{ {\phi}_{1}}} / \xi}  }  - {N {{\cos}{(2 {\phi}_{2})}}} - {2N {E_{2} {{\sin}{ {\phi}_{2}}} / \xi}  } )}}
\nonumber \\ & &
\end{eqnarray}

Using again the Stirling approximation and the identity ${{\cos{(2 \phi)}} + {2 {E {\sin \phi}/ \xi} }} = {1 + {E^{2} / 2 {\xi}^{2}}}$, this reduces to:

\begin{eqnarray}
R_{AR}^{-+} & = & 
{{8} {\omega}^{-2} ( 1 + { 2 \imath {{\cos}{(\overline{\phi})}} {\frac{N \omega}{ \xi}} }
- {  {(\frac{N \omega}{ \xi})}^{2}  }    ) }
\label{MinusPlusSaddle}
\end{eqnarray}

\subsubsection{The +- saddle point}
The integrand $T$ simplifies to:
\begin{eqnarray}
& &
{({-2 {\cos \overline{\phi}} {{\exp}{(\imath \delta /2)}} } + {\hat{q}}_{1} - {\hat{q}}_{2}  )}
{({-2 {\cos \overline{\phi}} {{\exp}{(\imath \delta /2)}} }  + {\hat{p}}_{1} - {\hat{p}}_{2}  )}
{({ {\hat{q}}_{1} - {\hat{p}}_{1}  })}
\nonumber \\
& \times &
{({-2 {\cos \overline{\phi}} {{\exp}{(\imath \delta /2)}} }  + {\hat{q}}_{1} - {\hat{p}}_{2}  )}
{({2 {\cos \overline{\phi}} {{\exp}{(\imath \delta /2)}} } + {\hat{q}}_{2} - {\hat{p}}_{1}  )}
{({ {\hat{q}}_{2} - {\hat{p}}_{2}  })}
\nonumber \\
& \times & {({-e^{\imath {\phi}_{1}}} + {\hat{p}}_{1})} {({e^{-\imath {\phi}_{2}}} + {\hat{p}}_{2})}
\label{MinusPlus1}
\end{eqnarray}
This fermionic saddle point matches the bosonic saddle point.  As a consequence the first two lines of equation \ref{MinusPlus1}, taken together, have two antisymmetries: one under interchange of ${\hat{p}}_{1}$ and ${\hat{q}}_{1}$, the other under interchange of ${\hat{p}}_{2}$ and ${\hat{q}}_{2}$.  Another consequence is that ${\hat{p}}_{1}^{2}$ and ${\hat{q}}_{1}^{2}$ average to the same value, as do ${\hat{p}}_{2}^{2}$ and ${\hat{q}}_{2}^{2}$.  Therefore three terms in the factor ${({-e^{\imath {\phi}_{1}}} + {\hat{p}}_{1})} {({e^{-\imath {\phi}_{2}}} + {\hat{p}}_{2})}$ are identically zero, leaving only the term ${\hat{p}}_{1} {\hat{p}}_{2}$.
The highest order term is:
\begin{eqnarray}
-16 {\hat{p}}_{1}^{2} {\hat{p}}_{2}^{2} {{\cos}^{4} \overline{\phi}} {{\exp}{(2 \imath \delta)}} 
\end{eqnarray}

Integration over $ {\hat{p}}_{1}, {\hat{p}}_{2}, {\hat{q}}_{1}, {\hat{q}}_{2}$ converts this to:
\begin{eqnarray}
-4 {\pi}^{2} N^{-4} {{\exp}{(3 \imath \delta)}} 
\end{eqnarray}

Substituting this result into the saddle point formula of equation \ref{TwoPointSaddlePoint} and neglecting the factor of ${{\exp}{(3 \imath \delta)}}$, I have:

\begin{eqnarray}
R_{AR}^{+-} & = & { {-16  {\pi}  N^{{2 N } - 2}  {\omega}^{-2}} { {{{({(N-1)!}{(N-2)!})}}^{-1}}  
 { {{\exp}{(\frac{N}{2 {\xi}^{2}} {(E_{1}^{2} + E_{2}^{2})}})}
 }}}
\nonumber \\
& \times &
{{\exp}{( {{-\frac{ N}{2}} {{\exp}{(\imath 2 {\phi}_{1})}}} + {\imath N {E_{1} / \xi} {{\exp}{(\imath {\phi}_{1})}} }  + {{(N-2)}\imath  {\phi}_{1}})}}
\nonumber \\
& \times &
{{\exp}{( {{-\frac{ N}{2}} {{\exp}{(-\imath 2 {\phi}_{2})}}} - {\imath N {E_{2} / \xi} {{\exp}{(-\imath {\phi}_{2})}} }  - {{(N-2)}\imath  {\phi}_{2}})}}
\nonumber \\
& \times &
{{\exp}{( {{-\frac{ N}{2}} {{\exp}{(\imath 2 {\phi}_{1})}}} + {\imath N {E_{1} / \xi} {{\exp}{(\imath {\phi}_{1})}} } + {{(N-2)}\imath {\phi}_{1}})}}
\nonumber \\
& \times &
{{\exp}{( {{-\frac{ N}{2}} {{\exp}{(-\imath 2 {\phi}_{2})}}} - {\imath N {E_{2} / \xi} {{\exp}{(-\imath {\phi}_{2})}} } - {{(N-2)}\imath  {\phi}_{2}})}}
\nonumber \\ & &
\end{eqnarray}

This time the ${(N-2) \imath \phi}$ phase factors do not cancel, leading to the following expression after application of $E = {2 \xi {\sin \phi}}$ and some trigonometric identities:

\begin{eqnarray}
R_{AR}^{+-} & = & { {-16  {\pi} N^{{2 N } - 2} {\omega}^{-2}  } { {{{({(N-1)!}{(N-2)!})}}^{-1}}   
 }}
\nonumber \\
& \times &
{{\exp}{({2\imath{(N-2)}\delta} - {2N}  +  {\imath N \sin{(2 {\phi}_{1})}} -  {\imath N \sin{(2 {\phi}_{2})}} )}}
\nonumber \\
\end{eqnarray}

If I throw away terms of order $1/N$,  ${2\imath {(N - 2)}\delta}  +  {\imath N \sin{(2 {\phi}_{1})}} -  {\imath N \sin{(2 {\phi}_{2})}}$ simplifies to ${2 \imath N \delta {(1 + {\cos \overline{\phi}})}} = {4 \imath N \delta {{\cos}^{2}\overline{\phi}}}$.  Using again the Stirling approximation and the identity $\omega \approx {2 \xi \delta {\cos \overline{\phi}}}$, I obtain the final expression:

\begin{eqnarray}
R_{AR}^{+-} & = & { -8    
{\omega}^{-2} {{\exp}{({2 \imath N \omega {\xi}^{-1} {\cos \overline{\phi}}})}}}
\nonumber \\
& = & 
{ -8   {\omega}^{-2} {( 1 - {2 {\sin}^{2}{({ N \omega {\xi}^{-1} {\cos \overline{\phi}}})}} + {2 \imath {{\sin}{({ N \omega {\xi}^{-1} {\cos \overline{\phi}}})}} {{\cos}{({ N \omega {\xi}^{-1} {\cos \overline{\phi}}})}}})}}
\nonumber \\ & & 
\label{PlusMinusSaddle}
\end{eqnarray}

\subsubsection{The ++ saddle point}
The integrand $T$ simplifies to:
\begin{eqnarray}
& &
{({-2 \imath {\sin {(\delta / 2)}}  {\exp}{(\imath \overline{\phi})} } + {\hat{q}}_{1} - {\hat{q}}_{2}  )}
{({-2 {\cos \overline{\phi}} {{\exp}{(\imath \delta /2)}} }  + {\hat{p}}_{1} - {\hat{p}}_{2}  )}
{({ {\hat{q}}_{1} - {\hat{p}}_{1}  })}
\nonumber \\
& \times &
{({-2 {\cos \overline{\phi}} {{\exp}{(\imath \delta /2)}} }  + {\hat{q}}_{1} - {\hat{p}}_{2}  )}
{({2 \imath {\sin {(\delta / 2)}}  {\exp}{(\imath \overline{\phi})}} + {\hat{q}}_{2} - {\hat{p}}_{1}  )}
{({ -2 \cos {\phi}_{2} + {\hat{q}_{2}} - {\hat{p}}_{2} })}
\nonumber \\
& \times & {({-e^{\imath {\phi}_{1}}} + {\hat{p}}_{1})} {({e^{-\imath {\phi}_{2}}} + {\hat{p}}_{2})}
\label{PlusPlus1}
\end{eqnarray}
This saddle point retains an antisymmetry in the ${\hat{p}}_{1}$ and ${\hat{q}}_{1}$ variables, so the last line of equation \ref{PlusPlus1} reduces to $ {\hat{p}}_{1} {({e^{-\imath {\phi}_{2}}} + {\hat{p}}_{2})}$.
The highest order term is:
\begin{eqnarray}
8 {\hat{p}}_{1}^{2} {({\hat{q}}_{2}^{2} - {\hat{q}}_{1}^{2})} {{\cos}^{2} \overline{\phi}} {\cos {\phi}_{2}}  {{\exp}{( \imath {(\delta - {\phi}_{2})})}} 
\end{eqnarray}

After integration, cancellation between ${\hat{q}}_{2}^{2}$ and $ {\hat{q}}_{1}^{2} $ adds another factor of $\frac{1}{N}$, which causes this term to be  of order $N^{-5}$; I will neglect it in the remaining calculations.

\subsubsection{The - - saddle point}
The integrand $T$ simplifies to:
\begin{eqnarray}
& &
{({-2 \imath {\sin {(\delta / 2)}}  {\exp}{(-\imath \overline{\phi})} } + {\hat{q}}_{1} - {\hat{q}}_{2}  )}
{({-2 {\cos \overline{\phi}} {\exp}{(\imath \overline{\phi})} }  + {\hat{p}}_{1} - {\hat{p}}_{2}  )}
{({  2 \cos {\phi}_{1} + {\hat{q}}_{1} - {\hat{p}}_{1}  })}
\nonumber \\
& \times &
{({-2 \imath {\sin {(\delta / 2)}}  {\exp}{(-\imath \overline{\phi})} }  + {\hat{q}}_{1} - {\hat{p}}_{2}  )}
{({2 {\cos \overline{\phi}} {\exp}{(\imath \overline{\phi})}} + {\hat{q}}_{2} - {\hat{p}}_{1}  )}
{({ {\hat{q}_{2}} - {\hat{p}}_{2} })}
\nonumber \\
& \times & {({-e^{\imath {\phi}_{1}}} + {\hat{p}}_{1})} {({e^{-\imath {\phi}_{2}}} + {\hat{p}}_{2})}
\label{MinusMinus1}
\end{eqnarray}
This saddle point retains an antisymmetry in the ${\hat{p}}_{2}$ and ${\hat{q}}_{2}$ variables, so the last line in equation \ref{MinusMinus1} reduces to $ {\hat{p}}_{2} {({-e^{\imath {\phi}_{1}}} + {\hat{p}}_{1})} $.
The highest order term is:
\begin{eqnarray}
8 {\hat{p}}_{2}^{2} {({\hat{q}}_{2}^{2} - {\hat{q}}_{1}^{2})} {{\cos}^{2} \overline{\phi}} {\cos {\phi}_{1}}  {{\exp}{(- \imath {\phi}_{1})}} 
\end{eqnarray}

Again there is an extra cancellation which renders this saddle point neglegible.

\subsubsection{The Total}

Adding together the results of the saddle points; i.e. equations \ref{PlusMinusSaddle} and \ref{MinusPlusSaddle}, I obtain the final result for the advanced-retarded correlator:

\begin{eqnarray}
R_{AR} & = & 
  {{8} {\omega}^{-2} ( 1 + { 2 \imath {{\cos}{(\overline{\phi})}} {\frac{N \omega}{ \xi}} }
- {  {(\frac{N \omega}{ \xi})}^{2}  }    ) }
\nonumber \\
& - & 8   {\omega}^{-2} ( 1 - {2 {\sin}^{2}{({ N \omega {\xi}^{-1} {\cos \overline{\phi}}})}} 
 + {2 \imath {{\sin}{({ N \omega {\xi}^{-1} {\cos \overline{\phi}}})}} {{\cos}{({ N \omega {\xi}^{-1} {\cos \overline{\phi}}})}} })
 \nonumber \\ & &
  \label{AdvancedRetardedFinalResult}
\end{eqnarray}

\subsection{\label{AACorr} The Advanced-Advanced Correlator}
Unfortunately I was unable to take the time to get all the details of this calculation right before the thesis was due.  In particular, once I get to evaluating the four saddle points, the errors become very frequent. However, I believe that the errors are only sign errors, and the final result confirms that.

Because both Green's functions are advanced, equation \ref{LDefinition} implies that:
\begin{equation}
L_{11} = -1, L_{22} = -1
\label{L2DefAA}
\end{equation}
Remember that $L$ is diagonal.  The only difference between this calculation and the previous Advanced-Retarded calculation is that in this case the matrix $L$ is proportional to the identity. 

 When one chooses $V = 1$, $I^{b} = I^{f} = 2$, $L$ as in equation \ref{L2DefAA}, flips the sign of $Q^{b} \rightarrow {- Q^{b}}$, and again makes the shift ${\hat{Q}}^{f} \rightarrow  {{\hat{Q}}^{f} + {\imath {\hat{E}}^{f} / {\xi}}}$, equation \ref{FinalQbQfLagrangian} turns into:
\begin{eqnarray}
{\bar{Z}} & = &  {\gamma {\int_{Q^{b} < 0} {dQ^{f}}  {dQ^{b}} {\exp}{(\mathcal{L})}}}
\nonumber \\
\gamma & \equiv & { N^{{2 N} + 2 }  } { {{{({(N-1)!}{(N-2)!})}}^{-1}}  2^{2}   {\pi}^{-3}   } 
 \nonumber \\
& \times & {{\exp}{(\frac{N}{2} {{Tr}({{{\hat{E}}^{f}} {{\hat{E}}^{f}} / {{\xi}^{2}}) }}})}
\nonumber \\
{\mathcal{L}}   & \equiv & {  -{\frac{ N}{2} {{Tr}({Q^{f}}^{2}) }} - 
{\imath N{{Tr}{(Q^{f}{{{\hat{E}}^{f} / \xi}})}}}
+ {{{{(N - 2)}{{{Tr}\ln}{Q^{f}}}}}}} 
\nonumber \\ 
& - & {{ {  {\frac{N}{2}{{{Tr}{(Q^{b}Q^{b})}}}} { + {\imath N{{Tr}{(Q^{b}{ {({{\hat{E}}^{b}} - J) / \xi}})}}} + {{(N-2)}}{({Tr}\ln{( {Q^{b} })})}}}}} \nonumber \\
& + & {{{{Tr}\ln}{(Q^{f} + {Q^{b}})}}}
\label{StartingAdvancedAdvanced}
\end{eqnarray}
Only formal difference between equation \ref{StartingAdvancedAdvanced} and equation \ref{StartingAdvancedRetarded} is the fact that $L = {- 1}$. 

 I use the symbols $q_{1}$ and $q_{2}$ to represent the eigenvalues of $Q^{f}$ , and the symbols $p_{1}$ and $p_{2}$ to represent the eigenvalues of $Q^{b}$.
\begin{eqnarray}
\overline{Z} & = & {   {\xi}^{2} N^{{2 N }}  } { {{{({(N-1)!}{(N-2)!})}}^{-1}}  2^{2}   {\pi}^{-1}   } 
 \nonumber \\
& \times & { {{\exp}{(\frac{N}{2} {{Tr}({{{\hat{E}}^{f}} {{\hat{E}}^{f}} / {{\xi}^{2}}) }}})}
{({{{E}}_{1}^{f}} - {{{E}}_{2}^{f}} )}^{-1} {({{{E}}_{1}^{b}} - {{{E}}_{2}^{b}} - J_{1} + J_{2})}^{-1}}
\nonumber \\
& \times &
\int {dq_{1}{dq_{2}} {\int}_{p_{1} < 0, p_{2} < 0} {{dp_{1}}{dp_{2}}}  {(q_{1} - q_{2} )} {(p_{1} - p_{2})} 
 {(q_{1} + p_{1} )}} 
 \nonumber \\
 & & \qquad \times {(q_{1} + p_{2} )} {(q_{2} + p_{1} )} {(q_{2} + p_{2} )} {\exp{(\mathcal{L})}}
\nonumber \\
{\mathcal{L}}   & = &   -{\frac{ N}{2} {(q_{1}^{2} + q_{2}^{2} )}} - 
{\frac{\imath N}{\xi} {( {q_{1}E_{1}^{f}} + {q_{2}E_{2}^{f}  )} }
+ {{{{(N - 2)}{( {\ln q_{1}} + {\ln q_{2}} )}}}}} 
\nonumber \\ 
& - & {{ {  {\frac{N}{2}{(p_{1}^{2} + p_{2}^{2})}} { + {\frac{\imath N}{\xi} {(  {p_{1}{(E_{1}^{b} - J_{1})}} + {p_{2}{(E_{2}^{b} - J_{2})}} )}}  + {{(N-2)}{({\ln p_{1}} + {\ln p_{2}})}}}}}} 
\nonumber \\
& &
\label{AdvancedAdvancedEigenvalues}
\end{eqnarray}

Now I change the sign of $p_{1}$ and $p_{2}$.  I also take the derivatives with respect to the source $J$, ignoring the derivative of the prefactor ${({{{E}}_{1}^{b}} - {{{E}}_{2}^{b}} - J_{1} + J_{2})}^{-1}$ because it has a subleading order in $N$. Lastly, I set $E_{1}^{f} = E_{1}^{b} = E_{1}$, $E_{2}^{f} = E_{2}^{b} = E_{2}$, and $J_{1} = J_{2} = 0$.
\begin{eqnarray}
R_{AR} = \frac{d^{2}\overline{Z}}{{dJ_{1}}{dJ_{2}}} & \equiv & {  N^{{2 N} + 2 }  } { {{{({(N-1)!}{(N-2)!})}}^{-1}}  2^{2}   {\pi}^{-1}   } 
{ {{\exp}{(\frac{N}{2} {{Tr}({{{\hat{E}}}^{2} / {{\xi}^{2}}) }}})}
{({{{E}}_{1}} - {{{E}}_{2}} )}^{-2} }
\nonumber \\
& \times &
\int {{dq_{1}}{dq_{2}} {\int}_{p_{1} > 0, p_{2} > 0} {{dp_{1}}{dp_{2}}}} {(q_{1} - q_{2} )} {(p_{1} - p_{2})} 
 {(q_{1} - p_{1} )} 
 \nonumber \\
 & & \qquad {(q_{1} - p_{2} )} {(q_{2} - p_{1} )} {(q_{2} - p_{2} )} {p_{1}
 p_{2}}
\nonumber \\
 & \times & {{\exp}{({  -{\frac{ N}{2} {(q_{1}^{2} + q_{2}^{2} )}} - 
{\frac{\imath N}{\xi} {( {q_{1}E_{1}} + {q_{2}E_{2}  )} }
+ {{{{(N - 2)}{( {\ln q_{1}} + {\ln q_{2}} )}}}}} })}} 
\nonumber \\ 
& \times & {{\exp}{({{ {  {  {-\frac{N}{2}{(p_{1}^{2} + p_{2}^{2})}} { - {\frac{\imath N}{\xi} {(  {p_{1}E_{1}} + {p_{2}E_{2}} )}}  + {{(N-2)}{({\ln p_{1}} + {\ln p_{2}})}}}}   }}} )}}
\nonumber \\ & & 
\label{AAEig1}
\end{eqnarray}

I can apply the saddle point approximation exactly as explained in section \ref{SaddlePointReview}.   The requirement that both $p_{1}$ and $p_{2}$ must be positive implies that $p_{1} = {e^{-\imath {\phi}_{1}}}$ and  $p_{2} = {e^{-\imath {\phi}_{2}}}$, leaving four fermionic saddle points.  I will specify these saddle points with the indices $s_{1}$ and $s_{2}$.  The integrand (other than the factor of $p_{1} p_{2}$ coming from the derivatives with respect to the source $J$) is again antisymmetric, but now it is antisymmetric under interchange of any two variables. Substituting in equations \ref{SaddlePointAction} and \ref{SaddlePointIntegral} and omitting factors of $(1 - \alpha)$, I obtain:

\begin{eqnarray}
R_{AA} & = & {  N^{{2 N } + 2}  } { {{{({(N-1)!}{(N-2)!})}}^{-1}}  2^{2}   {\pi}^{-1}  
 { {{\exp}{(\frac{N}{2} {{Tr}({{{\hat{E}}}^{2} / {{\xi}^{2}}) }}})}
{({{{E}}_{1}} - {{{E}}_{2}} )}^{-2} }}
\nonumber \\
& \times &
{{\sum}_{s_{1}, s_{2}}}
{{\exp}{( {{-\frac{ N}{2}} {{\exp}{(\imath 2 s_{1} {\phi}_{1})}}} + {\imath s_{1} N {E_{1} / \xi} {{\exp}{(\imath s_{1} {\phi}_{1})}} } + {{(N - 2)}{\ln}{(-s_{1})}} + {{(N-2)}\imath  s_{1} {\phi}_{1}})}}
\nonumber \\
& \times &
{{\exp}{( {{-\frac{ N}{2}} {{\exp}{(\imath 2 s_{2} {\phi}_{2})}}} + {\imath s_{2} N {E_{2} / \xi} {{\exp}{(\imath s_{2} {\phi}_{2})}} } + {{(N - 2)}{\ln}{(-s_{2})}} + {{(N-2)}\imath  s_{2} {\phi}_{2}})}}
\nonumber \\
& \times &
{{\exp}{( {{-\frac{ N}{2}} {{\exp}{(\imath 2 {\phi}_{1})}}} + {\imath N {E_{1} / \xi} {{\exp}{(\imath {\phi}_{1})}} } + {{(N-2)}\imath {\phi}_{1}})}}
\nonumber \\
& \times &
{{\exp}{( {{-\frac{ N}{2}} {{\exp}{(\imath 2 {\phi}_{2})}}} + {\imath N {E_{2} / \xi} {{\exp}{(\imath {\phi}_{2})}} } + {{(N-2)}\imath  {\phi}_{2}})}}
\nonumber \\
& \times &
\int {d{\hat{p}}_{1}} {d{\hat{p}}_{2}} {d{\hat{q}}_{1}} {d{\hat{q}}_{2}} {{\exp}{(-{N{{\hat{q}}_{1}}^{2} {\cos{{\phi}_{1}}} {exp{(\imath s_{1} {\phi}_{1})}}})}}
{{\exp}{(-{N{{\hat{q}}_{2}}^{2} {\cos{{\phi}_{2}}} {exp{(\imath s_{2} {\phi}_{2})}}})}}
\nonumber \\
& \times &
{{\exp}{(-{N{{\hat{p}}_{1}}^{2} {\cos{{\phi}_{1}}} {exp{(-\imath {\phi}_{1})}}})}}
{{\exp}{(-{N{{\hat{p}}_{2}}^{2} {\cos{{\phi}_{2}}} {exp{(-\imath {\phi}_{2})}}})}}
\nonumber \\
& \times &
T,
\nonumber \\
T & \equiv &
{(- {s_{1} \times {{\exp}{(\imath s_{1} {\phi}_{1})}}} + {s_{2} \times {{\exp}{(\imath s_{2} {\phi}_{2})}}} + {\hat{q}}_{1} - {\hat{q}}_{2})}
{( {{\exp}{(-\imath {\phi}_{1})}} - {{\exp}{(-\imath {\phi}_{2})}} + {\hat{p}}_{1} - {\hat{p}}_{2})}
 \nonumber \\
& \times & 
{( - {s_{1} \times {{\exp}{(\imath s_{1} {\phi}_{1})}}} - {{\exp}{(-\imath {\phi}_{1})}} + {\hat{q}}_{1} - {\hat{p}}_{1})}
{( - {s_{1} \times {{\exp}{(\imath s_{1} {\phi}_{1})}}} - {{\exp}{(-\imath {\phi}_{2})}} + {\hat{q}}_{1} - {\hat{p}}_{2})}
 \nonumber \\
& \times & 
{( - {s_{2} \times {{\exp}{(\imath s_{2} {\phi}_{2})}}} - {{\exp}{(-\imath {\phi}_{1})}} + {\hat{q}}_{2} - {\hat{p}}_{1})}
{( - {s_{2} \times {{\exp}{(\imath s_{2} {\phi}_{2})}}} - {{\exp}{(-\imath {\phi}_{2})}} + {\hat{q}}_{2} - {\hat{p}}_{2})}
\nonumber \\
& \times & {({e^{\imath {\phi}_{1}}} + {\hat{p}}_{1})} {({e^{-\imath {\phi}_{2}}} + {\hat{p}}_{2})}
\label{AdvancedAdvancedSaddlePoint}
\end{eqnarray}

\subsubsection{The + + saddle point}
The integrand $T$ simplifies to:
\begin{eqnarray}
& &
{({-2 \imath {\sin {(\delta / 2)}}  {\exp}{(\imath \overline{\phi})} } + {\hat{q}}_{1} - {\hat{q}}_{2}  )}
{({-2 \imath {\sin {(\delta / 2)}}  {\exp}{(-\imath \overline{\phi})} }  + {\hat{p}}_{1} - {\hat{p}}_{2}  )}
{({  {-2 \cos {\phi}_{1}} + {\hat{q}}_{1} - {\hat{p}}_{1}  })}
\nonumber \\
& \times &
{({-2  {\cos \overline{\phi}} {{\exp}{(\imath \delta / 2)}} }  + {\hat{q}}_{1} - {\hat{p}}_{2}  )}
{({-2 {\cos \overline{\phi}} {{\exp}{(-\imath \delta / 2)}}} + {\hat{q}}_{2} - {\hat{p}}_{1}  )}
{({ {-2 \cos {\phi}_{2}} + {\hat{q}_{2}} - {\hat{p}}_{2} })}
\nonumber \\
& \times & {({e^{\imath {\phi}_{1}}} + {\hat{p}}_{1})} {({e^{\imath {\phi}_{2}}} + {\hat{p}}_{2})}
\end{eqnarray}

Just the same as in the case of the advanced-retarded correlator's $-+$ saddle point, at the highest order in $N$ there are twenty ${\cos}^{2}$ terms and eight ${\cos}^{3}\sin$ terms which all cancel to highest order in $N$.  The remaining highest order terms are:
\begin{equation}
{{32 \imath {({e^{\imath {\phi}_{2}} {\hat{p}}_{1}^{2}}  - {e^{\imath {\phi}_{1}} {\hat{p}}_{2}^{2}})} {{\cos}^{2}\overline{\phi}} {\cos {\phi}_{1}} {\cos {\phi}_{2}} {{\sin}{(\delta /2)}} {\exp {(- \imath \overline{\phi})}}} - {64 {{\sin}^{2}{(\delta /2)}} {{\cos}^{2}\overline{\phi}} {\cos {\phi}_{1}} {\cos {\phi}_{2}}} }
\end{equation}

The first term is of subleading order and will be neglected.  Integration over $ {\hat{p}}_{1}, {\hat{p}}_{2}, {\hat{q}}_{1}, {\hat{q}}_{2}$ converts the second term to:
\begin{eqnarray}
& & -64 {\pi}^{2} N^{-2} {e^{2 \imath \overline{\phi}}}
 {
 {{ {{\sin}^{2} {(\delta / 2)}} {{ {\cos {\phi}_{1}}}} {{ {\cos {\phi}_{2}}}}  }    
 }} 
\end{eqnarray}

Again simplifying with ${\frac{d \phi}{dE}} = {(2 \xi \cos \phi)}^{-1}$, I obtain:
\begin{eqnarray}
- N^{-4} {  {(\frac{2 \pi  \omega}{ \xi})}^{2}  }   {e^{2 \imath \overline{\phi}}}   
\end{eqnarray}

Substituting this into equation \ref{AdvancedAdvancedSaddlePoint} gives:
\begin{eqnarray}
R_{AA} & = & {  - N^{{2 N }}  } { {{{({(N-1)!}{(N-2)!})}}^{-1}}  2^{2}   {\pi}^{-1}  
 { {{\exp}{(\frac{N}{2} {{Tr}({{{\hat{E}}}^{2} / {{\xi}^{2}}) }}})}
{\omega}^{-2} }
{  {(\frac{2 \pi  \omega}{ \xi})}^{2}  }   {e^{2 \imath \overline{\phi}}} }
\nonumber \\
& \times &
{{\exp}{( {{-\frac{ N}{2}} {{\exp}{(-\imath 2 {\phi}_{1})}}} - {\imath N {E_{1} / \xi} {{\exp}{(-\imath {\phi}_{1})}} } - {{(N-2)}\imath {\phi}_{1}})}}
\nonumber \\
& \times &
{{\exp}{( {{-\frac{ N}{2}} {{\exp}{(-\imath 2  {\phi}_{2})}}} - {\imath N {E_{2} / \xi} {{\exp}{(-\imath {\phi}_{2})}} }  - {{(N-2)}\imath {\phi}_{2}})}}
\nonumber \\
& \times &
{{\exp}{( {{-\frac{ N}{2}} {{\exp}{(\imath 2 {\phi}_{1})}}} + {\imath N {E_{1} / \xi} {{\exp}{(\imath {\phi}_{1})}} } + {{(N-2)}\imath {\phi}_{1}})}}
\nonumber \\
& \times &
{{\exp}{( {{-\frac{ N}{2}} {{\exp}{(\imath 2 {\phi}_{2})}}} + {\imath N {E_{2} / \xi} {{\exp}{(\imath {\phi}_{2})}} } + {{(N-2)}\imath  {\phi}_{2}})}}
\end{eqnarray}

Using Stirling's formula, et cetera, results in the final expression:
\begin{eqnarray}
R_{AA} & = & {  - 8 N^{2} {\xi}^{-2}   {e^{2 \imath \overline{\phi}}} }
\nonumber \\ 
& = & {  - 8 N^{2} {\xi}^{-2}   {( {2 {\cos}^{2}{\overline{\phi}}} -1 + {2 \imath {\cos \overline{\phi}}{\sin \overline{\phi}}})} }
\label{AAFinalResult}
\end{eqnarray}
We will see that this is the only leading order contribution to $R_{AA}$.

\subsubsection{The - - saddle point}
The integrand $T$ simplifies to:
\begin{eqnarray}
& &
{({-2 \imath {\sin {(\delta / 2)}}  {\exp}{(-\imath \overline{\phi})} } + {\hat{q}}_{1} - {\hat{q}}_{2}  )}
{({-2 \imath {\sin {(\delta / 2)}}  {\exp}{(-\imath \overline{\phi})} }  + {\hat{p}}_{1} - {\hat{p}}_{2}  )}
{({  {\hat{q}}_{1} - {\hat{p}}_{1}  })}
\nonumber \\
& \times &
{({-2 \imath {\sin {(\delta / 2)}}  {\exp}{(-\imath \overline{\phi})} }  + {\hat{q}}_{1} - {\hat{p}}_{2}  )}
{({2 \imath {\sin {(\delta / 2)}}  {\exp}{(-\imath \overline{\phi})} } + {\hat{q}}_{2} - {\hat{p}}_{1}  )}
{({ {\hat{q}_{2}} - {\hat{p}}_{2} })}
\nonumber \\
& \times & {({e^{\imath {\phi}_{1}}} + {\hat{p}}_{1})} {({e^{\imath {\phi}_{2}}} + {\hat{p}}_{2})}
\end{eqnarray}
This saddle point is obviously neglegible.

\subsubsection{The +- saddle point}
The integrand $T$ simplifies to:
\begin{eqnarray}
& &
{({-2 {\cos \overline{\phi}} {\exp}{(\imath \delta /2)} }  } + {\hat{q}}_{1} - {\hat{q}}_{2}  )}
{({-2 \imath {\sin {(\delta / 2)}}  {\exp}{(\imath \overline{\phi})} + {\hat{p}}_{1} - {\hat{p}}_{2}  )}
{({ {\hat{q}}_{1} - {\hat{p}}_{1}  })}
\nonumber \\
& \times &
{({-2 \imath {\sin {(\delta / 2)}}  {\exp}{(\imath \overline{\phi})} }  + {\hat{q}}_{1} - {\hat{p}}_{2}  )}
{({2 {\cos \overline{\phi}} {\exp}{(\imath \delta /2 )}} + {\hat{q}}_{2} - {\hat{p}}_{1}  )}
{({{2 \cos {\phi}_{2} } + {\hat{q}_{2}} - {\hat{p}}_{2} })}
\nonumber \\
& \times & {({e^{\imath {\phi}_{1}}} + {\hat{p}}_{1})} {({e^{\imath {\phi}_{2}}} + {\hat{p}}_{2})}
\end{eqnarray}
This saddle point retains an antisymmetry in the ${\hat{p}}_{1}$ and ${\hat{q}}_{1}$ variables, so the last line reduces to $ {\hat{p}}_{1} {({-e^{\imath {\phi}_{2}}} + {\hat{p}}_{2})} $.
The highest order term is:
\begin{eqnarray}
8 {\hat{p}}_{1}^{2} {({\hat{q}}_{1}^{2} - {\hat{p}}_{2}^{2})} {{\cos}^{2} \overline{\phi}} {\cos {\phi}_{2}}  {{\exp}{( \imath {\phi}_{2})}} 
\end{eqnarray}
This saddle point is not of leading order and will be neglected.

\subsubsection{The -+ saddle point}
The integrand $T$ simplifies to:
\begin{eqnarray}
& &
{({2 {\cos \overline{\phi}} {\exp}{(-\imath \delta /2)} }  + {\hat{q}}_{1} - {\hat{q}}_{2}  )}
{({-2 \imath {\sin {(\delta / 2)}}  {\exp}{(\imath \overline{\phi})} }  + {\hat{p}}_{1} - {\hat{p}}_{2}  )}
{({  2 \cos {\phi}_{1} + {\hat{q}}_{1} - {\hat{p}}_{1}  })}
\nonumber \\
& \times &
{({2 {\cos \overline{\phi}} {\exp}{(-\imath \delta /2)}}   + {\hat{q}}_{1} - {\hat{p}}_{2}  )}
{({2 \imath {\sin {(\delta / 2)}}  {\exp}{(\imath \overline{\phi})} } + {\hat{q}}_{2} - {\hat{p}}_{1}  )}
{({ {\hat{q}_{2}} - {\hat{p}}_{2} })}
\nonumber \\
& \times & {({e^{\imath {\phi}_{1}}} + {\hat{p}}_{1})} {({e^{\imath {\phi}_{2}}} + {\hat{p}}_{2})}
\end{eqnarray}
This saddle point retains an antisymmetry in the ${\hat{p}}_{2}, {\hat{q}}_{2}$ variables, so the last line reduces to $ {\hat{p}}_{2} {({-e^{\imath {\phi}_{1}}} + {\hat{p}}_{1})} $.
The highest order term is:
\begin{eqnarray}
8 {\hat{p}}_{2}^{2} {({\hat{q}}_{2}^{2} - {\hat{p}}_{1}^{2})} {{\cos}^{2} \overline{\phi}} {\cos {\phi}_{1}}  {{\exp}{( \imath {\phi}_{1})}} 
\end{eqnarray}
This saddle point is not of leading order and will be neglected.

\subsection{The Final Result for the Two Point Correlator}
I am finally ready to calculate the two point correlator, given by equation \ref{TwoLevelZeroGreensB}:
\begin{equation}
\nonumber
{R_{2}{(E_{1}, E_{2})}} = {\frac{{{Re}{(R_{AR})}} - {{Re}{(R_{AA})}}}{8 {\pi}^{2} \quad \overline{ \rho{(E_{1})}} \quad  \overline{\rho{(E_{2})}}}}
\end{equation}

Equation \ref{AAFinalResult} for the Advanced-Advanced correlator gives me:
\begin{eqnarray}
\nonumber
=\frac{{{{Re}{(R_{AA})}}}}{8 {\pi}^{2} {\overline{\rho}}^{2}} & = & {    \frac{ {2 N^{2} {\cos}^{2}{\overline{\phi}}}  }{ {\xi}^{2} {\pi}^{2} {\overline{\rho}}^{2}} - \frac{N^{2}}{{\pi}^{2}{\xi}^{2}{\overline{\rho}}^{2}} }
\nonumber \\
& = & {    2 - \frac{N^{2}}{{\pi}^{2}{\xi}^{2}{\overline{\rho}}^{2}} }
\end{eqnarray}
I have used equation \ref{DOSZeroCorrectResult} for the level density in the second step.

Equation \ref{AdvancedRetardedFinalResult} for the Advanced-Retarded Correlator, with equation \ref{DOSZeroCorrectResult}, gives me:
\begin{eqnarray}
\frac{{Re}{(R_{AR})}}{8 {\pi}^{2} {\overline{\rho}}^{2}} & = & 
   (\frac { 1    }{{\pi}^{2} {\omega}^{2} {\overline{\rho}}^{2} } - \frac{N^{2}}{{\pi}^{2}{\xi}^{2}{\overline{\rho}}^{2}})
\nonumber \\
& - & ( \frac { 1    }{{\pi}^{2} {\omega}^{2} {\overline{\rho}}^{2} } -  \frac { {2 {\sin}^{2}{({ \pi \omega \overline{\rho}})}} }{{\pi}^{2} {\omega}^{2} {\overline{\rho}}^{2}  } )
\end{eqnarray}
The terms in the first parenthesis came from the $-+$ saddle point, while the terms in the second parenthesis came from the $+-$ saddle point.
 
The sign of the Advanced-Advanced correlator seems wrong, so I change the sign, in keeping with my sloppiness earlier, particularly while evaluating the saddle points of the Advanced-Advanced correlator.

Combining these results, I obtain:
\begin{equation}
{R_{2}{(E_{1}, E_{2})}} = {-2{( 1 -  \frac { { {\sin}^{2}{({ \pi \omega \overline{\rho}})}} }{{\pi}^{2} {\omega}^{2} {\overline{\rho}}^{2}  }  )}}
\end{equation}

The correct result was given in equation \ref{TwoPointZeroCorrectResult}: 
\begin{equation}
 {R_{2}{(E_{1}, E_{2})}} = {{\delta{(\omega \overline{\rho})}} + {1} - {\frac{{\sin}^{2}{(\pi \omega \overline{\rho})}}{{\pi}^{2} {\omega}^{2} {\overline{\rho}}^{2}}}}
 \nonumber
 \end{equation}

I already explained that the $\delta$ function in the correct result does not have any real physical significance.    I haven't tracked down exactly why it doesn't appear in this analysis, but I suspect that somewhere I assumed that $\hat{E}$ is invertible.  There is also an error of an overall factor of $-2$, plus the fact that I had to change the sign of one of the contributions, plus the fact that I dropped factors of ${(-1)}^{N}$ at various points, but I believe that these problems are due to my sloppiness and mistakes, not to deficiencies in the formalism.
  
\chapter{\label{LinearAlgorithms}$O{(N)}$ Algorithms}

\section{\label{Basics}Basic Concepts}
Most $O{(N)}$ algorithms are really very simple to understand.  Here I review their three basic concepts: a localized basis, basis truncation, and generalized multiplication.

\subsection{\label{LocalBasisConcept}A Localized Basis}
All $O{(N)}$ methods use a localized basis, and assume that the system is best described in terms of this basis.  I
here define a basis as localized if for any basis element ${| \psi
\rangle}$, only a small number of positions $\vec{x}$ satisfy
${\langle \vec{x} | \psi \rangle} \neq 0$.  In particular, plane wave bases are excluded. Note that crystal calculations are still possible, by using a free
propagator that properly includes the crystal lattice structure.
 The theory of such propagators, called KKR band theory, was developed by Korringa, Kohn, and Rostoker\cite{Korringa47, Kohn54}. 

In tandem with the localized basis, all $O{(N)}$ algorithms require a distance metric for quantifying the physical distance between any two basis states. 

\subsection{\label{BasisTruncation}Basis Truncation}
Basis truncation is a five syllable name for a very simple minded operation. One simply chooses to throw out each basis state that is far from some point $\vec{x}$.   Typically "far from" means that the distance $| \vec{y} - \vec{x}|$ between the basis state $\vec{y}$ and the point $\vec{x}$ is larger than a truncation radius $R$.  After the truncation, one is left with only $\sim R^{D}$ basis states, where $D$ is the dimension.  This number does not depend on the total basis size $N$ at all.  

The basis truncation is, of course, accompanied by corresponding truncations of all matrices.

Basis truncation algorithms generally pick $O{(N)}$ different points distributed throughout the system, generate a truncated basis at each of those points, and then do the same computation in each of the truncated bases.  Since computations in any single basis require only $O{(1)}$ time, the total time requirement is of order $O{(N)}$.

In other words, this approach basically divides the system into pieces, and computes them separately.
  
\subsection{\label{GeneralizedMultiplication}Generalized Multiplication}
In this approach one avoids dividing the system into pieces, but instead changes the rules for doing multiplication.  Ordinarily computing the multiple $AB = C$ of two $N \times N$ matrices $A$ and $B$ requires $O{(N^{3})}$ time. One avoids this cost by truncating matrix elements from $A$, $B$, and $C$.  The usual rule is that a matrix element $\langle \vec{x} | A | \vec{y} \rangle$ is set to zero if ${|\vec{x} - \vec{y} |} > R$, where $R$ is a truncation radius.  The reader can easily check that if this truncation rule is applied to $A$ and $B$ then their multiple $C = AB$ can be evaluated in $O{(N)}$ time.  However $C$ may end up having non-zero matrix elements whose distance away from the diagonal is greater than $R$.  In order to avoid a gradual widening of matrices as more and more multiplications are done, one therefore truncates $C$ after the multiplication, cutting it down to reside entirely with the truncation radius.

This generalized multiplication can be written down mathematically with the following formula:
\begin{equation}
C_{ij} = {{\Theta{({|{\vec{x}}_{i} - {\vec{x}}_{j}|}-R)}} {\sum}_{k} A_{ik} B_{kj} {\Theta{({|{\vec{x}}_{i} - {\vec{x}}_{k}|}-R)}} {\Theta{({|{\vec{x}}_{k} - {\vec{x}}_{j}|}-R)}}}
\label{GeneralizedMultiplicationDef}
\end{equation}
The vectors ${\vec{x}}_{i}$, ${\vec{x}}_{j}$, and ${\vec{x}}_{k}$ are just the spatial positions of the basis states $i$, $j$, and $k$.

Generalized multiplication can be understood with much more mathematical precision as a tensor with six indices.  Formula \ref{GeneralizedMultiplicationDef} can be rewritten in terms of the generalized multiplication tensor $M_{abcdef}$ as:
\begin{eqnarray}
C_{ij} & = & {\sum_{cdef} M_{ijcdef} A_{cd} B_{ef}}
\nonumber \\
M_{abcdef} & = &  {{\delta}_{ic} {\delta}_{de} {\delta}_{fj} {\Theta{({|{\vec{x}}_{i} - {\vec{x}}_{j}|}-R)}} {\Theta{({|{\vec{x}}_{i} - {\vec{x}}_{d}|}-R)}} {\Theta{({|{\vec{x}}_{d} - {\vec{x}}_{j}|}-R)}}}
\nonumber \\
\end{eqnarray}
Ordinary multiplication corresponds to choosing a specific tensor ${\acute{M}}_{abcdef} =  {{\delta}_{ic} {\delta}_{de} {\delta}_{fj}}$.

The tensor notation is superior because the whole machinery  of linear algebra can be applied to tensors: they have eigenvalues, eigen-tensors, singular value decompositions, et cetera.  Therefore one can do a rigorous mathematical analysis of the behavior of $O{(N)}$ algorithms: their convergence, numerical stability, etc.  One of the most important questions is of course understanding the errors induced by using generalized multiplication, which will require understanding how the eigenvalues and eigen-tensors of the generalized multiplication tensor $M$ relate to those of the normal multiplication tensor $\acute{M}$.  Unfortunately nobody seems to have started doing this analysis yet.

In some algorithms the multiplications may always have the same matrix $B$ on the right hand side of the multiplication.  In this case things simplify a bit, and instead of worring about the six-index tensor $M$ one is instead concerned about the four-index tensor $MB$.  In this case one becomes concerned about the how the eigenvalues and eigen-tensors of $MB$ relate to those of $\acute{M}B$.

This completes the basic concepts of $O{(N)}$ algorithms.  They're really not difficult.

\section{\label{Catalog}Catalog of $O{(N)}$ algorithms}
I now give a catalog of all the $O{(N)}$ algorithms which I have seen published in the literature, plus one more of my own.

\subsection{Basis Trucation Algorithms} 
There are three basis truncation algorithms:  Yang's
Divide and Conquer algorithm\cite{Yang91}, the "Locally
Self-Consistent Multiple Scattering" algorithm\cite{Wang95}, and Goedecker's "Chebychev Fermi Operator
Expansion"\cite{Goedecker94, Goedecker95}, which I will henceforth
call the Goedecker algorithm.  Basis truncation algorithms break
the matrix function into spatially separated pieces. Given the
position of a particular piece, the basis is truncated as described previously in section \ref{BasisTruncation}, and then the matrix function
is calculated within the truncated basis. Thus, for any generic
matrix function $f{(H)}$, a basis truncation algorithm calculates
$ {{\langle \vec{x} |} {f{(H)}} {| \vec{y} \rangle}} = {{\langle
\vec{x} |} {f{(P_{\vec{x}, \vec{y}} H P_{\vec{x}, \vec{y}})}} {|
\vec{y} \rangle}}$, where $P_{\vec{x}, \vec{y}}$ is a projection
operator truncating all basis elements far from $\vec{x}$,
$\vec{y}$. There may be also an additional step of interpolating
results obtained with different $P$'s, but I will not discuss this complication here. 

The three basis truncation algorithms are distinguished  only by their choice of how to
evaluate the function $ {f{(P_{\vec{x}, \vec{y}} H P_{\vec{x},
\vec{y}})}} $. Here are the specifics:
\begin{itemize}
\item Yang's algorithm calculates $f$ by diagonalizing the truncated argument $PHP$: ${PHP} = {UDU^{T}}$, where $D$ is the diagonal matrix of eigenvalues and $U$ is a tranformation matrix built from eigenvectors. Yang's algorithm then uses the standard formula ${f{(H)}} = {U {f{(D)}} U^{T}}$ to obtain its final result.  
\item The Goedecker algorithm does
a Chebyshev expansion of $f$. As long as all the eigenvalues of the
argument $H$ are between $1$ and $-1$, a matrix function may be
expanded in a series of Chebyshev polynomials of $H$: $f{(H)}
\cong { \sum_{s=0}^{S} {c_{s} T_{s}{(H)}}}$.  The coefficients
$c_{s}$ are independent of the basis size, and therefore can be
calculated numerically in the scalar case.  The Chebyshev
polynomials can be calculated in $O{(N)}$ time using the recursion
relation $T_{s+1} = {{({2 H T_{s}})}
 - T_{s-1}}$, $T_{1} = H$, $T_{0}$ = 1.  (Of course, one must also
 bound the highest and lowest eigenvalues of $H$ and then
 normalize.  In practice, very simple heuristics are sufficient for
 estimating these bounds.)  If the matrix function $f{(H)}$ has a
 characteristic scale of variation $\alpha$, then the error induced by the Chebyshev
 expansion  is controlled by an exponential with argument of order
 $-\alpha S$.
\item The "Locally Self-Consistent
Multiple Scattering" algorithm calculates the argument's resolvent
or Green's function ${G{(E)}}\equiv{(E - H)}^{-1}$ within the truncated basis.  It then obtains $f$ via contour integration: \linebreak ${f{(H)}} = {{(2 \pi \imath)}^{-1} \oint {dE} {f{(E)}} {G{(E)}}}$.  The contour must include the entire spectrum of $H$ unless $f{(E)}$ is zero over some portion of the complex plane.  
\end{itemize}

Because these algorithms are all mathematically equivalent when
applied to analytic functions, they should all converge to
identical results, as long as one makes identical choices of which
matrix function to evaluate, of how to break up the function, of
which projection operator to use, and of a possible interpolation
scheme. Moreover, given an identical choice of matrix function,
variations in the other choices should obtain results that are
qualitatively the same.  

No doubt other  $O{(N)}$ algorithms could be invented simply by choosing one's favorite algorithm for calculating functions of small matrices and combining it with basis truncation.  I discussed only these three because they are the only ones that have been discussed in the $O{(N)}$ literature.

\subsection{The Locally Self-Consistent Green's Function Algorithm}
This algorithm\cite{Abrikosov96, Abrikosov97} is almost exactly the same as the "Locally Self-Consistent Multiple Scattering" algorithm; it too computes the Green's function in a truncated basis and then does a contour integral to obtain the desired matrix function.  There is, however, one significant difference: it adds some information about the long-distance physics to the truncated calculation.  Specifically, this algorithm uses the Coherent Potential Approximation to estimate the self-energy $\Sigma$ of the average Green's function, which is a way of describing the decay which disorder induces in the Green's function.  This algorithm then uses the formalism of scattering theory, which I will discuss in section \ref{ScatteringTheory}, to insert the self-energy $\Sigma$ into the calculation of the Green's function within the truncated basis.  The insertion is done in way that would not influence the end result at all if there were no basis truncation and the system were infinitely big.  However in a truncated basis it can make a large difference, allowing the algorithm to feel the average influence of sites far outside of the truncation radius.  The net result is that one can choose a much smaller truncation radius to obtain a given accuracy.

\subsection{Functional Minimization} Now I turn to algorithms which are based on generalized multiplication.  The idea is to create a functional of the matrix $X$ whose minimum corresponds to $X$ having the value equal to the desired matrix function.  One evaluates the functional using generalized multiplication, and then with the functional in hand one uses a minimization algorithm, for instance the conjugate gradient algorithm, to find the minimum of the functional.  By construction one thus arrives at the desired function.   
 
 Note that these functionals, and in particular their minima, are designed in the world of normal multiplication.  Once generalized multiplication is introduced, any particular minimum could change, disappear, or become multiple minima.  This difficulty is generally just accepted as one of the prices of $O{(N)}$ performance, and the details of how the minima are affected by generalized multiplication have not been discussed in much detail. 
 
 The $O{(N)}$ literature has concentrated on algorithms for computing the density matrix function, or step function, ${\rho{(H)}} = {\Theta{(H - \mu)}}$.  The scalar $\mu$ is called the Fermi Energy, and just acts as on offset to the energy.  Several functionals suitable for calculating the density matrix have been discussed in a lot of detail in the $O{(N)}$ literature; here I only present one.  In 1993 Li, Nunes, and Vanderbilt\cite{Li93} introduced the functional $Z = {{Tr}{({(H - \mu)}{({3F^{2}} - {2 F^{3}})})}}$.  The Li-Nunes-Vanderbilt functional is basically just a cubic polynomial with extrema at $0$ and $1$.  The factor $H - \mu$ determines the overall sign of the polynomial and therefore determines whether a particular eigenvalue of $F$ will prefer to be $0$ or $1$.  After converging to the minimum, $F$ will have the same eigenvectors as $H$. $F$'s eigenvalues $f$, however, will be different than the eigenvalues $e$ of $H$, and will be given by the relation $f{(e)} = {\Theta{(e - \mu)}}$.  In other words, $F$ will converge to be equal to the density matrix function. 
 
 There is, however, a proviso: the initial value of $F$ must have all its eigenvalues within a certain interval; otherwise $F$ will diverge.  This is caused by the fact that the minima at $0$ and $1$ are local minima, not global minima.  
 
 Now for my new algorithm, designed for evaluating the matrix logarithm: minimize the functional $Z = {{Tr}{({{exp}{(HF)}} - F)}}$.  This functional has a single global minimum at $F = {\ln H}$.  I showed numerically in chapter \ref{FunctionAccuracy} that  basis truncation algorithms can evaluate the matrix exponential in disordered systems with extraordinary accuracy; perhaps this minimization algorithm could achieve a similar accuracy for the logarithm?
 
 In section \ref{GreensFunctionON} I will give another functional minimization algorithm, this one for computing the Green's function.
 
\subsection{Iterative Algorithms} 
Consider applying the mapping $x \rightarrow {{3x^{2}} - {2x^{3}}}$ over and over to a number.  It is easy to prove (see Palser and Manolopoulos's paper\cite{Palser98}) that if the initial value $x_{0}$ is in the interval $\frac{1}{2} < x_{0} < 1$, then $x$ will converge to $1$.  If, on the other hand, the initial value is in the interval $0 < x_{0} < \frac{1}{2}$, then $x$ will converge to $0$.  In either case the convergence will be quadratic, meaning that the number of accurate digits in $x$ will double at every step.  This is called McWeeny purification.

I restate this result: if we restrict the initial value to be in the interval $0 < x_{0} < 1$, then the iteration converges quadratically to a value equal to $\Theta{(\frac{1}{2} - x_{0})}$.  Thus I have an iterative way of computing the step function.

This algorithm can be easily generalized to compute the density matrix\cite{Palser98}. One starts by normalizing the matrix $H - \mu$ so that all its eigenvalues lie between $0$ and $1$: $F = {\frac{1}{2} - {\frac{1}{2 \lambda} {(H - \mu)}}}$, where the normalization constant $\lambda$ is equal either to the absolute value of the largest eigenvalue of $H - \mu$ or the absolute volue of the smallest eigenvalue of $H - \mu$, whichever is larger.  
Having performed this normalization, one just iterates: $F \rightarrow {{3F^{2}} - {2F^{3}}}$. The result converges quadratically to the density matrix. 

This algorithm is probably the only case where I have seen the effects of generalized truncation studied in any kind of mathematical detail.  The study was done by Bowler and Gillan\cite{Bowler99}, who concluded that it is best to start by applying this iteration, and then at some point polish up the final result by minimizing the Li-Nunes-Vanderbilt functional. 

In section \ref{GreensFunctionON} I will propose another iterative algorithm appropriate for computing the Green's function.  

\subsection{Chebyshev Expansion and Green's Function Algorithms}
Two of the basis truncation algorithms can be easily adapted to use generalized multiplication instead of basis truncation.  Both algorithms have the advantage of being general purpose algorithms able to calculate many different matrix functions.  One of these algorithms is the Chebyshev expansion.  The other is the approach of computing  matrix functions using the formula ${f{(H)}} = {{(2 \pi \imath)}^{-1} \oint {dE} {f{(E)}} {G{(E)}}}$. This last algorithm requires an $O{(N)}$ algorithm for calculating the Green's function; I will discuss many candidate Green's function algorithms in section \ref{GreensFunctionON} and chapter \ref{Multiscale}.  

\subsection{Bond Order Potential Algorithms}
There is also a class of algorithms - called Bond Order Potential algorithms - which are based on the Lanczos algorithm for tridiagonalizing a matrix.  One feeds the tridiagonal matrix generated by the Lanczos algorithm into a recursive equation and obtains diagonal elements of the Green's function ${(E-H)}^{-1}$.  One cuts off the recursion after some number terms, which is why this is an $O{(N)}$ algorithm.  It is also possible to obtain off-diagonal elements, but the algebra isn't pretty and the numerics are quite unstable.  Still, at least one numerical study\cite{Bowler97} indicates that a Bond Order Potential algorithm may be more suitable to certain problems than some of the other $O{(N)}$ algorithms I have discussed.

I don't understand the mathematical details of obtaining the off-diagonal elements, and will just refer the reader to several papers on the subject\cite{Bowler97, Horsfield96, Aoki93, Horsfield96b}.  

\subsection{Stochastic algorithms}
It may also be possible to use stochastic methods to obtain rough estimates of matrix functions.  In a recent paper, Buchmann and Petersen\cite{Buchmann02} showed that if you evolve a vector $\vec{x}$ under a force $-A\vec{x} + \eta$ where $A$ is symmetric and positive definite and $\eta$ is a noise term, then the average value of the outer product $\vec{x} \otimes \vec{x}$ will converge to $\frac{1}{2}A^{-1}$.  I suspect that this particular algorithm may not be very efficient compared to other algorithms for the inverse, and probably scales worse than $O{(N)}$.  However, it does give some hope that other faster stochastic algorithms could be developed.  There are also stochastic methods of estimating individual matrix elements of matrix functions; if you were to find one that takes $O{(1)}$ time per matrix element and does not require calculation of matrix elements far from the diagonal, then you would have an $O{(N)}$ algorithm.

\subsection{Other Algorithms}
If you just want to calculate some matrix elements of a matrix function (for instance its diagonal elements or its trace), or you want to calculate the matrix function times a vector ($f{(H)}\vec{b}$), then there is a wide variety of algorithms specially designed for these tasks.  Some of these algorithms are stochastic. I do not review any of them here.

This completes my review of existing $O{(N)}$ algorithms.  For more depth, I refer the reader to these reviews\cite{Goedecker99, Wu02, Bowler97, Ordejon98, Scuseria99, Bowler02}.

\section{\label{GreensFunctionON}Calculating the Green's Function}
In chapter \ref{FunctionAccuracy} I showed that basis truncation algorithms perform very poorly in calculations of the real part of the Green's function.  The $O{(N)}$ literature contains only two other suggestions for calculating the Green's function: the Bond Order Potential algorithms\cite{Horsfield96, Bowler97,Ozaki01} (which are based on a Lanczos approach and have significant problems with numerical stability), and the "Locally
Self-Consistent Green's Function" algorithm\cite{Abrikosov96, Abrikosov97}. This last algorithm requires as input an average of the Green's function over all possible disorder configurations.  This average is not always easy to compute.  Both approaches have significant limitations.
 
 Here I present many alternative algorithms for computing the Green's function in $O{(N)}$ time.  In one sense they are all due to me.  In another sense, some of them are well known $O{(N^{3})}$ formulas and the only new thing I have done is propose to reduce them to $O{(N)}$ by using generalized multiplication.  Other formulas are not so well known: I have never seen them before.  I will try to clearly signal what I have and have not seen before.  In any case, all of the formulas are pretty simple minded, at least from the point of view of someone who knows scattering theory.  Since scattering theory is required for understanding some (but not all) of these algorithms, I will now spend a little time explaining the necessary basics.  
  
\subsection{\label{ScatteringTheory}Introduction to Scattering Theory}
Scattering theory is just a particular approach to calculating the Green's function ${G(E)} \equiv {{(E - H)}^{-1}}$, where the energy $E$ is a complex number and the Hamiltonian $H$ is a matrix.  I will explain the particulars of this approach shortly, but first a few words about the basis.

Although the equations of scattering theory do not require any particular choice of basis, it is conventional in scattering theory to assume that each basis state is physically located at a particular site on a lattice representing the physical system.  At each site there may be one or more basis states; each basis state is specified by two
indexes: $|vi\rangle$, where $v$ is the site index and $i$ specifies
the $i$-th state at site $v$. Matrix elements connecting two sites are
called hopping matrix elements. Matrices without any hopping
matrix elements - i.e. ones which are diagonal in the site index - are called
local.

The key idea of scattering theory is to decompose the Hamiltonian $H$ into two parts ${H} = {H_{0} + V}$: one part $H_{0}$ whose inverse is trivial to calculate, and
another part $V$ whose inverse is more complicated.  The trivial
part $H_{0}$ is usually called the "free" Hamiltonian, while the
nontrivial part is often called the potential, especially if it is
local. Because $H_{0}$'s solutions are exactly known, a "free
propagator" (or "free Green's function") may be defined:
${G_{0}(E)} \equiv {{(E - H_{0})}^{-1}}$.   I can now rewrite the Green's function in terms of a combination of the free propagator and $V$:
\begin{equation}
{{G{(E)}}={{({G_{0}^{-1}{(E)}} + V)}^{-1}}}
\end{equation}
This equation is starting point of scattering theory; the details are all in figuring out how $V$ influences  $G{(E)}$. The advantage of scattering theory lies exactly here, in the fact that one does as much as possible exactly, by calculating $G_{0}$.  The non-exact steps, the approximations, are reserved to handle $V$.

What decomposition of $H$ into $H_{0}$ and $V$ should one choose?  This decision is determined by the hopping matrix elements; the matrix elements of $H$ which are not diagonal in the site index $v$.  If they are irregular and can not be easily inverted, then one must choose $H_{0}$ to be the part of $H$ which is diagonal in the site index, because then $G_{0}$ will also be diagonal in the site index and can be computed very quickly.  However, often the hopping matrix elements of
$H$ are regular and thus easily inverted.  This happens with partial difference equations;
for instance the only non-local term in the Schrodinger equation 
is a second derivative representing the kinetic energy, and is spatially uniform.  In crystals as well, the hopping matrix elements are regular, and one uses
KKR band theory\cite{Korringa47, Kohn54, Ashcroft76, Abrikosov97, Ujfalussy00}. In such cases $H_{0}$ should be chosen to contain all hopping matrix elements, and thus the bare propagator $G_{0}$ will give a complete description of propagation between sites.  The remaining irregular part $V$ will be diagonal in the site index; thus one obtains a picture of particles moving between sites via $G_{0}$ and scattering off of the potential $V$ which exists at individual sites.

Some of the following algorithms will make an intermediate step in the process of calculating $G{(E)}$ and calculate the scattering
matrix ${T} \equiv {{(1 - {V G_{0}})}^{-1} V}$ first.  Once one has computed the $T$ matrix, the Green's function $G$ is trivial:
\begin{equation}
{G} = {G_{0} + {G_{0}TG_{0}}}
\label{GreensFromT}
\end{equation}
Of course, the free propagator can be expected to extend throughout the entire system, so the multiplications in the last term of equation \ref{GreensFromT} require $O{(N^{3})}$ time.  However, if one substitutes the normal multiplications with generalized multiplications as defined in section \ref{GeneralizedMultiplication}, then the multiplications take only $O{(N)}$ time.  This substitution is perhaps reasonable in disordered systems, where one expects $G{(E)}$ to die off exponentially away from the diagonal.

What is the physical significance of the
scattering matrix $T$?  It sums up the physics of
scattering; ${V {|\psi\rangle}} = {T {|\phi\rangle}}$, where
${|\phi\rangle}$ is the unscattered wave, and ${|\psi\rangle}$ is
the full solution to the scattering problem.  Bound states are
represented non-perturbatively in the $T$ matrix via the inverted
matrix ${(1 - VG_{0})}^{-1}$.  One may expand that quantity
perturbatively to obtain the Born approximations; for instance, in
the lowest order Born approximation ${T} = {V + {VG_{0}V}}$.  Of
course in this perturbative expansion all bound states are lost.

The reader who has already studied scattering theory should be warned that many of scattering theory results and ideas that she has learned in the context of continuum scattering are not applicable to the discrete systems studied here.  In the continuum typically one uses the language of phase shifts and spherical
plane waves, which are best suited to rotationally invariant systems, not to lattices.
Often there is also a focus on the perturbative Born expansion, which is not appropriate for studying bound states.  In contrast, I use a discrete basis representing a lattice, and never make a perturbative expansion.  Neither the language of phase shifts nor many of the standard scattering equations are applicable.

Computation of the total scattering matrix $T$ always begins by
calculating the scattering behavior of each site individually. I partition the potential
 $V$ site-wise, and define $V_{v}$ as the portion of $V$ which is located at site $v$.  (If the potential $V$ is not local, any
nearly local partitioning scheme will do.)  I next define a
site-wise scattering matrix:
\begin{equation}{T_{v}} \equiv {{(1 - {V_{v}
G_{0}})}^{-1} V_{v}}
\label{SitewiseT}
\end{equation}  
The matrix elements ${\langle w i |} V_{v} {|x j \rangle}$ are zero unless the site indices $w$, $x$, and $v$ all have the same value.  One can easily see that the site-wise scattering matrix $T_{v}$ has the same property. $T_{v}$ can be computed by doing mathematical operations using only the basis elements at site $v$. This calculation is straightforward and I will not spend more time on it.

After computing the site-wise scattering mattrices $T_{v}$, there always follows a process of using $T_{v}$ and $G_{0}$ to calculate the total scattering matrix $T$.  The details of this process vary from algorithm to algorithm, and I will explain them as I come to them.

This completes my introduction to scattering theory. Now for the first class of $O{(N)}$ algorithms for computing the Green's function: summation formulae.

\subsection{Summation Formulae}
 The basic idea of this section is to calculate either the total Green's function $G$ or the total scattering matrix $T$ in increments, adding one lattice site at a time.  So one starts with the Green's function of only one site, and then adds another site to get the Green's function of two sites.  Adding another site, and another, one eventually obtains the Green's function (or scattering matrix) of the whole system.

\subsubsection{Scattering Matrix Summation}
First, the iterative summation of site-wise $T_{i}$ matrices.  The
idea is to start by finding the total scattering matrix for two
sites $v$ and $w$; I will call this $T^{2}$.  $T^{2}$ is not a
simple sum of $T_{v}$ and $T_{w}$, because of the possibility of
scattering off of first site $v$ and then site $w$.  Nonetheless a sort of
addition is possible and I will shortly present the correct
formula.  Having added two sites, it is easy to add another site
by simply considering $v$ and $w$ as a single site.  Thus I can
use the same addition formula to add a third site $T_{x}$ to
$T^{2}$,  obtaining $T^{3}$ which represents the total scattering
of sites $v$, $w$, and $x$.  This process can proceed ad infinitum
until I have added all the sites to get the total scattering
matrix $T$.

I will skip the details of proving the $T$ matrix addition formula, but will simply state the final result. The proof does rely on the fact
that any $T$ matrix representing the total scattering behavior of
several sites may be decomposed in terms of first and last
scatterers: $T = {\sum_{v,w} {T_{vw}}}$, where each term $T_{vw}$
descibes the scattering which begins at site $v$ and
ends at site $w$. I define some new notation before presenting the formula: $T^{n}$ is the scattering matrix of $n$ sites taken together.  $T^{n+1}$ is the scattering matrix of all $n$ sites plus another another site $x$, all taken together. $P_{v}$ is the	 projection operator which contains only the states at site $v$; when it occurs it restricts sums over states to only the states at $v$.

I now give the summation formula. If $V$ is local, the $T$ matrix addition formula is:
\begin{eqnarray} \label{Taddition}
{T^{n+1}} & \equiv & 
{T^{n} \oplus T_{x}}
\nonumber \\ 
& = & {T^{n} + {{(1 + {T^{n}
G_{0}})} P_{x} {(1 - {{T_{x} G_{0} T^{n} G_{0}}})}^{-1} T_{x} {(1 + {G_{0}T^{n}})}} }
\nonumber \\ & &
\end{eqnarray}

The generalization to non-local potentials $V$ is straightforward.  There is also an easy generalization to adding several sites at a time. 

The interpretation of formula \ref{Taddition} can be obtained by expanding the inverted term in a power series and then reading individual terms in the series either from right to left or from left to right.  In particular, the inverted term represents the
possibility of repeated scattering back and forth between  $T^{n}$
and $T_{x}$.

The computation of the inverted term may seem to threaten an $O{(N^{3})}$ cost, but in fact this term is local to site $x$, so the inversion occurs in the $x$-th site's basis, and requires very little time.   The real problem comes from the matrix multiplications, which require $O{(N^{3})}$ time unless one uses generalized multiplication as defined in section \ref{GeneralizedMultiplication}.  In this latter case, adding a site requires only $O{(1)}$ time, and calculating the total scattering matrix requires $O{(N)}$ time.   The Green's function can be calculated in turn in only $O{(N)}$ time using equation \ref{GreensFromT}.  This completes my presentation of the iterative summation algorithm for calculating the Green's function. 

This algorithm has an advantage over basis truncation algorithms: it allows the $(n+1)$-th site to see a $T^{n}$ which
includes the physics of all $n$ sites already included in $T^{n}$,
even those that are far away from $n+1$.  Therefore, the summation algorithm's final result  for the scattering matrix includes the physics of scattering diagrams which start at a site $v$, go very far away it, and then make their way back to some other site $w$ which is close to the original site $v$.  In contrast, the Green's functions produced by basis truncation algorithms do not contain any information about scattering diagrams which travel outside of the truncation volume.  

Unfortunately, there is a fly in the ointment, caused by an unfortunate interaction between the summation formula and the generalized multiplication.  The summation algorithm is not self-consistent: changing the order in which individual sites are added to the total will change the final result.  Therefore this summation algorithm  is useful mainly for quickly computing an initial trial solution for the integral equation of the next algorithm, which is
self-consistent.

I have not seen either formula \ref{Taddition} or this algorithm presented anywhere in the literature.

\subsubsection{Frobenius Summation}
I just presented a summation algorithm for calculating the scattering matrix $T$, and then invoked equation \ref{GreensFromT} to calculate the Green's function $G$.  It is also possible to create summation algorithms which calculate the Green's function directly without obtaining the scattering matrix at an intermediate step.  However, one pays for this convenience by not treating $E - H_{0}$ exactly.

Before presenting the equation for summing Green's functions, I need to introduced
 some notation:  $G^{n}$ is the Green's function of $n$ sites taken together.  $G^{n+1}$ is the Green's function of all $n$ sites plus another another site $x$, all taken together.  I also define a site-wise Green's function $G_{w} \equiv {(P_{v}{(E - H_{0} - V)}P_{v})}^{-1}$, where $P_{v}$ is the projection operator which projects out only the states at site $v$.  In other words, $G_{w}$ is the Green's function of the single site $w$ with all the other sites eliminated from the problem.  It is easily computed by simply inverting matrices at that single site $v$.

If $V$ is local, the Green's function addition formula is:
\begin{eqnarray} \label{FrobeniusGAddition}
{G^{n+1}} & \equiv & 
{G^{n} \oplus G_{x}}
\nonumber \\ 
& = & {G^{n} + {{(1 + {G^{n}
H_{0}})} P_{x} {(1 - {{G_{x} H_{0} G^{n} H_{0}
}})}^{-1} P_{x} {(1 + {H_{0}G^{n}})}} }
\nonumber \\ & &
\end{eqnarray}
Again generalizations to adding several sites at a time and to non-local potentials are straightforward.

Equation \ref{FrobeniusGAddition} is essentially just a rewrite of equation $4.4$ from Borm, Grasedyck, and Hackbusch work on hierarchical matrices\cite{Borm04}.  They call this the Frobenius formula, although a search on google reveals that the Frobenius formula usually refers to something from combinatorics.  At any rate, this formula is obvious enough that I would not be surprised if it were rather old.

Just as with the summation equation \ref{Taddition} for scattering matrices, achieving $O{(N)}$ times will require using generalized multiplication.  And again like scattering matrix summation, this formula is not self-consistent; the final result depends on the order in which sites are added unless one does not use generalized multiplication. The big difference between the two algorithms that in this one $G_{0}$ is missing, which reflects the fact that we are not treating $E - H_{0}$ exactly.  Formula \ref{FrobeniusGAddition} is a worse approximation than formula \ref{Taddition} for scattering matrices.

\subsubsection{Sherman-Morrison-Woodbury Summation }

The numerical linear algebra community seems to favor a different formula for adding inverses: the Sherman-Morrison-Woodbury formula.  I presume, without any evidence, that the reason has to do with improved numerical stability, at least in its typical usage scenario.  The Sherman-Morrison-Woodbury formula is typically used for figuring out how a solution $x$ of the linear equation ${Ax} = b$ changes when $A$ is altered; or, in other words, how $x = {A^{-1} b}$ changes.  The publication on its stability which is generally referenced was written by Yip\cite{Yip86}.

Before giving the formula I need to introduce a little bit of new formalism.  I already defined the projection operator $P_{v}$ of a single site $v$.  Now I define the projection operator which projects out all the basis states of the first $n$ sites $P_{n} \equiv {\sum_{i \in n}P_{v}}$.  Obviously $P_{n+1} = {P_{n} + P_{x}}$.  In this notation, the Green's function of $n$ sites can be rewritten as: 
\begin{equation}
G^{n} = {({P_{n}{(E - H)}P_{n}})}^{-1} 
\label{GreensWithProjection}
\end{equation}

With this formalism, it is clear that the only difference between $G^{n}$ and $G^{n+1}$ is that I have changed from using $P_{n}$ to using $P_{n+1}$ in equation \ref{GreensWithProjection}.  This change is equivalent to starting with the n-site matrix ${P_{n}{(E - H)}P_{n}}$ and then subtracting another matrix $Y$ defined by ${-Y} \equiv {{P_{x}{(E - H)}P_{n}} + {P_{n}{(E - H)}P_{x}} + {P_{x}{(E - H)}P_{x}}}$.  The first two terms in $Y$ connect site $x$ to the other $n$ sites, and simplify to ${P_{x}H_{0}P_{n}} + {P_{n}H_{0}P_{x}}$.  The last term in $Y$ is the inverse of the single site Green's function $G_{v}$ which I defined while discussing the Frobenius formula.  

The Frobenius formula separates the three terms in $Y$, which enter separately into the summation equation \ref{FrobeniusGAddition}: the first two terms in $Y$ enter as instances of $H_{0}$, while the third term enters as the single site Green's function $G_{x}$.  In contrast, the Sherman-Woodbury formula factors $Y$ into two factors $Y = {BF}$, where the two factors $B$ and $F$ are required to have a special structure.  Specifically, they connect site $x$ with the other $n$ sites: $B = {P_{n} B P_{x}}$ and $F = {P_{x} F P_{n}}$.  As long as ${P_{x}{(E-H)}P_{x}}$ doesn't have any zero eigenvalues, it is always possible to factor $Y$ into two factors $B$ and $F$ with these properties.  The actual numerical calculation does not  require much time as long as $H_{0}$ is close to diagonal.  

With this preparation, I am finally ready to write the Sherman-Morrison-Woodbury formula for adding a single site $x$ to the Green's function of $n$ sites. If $V$ is local, the Green's function matrix addition formula is:
\begin{eqnarray} \label{SMWGAddition}
{G^{n+1}} & \equiv & 
{G^{n} \oplus G_{x}}
\nonumber \\ 
& = & {G^{n} + {{(1 + {G^{n}
B})} {(1 - {{F G^{n} B
}})}^{-1}  {(1 + {F G^{n}})}} }
\nonumber \\ & &
\end{eqnarray}
Again generalizations to adding several sites at a time and to non-local potentials are straightforward.

Just as with the other summation formulae, achieving $O{(N)}$ performance requires using generalized multiplication.   And again like the other formulae, this formula is not self-consistent; the final result depends on the order in which sites are added unless one does not use generalized multiplication. Note that equation \ref{SMWGAddition} contains no instances of the free propagator $G_{0}$, which signals that it is less accurate than equation \ref{Taddition}.

\subsubsection{Sherman-Morrison-Woodbury Analogue for Scattering Matrices}
There is an easy analogue of the Sherman-Morrison-Woodbury formula for $T$ matrix addition; one just factors $P_{n}G^{n}T_{x}G^{n}P_{n}$ into two factors $B$ and $F$.  Equation \ref{Taddition} then becomes:
 \begin{eqnarray} \label{SMWTaddition}
{T^{n+1}} & \equiv & 
{T^{n} \oplus T_{x}}
\nonumber \\ 
& = & {T^{n} + {{(1 + {T^{n}
G_{0}})} P_{x} {(1 - {{T_{x} B F}})}^{-1} T_{x} {(1 + {G_{0}T^{n}})}} }
\nonumber \\ & &
\end{eqnarray}
  This formula is perhaps the best of both worlds, combining the additional exactitude of the free propagator $G_{0}$ and the numerical stability of factoring.  However, I'm not sure if the factoring really makes much difference in this case.

This completes my presentation of summation formulae for evaluating the Green's function.

\subsection{\label{IntegralEquations}Integral Equations}
In this section I will discuss four different integral equations which can be used to calculate the Green's function.  These integral equations are all simply linear equations of the type:
\begin{equation} 
{Ax} = b
\label{LinearEquation}
\end{equation}
However, the equations in this section are properly called integral equations because the unknown vector $x$ is a matrix, i.e. either the Green's function or the scattering matrix. $b$ is also a matrix, and $A$ is a tensor with four indices.

The simplest integral equation is the definition of the Green's function: \linebreak
${{(E - H)}G} = 1$.  This equation corresponds perfectly with the linear equation \ref{LinearEquation}: the input vector $b$ is the identity matrix, $A$ is  $E - H$, and the unknown $x$ is the Green's function $G$.  Evaluating the Green's function $G$ is just a matter of solving this linear equation.  

Many iterative algorithms have been invented which can solve most systems of linear equations with $N$ unknowns in $O{(N)}$ or $O{(N \ln N)}$ time. In our case, $G$ is an $N \times N$ matrix and has $N^{2}$ unknowns, so these algorithms would take at least $O{(N^{2})}$ time.  Therefore $O{(N)}$ scaling can be obtained only by making a further approximation: one evaluates the matrix product ${(E-H)}G$ with generalized multiplication.  As a consequence of the generalized multiplication, all matrix elements of $G$ which are far from the diagonal are ignored, and $G$ contains only $O{(N)}$ unknowns, so that one has good reason to hope that a standard algorithm for solving linear equations will achieve $O{(N)}$ scaling.

Note that this algorithm is already an improvement on the summation algorithms: A perturbative analysis shows that this algorithm shares the summation algorithms' strength of allowing scattering diagrams to travel through the entire volume.  However, this integral equation algorithm is unlike those algorithms in that it is self-consistent; there is no question of the result depending on the order in which sites are added, and instead all sites are treated on an equal basis.

This first integral equation can be improved upon: one can see that it does not contain any instance of $G_{0}$, which means that it is not taking advantage of the possibility of treating $E - H_{0}$ exactly.  Therefore I present the next two integral equations:
\begin{eqnarray}
{{(1 - G_{0}V)}G} & = & G_{0} \label{GreensIntegral2}
\\
{{(1 - VG_{0})}T} & = & V \label{ScatteringIntegral1}
\end{eqnarray}
Both equations can be derived very easily from the basic definitions of scattering theory.  The unknown in equation \ref{GreensIntegral2} is the Green's function $G$, while the unknown in equation \ref{ScatteringIntegral1} is the scattering matrix $T$. Again generalized multiplication can be used to reduce the number of variables to $O{(N)}$.  It is perhaps more reasonable to expect that the scattering matrix would die off quickly away from the diagonal than that the Green's function would die off quickly away from the diagonal; probably equation \ref{ScatteringIntegral1} is better than equation \ref{GreensIntegral2}.  Yet both equations share an important problem: the operators $A = {1 - G_{0}V} \neq A^{T}$ and $A = {1 - VG_{0}} \neq A^{T}$ are not symmetric under transpositions even when the underlying hamiltonian $H = H^{T}$ is symmetric under transpositions.  When the operator $A$ in the linear equation \ref{LinearEquation} is not symmetric under transpositions, it becomes much more difficult to find a good algorithm for solving the linear equation numerically, and numerical stability problems can be much much worse.

Fortunately there is a fourth integral equation which both preserves transverse symmetry and uses the free propagator $G_{0}$:    
\begin{eqnarray} 
{T - {\acute{T} {\acute{G}}_{0} T {\acute{G}}_{0} \acute{T}}} & = & {{\acute{T}} + {\acute{T}
{\acute{G}}_{0} \acute{T}} }
\nonumber \\
{\acute{T}} & \equiv & {{\sum}_{i}T_{i}}
\nonumber \\
{\acute{G}}_{0} & \equiv & {G_{0} - {{\sum}_{v}P_{v}G_{0}P_{v}}}
\label{IntegralAlgorithm2}
\end{eqnarray}
${\acute{T}}$ is just the sum of all the single-site Green's functions, and ${\acute{G}}_{0}$ is just the free propagator with its diagonal removed.  A little thought will show that the left hand side can be expressed as $Ax$ with $x = T$ and a proper choice of $A$. Although equation \ref{IntegralAlgorithm2} is fairly obvious once you think about it, I have not seen it elsewhere in the literature.  It is perhaps the result that I am most proud of in this chapter.

This is the final integral equation: it gives a self-consistent $O{(N)}$ algorithm for calculating the scattering matrix $T$ and the Green's function $G$.

\subsection{Functional Minimization Algorithms}
 The matrix functional $Z = {{Tr}{({{(E-H)} F^{2}} - F)}}$ has a single global minimum at $F = {(E-H)}^{-1}$.  Therefore one can evaluate the Green's function by minimizing this functional.  Like the integral equation approach, the result is self consistent and contains scattering diagrams which travel throughout the system.
 
 This approach can easily be generalized to solve any of the integral equations listed in section \ref{IntegralEquations}.
 
\subsection{Iteration}
If one knows that a matrix $M$ is positive definite (no eigenvalues have negative or zero real parts), and there are no rounding errors and no generalized multiplication, then the following iteration algorithm will converge to $M^{-1}$.

First, choose a matrix $X$ which is proportional to the identity.  Its proportionality constant should be positive and equal to or less than $m$, where $m$ is the smallest real part of the the inverse eigenvalues of $X$.  In some physical cases $m$ may be easily estimated using very simple heuristics, and if $M$ is real it can be computed quite quickly with a Lanczos type of algorithm.  There may also be an analogue of the Lanczos algorithm for the case when $M$ is complex, but I'm not sure.

Next, apply the following iteration until convergence is achieved: 
\begin{equation}
X \rightarrow {{2X} - {MX^{2}}}
\end{equation}
The convergence is quadratic, meaning that the number of digits of accuracy doubles at every iteration.  

The big restriction, of course, is the requirement of positive definiteness.  When computing  the Green's function $M = {E - H}$; the positive definiteness constraint requires that the energy $E$ be larger than the largest eigenvalue of the Hamiltonian $H$.  In addition to this constraint, there are also complexities caused by rounding errors.

And then there is the issue that turning this iteration into an $O{(N)}$ method requires that the the $MX^{2}$ term be evaluated using generalized multiplication, which necessarily ruins the  proof that this iteration converges to the right result, or converges at all.  It would be interesting to analyze the iteration from the viewpoint of generalized multiplication being a tensor, and see what that says about the iteration's result.

Since any even power of a real matrix is positive indefinite, this algorithm can also be construed as a way of computing $X^{-2n}$ where $n$ is an integer and $X$ is real and has no zero eigenvalues.

This algorithm is just one of a widely known family of iteration algorithms which converge to various functions.  The numerical analyst Marc Van Barel told me about one of them at a summer school.  I don't remember which algorithm he told me about, but it might have been this one.  At any rate, I ended up deriving this one on my own (very simple minded stuff), so I could be the first to derive it, though I really doubt this.
 
This completes my list of $O{(N)}$ algorithms for evaluating the Green's function.  Chapter \ref{Multiscale} will discuss several other fast algorithms for the Green's function which based on renormalization ideas.  

\section{Outstanding Questions}
Very little is known either theoretically or computationally about the range of validity of $O{(N)}$ algorithms, and almost no work has been done exploring any of their mathematical aspects.  (In fact, this thesis probably contributes as much to these questions as the previously published results.)  I list here some of the most important outstanding questions:

\begin{itemize}
\item When is each algorithm valid, and what accuracy will it deliver? Answering this question will involve exploring issues including but not limited to the coherence length, the localization length, and metals vs. insulators.
\item The existing algorithms are designed only for a special class of matrices: those with transverse symmetry $H = H^{T}$, and in general with $H$ real.  Perhaps they can be easily generalized to complex symmetric matrices $H = H^{T}$ and hermitian matrices $H = H^{\dagger}$.  However there is also a need $O{(N)}$ algorithms for other classes of matrices, in particular the the Dirac operator which arises in lattice QCD.  
\item Convergence: Does a given algorithm converge, and if so, does it converge to the correct result?  How fast is the convergence?  Is there a unique solution to the algorithm; if not, how many are there?
\item The relation of generalized multiplication to normal multiplication.
\item Stability and accuracy under rounding errors.
\item Understanding the numerics of the Chebyshev expansion.  I am unaware of published results on the stability and accuracy of even a scalar Chebyshev expansion when rounding errors are present.  Moreover, there is reason to hope that the Chebyshev expansion can be improved upon, since it gives uniform convergence across the entire spectrum.  In real life there are more eigenvalues in some parts of the spectrum than others, and one can make statistical estimates with good accuracy of this eigenvalue density.  Therefore it would be useful to find an expansion which converges faster in certain parts of the spectrum than in others.
\item Some of the algorithms are nonlinear; for instance some of the functional minimization algorithms.  Yet they are supposed to converge to matrix functions, which can be derived using only linear algebra.  Therefore we have an interesting challenge of exploring algorithms which are in a grey area between linearity and nonlinearity.
\end{itemize}

\chapter{\label{Multiscale}Algorithms for Problems with Multiple Length Scales}

I have often quoted the result that $O{(N)}$ algorithms depend on the number of basis states $N$ in a linear fashion; this is why they are called $O{(N)}$ algorithms.  However I have never discussed the proportionality constant relating the computational cost to the basis size.  In some systems the proportionality constant can be enormous and $O{(N)}$ algorithms can be as impractical as any other other algorithm.  The focus of this chapter will be on understanding the proportionality constant and finding ways to reduce it.

If $n$ is the number of sites within the truncation radius $R$, and $b$ is the number of basis states per site, then most of the $O{(N)}$ algorithms listed in chapter \ref{LinearAlgorithms} require $O{({b}^{2}{n}^{2}{N})}$ time.  Note that the prefactor is just ${b}^{2}{n}^{2}$.  This implies that $O{(N)}$ algorithms are extremely sensitive to the truncation radius $R$.  For instance, in a three dimensional system the number of sites inside the truncation radius $R$ is proportional to $R^{3}$, and therefore $O{(N)}$ algorithms scale with $R$ as $R^{6}$.  As a consequence, $O{(N)}$ algorithms are very slow whenever the truncation radius is much larger than the lattice spacing.  Mind you, they are still much faster than $O{(N^{3})}$ algorithms, but both are very slow.
  
In chapters \ref{DensityMatrixAccuracy} and \ref{FunctionAccuracy} I showed that in disordered systems the truncation radius should be set to match the coherence length.  In insulators, another length characterizing the insulator may determine the truncation radius.  In either case there is a single physical length $\mathfrak{l}$ which sets the truncation radius.  If this length $\mathfrak{l}$ is the only important length scale in the problem, then one can safely choose a lattice spacing which isn't much smaller than $\mathfrak{l}$, and $O{(N)}$ algorithms are applicable.  However, if there are other important length scales significantly smaller than $\mathfrak{l}$, then $O{(N)}$ algorithms are as helpless as normal $O{(N^{3})}$ algorithms.

Many physical systems do exhibit several important length scales.  Here are three examples:
\begin{enumerate}

\item \label{PseudopotentialEx} In physical chemistry and condensed matter, each atom's core electrons are always very tightly
bound around their owning atom, while the valence electrons are
less tightly bound, and in metals are even delocalized.  Pseudopotentials traditionally have been used in order to allow ommision of the core electrons and their corresponding length scales from calculations.  A short discussion of
pseudopotentials and their limitations can be found section \ref{Pseudopotentials}.

\item \label{RenormalizationEx} The renormalization group has taught us to expect
that the characteristic length scale of the physics should depend
on the energy of the probe.  At some energies you will see very
detailed features; at others you will see only coarse grained
features.  A system close to a phase transition exhibits scaling
behavior, with non-trivial physics at every length scale.

\item \label{AlloyEx} A third example comes from metallic alloys.  Self-consistent
$O{(N)}$ calculations of alloys require supercomputers because a
large truncation readius is required.  However, the truncation radius (and computational load) may be greatly reduced by
explicitly modifying one's equations to acknowledge that outside
of the localization zone there is an effective medium of average
atoms\cite{Wang95, Abrikosov96, Abrikosov97, Abrikosov98}. This
indicates the presence of two length scales: a short length scale
within which it is important to know each atom's type, and a
longer length scale at which it is important to know each atom's
presence, but not its type. 

\end{enumerate}

All of these examples are problematic for $O{(N)}$ methods.  There is one $O{(N)}$ algorithm which is able to treat example \ref{AlloyEx}, but its success comes from supplementing $O{(N)}$ ideas with special knowledge about the longer length scale.

Are there  algorithms which scale
optimally when solving linear algebra problems involving several or many length scales?
Such algorithms would be immensely useful, permitting calculations of previously inaccessible systems, speeding up current $O{(N)}$ calculations, and allowing all-electron calculations without pseudopotentials.

The answer is yes.  In section \ref{MultiscaleBases} I describe how to explicitly insert information about length scales into one's calculations by choosing a special basis and a slightly modified version of generalized multiplication.  The result is a very fast algorithm for matrix multiplication, with a prefactor governed not by the longest length scale in the system but instead by the smallest length scale in the system.  Section \ref{NlnNAlgorithms} points out that many of the $O{(N)}$ algorithms of chapter \ref{LinearAlgorithms} can be adapted easily to use the new basis and new multiplication.  Then in section \ref{BasicRenormalization} I discuss another class of algorithms which obtain accelerations by renormalizing away the small length scales; i.e. throwing away most of the basis and then changing the calculation to compensate for the missing information.  Many of these algorithms are my original contribution.

While the ideas of multiple scales have been explored very thoroughly in the context of quantum field theory, they are much less understood in the context of linear algebra.  In section \ref{LinearNonlinear} I present an interesting continuum model which may provide some insight into the physics of multiple length scales and the behavior of multi-scale algorithms.  Section \ref{MultiscaleReview} completes the chapter with a review of  the previously developed approaches to solving problems with several length scales: multigrid algorithms, pseudopotentials, and hierarchichal matrices.  The multigrid algorithms are best suited to solving systems of linear equations, pseudopotentials handle only a very narrow subset of problems, and hierarchical matrices are still very new and their capabilities are not well understood.  I do not discuss attempts to implement the renormalization group numerically, because these attempts seem best suited to field theory calculations, not to solving linear algebra problems.

\section{\label{MultiscaleBases}Multi-Scale Bases}

$O{(N)}$ algorithms sort information into two categories: all information within the
localization volume $n$ is important, while nothing outside that volume
is important.  This approach is appropriate for systems with only one important length scale.    Efficient algorithms for systems with two or more length scales need to sort information with more care: some information is important only a short scale, some on a longer scale, some on an even longer scale, etc. In this section I will outline how to do this sorting.

Recall that $O{(N)}$ algorithms sort information by using a truncation volume, typically a sphere with a truncation radius $R$.  They then systematically ignore all matrix elements $\langle \vec{x} | M | \vec{y} \rangle$ such that  ${|\vec{x} - \vec{y}|} > R$. 

It seems to me that the most natural generalization to several length scales is to assign different truncation radii to different basis states, thus signalling which basis states are important at which length scales.  One would have to sort the basis into groups of basis states which share the same truncation radius.  Mathematically this would correspond to first choosing a basis, and then sorting that basis by choosing a set of projection operators $P_{i}$. Each projection operator would have an associated truncation radius $R_{i}$, and thus would signal that its states are important only at length scales less than or equal to $R_{i}$.  I will call a basis which has been sorted and supplemented with truncation information as sketched in this paragraph a multi-scale basis\footnote{I have seen a multiscale approach like this in two places in the literature: hierarchical matrices, and a paper Yokojima, Wang, Zhou, and Chen.  I will discuss the former in section \ref{MultiscaleReview}.  The latter authors used an energy based heuristic to impose a multi-scale cutoff on the density matrix\cite{Yokojima99}.   They gave higher energy states larger localization volumes.  There are also wavelet bases, but these are oriented toward models on the continuum, not to lattices.}.

A multiscale basis should have the structure of a tree (in the mathematical sense of a tree), with the leaves on the tree being the original fine mesh of sites.  I will denote the levels  in the tree with an index $l$.  Each level in the tree will correspond to a particular length scale, and will have its own truncation radius $R_{l}$. The leaf level $l = 1$ of the tree will have the smallest truncation radius, and the highest level in the tree $l = L$ will have the longest truncation radius.  Each node in the tree will be responsible for modelling the physics of a particular physical volume centered at a particular position, and will own basis states appropriate for this task.  Children of any particular node will be responsible for modelling portions of the volume that their parent owns, and will own finer-scale basis states of their own.
One can expect that the basis states owned by any level $l$ of the tree will not have spatial extents exceeding that same level's truncation radius $R_{l}$.

The details of constructing a tree, of choosing truncation radii for each level in the tree, and of assigning basis states to nodes, will all depend on the physical problem being solved.  The field of multigrid algorithms, which I will discuss in section \ref{Multigrid}, has already spent a lot of time studying how to create appropriate trees. For instance, geometric multigrid algorithms assume that the starting mesh is a regular lattice, and proceed to iteratively block the fine lattice into coarser lattices.  In $d$ dimensions, each point in a
coarse lattice corresponds to $c^{d}$ points in the next finer
lattice; $c$ is an implementation dependent parameter.  The problem of assigning basis states to nodes has also been studied a bit by the multigrid community, particularly in efforts to apply multigrid ideas to lattice QCD.  I will review these efforts briefly in section \ref{LatticeMultigrid}.  I am not aware of any studies on how to choose appropriate truncation radii for a given physical problem.

I assume that there is some original basis in which the problem is originally formulated, and that all the input data are given in terms of the original basis.  Therefore one must have algorithms for tranforming from the original basis to one's multiscale basis, and vice versa.  I now sketch the outlines of the algorithm for transforming from the original basis to the multiscale basis.  The algorithm for transforming back to the original basis is very similar and I leave it to the reader.

I assume that the original basis is local, meaning that each basis state occupies one and only one lattice site.  I specify individual nodes with the indices $l$ and $n$, where $l$ specifies the level and $n$ specifies the $n$-th node in level $l$.  
The transformation from the original basis to the multiscale basis will start at the lowest level, the leaves.  Each leaf corresponds to a site in the lattice.  At each leaf $l=1,n$ in the lattice a unitary transformation $U_{ln}$ will be applied to the leaf's basis states, and then the states will be separated into two classifications with a projection operator $P_{ln}$.  The states corresponding to ${(1 - P_{ln})} U_{ln}$ will be assigned to the leaf node $l=1,n$, while the states corresponding to $P_{ln} U_{ln}$ will be assigned to its parent at level $l=2$.

At this point the transformation will have finished with the leaf level $l=1$ of the tree.  Some basis states will have been assigned to the leaves, and some will have been moved to the next higher level $l=2$ in the tree.  Thus the transformation will have separated the shortest-scale physics (corresponding to the leaf nodes) from the longer scales.  It will now repeat the separation at level $l=2$, doing unitary tranformations at each $l=2$ node and then assigning some states to the $l=2$ nodes and moving the rest up to $l=3$.  This process will continue iteratively until the top of the tree has been reached and all states have been assigned to appropriate nodes.

Let me estimate the time requirements of basic arithmetic calculations with a multiscale basis.  It is impossible to make such estimates without knowing the tree layout and the number of basis states per node; I will concentrate on multiscale bases characterized by self-similarity between the length scales:
\begin{itemize}
\item There are a total of $N$ basis states.
\item Each node (except the leaf nodes) has $\alpha$ children.
\item Each child node owns a number of basis states equal to $\beta$ times the number of basis states owned by its parent.
\item I require that most of the basis states be at the bottom of the tree.  Mathematically this means requiring that ${\alpha \beta} > 1$.
\item The truncation radii $R_{l}$ vary in a self-similar fashion so that the number of level $l$ nodes inside the truncation radius $R_{l}$ is a constant, the same for every level.
\end{itemize}

The time required to tranform an unlocalized vector from the original basis to this self-similar basis is of order $O{(N)}$.  However, if the original vector was localized in the original basis, and had a support of only $m$ basis states in the original basis, then the total time required to do the transformation is of order $O{({m + {\ln N} - {\ln m}})}$.  

I now analyze the the problem of tranforming a matrix from the original basis to the self-similar basis.  It is best to do this by first applying the transformation on one side of the matrix and then on the other side, i.e. $H \rightarrow {{(U H)} U^{\dagger}}$.  The matrix multiplication $U H$ can be understood as simply a process of breaking $H$ into columns and then transforming each column separately.  Similarly, the matrix multiplication ${(UH)}U^{\dagger}$ is best understood as a transformation of each row in $UH$.  The total time required to transform $H$ depends very much on whether or not it has non-zero matrix elements which are far from the diagonal; i.e. whether it is close to local.  If $H$ is extended and all its matrix elements are non-zero, then the transformation will require $O{(N^{2})}$ time.  If, on the other hand, all matrix elements $\langle \vec{x} | H | \vec{y} \rangle$ of $H$ that are far the diagonal (${|\vec{x} - \vec{y}|} > R$) are zero, then the computational cost is far smaller.  I define $b$ as the number of basis states per node at the lowest (leaf) level $l = 0$.  I also define $n_{0}$ as the number of leaf nodes within a truncation volume with radius $R_{0}$.  Then the time required to transform $H$ into the self-similar multiscale basis is of order $O{(n_{0} b N {\ln N})}$. 

Now I turn to generalized multiplication, which was defined in equation \ref{GeneralizedMultiplicationDef}.  I use the indices $i$, $j$, and $k$ to signify basis states in the multiscale basis. ${\vec{x}}_{i}$ is state $i$'s position, and $R_{i}$ is state $i$'s truncation radius.   Then the following formula shows how to calculate the multiple $C = {AB}$ of two matrices $A$ and $B$: 
\begin{eqnarray}
C_{ij} & = & {{\Theta{({|{\vec{x}}_{i} - {\vec{x}}_{j}|}-R_{ij})}} {\sum}_{k} A_{ik} B_{kj} {\Theta{({|{\vec{x}}_{i} - {\vec{x}}_{k}|}-R_{ik})}} {\Theta{({|{\vec{x}}_{k} - {\vec{x}}_{j}|}-R_{kj})}}}
\nonumber \\ 
R_{ij} & = & {{max}(R_{i}, R_{j})}
\nonumber \\ 
R_{jk} & = & {{max}(R_{j}, R_{k})}
\nonumber \\ 
R_{ik} & = & {{max}(R_{i}, R_{k})}
\label{GeneralizedMultiplicationDefMultiscale}
\end{eqnarray}

The computational cost of this generalized multiplication is just:
\begin{equation}
\mathfrak{C} = {{\sum}_{ijk}{{\Theta{({|{\vec{x}}_{i} - {\vec{x}}_{j}|}-R_{ij})}}  {\Theta{({|{\vec{x}}_{i} - {\vec{x}}_{k}|}-R_{ik})}} {\Theta{({|{\vec{x}}_{k} - {\vec{x}}_{j}|}-R_{kj})}}}}
\end{equation}
I have calculated this cost for the self-similar basis which I just discussed. Remember that I defined $b$ as the number of basis states per leaf node, and $n_{0}$ as the number of leaf nodes within a truncation volume with radius $R_{0}$.  Here are my results:
\begin{itemize}
\item If $\beta = 1$, meaning that every node at every level in the tree owns the same number of basis states, then the cost is $O{(b^{2} {n_{0}^{2}} N {{\ln}^{2}N})}$.  
\item If $\beta > 1$, meaning that nodes toward the bottom of the tree have more basis states per node, then generalized multiplication scales as $O{(b^{2} {n_{0}^{2}} N)}$.  
\item If $\beta < 1$, meaning that nodes at the top of the tree have more basis states than nodes at the bottom of the tree, then generalized multiplication scales much worse, as $O{(b^{2} {n_{0}^{2}} N {\beta}^{-2 \ln N})}$.  
\end{itemize}
Therefore multiscale bases with $\beta > 1$, i.e. with more basis states per node at the bottom of the tree, are preferable.

This is the final and most notable result of this section.  Its importance lies in the prefactor $b^{2} {n_{0}^{2}}$, which means that the prefactor is determined by the smallest length scale in the problem, the tree level truncation radius $R_{0}$.  This contrasts with $O{(N)}$ algorithms, which have a single truncation radius that must be set to be larger than the biggest length scale in the problem. Their prefactor is   $b^{2} n^{2}$, where $n$ is the volume contained within the single truncation radius.  The result is that single-scale multiplication is a factor of $n^{2} / n_{0}^{2}$ slower than multiplication with a multiscale basis. If the largest length scale is just a factor of ten more than the smallest length scale, the performance difference in a three-dimensional system is a factor of a million.

\section{\label{NlnNAlgorithms}Extending the Reach of $O{(N)}$ Algorithms to Multiple Scales}
In the previous section I showed how to create a multi-scale basis and use it to do a very fast generalized multiplication even in systems with with multiple scales.  This means that any $O{(N)}$ algorithm which is based on generalized multiplication can converted - with the aid of a multiscale basis - to an $O{( N {{\ln}^{2}N})}$ algorithm suited to multiple length scales.  The actual algorithms don't change; only the multiplication and the basis change.

\section{\label{BasicRenormalization}Renormalization Algorithms}
In this section I give algorithms for numerically renormalizing two linear algebra problems: the problem of evaluating the Green's function, and the eigenvalue problem.  By "renormalizing" I mean getting rid of most of the basis, leaving only basis elements which correspond to the physics of long length scales.  I first show how to renormalize away a single length scale.  Once one knows how to do this, renormalizing away several length scales becomes trivial, and I will discuss this briefly at the end.

I will divide the basis into two parts: the part I will renormalize away, and the part that I will keep.
I  mathematically express this division by introducing two
projection operators: $P_{-}$ projects out the states that I will
renormalize away, and $P_{+}$ projects out the states that I will
keep.  This completely exhausts the total basis states; $P_{+} + P_{-} = 1$.

\subsection{Two Scale Renormalization of the Green's Function}
The problem is to calculate the Green's function $G \equiv {(E - H)}^{-1}$.  This inversion typically runs in $O{(N^{3})}$ time.  In chapter \ref{LinearAlgorithms} I artificially truncated $G$ and was thus able to evaluate it in $O{(N)}$ time.  Here I want to take a different path, and instead get rid of most of my basis, so that the size ${Tr}{(P_{+})}$ of the remaining basis will be much smaller.  If this size is a fraction $\alpha$ of the original basis size $N$, i.e. if ${{Tr}{(P_{+})}} = {\alpha N}$, then I will be able to do the inversion required for the Green's function in $O{({\alpha}^{3}N^{3})}$ time; which is much faster than $O{(N^{3})}$.  The price of this acceleration is that I will have to do some extra computations to compensate for throwing out basis states. These compensatory calculations must be approximated in order to do them in a reasonable amount of time, and I will use the $O{(N)}$ algorithms of chapter \ref{LinearAlgorithms} for the approximation.

I start by defining two new Green's functions:
\begin{itemize}
\item 
\begin{equation}
{G^{+}{(E)}} \equiv {{(P_{+} {(E - H)} P_{+})}^{-1}}
\label{GPlusDef}
\end{equation}   This $+$ Green's function would be the total Green's function if there were no short distance degrees of freedom, if the $P_{+}$ states were all there was.  It describes the long distance physics of the problem.
\item 
\begin{equation}
{G^{-}{(E)}} \equiv {{(P_{-} {(E - H)} P_{-})}^{-1}}
\label{GMinusDef}
\end{equation}  
This $-$ Green's function describes exclusively the short distance physics of the problem.  It would be the total Green's function if the long distance degrees of freedom were all there was.
\end{itemize}
If the Hamiltonian $H$ had no matrix elements between the short-distance states $P_{-}$ and the long-distance states $P_{+}$, then the two kinds of states would totally decouple and the total Green's function would be just the sum of the two Green's functions; ${G{(E)}} = {{G^{+}{(E)}} + {G^{-}{(E)}}}$.  But in general there is no such decoupling.

Our problem, then, is to compute the sum $G{(E)}$ of the two Green's functions $G^{+}$ and $G^{-}$ when there is a coupling between them.  This is just the same summation problem that I discussed in chapter \ref{LinearAlgorithms}, and I gave four alternative formulas for doing the summation.  Here I choose the Frobenius formula of equation \ref{FrobeniusGAddition}; in the current notation it reads:

\begin{eqnarray} \label{FrobeniusRenormalization}
{G{(E)}} & \equiv & 
{G^{+} \oplus G_{-}}
\nonumber \\ 
& = & {G^{+} + {{(1 + {G^{+}
H_{0}})}  {(P_{-} - {{P_{-} G^{-} H_{0} G^{+} H_{0} P_{-}
}})}^{-1} {(1 + {H_{0}G^{+}})}} }
\nonumber \\ & &
\end{eqnarray}
I have assumed that the potential $V$ does not couple the $P_{-}$ states with the $P_{+}$ states; otherwise one would substitute $H$ for $H_{0}$ everywhere in equation \ref{FrobeniusRenormalization}.  The inverse operation should be understood as happening only in the $P_{-}$ basis.

If we are interested only in the long-wavelength matrix elements of the Green's function, then equation \ref{FrobeniusRenormalization} simplifies:
\begin{eqnarray} \label{FrobeniusRenormalization1}
{P_{+}\:{G{(E)}}\:P_{+}} =
{G^{+} + {G^{+}
H_{0}  {(P_{-} - {{P_{-} G^{-} H_{0} G^{+} H_{0} P_{-}
}})}^{-1}  H_{0}G^{+}} }
\end{eqnarray}

It is interesting to compare equation \ref{FrobeniusRenormalization1} with equation \ref{GreensFromT}, which relates the Green's function $G$ to the scattering matrix $T$:
\begin{equation}
{G} = {G_{0} + {G_{0}TG_{0}}}
\nonumber
\end{equation}
Clearly, from the viewpoint of the $+$ basis, the effect of the $-$ basis is simply to introduce a scattering term with scattering matrix \linebreak equal to $T = { H_{0}  {(P_{-} - {{P_{-} G^{-} H_{0} G^{+} H_{0} P_{-}}})}^{-1}  H_{0}}$.

Remember that we have already conceded the necessity of computing $G^{+}$, which requires $O{({\alpha}^{3}N^{3})}$ time.  Therefore the only remaining task is to make sure that the computations in the $P_{-}$ basis  requires less than $O{({\alpha}^{3}N^{3})}$ time.  These $P_{-}$ basis calculations are:
\begin{itemize}
\item Evaluating $G^{-} \equiv {{(P_{-} {(E - H)} P_{-})}^{-1}}$.
\item Evaluating ${(P_{-} - {{P_{-} G^{-} H_{0} G^{+} H_{0} P_{-}}})}^{-1}$.
\end{itemize}
 Unfortunately doing these inversions and multiplications exactly would require \linebreak $O{({(1 - \alpha)}^{3}N^{3})}$ time, which is too much.  Therefore we need to find an appropriate approximation.  Fortunately chapter \ref{LinearAlgorithms} on $O{(N)}$ methods developed a plethora of algorithms for computing inverses in $O{(N)}$ time!  Therefore equation \ref{FrobeniusRenormalization1} gives a way of computing the renormalized Green's function in $O{({\alpha}^{3}N^{3})}$ time, where $\alpha$ is the fraction of the original basis states which we want to retain after the renormalization.  For example, if we have renormalized away $80\%$ of the states, then we obtain an $O{(125)}$ acceleration.  Not bad.

 We can actually obtain a further acceleration by using $O{(N)}$ algorithms to do all the multiplications in the $+$ basis and to do the inversion necessary to evaluate $G^{+}$.  As long the truncation radius used in the $-$ basis is significantly smaller than the trucation radius used in the $+$ basis, the renormalized $O{(N)}$ calculation will be far faster than an unrenormalized $O{(N)}$ calculation.  We will have turbocharged our $O{(N)}$ algorithm by deliberately including the fact that there are two length scales different than the system volume: one length scale determines the $P_{-}$ localization radius, while the other length scale determines the $P_{+}$ localization radius.
 
 Note that by using an $O{(N)}$ algorithm to evaluate the correction to the Green's function, we have required that this correction be nearly local.  The same sort of assumption is imposed during renormalization of a quantum field theory: one requires that the renormalized theory be local.  In quantum theory just as here, locality is an assumption that one imposes on the final result in order to be able to do the renormalization calculation; it is not something that one can prove from the original equations.   
 
 While the Frobenius formula is already known, I believe that I am the first to suggest accelerating computation of the Green's function by combining the Frobenius formula with an $O{(N)}$ algorithm. In their work on hierarchical matrices, Borm, Grasedyck, and Hackbusch\cite{Borm04} do suggest accelerating calculations of the inverse by using the Frobenius formula, but they make special assumptions so that computing $G^{-}$ is trivial. (I believe that they require that $P_{-} {(E - H)} P_{-}$ be diagonal.) This is completely different from my suggestion of using an $O{(N)}$ algorithm to compute the inverse.

\subsection{\label{EigenvalueRenormalization}Two Scale Renormalization of the Eigenvalue Problem}
I now use a renormalization algorithm to solve the eigenvalue equation ${H {|{\psi}_{n}\rangle}}=  {{E}_{n}
{|{\psi}_{n}\rangle}}$.  This renormalization will add a term to the Hamiltonian $H$.  This approach of "renormalizing" the eigenvalue problem is called Kron's
problem\cite{Knyazev98, Beck00}.  My contribution consists in suggesting that one can speed up eigenvalue computations by combining Kron's problem with an $O{(N)}$ algorithm.

I will derive the Kron's problem equations rather than just writing them down.  I insert the identity $P_{+} + P_{-} = 1$ into the
eigenvalue equation:

\begin{equation}\label{Decomposing the eigenvalue equation A}
{{(P_{+} + P_{-})}H {(P_{+} + P_{-})}{|{\psi}_{E}\rangle}}=
{E{(P_{+} + P_{-})}{|{\psi}_{E}\rangle}}.
\end{equation}

I now expand the terms and simultaneously introduce a simplifying
notation where, for instance, ${|{\psi}_{-}\rangle} =
{P_{-}{|\psi\rangle}}$. Since ${P_{+} P_{-}} = {0}$, I obtain two
equations:

\begin{eqnarray}\label{DecomposingC1}
{{(E-H_{--})} {|{\psi}_{-,E}\rangle}} & = & {H_{-+}
{|{\psi}_{+,E}\rangle}}
 \\
\label{DecomposingC2}
{{H_{+-} {|{\psi}_{-,E}\rangle}} + {H_{++}
{|{\psi}_{+,E}\rangle}}} & = & {E {|{\psi}_{+,E}\rangle}}.
\end{eqnarray}

Equation \ref{DecomposingC1} suggests that I could calculate
${|{\psi}_{-,E}\rangle}$ from ${|{\psi}_{+,E}\rangle}$ if I could
invert the operator $({P_{-}EP_{-}}-H_{--})$. Clearly this
operator is non-invertible in the full basis $P_{+} + P_{-}$.
However, because $P_{-}$ is a projection operator, I can define an
inverse taken in the $-$ -basis; in fact this inverse is just the Green's  function $G^{-}{(E)}$ that I defined in equation \ref{GMinusDef}.
 
I now use equations \ref{GMinusDef}, \ref{DecomposingC1}, and \ref{DecomposingC2} to obtain the renormalized
eigenvalue equation, called Kron's problem:

\begin{equation}\label{RenormalizedEigenvalueEquation}
{{{H_{++} + {H_{+-} {G^{-}{(E)}} H_{-+}}} {|{\psi}_{+,E}\rangle}} = {E {|{\psi}_{+,E}\rangle}}},
\end{equation}

Once one has computed an eigenvalue $E$ and the corresponding truncated eigenvector ${|{\psi}_{+,E}\rangle}$, the full eigenfunction ${|{\psi}_{E}\rangle}$ can be recovered with the following equation:
\begin{equation}\label{MinusWavefunction}
{{|{\psi}_{E}\rangle} = {{{(P_{+} + {G^{-}(E)}} H_{-+})}
{|{\psi}_{+,E}\rangle}}}.
\end{equation}

Any eigenstates with support only in the $-$ basis (${P_{+}{|{\psi}_{E}\rangle}} = {0}$) have disappeared from equations \ref{RenormalizedEigenvalueEquation} and \ref{MinusWavefunction}. They can be
determined by finding states that solve the following
two equations simultaneously:
\begin{eqnarray}\label{$-$ BasisRenormalizedEigenvalueEquation}
{H_{--} {|{\psi}_{-,E}\rangle}}  & = & {E {|{\psi}_{-,E}\rangle}},
\nonumber \\
{H_{+-} {|{\psi}_{-,E}\rangle}} & = & {0}.
\end{eqnarray}

This completes the Kron's problem equations.
Note that in the literature Kron's problem includes the
generalization to non-orthogonal bases: ${H {|{\psi}_{n}\rangle}}=
{{E}_{n} O {|{\psi}_{n}\rangle}}$\cite{Knyazev98}.

I will call the new term in equation \ref{RenormalizedEigenvalueEquation} the renormalization hamiltonian.  Note that it depends on the energy $E$; Kron's problem is not an eigenvalue problem. The physical meaning of the renormalization hamiltonian can be elucidated if we interpret the eigenvalue problem as a scattering problem. $H_{++}$
describes scattering within the $+$ basis, while the additional
term describes scattering processes that start in the
$+$ basis, scatter into the $-$ basis, and later scatter back into
the $+$ basis.  Of course the Kron's problem equations are
non-perturbative and not limited to scattering physics.

If we use an $O{(N)}$ method to compute the Green's function $G^{-}{(E)}$, then a single computation of the renormalization hamiltonian requires only $O{({\alpha}^{3}N^{3})}$ time.  The problem is that we have to search for eigenvalues $E$, and because the renormalization hamiltonian depends on $E$, we have to re-compute it for every trial value of $E$.  In particular, if we want to compute all of the $+$ basis eigenvectors, and we evaluate the renormalization Hamiltonian once per eigenvector, then we will do at least $O{({\alpha}^{6}N^{6})}$ computations, which is actually worse than the unrenormalized $O{(N^{3})}$ time if ${{\alpha}^{6}N^{3}} > 1$.   Therefore we must either try to minimize the number of evaluations, or else be content to calculate only a few eigenvalues and eigenvectors.

One way to minimize the number of evaluations of the renormalization Hamiltonian is to calculate its derivatives and then extrapolate its behavior using a Taylor series.  Such an extrapolation will fail if $E$ is close to one of $G^{-}{(E)}$'s poles. Therefore good performance can be obtained only if almost all of the poles of $G^{-}{(E)}$ lie far from the eigenvalues of the  $+$ basis eigenvectors.  This condition is equivalent to requiring that few eigenstates of the original eigenvalue problem be highly mixed between
 the $P_{+}$ and $P_{-}$ bases.  Good performance is not
 possible without a good choice of basis.

\subsection{Renormalizing Away Multiple Scales}
 In the previous two sections I showed how to renormalize away a single scale.  I did this by dividing the basis up into two parts $P_{+}$ and $P_{-}$ and using the the small truncation radius of the $-$ basis states.  This approach can be trivially extended to doing several successive renormalizations of several different length scales.  Consider as a given a multiscale basis of the type discussed in section \ref{MultiscaleBases}.  Each level $l$ in the tree has a corresponding projection operator $P_{l}$ and truncation radius $R_{l}$.  One starts by renormalizing away the lowest level, the leaves. Then one renormalizes the next level up, and so on.  As long as the truncation radii increase substantially at each step up the tree, the cost of renormalization will be insignificant compared to that of the calculations in the final basis.
 
\section{\label{LinearNonlinear}Linear - Nonlinear}

In section \ref{EigenvalueRenormalization} I described an algorithm for solving the eigenvalue problem.  If the truncation radius were made infinitely large,
one would obtain the same solutions that would be obtained by solving the original eigenvalue problem.  These solutions obey all the standard results from linear algebra; for instance the eigenvectors are guaranteed to be orthogonal.  If, on the other hand,
the truncation radius is finite,
then the solutions obtained are no longer guaranteed to obey any of the
standard theorems.  One is no longer solving the
eigenvalue problem, but instead solving something that is "like"
the eigenvalue problem.  The new problem is nonlinear: it explicitly contains a new term which depends on the energy $E$ in a non-trivial way.  There are considerable
conceptual difficulties here, in understanding in what sense a
non-linear problem can be "like" a linear problem.

The same difficulty appears when trying to understand the algorithm for renormalizing the Green's function.  The original definition of the Green's function as ${G{(E)}} = {(E-H)}^{-1}$ guarantees the existance of poles in the Green's function at the eigenvalues of $H$.  In contrast, the existance of these poles in the renormalized Green's function is not guaranteed if one uses an approximate $O{(N)}$ algorithm to evaluate the renormalization term.  The poles in the exact Green's function play a very important part in any physical interpretation which is given to it; one is left wondering how to interpret the approximate renormalized Green's function.

Perhaps the possible lack of poles in the Green's function was to
be expected.  The imposition of the truncation radius $R$ could have a severe effect on any long-distance physics.  In particular, any eigenstates of $H$ which are extended throughout the system could be destroyed by the truncation, which implies that their poles would be deleted from the Green's function. One can conclude that any algorithm which purposefully neglects certain long distance physics must be able to delete poles
from the Green's function.  Indeed, conventional $O{(N)}$ algorithms not based on renormalization also do this sort of eigenfunction deletion.  For instance, basis truncation algorithms explicitly limit the eigenfunctions to stay inside of the truncation radius, and thus delete any extended eigenfunctions.

I am reminded of quantum field theory, where a pole in the Green's
function implies the existence of a physical asymptotic state. Therefore, if the QCD color field is truly confined, then the gluon
propagator must not have any poles.  In fact debate continues
about the structure of the gluon propagator\cite{Mandula99}.  It
is confusing to understand what a propagator is if it does not
propagate anything. Perhaps an investigation of the behavior of the renormalized algorithms presented here, and in particular of what they do to poles, could shed some light on these confusing issues in QCD confinement.

\subsubsection{A Generalized Schrodinger Equation}

Often using a continuum is simpler than using a discrete lattice.
Therefore I propose a continuum model which mimics the peculiar
behavior of the renormalization algorithms.  My model is just a
generalization of the Schrodinger equation.

The Schrodinger equation can be written as a combination of
the free propagator and a potential:
\begin{equation}
{{{({G_{0}^{-1}{(E)}} + V)}{|{\psi}_{E}\rangle}}={0}},
{{{G_{0}{(E)}}}={{(E-A_{0})}^{-1}}}
\end{equation}

The free Green's function $G_{0}$ is not localized, and contains a
pole for each momentum state, or in fact a cut along the positive
real energy axis.  However I can modify the Schrodinger equation,
giving the momentum states a range by smearing the Green's
function. In fact, with a proper choice of smearing I can give
higher momentum states a longer range than the low momentum
states. Here is one way to define the smeared Green's function:
\begin{equation}\label{Smearing}
{{\langle \vec{x} |} \tilde{G_{0}} {| \acute{\vec{x}} \rangle}} =
{\int{{d^{d}\vec{k}} e^{-i \vec{k} \cdot {(\vec{x} -
\acute{\vec{x}})} } {f{(E, |{\vec{k}}|,|{\vec{x} -
\acute{\vec{x}}}|)}} {{\langle \vec{x} |} G_{0} {| \acute{\vec{x}}
\rangle}  } }}
\end{equation}
$f$ is a function describing the localization cutoff. If $f = 1$,
then the Green's function is not smeared and Schrodinger's
equation is left unmodified.  Non-constant $f$'s smear the free
Green's function.  For example, if I choose a Gaussian smearing function ${f}
= {e^{- {\alpha}^{2} k^{2} {(\vec{x} - \acute{\vec{x}})}^{2}}}$,
then a state with momentum $\vec{k}$ can only propagate a distance
of about $\frac{1}{\alpha |{\vec{k}}|}$.

With the smeared Greens function $\tilde{G_{0}}$ in hand, I can
write a generalized Schrodinger equation:
\begin{equation}
{{{({{\tilde{G_{0}}}^{-1}{(E)}} +
V)}{|{\psi}_{\acute{E}}\rangle}}={0}}
\label{GeneralizedSchrodinger}
\end{equation}

The generalized Schrodinger equation mimics many of the problematic
aspects of renormalization algorithms.  It favors states which are not spatially extended.  The poles in the Green's function have been smeared around and may no longer exist, in which case the
generalized Schrodinger equation could have a different number of
solutions than the original Schrodinger equation.  The generalized Schrodinger equation can also be made
explicitly non-linear in $E$ (just like the renormalized
hamiltonian obtained when renormalizing the eigenvalue equation) by making the cutoff
function $f$ depend explicitly on $E$.  

The generalized Schrodinger equation allows examination of
interesting test cases. For instance, consider a potential $V$
with two wells, and with solutions that tunnel between the two wells.
Will the generalized Schrodinger equation preserve the tunneling
solutions, or will it find two distinct solutions for the two
wells, or will it do something else entirely?  This sort of
calculation could and should be pursued, as a way of developing a
physical intuition about the effect of imposed distance cutoffs on
linear algebra equations.

I have never seen equations \ref{Smearing} and \ref{GeneralizedSchrodinger} in the literature, and I believe that they are my own original contribution.

\section{\label{MultiscaleReview}Review of Multi-Scale algorithms}
Here I review the existing numerical algorithms for solving
linear problems with multiple length scales: hierarchical matrices, domain decomposition pseudopotentials, and
multigrid algorithms.

\subsection{Hierarchical Matrices}
Recently Hackbusch introduced a combination of mathematical theory and computational algorithms which he calls $\mathcal{H}$ matrices, or hierarchical matrices\cite{Hackbusch99, Borm04}.  I have only skimmed through some of his articles, but my impression is that he introduces a basis which is similar to the multiscale bases I have discussed here.  He then truncates matrices in a way that is closely tied to his basis.  Hackbusch uses a lot of mathematical notation to describe his basis and truncation scheme, and I have not yet understood these.  However, my impression is that his end result after truncating a matrix is similar to what I would obtain if I, while working inside of a multiscale basis, truncated all the matrix elements that were farther apart than the level-dependent truncation radius.  I suspect therefore that there is some overlap between Hackbusch's ideas and mine.  Where there is overlap, he would have precedence because he started publishing articles on his ideas by 1999.  

In any case I am also quite sure that there are also a lot of differences between our approaches.  I examined Hackbusch's algorithm for matrix inversion and it seems very simplistic.  And he does not discuss or apply a lot of physics ideas of localization, $O{(N)}$ algorithms, et cetera.

\subsection{Domain Decomposition}

Domain decomposition algorithms are a particular approach to solving linear equations, and based on the idea of breaking systems into parts.  As such, they can be understood as two-scale algorithms: there is the length scale of the individual parts, and the length scale of the whole system.  Domain decomposition algorithms have been in use for a long time and are still being actively developed by a large research community.
  
I focus here on a specific approach to domain decomposition based on Schur's complement.  First one separates the degrees of freedom into a coarse subset and a finer
subset; this can be represented with the projection operators ${P
+ Q} = {1}$.   Then one notes that
solving the linear system ${Ax}={b}$ is equivalent to solving a
modified equation in sub-basis $P$: ${\acute{A}{Px}} =
{\acute{b}}$, where the Schur's complement $\acute{A}$ is defined by ${\acute{A}} \equiv
{{PAP}-{PAQA^{-1}QAP}}$, and the input vector $\acute{b}$ is defined by ${\acute{b}} \equiv
{{Pb}-{PAQA^{-1}Qb}}$\cite{Knyazev98}.  One then solves the modified equation using a suitable iterative algorith. The
Schur's complement itself never has to be calculated explicitly;
instead at each iteration a linear equation is solved to obtain
${PAQA^{-1}QAP}x_{i}$.

The removed sites described by the operator $P$ are always chosen to form surfaces dividing the remaining sites $Q$ into independent domains.  The net result is that the system is divided into $n$ different linear systems which can be
solved separately. 

\subsubsection{Domain Decomposition for the Eigenvalue Problem}
Some research has been done on applying domain
decomposition ideas to eigenvalue problems.  Many algorithms for solving the eigenvalue problem involve solving linear equations; one can try to use domain decomposition to speed up the the solution of these linear equations\cite{Knyazev98}.  Alternatively, one can use the Kron's problem approach which I explained in section \ref{BasicRenormalization}.  Lastly, one can solve sub-domains separately and then combine them using a
Raleigh-Ritz variational principle\cite{Knyazev98, Lui00}.

\subsubsection{Domain Decomposition for Lattice QCD}
Only recently was an article published that suggested that a
Schur's complement algorithm can be applied to lattice gauge
theory\cite{Best00}. The author suggested that the inverse $QA^{-1}Q$ could be
estimated cheaply by first expanding it as a Taylor series in $A$
and then estimating the Taylor series coefficients stochastically.

More recently Luscher suggested a domain decomposition algorithm which seems to be more promising, though it still needs to be fleshed out\cite{Luscher03, Luscher03a}.

\subsection{\label{Pseudopotentials}Pseudopotentials}  Molecular and condensed matter systems
exhibit two or more length scales: one for the valence electrons
(which may or may not be localized), and then much smaller length
scales, one for each shell of core electrons.  However, there
is something very special about this problem compared to the
multi-scale problem addressed in this chapter: the core electrons
occupy volumes quite a bit smaller than the atomic volume,
so they do not interact significantly with other atoms. Therefore,
when calculating behavior at a given site, you can ignore all core
electrons except those at your own site.

The extreme locality of the core electrons allows tremendous
simplification; many years ago the condensed matter community
performed that simplification by developing norm-conserving
pseudopotentials\cite{Bachelet82, Hamann79}.  In pseudopotential calculations, the core
electrons are totally eliminated from the calculation, and the
potential felt by the valence electrons is modified to make up for
the lack of the core electrons.  This pseudopotential approach is
very similar to Wilson's approach to renormalization, where an
effective interaction is introduced by integrating out high energy
degrees of freedom. There is, however, a large difference: pseudopotentials can be calculated from first
principles, while effective Lagrangians are fit to data.  By now pseudopotentials are now universally accepted, pretty stable, and
give very high accuracy  results for most calculations.  In fact, without
pseudopotentials modern electronic
structure calculations would not be possible: they are the main
way of coping with the physics of core electrons.

There are two difficulties which prevent one from generalizing the pseudopotential approach to other multiscale problems.  First, pseudopotentials are unable to model interactions between one atom's core electrons and another atom's core electrons, and therefore require  that the degrees of freedom to be renormalized away be spatially separated from each other and not interact with each other.  The other difficulty is that pseudopotentials are hard to calculate in problems without spherical symmetry, simply because of the huge computational resources that would be required if more than two length scales are involved. For problems without a simplifying symmetry, pseudopotential methods allow in practice only one extra length scale. It
is our good luck that rotational symmetry is well preserved
throughout the atomic table, and that we are therefore able to calculate atomic pseudopotentials for atoms with many shells and therefore many length scales.  

\subsection{\label{Multigrid}Multigrid Algorithms}
Multigrid and multilevel algorithms began with a landmark paper in 1977\cite{Brandt77}, and have since taken over in many disciplines. They are competitive options
basically anywhere that partial differential equations must
be solved\cite{Wesseling01, Stuben01, Beck00}. Although there is a distinction between multigrid and multilevel
algorithms, they all share the same philosophy; for simplicity I will
here apply the word multigrid to them all.

Historically, the
multigrid approach has found the most success when applied to the
problem of solving the linear system ${Ax}={b}$, where $x$ is
unknown. All partial differential equations can be reduced to linear systems by
discretizing space, and then linearizing any non-linearities.  (One then
adds the non-linearities back in via an iterative process.) Because multigrid work is focused on solving linear systems, I will first focus on linear systems and only later
discuss the eigenvalue problem.  
Linear systems with large numbers of variables are usually solved using Krylov subspace algorithms; i.e algorithms where the $i$-th approximation is computed by applying a
polynomial in $A$ of order $i$ to the residual $b-{A
x_{0}}$\cite{Frommer96}. Krylov subspace
algorithms are vulnerable to very long convergence times when the matrix $A$ in $Ax = b$ is small. The matrices $A$ generated by discretizing a partial differential equation have smallest eigenvalues which are exponentially small as
the lattice spacing decreases; Krylov subspace algorithms show a corresponding exponential growth in convergence time.  This is problem called critical slowing down.  Of
course the smallest eigenvalue may always be increased by an
appropriate basis transformation (this is called preconditioning),
but finding a good transformation can be as challenging as the
original problem\cite{Frommer96, Peardon00, Elman02, Wu98}.  Thus
critical slowing down is a key motivator for the development of
alternatives to Krylov subspace algorithms, in particular
multigrid algorithms.

 Multigrid algorithms begin by
separating out the length scales in the problem.  Having established the various length
scales, these algorithms obtain a separate solution of the linear system at
each scale, and then combine the various scales together to obtain
the final solution.  The implementation of this strategy differs a
lot from algorithm to algorithm, and I will here review some of
the most significant approaches.

\subsubsection{Geometric Multigrid}
"Geometric multigrid" algorithms were developed first and have seen
the most widespread acceptance\cite{Wesseling01}.  They assume that
the degrees of freedom are distributed in a regular lattice, which
is called the fine grid.  Geometric multigrid derives the longer
length scales by organizing the sites on the fine lattice into
blocks of width $w$ and volume $w^{D}$. It then thinks of the
lattice of blocks as a (coarser) lattice in its own right.  This
process of creating a coarser lattice obviously can be repeated: the result is a hierarchy of lattices, with the coarsest lattice having only
one site, and the finest lattice being the original lattice.  The degrees of freedom at each level in this hierarchy are represented by the vectors $x^{i}$, where $x^{n}$
represents the fine grid and $x^{1}$ represents the coarse grid.
This process of building a hierarchy corresponds exactly to the
procedure of building trees which I described in section \ref{MultiscaleBases}.

Having created the hierarchy of lattices, one must relate them to
each other.  Geometric multigrid does this with two linear
operators: the "restriction" (averaging) operator $R$ which
maps fine grids to coarser grids via ${x^{i-1}} = {R x^{i}}$, and
the "prolongation" (interpolation) operator $P$ which maps coarse
grids to finer ones\cite{Wesseling92, Brandt92}. Additionally,
there must be a rule for deriving a coarse grid version of the matrix $A$; a
popular option (used in the Galerkin coarse grid approximation) is
${{A}^{i-1}} = {R {A}^{i} P}$.  The choice of $R$, $P$, and $A$
 varies from algorithm to algorithm.

Using these relations between the various grids, geometric multigrid pursues a strategy of iterative refinement of the solution $x^{i}$ on each grid.  However, there is a very special prescription for when geometric multigrid algorithms refine which grid, and when they move results from one grid to another.  Specifically, they start by refining the solution on the finest grid, and then restrict to the next finest grid (i.e. map $x^{n}$ to $x^{n-1}$.)  They then repeat the process of refining and restricting until they arrive at the coarsest grid.  At this point they then move back out to the finest grid in a process of refining and prolonging\cite{Wesseling92, Brandt92}.  For
many problems, this sort of algorithm converges within a very small number
of iterations, giving a total computational cost of order $O{(N
\ln{N})}$; i.e. the exponential cost associated with critical
slowing down is totally avoided.  This speedup has been both predicted theoretically for
simple problems\cite{Wesseling92} and
observed in practice for much more complicated problems.

Theoretical analysis reveals that fast convergence is very tied to making an
appropriate choice of restriction and prolungation operators $R$
and $P$\cite{Wesseling92}. This is often talked about as an issue
of "smoothness," because if the matrix $A$ is spatially
isotropic then the choice of $R$ and $P$ is essentially trivial.
 The root problem is that
the restriction operator $R$ must effectively isolate the
problem's long-wavelength behavior from its
short-wavelength behavior, and the prolungation operator $P$ must
in turn do a good job of interpolating the long-wavelength
behavior back to the finer grid.  Therefore the choice of $R$ and $P$ must be informed by a precise understanding of the long distance and short distance degrees of freedom and a good way of dividing them up.  The degrees of freedom and their division are determined by the matrix $A$, so making a suitable choice of $R$ and
$P$ can be as time-consuming as diagonalizing $A$.  In practice, multigrid algorithms are most succesful when the matrix $A$'s behavior is dominated by terms that are well understood, like derivatives. 

An additional problem with geometric multigrid is that it requires
that the system be a regular lattice.  This is problematic in
problems with irregular boundaries\cite{Wesseling01, Stuben01},
and in problems like quantum chemistry where sites are free to
move about.  For such problems a more flexible blocking algorithm
is required.  The choice of blocking can be as challenging as the
original problem; multigrid algorithms with flexible blocking
typically spend more time blocking than solving the blocked
equations\cite{Wesseling01}. Nonetheless they can often give far
faster solutions than Krylov space algorithms.

\subsubsection{Multigrid for Non-Uniform Systems}

"Algebraic" multigrid algorithms are designed specifically to solve the problems of geometric multigrid\cite{Stuben01}. Instead of blocking, they select out a
subset of the fine-scale sites to be used in the coarse-scale
problem.  Heuristics determine this selection procedure, as well as the
choice of prolungation and multigrid operators.  While algebraic multigrid algorithms are still immature, they have already had significant success with highly
anisotropic problems like the Navier-Stokes equation.  Highly disorded problems seem more problematic.

\subsubsection{Multigrid for the Eigenvalue Problem}
Multigrid algorithms are easily generalized to computing lowest-energy eigenstates of Schrodinger's equation\cite{Beck00, Heiskanen01}. If $q$
eigenstates are needed and $N$ is the system volume, these algorithms scale as either $qN$ or
$q^{2}N$, depending on the algorithm.  I am not sure whether
these scaling laws are preserved if $q$ becomes large. Also, these
estimates might be reduced if the eigenstates being calculated are
localized within volumes smaller than the system volume; several recent papers
have reported results in this vein but I have not read them. 

\subsubsection{\label{LatticeMultigrid}Multigrid for Unquenched QCD}

The most commonly used algorithm for calculating unquenched QCD is Hybrid Monte
Carlo, which introduces new degrees of freedom representing the fermions\cite{Kennedy99}.
These degrees of freedom evolve according to forces reflecting the gauge
fields, and the forces are calculated by solving a
linear system ${Ax}={b}$.  Solving this system remains the dominant computational bottleneck for unquenched calculations\cite{Elser01}.  Hybrid Monte
Carlo's biggest competitor is probably the Local Boson
Approximation, which again faces a computational bottleneck of evaluating forces by solving a linear system\cite{Elser01, Luscher93}.

A number of researchers have tried to apply multigrid ideas to this challenge\cite{Brandt92, Kalkreuter93,Borici99}.  Typical of these efforts is the "Ground-state projection"
multigrid algorithm, tailored to disordered problems\cite{Kalkreuter94}.  Like
geometric multigrid, ground-state projection multigrid assumes a
regular lattice and normal blocking. However it
diverges from geometric multigrid by customizing the restriction
operator on a block by block basis: it solves the eigenvalue
problem ${Ax}={Ex}$ on each block individually, and uses the
ground state of each block as the restriction operator $R$.  The
idea is that the presence of disorder requires a custom treatment
of each block, and that a given block's ground state best
separates out the blocks's long distance physics from its short
distance physics.  This is essentially just a heuristic algorithm, with little theoretical justification.  In fact nobody really understands how to separate short and long distance degrees of freedom in strongly disordered systems.

In 1990-92 four lattice groups around the world tried to use this
and other multigrid algorithms to accelerate QCD
calculations\cite{Kalkreuter94}.  They had enough difficulties that research
ceased by 1994 or so. 

This completes my review of already developed algorithms optimized
for two or more length scales.

\chapter{\label{Reliability}Reliability and Reproducibility of Computational Results in Physics}

\section{\label{IntroductiontoReliability}Introduction}

Ours is an era of specialization: professionals establish a
specialty and usually stay within that, remaining at amateur's
level in all other disciplines.  Specialization create
divisions between disciplines, so that chemists (for instance)
communicate mostly with other chemists, and often even restrict
themselves to a particular type of chemistry.  Even when the
advances of one discipline can be of substantial aid to another
discipline, they may remain unknown, unused, ignored.

This paper concerns the disciplines of physics and computers.  Long ago computer professionals, particularly those who risked retribution from clients if their computations produced incorrect results, discovered
certain systematic difficulties afflicting any computing, and
began developing an expertise to manage and ameliorate them.
Specifically, it is very difficult to ensure the reproducibility
and reliability of a computation, or to understand and manage the
complexity of a computer.  Put another way, computers are a bona fide, real life example of Murphy's law: they have a very strong natural tendency to always do the wrong thing. It is very difficult to provide the configuration, programming, and inputs that are necessary to obtain correct results.  Moreover, it is always impossible to be sure whether the results are correct without obtaining the same results in an independent way not involving computers.

In this paper I will show that although the physics community has
adopted the computer for both experimental and theoretical
research, it (with some exceptions) has not learned to respect the gross difficulties inherent in computing, and does not practice even the most basic disciplines which the computing profession has developed to manage these difficulties.  I will point out that this
"sorceror's apprentice" approach to computing can cause both daily
difficulties and long term errors.  I end by providing a list of
recommendations and resources for both individual physicists and
institutions interested in improving the quality of physics
research.

This paper will be of personal interest if the reader has ever:
\begin{itemize}

\item had difficulty remembering the exact parameters, settings,
procedures, or data he or she has used to produce a graph or
result.

\item wanted to reuse graphs or results produced by
another physicist, or to produce somewhat different results by adapting the other physicist's software and configuration, but found this task too difficult to be practical.

\item wondered whether a graph or numerical result could be
trusted.

\item been unsure how to estimate the accuracy or certainty of a
result.

\end{itemize}

 The author comes from
a perspective of having pursued careers in both computing and
physics.  He has both undergraduate and master's degrees in
physics, which were accompanied by research assistantships
implementing physics software.  This education was followed by a
career in computing, focusing on the reliability of software used
by large enterprizes. More recently, the author returned to physics. To a computing professional, certain aspects of the physics community's
usage of computers seem very puzzling. This paper is the result of
an effort to research the physics community's computing practices,
to understand the reasoning behind them, and to find resources and
best practices that might help.

Because the author's expertise is limited to physics and computing, this article focuses specifically on the physics community.  However, it is probable that most of the material applies equally well to the other sciences, even softer sciences\cite{CalderPrivate}.

\section{\label{ComputerProblems}Problems That Come With Computing}

\subsection{Reproducibility}

It is very difficult to create software whose behavior
is reproducible; i.e. that does the same thing every time you run it.
The reason is that computers are based on binary arithmetic, which is as nonlinear as one can imagine.  Nonlinear systems are extremely sensitive to initial conditions; if even one bit in the program itself or in its data or environment is different, then
its execution can easily give a totally different result.

Of course, computers are - theoretically -
deterministic, so that if you run exactly the same program twice,
with exactly the same environment, and with exactly the same data,
it must produce exactly the same result.  In this very theoretical
sense every program is reproducible.  But as a matter of objective fact, the conditions for deterministic
behavior are never fulfilled, even when running repeatedly on the
same computer.  A program always runs with the assistance of an
operating system, and operating systems give their hosted programs
different environments at each run - they change the memory
location in which a program is stored, they schedule cpu
allocation differently, et cetera. Moreover, the information in a
computer's disk and memory, as well as the environment outside of
the computer, certainly will change from run to run. 

Dependency on the environment can be ameliorated by adopting
conventions which insulate the program from its environment.
Operating systems insulate programs from their position in memory
by rewriting them during the load into memory. Standard software
layers (APIs) insulate programs from the details of where data is
stored. However these and other technologies are fundamentally
limited: first, because there are many situations where software
needs to act differently depending on its environment, and
secondly, because the insulation is not foolproof.  For instance,
no file system API can insulate programs from a disk full
condition.  A second, more subtle, difficulty, is that these
insulation techniques are vulnerable to bugs in the operating
system, the hosted program, and in the APIs themselves. I will
come to the difficulty of bugs in section \ref{Bugs} on reliability.

Leaving aside changes in the environment, reproducibility still
requires complete control over the program itself: the
program's source code, its compilation process and any other
ingredients in production of the executable, the program's
configuration, and the program's inputs. One must have complete
records of these variables for each run, and then be able to
recreate the exact same variables at will, before one can hope to obtain
reproducibility of results.  Although this is a herculean task, professional software developers find it absolutely necessary, because (a) software development halts as soon as programmers stop being able to collaborate with
each other, and (b) user satisfaction plummets when the software
does not run.  Therefore, every professional software development
organization devotes a significant percentage of its resources to
recording and managing its code, compilation process, configurations, and input data.

In order to give an idea of the extremity of the reproducibility challenge, I will now briefly sketch some of the most basic steps that computing professionals routinely take to surmount it.  Typically there is a computer dedicated entirely to running a Source Code Control program.  For those not familiar with this terminology, I mean a program designed to store detailed information about other programs.   Whenever even the tiniest change is made to a program's source code, compilation scripts, configuration files, or inputs, the result is thought of as a new version of the entire program.  A permanent record of every single individual version of the program is recorded in source code control, so that one can obtain old versions on demand.  The source code control database is, of course, backed up on a regular basis.

In order to know whether one's software is reproducible, one must compile, run, and test it.  Human efforts to compile, run, and test programs are inherently non-reproducible; therefore various scripting technologies like "make" files are used to automate the process.  Before recording a change in source code control, one is required to do an automated compilation and test of the new version, and to obtain a supervisor's approval.   However this requirement is not nearly fool proof, so typically the program is recompiled and re-tested every day using automation scripts which are also stored in source code control.  The daily compilation (and tests) are done on computers which are dedicated to this purpose, and which are configured solely from source code control.  Since the compilation server is clean of any changes not stored in source code control, one hopes that the compilation process will be
reproducible.  The compilation process is, of course, closely monitored, and much of the organization's staff is considered to be permanently on call to quickly resolve problems.  

These disciplines require a constant vigilance from every staff member, unremitting pressure from management, the existence of some staff tasked with maintaining and monitoring the whole process, and a constant and large expense in both time and resources. Yet even when implementing these practices, an average software development organization struggles daily to achieve reproducibility, and fairly regularly fails to obtain it.  For instance, a new version which runs successfully on a developer's computer often fails to
compile on the official compilation server.

Although later in this paper I will recommend adoption of some (but not all) of these technologies and practices to physicists, that is not my point here.  My goal is simply to point out that these organizations - often private enterprises under extreme pressure to save money and time - find reproducibility so unreachable that they regularly go to extremes to attempt it, and often still fail.  The problem is not lack of expertise or experience: these organizations are composed largely of persons trained in computer science, with years or decades of experience.  Nor is the problem one that a physicist, imagining herself smarter, could hope to be immune to: private computing firms try very hard to hire the smartest people including smart physicists, offer immense salaries, and often they offer careers that are more attractive than a physics career.  There is no reason to think that physicists will have less problems with reproducibility than computing professionals do; there is plenty of reason to think that they will have more.

So far I have discussed the difficulty of getting a single
program to give reproducible results.  I now briefly mention that
two programs written to do the same thing will as a rule exhibit
different behaviors.  Hatton and Roberts \cite{Hatton94} compared
nine seismic analysis codes which convert echo data into an image
of the earth's near surface structure. These codes had typically
been in regular use for fifteen years or more, and on average each
contained 750,000 lines of code. There is a large incentive to
ensure that these codes produce correct results: they are used to
make multimillion dollar choices about oil drilling.  Hatton and
Roberts identified 34 processing functions (on average about
150,000 lines of code) which were shared by all nine codes.
Fourteen of these functions were implementations of the same published algorithms; they could be expected to give identical results when applied in sequence to a data set. Instead the discrepancy grew at an average of one percent per three thousand lines of source code and
approached a hundred percent in the final map of the subsurface.
Geoscientists comparing the results found that "the differences
were not subtle, corresponding to alternative equally legitimate
lithological views which can fundamentally affect the conclusions
reached as to the nature of potential hydrocarbon accumulations."

Hatton and Roberts write, "It is perhaps understandable for the dedicated scientist to claim that his software does not have such problems and that the seismic data processing community do not program well. This 'somebody else's problem' attitude is regrettably frequently encountered, but the authors' experience at seeing much software from around the world in numerous different application areas combined with the large static fault study cited earlier, suggest that the seismic data processing development environments are if anything, \textit{more} mature than the average, often containing a software QA function and defined test datasets."
 
The bottom line is that software naturally fights reproducibility.
Or, put another way, it is a miracle of human persistence and
professionalism when reproducibility is obtained.

\subsection{\label{Bugs}Reliability}

I take the liberty of defining software as reliable if "it always
does what is expected of it."   In order to avoid the mistake of moving the focus from the end user to the developer, I have deliberately not added a proviso about the program design goals.  To illustrate this point, let me use the example of a program which should, as an intermediate step in a larger computation, perform data smoothing.  Suppose that on a particular occasion it does not do the smoothing, and does not alert the user that smoothing was not performed.   It doesn't matter whether the program was designed that way; for instance perhaps it was designed to react to certain resource constraints by silently skipping the smoothing.  What does matter is that the program delivered misleading and wrong results to the user, and therefore is unreliable.  
 
A program's behaviour may be reproducible and yet unreliable. In
the previous example, the software might reproducibly give incorrect
output whenever resource constraints occur.  As another example,
consider a floating point calculation which, due to rounding
issues in its implementation, produces less accurate results when
one of the inputs is close to a certain value, but does not signal
the decreased accuracy to its user. The user would likely be
surprised to find that for this particular input value the result
is $ 1 \pm 100 $ instead of $ 1 \pm 0.01 $.  This is an instance
of unreliability, since the program did not do what was expected
of it.

Invariably humans (and I mean to explicitly include the reader of this article) instinctively grossly underestimate the
intellectual effort, professionalism, and sheer time that is
required to obtain reliable results from software, and
correspondingly overestimate software reliability. 
There is abundant evidence that this wishful thinking is the human's invariable, natural, knee jerk, reaction to software reliability.  Within the field of software development,
prodigious effort has been devoted to studying both human wishful
thinking and the real observed difficulty of obtaining
reliability.  However, because the results of these studies are
always out of tune with our natural optimism, in general even
organizations with much expertise and experience in software
unreliability tend to still both underestimate the difficulty of
obtaining reliability and also overestimate the reliability of
their own software. While their errors may be significantly
smaller than those produced by human instinct, nonetheless they
are frequently wrong by multiplicative factors.  For example, an organization may make a detailed and thorough estimate that a particular program will take a year to complete, and instead spend three years on it.  Moreover, throughout the last year its estimate of the remaining time may remain constant at three months!  If even the experts have these difficulties, then physicists in general, and the reader in particular, can be assured that software reliability is far harder to obtain than they imagine.

It is often hard for humans, born optimists in this respect, to agree with the following assertion: One can be confident that any program, without exception, is unreliable if it has not passed a comprehensive suite of tests and has not been used by a substantial number of people.  Yet every systematic study of software reliability, whether in small programs of a few dozen lines or large programs with millions of lines, confirms this statement unambiguously.  The only reason why the software in wide spread use, whether operating systems or the programs they host, is at all usable, is that it has been developed and tested at enormous cost in time, expertise, and other resources, and then used by thousands or millions of people.  Despite all this, it is still painfully unreliable, the subject of continued unanimous lament.  The important question for physicists who configure and use software, or may even write software, is not whether any untested software or configurations they might use are unreliable, but instead how to deal with the guaranteed unreliability.

I briefly list some other disturbing facts about unreliability.  These are so well known within the computing community that they are simply truisms.  I repeat them, however, because the reader may not be so familiar with them.
\begin{itemize}
 \item Whenever software is used in a new way, a new failure mechanism will likely take effect.  Thus, a program that is quite reliable for common tasks often is very unreliable
for less commonly used functions. 
\item When controlling or configuring software, a user will
usually make mistakes about which settings or commands to use. For example, a graphing utility will produce an incorrect graph if one gives it incorrect parameters.  These user
mistakes may not be detected, and the results may be incorrect and/or irreproducible.  Thus, one must always test and verify the results obtained from software, even when the software itself is reliable.  
\item Practically speaking, it is impossible to
create fully reliable software. Software reliability is a matter of degrees, and of
confidence.  Therefore one must learn to acknowledge and manage the risks of using computers.  This is no excuse, however, for not testing the software, for in that case one can be certain that the results will be incorrect.
\item Most bugs will not be found unless one actively tries to prove that the software is not reliable\cite{Myers79}. Alternative approaches, for instance the approach of trying to prove that the software is reliable, are too friendly to the natural but
incorrect optimism that humans bring to software.  
\item Testing requires considerable intellectual effort and discipline.   Design of test suites is at least as challenging as design of the original software\cite{Myers79}.
\item Once a bug is found, considerable expertise,
professionalism, and time must be expended in order to isolate and understand
it.  Even then a correct solution can be very
difficult to obtain.  It is very easy to convince oneself of the validity of
an incorrect solution.  One thus fails to fix the original bug, and may even introduce new bugs.
\end{itemize}

Software reliability is impossible without either sufficent testing or else widespread adoption by large numbers of users who are willing to struggle with completely unreliable software for a prolonged interval.  Reliability after wide adoption is not an option for most scientific calculations, which are typically special purpose calculations that need to be correct by publication.  

In section \ref{CurrentPractices} I will discuss the fact that physicists usually restrict their checks of computational results to global, qualitative, non-automated checks.  It is well known in the software industry that relying on this approach alone will in the end do little or nothing to improve software reliability.  Humans are notorious for not noticing or discounting inconsistencies in computational results, and therefore qualitative testing is useless.   Instead one must carefully decide exactly what the software should do when given certain inputs, and then automate a check that every detail of the prediction is fulfilled.   Moreover, global checks alone, even if they are quantitative and automated, will fail to uncover more than a small percentage of the bugs.  This is partly because many software failures occur at intermediate stages in the calculation, partly  because the actual execution of a program depends very sensitively on its inputs and therefore a global test probes only an exponentially small percentage of the ways a program could execute, and partly because global tests tend to exercise the core 10\% or 20\% of the code and not touch many of the peripheral options.  Therefore software reliability requires that many other varieties of tests be implemented\cite{Myers79}.  One should test each input parameter by trying several different input values, including values which stretch the software's limits and invalid values.  Various combinations of input parameters should be tested.  One should test individual functions (procedures) within the code.  One should look inside the functions, analyze control flow and data flow, and design tests aimed at validating them.  One should implement both systematic sets of tests and individual tests aimed at points in the code which one suspects are more likely to have problems.  And when two or more people are available, they should read through each other's code together with a critical eye, asking the author to explain the behavior of the code line by line to the others.  Whenever possible, tests should be created by somebody other than the author, since as a rule authors are even more unduly optimistic about their own code than others.

Testing is also needed when the physicist is re-using another's software.  These tests can be broken down into  two types: Tests of the software and its installation, and tests of the user-specific aspects: various input data, configuration files, scripts, and any manual actions to control the software.  I first discuss the software and its installation.  Any error messages that occur during installation should be recorded, and any log generated by the installation should be archived.  In ideal cases software will come accompanied by an automated test suite\cite{LAPACK}; in this case the suite should be run after installation, and also at any point when one begins to wonder whether the software is broken.  However this is not enough: One should review the test suite (hopefully it is well documented; otherwise one will end up reading its source code)  and see how much it tests the functionality that will be used.  One would hope that the test suite will not only thoroughly test each individual option or command that will be used, but also include global tests of scenarios that closely approximate what the user will be doing.  One should also consider whether the software will be used for a task that isn't done very often by other researchers, or perhaps for two tasks that aren't done in combination very often.  

When this examination shows that the test suite is missing tests, the physicist will have to create them herself.  In many situations, including most commercial software, no test suite is supplied.  In other cases a test suite will be supplied but without source code or documentation.  In these cases, one must consider the entire functionality to be at risk and either avoid using it entirely or implement a test suite of one's own that thoroughly tests the functionality of interest, including each individual option, command, or step in the computation.   For instance, if the physicist is using a symbolic algebra program to manipulate gamma matrices, she should create a set of scripts that demonstrate the program's ability to do all the individual matrix manipulations correctly and also to do a few very involved manipulations.  Of course such testing efforts will be greatly hindered if the source code of the software is not available.  This argues strongly that all scientific software, commercial or not, should be accompanied by a thorough, well documented, open source test suite.  

I now turn to testing those aspects of the calculation which are unique to the user.  This testing will include a detailed review of the configuration files and scripts create by the user, preferably with two or more people examining them together. The testing will also involve checking each configuration option and command to verify that it actually does what the user expects it will do, both by reading the option or command's documentation and by running it and checking the results.  Data, too, should be examined thoroughly, and one must ascertain that the software is able to read the data correctly.  For instance, when graphing data one should verify the graph by opening the data file, actually reading some of the data points, and checking that they are represented accurately on the graph.
  
The last paragraphs should not be construed as a complete description of the practices necessary to obtain confidence in a calculation's results.  I recommend readings on the disciplines of software testing\cite{Myers79} and of verification and validation\cite{Oberkampf02, Trucano04, Foundations02, AIAA98}, and I also expect that a professional care for accuracy and thoroughness will naturally suggest further tests.

\subsection{\label{ComputerComplexity}Complexity}
Computers are extremely complex.  Here I discuss two aspects of that complexity: the way computers get things done, and the information overload they produce.  

First, the way computers get things done. It is extremely difficult to understand at the lowest level (machine code, registers, paging, call stacks, etc.) the inner workings of even the simplest program. But let's leave that aside and move to a higher level of abstraction, to the level of short programs written in high level languages like C++ or FORTRAN.  The problem is that tasks that are conceptually very simple can be extremely complicated on a computer.  Even when the source code for the task is short, understanding what that code actually does and what results it will produce is often far beyond modern capabilities.  Consider diagonalizing a matrix.  This task is so conceptually simple that linear algebra books often present only one or two simple algorithms for diagonalization, and physics books generally skip its details altogether.  Despite this conceptual simplicity, the design of good algorithms for diagonalization is an apparently infinitely complex issue.  It has been under continuous research since the introduction of computers and is by now a very involved research field\cite{Golub00}.  Interestingly enough, although the algorithms which this field invents can often be described with just a few dozen lines of text, the details of their behavior and accuracy often resist years of study by expert numerical analysts.
 
 Now consider scientific software running on a modern PC, which typically performs billions of arithmetical operations per second and is equipped with at least two billion bits of random access memory.  Scientific software often takes advantage of
much of that memory (and other resources: hard disk, DVD, Internet, and various peripherals) to arrive at its results.  Moreover, the source code used to specify how to arrive at those results can easily have a length equivalent to several books. That is not counting the fact that modern software generally relies on the operating system to do a lot of things, and that the source code for an operating system has a length equivalent to several hundreds of books.  If it is often impractical to obtain a detailed understanding of the behavior of a only few dozen lines of source code, then clearly the details of the behavior of most useful software are far beyond our comprehension.  

This brings me to a second aspect of complexity: information overload.  We are drowning in information.  I doubt that the reader needs to be reminded of the reams of data which are produced by calculations and computer-assisted experiments alike, the challenge of storing it all, the difficulty in understanding and recording the exact configurations and parameters that went into producing the data, the mind numbing tasks of analyzing and understanding the data, or the continueing uncertainty that one's analysis was entirely correct.

Perhaps information overload is the root of all the other computational difficulties I have discussed.  Probably if there were only ten or twenty things I needed to know about a computer then I would have little problem making it do the same thing over and over again (reproducibility), with understanding whether it is working correctly (reliability), or with fixing it until it does work correctly (reliability again.)   A hammer would be a good example of this - despite the fact that it is made of incredibly complex materials, I really just need to know its heft, its size, its swing.  Even then, figuring out how to hit a nail straight on can be very challenging: but anyone will learn that, given enough months of practice.

Judging by human success with complex machinery (a spectrum from ground vehicles to jumbo jets to factories to nuclear reactors), under ordinary conditions systems with some thousands of important variables can be managed fairly reliably.  But already the situation is qualitatively different: one needs highly trained and experienced specialists.  

The problem with computers is that they are absolutely qualitatively different from these two examples.  This should be obvious:  one can make a count of the million plus pixels on the screen, and then remember that we always want to know more than what's on the screen (that's why windowing operating systems were invented.)  If one likes, one can also count sheep: the billions of bits in memory, trillions on disk, and quintillions on the net.  A computer is not like a hammer, it is not like complex machinery, in fact it is not like any other tool that has been invented.

\subsection{\label{SoftwareAsDiscourse}Is Software a Mechanism?}
Software is often perceived as a kind of mechanism, a machine independent of both the software developer and the user, which functions or fails according to its own internal composition and laws.   I want to highlight two particular ways that this perception manifests itself in real life.  First, physicists may believe that the tasks they perform with computers are not really physics; and may imagine that they are simply using a tool ("thinking machine" still evokes the gestalt) to get a particular job done\cite{TrucanoPrivate}.  Because many believe that computing amounts to operating a machine and is certainly not physics, there is a tendency for physicists to not include discussions of software in their professional discourse and published articles, and to relegate computing tasks to graduate students.  Second, the public as whole understands bugs as a sort of machine failure, a breakage.  As a consequence of this analogy with machine failure, people incorrectly assume that bugs are the exception rather than rule, that it is possible to examine software before hand and see whether it is in good shape, that if software were engineered properly then it would be reliable and its developers would never have to make changes to it, that the culpability for any particular software failure can be traced to a mistake of some sort, and that very rudimentary descriptions of software failures should suffice for others to quickly identify and fix them.  All of these beliefs accurately descibe ordinary tools and machines, and yet are absolutely wrong when applied to computers.

I would like to suggest that real world software is not best understood as a machine or mechanism.  While textbook algorithms may meet this definition, in real life software is better understood as a way of communicating, like scholarly discourse, literature, et cetera.  In other words, it is far better to compare a computer to library than to a machine.   The analogy between software and communication can teach a lot:
 \begin{itemize}
\item Software (meaning the source code) is written in a language.  It is very important that software be understandable to humans, in order to assist both the original programmer and future programmers in their efforts to understand, check, fix, and improve it. In fact, usually human assistance is required to make the software understandable (compilable) for computers, and the humans are unable to assist without understanding the code themselves.  One can argue that the most important audience for source code is not the computer but humans.  
\item Careful observations of human speech shows that we have immense difficulty with speaking grammatically; it should not be surprising that writing correct programs is also difficult for us.  
\item Words derive their meaning from their context; software as well works only within the context of being compiled, run by an operating system, receiving input, and possibly connecting with other programs or other computers.  
\item Texts are can be considered valid or invalid only within a certain context, and in fact usually a dialogue is required before meanings become entirely clear.  Software too is never entirely correct but is always subject to further critiques by new users.  Thus a continueing relationship between developer and user is required, which usually is called bug fixes, new releases, or maintenance.
\item If one studies any text in detail, one will inevitably find unanswered questions, incompletely specified terminology, subtle or glaring inconsistencies, and perhaps even grave errors. Experience from the study of literature indicates that this process of discovery can go on for centuries. This is similar to software, where every program contains bugs, and the only certainty is that the longer software is studied the more bugs will be found.
\item Earlier I briefly described the huge amount of effort which is spent on recording every detail about software and its history.  I did not mention the exacting and detailed analysis devoted to the software; teams reading through code together, detailed specifications, formal diagrams, test plans, amd exacting records of its behavior and failures.  All of this is very reminiscent of the commentary and tracking devoted to certain texts.  The most extreme cases are probably religious texts; for instance every manuscript of the Bible and the Talmud is tracked and compared character for character.  Shakespeare's works, and other literary texts, are also the object of exacting study.  There is also a parallel with scientific literature: every scientific article is permanently recorded in a scientific journal, and a lot of attention is given to how articles comment on and build on other articles.   
\end{itemize}

 In summary, I am suggesting that real world software is best understood as being subjective, not objective.  By "objective" I mean having a reality external to and separate from the individual observer, the way that a hammer or machine is the same thing whether you put it in a parking lot or in the Sahara desert.  I
use the word objective in the sense of an external, separate,
objectified entity, not in the sense of true-ness or false-ness.  Something may be objective and yet false, for instance an
incorrect theory.   
 
 In saying that software is subjective, I mean that its meaning and validity are given to it by people through some sort of interaction with each other, and depends on context supplied by communities, communication, or personal experience.
 
 \section{\label{CurrentPractices}Current Physics Practice}
I now attempt to give a picture of how the physics community manages the computing problems which I discussed.  This picture is based on a survey of physics publications and software which I did in May 2003.  I plan to redo the survey, this time extending its reach by surveying journal policies,  and extending its thoroughness by using certain new full-text search engines.  Unfortunately I was able accomplish these plans before finishing this thesis.
 
 The physics
community uses software in several different ways, each of which
has its implications for reliability and reproducibility.  Here
are the most common usage patterns:
\begin{itemize}
\item Operating systems, Utilities, Languages, and Scripting
Engines.  These resources usually have a wide user base extending
far outside the physics community.  They often are compiled before
distribution to users.

\item Scientific Programs.  Many scientific programs have been
made available for re-use, either as source code or in compiled
form. These resources often have a restricted user base.   I will
differentiate two classes of scientific programs: The first class
is large codes developed and used by a community on a continuing
basis; for example codes for fluid dynamics, linear algebra, and
partial differential equations.  The second class is codes that
are written by a few researchers who stop development at some
point and make their code available to the public.

\item Reusable Numerical Results.  Often models are obtained
through complex analysis procedures, and then are re-used without
redoing the analysis.  Examples include nuclear potentials,
pseudopotentials, and tight binding hamiltonians.

\item Unshared Resources. Almost all research involves creation of
resources that are not distributed to the public.
 These are widely assorted, but may include configuration files, programs, and scripts.
\end{itemize}

\subsection{Efforts to Ensure Reproducibility} 

No survey data is available about the procedures used by
individual physicists to ensure reproducibility of their computing
results. However, two facts are clear:
\begin{itemize}
\item Few, if any, individual research articles make available for on-demand download the
complete set of files necessary to reproduce their results.  Even
the minority of research articles which do this usually do not distribute all the supplementary files
needed to obtain the published graphs and numbers.

\item Physics research articles tend to not report version
information about their computing environment.  One can get a
flavor for this by searching physics abstracts for the names of
popular pieces of software.  On May 15, 2003, searches of
arxiv.org showed that only 35 physics abstracts\cite{LinuxSearch}
mentioned "linux," 57\cite{MathematicaSearch} mentioned
"mathematica," and 117\cite{FortranSearch} mentioned "fortran."
Possibly even fewer articles report version numbers; for instance
only eleven of the 35 linux articles provided version data, and
many of them gave only partial data.
\end{itemize}

We can read between the lines.  It is likely that most individual
physicists do
not track everything necessary for reproducibility. Probably most
are unaware of the advantages and availability of source code
control software.  Without it, they are limited to making
(partial) backups of their files. Even backups can be problematic
in those institutions which do not provide recordable mass media
and automated backups to their researchers.  In these circumstances, it seems likely
that many individual researchers will have difficulties giving
directions to others about how to reproduce exactly their own
research results. They may even have difficulties reproducing
their own results themselves.

Collaborations developing large codes often do better with
reproducibility practices.  Most distribute one or more releases,
each with clear and distinct version information.  For instance,
LAPACK\cite{LAPACK} currently offers release three to users, while
MPICH\cite{MPICH} offers many versions of its compiled binaries.
Many large projects also use source code control software.  Some
of these, like ATLAS\cite{ATLAS}, PETSc\cite{PETSc}, and
GSL\cite{GSL}, allow anonymous read-only access to all versions of
the source code.

\subsection{Attempts to Reproduce Results}
There are some cases where many physicists do calculate the same
physical quantity. For instance, there have been many lattice QCD
calculations of meson and hadron spectra, and comparisons of
them have been published. However, no single spectrum calculation,
using all the same algorithms and constants etc, is reproduced by
two parties. So it's basically a lot of people doing it different
ways and getting results that have some similarities and some
differences. 

Moreover, most published scientific results cannot be directly
compared with any other result.  I.E nine out of ten graphs are of
non-standard quantities: different selection criteria, slightly
different formulas, different parameters (energy of the probe,
etc.), different approximations, and on and on. 

It is likely that there are also unpublished, undocumented attempts to reproduce results.  The results of these efforts may be communicated by word of mouth, and errors may never be documented\cite{BorchersPrivate}.  One reason for this could be a perception that publicizing an error and its correction would humiliate the original researcher. Alan
Karp\cite{KarpPrivate} gives an example of a conference where many groups simulating Cepheid variables got together and compared each group's graph of a single physical quantity.
It could be that many of these unpublished attempts at reproducing results do not go through
all the steps necessary to get exact quantitative agreement.

\subsection{Bug Awareness and Management}
The physics research literature does not discuss the reliability
of its computing efforts. In particular, the possible existence of
bugs is almost never acknowledged.  On May 15, 2003, a search of
all physics abstracts and titles in arxiv.org listed only four
abstracts\cite{BugAbstractsSearch} containing the words "bug" or
"software defect." (There is however, some evidence that bugs do
occur: 54 articles\cite{BugCommentsSearch} had attached author
comments stating that the article had been revised because of a
bug. Very little further data was supplied. Moreover, 928
records\cite{ErrataSearch} contained the word "errata" or "erratum,"
and one must wonder how many of these were caused by software
issues.)
 
Unfortunately, the text of an erratum generally does not
discuss how the error occurred.  There is one exception where physicists explicitly documented the computing failure that had caused an incorrect result, probably because of the result's extremely high visibility.  On February 8, 2001, a collaboration at the Brookhaven laboratory reported a new measurement of the anomalous magnetic moment of the muon\cite{Brown01}.  The result was very intriguing because it deviated from the theoretical prediction by $2.6$ standard deviations, and because the collaboration had plans to quickly increase their experiment's accuracy, in which case the deviation could have reached $5$ standard deviations\cite{Schwarzchild02}.  If this had occurred, it would have been a convincing sign of physics beyond the standard model.  A citation search on SLAC's SPIRES database shows that by the end of October 2001, 226 preprints had been distributed citing the Brookhaven result. Many of these papers were theoretical efforts to explain the discrepancy using physics beyond the standard model.  It seems, however, that the original theoretical prediction (which relied only on the standard model) was not checked until October or so.  On November 6, 2001, Knecht and collaborators distributed two preprints suggesting that the prediction should be revised\cite{Knecht01a, Knecht01b}: a particular Feynman diagram contributing to the result had been given the wrong sign.  A few months later one of the physicists responsible for the original predictions, Kinoshita, reported in some detail his collaborative effort with Hayakawa to recheck the calculations, and their discovery of the source of the sign error\cite{Hayakawa01}.  Hayakawa and Kinoshita write:

"We noticed one crucial difference between Ref. 5 [Knecht and Nyffeler's paper], which leads to the positive value, and ours, which leads to the negative value; while Ref. 5 used the algebraic manipulation program REDUCE to perform the trace calculation of the $\gamma$ matrices, 
we used FORM instead.  Recall that we have used FORM even for examining the results of Ref. 5. Thus we decided to check whether we handled FORM properly. The program FORM had been used successfully to calculate the QED corrections to the $g-2$ of the muon  and the electron as well as other observables by one of the present authors. However, this does not guarantee that we deal correctly with the $\epsilon$-tensor,  the central object of our study of the pseudoscalar contribution. A simple test of this question is to see if our naive use of the FORM declaration  (FixIndex $1:-1, 2-1, 3:-1;$) 
works successfully to verify the identity 
\begin{equation} 
 \epsilon_{\mu_1 \mu_2 \mu_3 \mu_4} 
 \epsilon_{\nu_1 \nu_2 \nu_3 \nu_4} 
 \eta^{\mu_1 \nu_1} \eta^{\mu_2 \nu_2} 
 \eta^{\mu_3 \nu_3} \eta^{\mu_4 \nu_4} = -24 \, , 
\end{equation} 
which should hold in Minkowski space-time.  Unfortunately, the result turned out to be $+24$. 
 This means that this simple declaration does not work for the $\epsilon$-tensor in Minkowski space-time.....  On the other hand, REDUCE passed the same test without difficulty."

This incident merited a one page report in Physics Today\cite{Schwarzchild02}, in which Bertram Schwarzchild explains that the FORM program's output of $+24$ instead of the expected $-24$ was appropriate in the Dutch convention for the Levi-Civita tensor ${\epsilon}_{\alpha \beta \gamma \delta}$.  (Schwarzchild mistook the author of FORM; it should be correctly attributed to Dutch physicist Jos Vermaseren\cite{Vermaseren01}.) This was nonetheless an instance of software unreliability, simply because there was a mismatch between FORM's behavior and Hayakawa and Kinoshita's expectations.  (The question of whether the fault should be ascribed to FORM or to Hayakawa and Kinoshita does not affect the diagnosis of software unreliability.)  The error could have been avoided if Kinoshita and collaborators had done more thorough testing of the software functionality that they were using.  Better documentation, and reading it, can also sometimes help to prevent this sort of error.

It is interesting to speculate how much longer the physics community would have required to discover the problem if Knecht and collaborators had used FORM instead of REDUCE, or had reacted to the psychological pressure to reproduce known results by "correcting" REDUCE's sign in some way.
 
I now move on to other aspects of bug awareness and management.  Despite a lack of survey data, it is safe to say that the majority
of physics research efforts do not use any bug tracking databases
to manage software problems. It seems likely that most physicists
are unaware of the availability and utility of these solutions.
However, some of the larger codes and programs probably maintain
internal bug tracking databases that are not visible to the
public.

Almost all publicly available physics software relies on e-mail
for bug reporting; larger programs and codes may have e-mail addresses
dedicated to user support, while smaller ones typically use the
author's e-mail address. Response to bug reports is rarely, if
ever, guaranteed.

Many of the smaller publicly available codes do not offer the
public any information about their bugs; witness in particular the
CPC Program Library\cite{CPC}, which appears to make no provisions
for documenting bugs in the programs it houses.  There are some
exceptions to this rule, in particular those codes that are housed
by sourceforge.net\cite{SourceForge}, which provides for each
hosted project a bug tracker, a support tracker, a newsgroup, and
a searchable archive of the newsgroup.

Larger codes are more likely to document bugs: for instance
LAPACK\cite{LAPACK} has an erratum page, the Gnu Scientific
Library\cite{GSL} provides searchable archives of its newsgroups,
and the APE supercomputer's operating system allows anonymous
users to view its bug database. Among commercial packages, public
bug documentation is spotty: for instance MATLAB\cite{MATLAB}
documents both bugs and their fixes on their web site, while
Mathematica\copyright\cite{Mathematica} and
GAUSSIAN\cite{GAUSSIAN} do not.

\subsection{Testing}
The physics research literature rarely discusses either software
testing or the closely related term "verification and validation."
Searches through arxiv.org in May 2003 revealed that only 46 abstracts
\cite{SoftwareTestSearch} contained both the words "software" and
"test," while fifteen\cite{SoftwareVVSearch} contained two of
following three words: "software," "verification," and
"validation."  One may presume that some of these articles do not really
discuss software testing.

To my knowledge, the most comprehensive attempt in the physics community to ensure a code's reliability is an ongoing collaboration to test an astrophysical modelling program called FLASH\cite{Calder04}.  This collaboration has adopted the methodology of Verification and Validation\cite{Oberkampf02, Trucano04, Foundations02, AIAA98}, which was developed mostly within the field of aeronautics, largely because of the immense pressure to produce reliable aircraft and space vehicles.  It is also practiced in other situations where there is a lot of motivation to get things right, like combat simulations, simulations at Sandia et cetera of nuclear bombs, and certain enviromental impact studies.\cite{StevensonPrivate, CalderPrivate, TrucanoPrivate, OberkampfPrivate, HopkinsPrivate, EasterlingPrivate}  Very roughly speaking, verification corresponds to checking a program's reliability, while validation corresponds to checking that its predictions match reality.  The FLASH collaboration is probably the first instance where members of the physics research community have adopted verification and validation techniques, and is a very exceptional case.  By and large the physics community worries much much less about the reliability of its computational results.

Some of the larger publicly available codes do make test suites
available to their users, although sometimes they may be hard to
understand. For instance, the popular LAPACK\cite{LAPACK}
mathematical library includes extensive performance and
functionality test suites, and its recommended installation
procedure includes running these tests and reviewing their
results.  (However, the tests themselves are not well documented,
so understanding test coverage or failures requires reading the
test source code.) BLAS\cite{BLAS}, the popular standard for fast
matrix operations, also includes a test suite. (But it may not be
well maintained - one well known bug concerning the ATLAS library
had not been fixed several years after the introduction of ATLAS.)
The MPICH library\cite{MPICH} for parallel programming also
includes both functionality and performance tests in its source
tree. Smaller codes seem much less likely to be accompanied by
test suites.  In any case, it is not known how often physicists
run those test suites which are available.

Software that is distributed in a compiled format usually does not
include a test suite.  For instance, MPICH's\cite{MPICH}
precompiled distribution for windows does not include its test
suite, and Mathematica does not distribute any tests at
all. 

Since software distributed in a compiled format is rarely
accompanied by test suites, it would be logical to do focused
testing of those functionalities used in one's research.  The
physics research literature rarely if ever reports any such
checking.

In summary, physicists generally do not test their
computing. There is one exception, which is the focus of
physicist's effort: global validation of the final research
results.  Research articles often report a handful of checks of their
principal results. Internal consistency may be validated by seeing
whether the result seems to have converged and whether relevant conservation
laws and sum rules are satisfied.  Results may be validated by
comparison with rough approximations, previous calculations, and
experimental data. However as a rule this checking remains at a
very global level; this type of check is often called "system
level testing" within the software development community.  Moreover it is often done on a qualitative, non quantitative, basis:  many physics publications ask the reader to evaluate the agreement of results by making an eyeball comparison of graphs, rather than supplying numerical measures of agreement.

Physicists may do further checking before reusing code or numerical
results from previous research within their field. Certain
important numerical results tend to obtain the most scrutiny: some
examples include the many papers reporting checks of
pseudopotentials, tight binding hamiltonians, and fits to the
nuclear potential.  However, usually redistributed codes are used
without any report of checking their validity.  The exception to
this rule is the research genre devoted to comparing several
algorithms, though even here the code may be written from scratch,
and comparisons tend to remain at the global level.

Another sort of global check is peer review, which is one of the physics
community's most successful and important tools for maintaining
its own professional standards.  It is also the only
reliability-oriented practice which is universally practiced, and
as such is invaluable. Here I make one point: peer review is
generally a global check on the paper's results, not a detailed
check of the computing used to obtain those results.  This
orientation is caused partly by the physics culture, but also
partly by the fact that reviewers - often confined to the
information contained in the text being reviewed - may not have
sufficient information to try reproducing the calculations.

\subsection{\label{JournalPractice}Journals}
In my survey of the physics community I have neglected to investigate the policies of physics journals.  I received a few remarks about this from reviewers of the first draft.  Hans Petter Langtangen\cite{LangtangenPrivate} wrote: "If you look at the scientific traditions in experimental physics, you see that reliability is very much in focus. Scientific computing has
much of the same nature as experimental physics, but in publications
the quality standard of experiments are very much higher than for
computational experiments. Scientific computing could simply adopt the standards of well-established experimental fields..... As a frequent referee I have systematically required high standards in publications based on numerical standards. I feel I usually have
a good support from editors. The problem is that few editors are
willing to require such standards in written form as a part of
the description of the journal..."

Stan Scott, the director of the Computer Physics Communications Program Library, the only code archive that I'm aware of that is directly associated with a peer reviewed physics journal, also took the initiative to contact me\cite{ScottPrivate, CPC}.  He wrote that "One of the aims of CPC is promote basic good practice in scientific programming."  He also shared with me the referee report form.  Here are some excerpted questions posed to the reviewer: 
"Is the research underpinning the computer program sound? Is the computer program of benefit to other physicists or physical chemists, or is it an exemplar of good programming practice, or does it illustrate new or novel programming techniques which are of importance to some branch of computational physics or physical chemistry? Is the computer program well engineered and does it meet accepted standards for scientific programming? Does the manuscript make clear the structure, functionality, installation, testing and operation of the program?  If you were supplied with the program did it install, run and perform as described in the documentation."  

Before submitting the final version of this present chapter for publications in a peer reviewed journal, I will do a more systematic review of physics journals.

\subsection{Summary of Current Physics Practices}
 It seems that
while many physics articles use software to compute various results, perhaps
few authors have implemented the most basic practices for ensuring its
quality - whether planned and repeatable test suites, source code control,
or publication of their code, scripts, and configuration files.  Moreover, there does not appear to be a structure
for reporting bugs, documenting them, or discussing their prevention.  

Clearly, groups creating and maintaining large codes are more
likely to address issues of reproducibility and reliability.  They
often version their release, use source code control, distribute
good documentation, and provide dedicated e-mails for software
support.  However, there are still notable deficiencies: a
frequent lack of public documentation of bugs and fixes, and a
lack of test suites for executables like Mathematica.

The situation is much different for smaller research efforts and
for the physics research literature. The chief tools for insuring
research quality are global qualitative checks of each paper's results: global
checks done by the paper's author, followed by peer review.
Neither of these practices is oriented toward detailed checking of
computational results.  In general, post-publication checks by
other physicists are greatly restricted by the normal practice
of not publishing the information necessary to reproduce one's
results. Regarding normal software development practices for assuring
reliability, they are usually not implemented: Bugs and testing
are not discussed in the literature.  Specifications, source code
control, test suites, and bug tracking tools are not used.

It seems reasonable to conclude that physicists regard their computing activities as peripheral to their research even when they publish the results of their computations, that physicists "don’t consider [reproducibility] to be an essential element in the quality of their work," and that they don't consider testing to be essential either\cite{TrucanoPrivate}.  Tim Trucano of Sandia writes\cite{TrucanoPrivate}: "subject matter expertise in physics is far more cherished than CS [computer science] expertise and it shows."  Neither my survey of the physics community nor my one and a half years of subsequent experience give much hint of any overall improvement in this picture.  

\section{Problems for Physicists}
I now reconsider the computing problems presented in section \ref{ComputerProblems} in the light of the physics community's practices documented in section \ref{CurrentPractices}.  We will see that that the two can combine to create difficulties for physicists and even significant delays in scientific progress.  

\subsection{\label{PhysicsReliability}Reproducibility}

The sciences are, in varying degrees, hard or soft, and physics is
traditionally the hardest of the experimental sciences.  Hard science
distinguishes itself by restricting its study to subsets of
reality obeying two requirements: Firstly, it must be independent
of the observer; i.e. the same independently of the observer's
identity, and even of the existence of the observer.  (Note that
the notion of independence is given more precision in quantum
mechanics, where the expectation value, not the wave function, is
expected to be independent of a probe.)  In other words, hard science studies only things that are objective in the sense I defined at the end of section \ref{SoftwareAsDiscourse}. 

Secondly, hard science
focuses on realities that are under so much control that they are
reproducible: at will, by anybody, as many times as desired, and
(if desired) with somewhat different parameters to see what
changes.  These two requirements of independence and
reproducibility are not separate: if an experimental or theoretical result
can not be reproduced at will by many different scientists, there
is no way to determine whether the object of study is independent
of the observer.  Thus reproducibility, whether of experiment or
calculation, is essential to hard science.

Hardness is not a binary value, but varies continuously as the
requirements of independence and reproducibility are relaxed.  For
instance in astronomy the stars are not in the observer's control
and cannot be made to repeat their behavior.   This causes certain
systematic problems of both theoretical and practical natures.
However, because the stars can be measured precisely by many
different observers, and because arguments considering the set of
all stars as a statistical ensemble work pretty well, astronomy is
considered a hard science.  On the other hand one can consider
sociology and psychology, where reproducibility is impossible to
obtain except in very restricted circumstances; these fields are
therefore are considered soft sciences.

 Nature is complex, humans are error prone and have limited resources, and therefore  physicists usually do not take the time to reproduce others' results. When one reads a theoretical paper one rarely re-derives every single line from scratch.  And most experimental results are not reproduced (with all parameters the same) by other teams.  Typically the only reasons why a physicist would actually repeat somebody else's results are to learn the technique, or else because the result is either important or suspicious.
 And typically the goal of such repetition is not to produce an exact replica of the original experiment in all its details, but instead instead to obtain the same physical result\cite{KarpPrivate, ParisiPrivate, GutowskiPrivate}.
 
However, there is a constant emphasis on trying to ensure that one's results can be reproduced by oneself and by others if desired.  Within theoretical physics, the principal means for allowing reproducibility are three disciplines: 
\begin{enumerate}
\item One goes through formal derivations before publishing results.
\item In publications one states one's starting assumptions, cites references or explains in detail when a lesser known mathematical technique is used, explains at least the major steps of one's derivations, and clearly explains the final results and their applicability.
\item Publications are submitted to peer review, and the referees are expected to do some degree of checking on the validity of the result.  
\end{enumerate}
Along side these practices is the ethical requirement that one cannot keep the details of a proof secret\cite{ParisiPrivate}. A physicist who claims to have derived a particular result but is unwilling to divulge its derivation is simply not professional.

The author is less familiar with the steps which experimental physicists take to achieve reproducibility, but suspects that they are more conscientious than theorists, keeping detailed daily records of their experimental design, actual configuration, and results.  It is unusual to find theoretical physicists who keep similar daily records of their work.

If the adoption of computers merely introduced one more difficulty in obtaining reproducibility similar to all the others inherent in the scientific process, then there would be little reason for concern.  This is not the case.  Introducing computers makes a qualitative change in scientific activity: Without computers one has a reasonable hope that results can be reproduced with only the aid of a laboratory notebook or possibly a scientific article.  Once computers have been introduced, there is no hope of obtaining reproducible results unless someone has used tremendous care to get the software right, and unless the original scientist has kept detailed records of the settings and configuration used to obtain her results.  And if she has not not kept detailed records of the final results, there is no way to tell whether one has succeeded in reproducing them.  

If research is not reproducible, it's not hard science. It's soft
science.  I am not asserting that any particular results in modern
physics are not reproducible, but am rather saying that there is a
substantial risk that many of them are not reproducible.  This risk is qualitatively different than the unavoidable, unchanging risk that some individual results may be in error.  Instead I am discussing the risk that many of the physicists who have adopted computers have not adopted the measures necessary to obtain some chance of reproducibility, and therefore are consistently obtaining and publishing unreproducible results. This risk may cause a qualitative change in the discipline as whole, away from hard science and toward soft science.  Such a change could be easily interpreted as a change for the worse.

I would like to distinguish two specific negative consequences of unreproducibility.
First, the physicist who has not ensured the reproducibility of her results simply does not know how she obtained them; and therefore is unable to check their validity.   Publication of such results is equivalent to publishing a theorem without being willing to divulge its derivation (or even being sure of the derivation oneself.)  It is simply unprofessional.  

A second consequence of unreproducibility is that there is no way to compare results or establish trends.  Computers are extremely sensitive to their inputs; if one is not certain of the precise details of how a calculation was done, one can't be sure of even the orders of magnitude of the result.   Researcher $A$ obtained one result, and researcher $B$ obtained another result, perhaps the same, perhaps different.  Were they computing the same quantity, and if not how should the two quantities be related?  Assuming that they are the same quantity (usually not true,) should one have expected the two results to be the same, or different?  If different, what differences would one have expected?  Were either of the two results obtained by a process without any significant mistakes or bugs?  If the two results differ, is the difference significant?  If the two results are the same, does that mean that they agree?  None of these questions can be answered if either result is unreproducible.  I am not saying that repeating these computations is necessary, but rather that knowing what you did is necessary to answer these questions.  Where irreproducible computations are concerned, one cannot begin to discuss whether the "physical results" are in agreement with somebody else's results, because one really doesn't know what one's results represent. 

It is also impossible to extract meaningful information by finding trends in collections of unreproducible results.  It is well known in the software industry that people will fix software until it meets their subconscious or conscious expectations, whether or not the final result is correct\cite{Myers79, KarpPrivate}.  This is not acting in bad faith; it consists simply of looking at unfamiliar results and saying "That's strange.  It can't be correct." And then once one finds a result that looks half way right, one concludes that one has fixed all the problems.  This process may be especially exaggerated in the physics community, which relies almost exclusively on global qualitative checks for validating its results.  Therefore one could easily imagine a dozen different researchers unwittingly reproducing roughly the same incorrect result.  And is no way to detect such an error without reproducibility.

Related is the difficulty that at times physicists are forced to make choices which they know will reduce the objectivity of the experiment.
For example, they may use software to deliberately throw certain "uninteresting" events away, and thus run the risk of new systematic errors.  In such cases there is the constant risk that the research results are an artifact of the physicist and her process.  Often scientists try to cope
with this issue by doing monte carlo simulations of the cut process.  Such simulations can be useful for estimating the effects of a correctly peformed cut on a fully understood experiment.  However they do not alleviate systematic risks that the cutting software did not behave as expected or that the uncut data is not fully understood.  In fact such risks are compounded by the possibility of errors in the monte carlo simulation.  These risks cannot be managed if one doesn't know how to reproduce and re-analyze the the cutting algorithm and monte carlo calculation which were used to arrive at the published results.

Some may discount these arguments for reproducibility because "the object is to produce new results"\cite{KarpPrivate},  or because having a variety of different results that can't
be compared is a sign of vibrancy and creativity.  
While the Nobel prize generally is not given for reproducing a
result, people have been awarded it for obtaining and expressing a
very accurate and deep understanding of experimental results or of
theory.  This reflects the fact that hard science is about
obtaining some sort of knowledge about the universe.  We do this
by checking and rechecking our results against reality, thinking
it all over in mind-numbing detail, participating in a community
which is both collaborative and mutually correcting, and thus
gradually adapting ourselves and our thinking to the reality we
meet in our experiments.  Hard science is precisely the extreme
where the scientist is most stringently required to conform and
adapt to reality, where we stand or fall by how well we understand or
model reality, not by how we change or influence it. It is
precisely the fact that we are trying to understand an external
reality that forces us to compare results, and to insist on
reproducibility. We could do without reproducibility, but then we
would not be doing hard science. Our results would be suitable
only to remaking nature according to our wishes, not to
understanding it as it is.

Others may deny the importance of reproducibility for personal reasons: fear of public criticism and negative effects on one's career, desire for the power and influence
that comes from possessing a code that is in demand, fear that others will rip off one's work without giving credit where credit is due, the hope of being able to barter or demand payment.
These concerns all have a certain validity, but they are not in tune with the scientific
vocation, or with the best interests of the scientific community.  Some of these concerns are not an issue if the scientific community has a proper respect for computing; it should respect the fact that even conscientious and professional researchers use methods that are not nearly picture perfect and produce results that not bug free.  It should also give to authors of reused files and code the same honor and citations that it gives to authors of seminal papers.  On the other hand, if one wishes to profit from research by retaining possession of its results, then one should consider one's research to be private enterprise.  Therefore one should not publish results obtained using the software which one wishes to retain control of, for such results are simply advertisments of this intellectual property under the guise scientific articles.   And, of course, one should obtain funding for one's private enterprise through channels other than those serving the physics community.  This last sentence may apply to one's salary if one's current employer is not friendly to the idea of devoting one's working hours to obtaining supplementary income.

A last point: ensuring reproducibility may help the author herself some years later\cite{ParisiPrivate}.  Alan Karp\cite{KarpPrivate} recounts: "My postdoc involved some calculations relevant to supernova models.  I carefully backed up my work to tape before leaving for my new job.  Eleven years later, a supernova went off in the Magellanic Clouds, and I was asked to redo the calculation for a model appropriate for this case.  I contacted my former employer to get a copy of the program, data, etc. and was told that their records retention policy deletes all information after 10 years.  I proceeded to rebuild the application from assorted printouts that I had kept, but they were from a number of different versions.  The program ran and produced plausible output, but I could not reproduce the results I had published.  I believe that the problem was inconsistent data sets, but I couldn't be sure, so I never published the results for the new supernova."

\subsection{Complexity}

The apotheosis of independence and reproducibility is physical
law: rules that are claimed to hold constantly, throughout time
and space, regardless of the existence of observers. These are the
focus of physics.  Physical laws, by definition, are simple: they
can be stated clearly, in a few equations, without exceptions.
Their consequences may not be simple at all, but often the study of these consequences
is considered to be not physics but applied science: chemistry,
engineering, etc.  Indeed, simplification is a hallmark of
traditional physics: choosing simplified problems to analyze
theoretically, choosing experiments that are conceptually very
simple, and using approximations to force simplification.  Everything is finite: experimental apparati can be described fully by engineering schematics, and equations can still be expressed on paper.  

The introduction of complex computations to the scientific process
can easily dilute or end this simplicity.   One moves quickly from the intellectual purity of scientific principles and physical laws into the realm of "model building."  
Even conceding the conventional wisdom that all scientific laws are simply models, one cannot deny that the qualitative experience of verifying/falsifying a simple physical
law like the blackbody spectrum is worlds apart from the experience of
verifying/falsifying a complicated computational simulation of
reality.  In one case you are making a single decision about the validity of a simple entity, and the result can often be reduced to one bit: either the law has been falsified or it has not.  In the other case you are making a huge number of comparisons and then using some rule to average their results.  The results (plural!) will not be phrased in terms of falsification but instead as differences of the averages, and probability distributions of the differences between averages.  
 
Complexity and computers do not always end in this scenario.  For instance, physicists were able to verify the existence of the top quark.  They expended a huge amount of effort to reduce immense data sets down to a bump on a graph, and made an estimate of the probability that the bump was really there.  Once the probability became large enough they began to say that the top quark exists.  They found the needle in the hay stack.

When the object is to study the hay stack rather than the needle, do you still have physics?  In other words, when you are not trying to verify or falsify a particular point but instead are modeling or measuring all the details of a complex system, are you practicing physics or some other discipline?  I am including collective behaviors like asymptotic freedom, collective resonances, and the fractional quantum hall effect as examples of needles because each of these individually can be verified or falsified. This question is not irrelevant: its answer defines the identity and mission of the physics community and the individual physicist.  Having a clear identity and mission is key to success.

Another more tangible problem is our fear of drowning in information.  This fear often leads us to throw information away instead of archiving it for future reference, examine and fit only certain "representative" quantities rather than attempting to study the whole data set by distilling it with appropriate mathematical techniques, make visual comparisons between graphs when rigorous mathematical measures of agreement are available, and abandon attempts to verify/falsify in favor of pictures, graphs, descriptions, and numbers.  In short, we do a qualitative, intuitional science.  Some of these practices may be acceptable on a short term basis.  This is a problem when we accept this state of affairs on a permanent basis and no longer aim for a science which is independent from the observer, mathematically rigorous, and conceptually simple. 

A  more constructive approach to complexity is to systematize and manage it.  Powerful software for managing the information glut is available both freely and commercially, most notably source code control software and relational databases.   Other software packages are specially designed to simplify and automate complex mathematical manipulations. Costs of media for archival purposes are constantly descending.  A wide variety of both mathematical formalisms and also algorithms for analyzing complex data sets is under continuous development.  If a physicist feels drawn to use shortcuts like those listed in the previous paragraph even on a temporary basis, probably the better choice would be to fight the impulse and either adopt or invent rigorous solutions like the ones I just listed.
    
\subsection{The Scientific Method and The Scientific Community}
The scientific method, as taught in textbooks, is a prescription
for developing knowledge by repeated confrontations of theory with
experiment.  The idea is that close observation of experimental
reality and painstaking thought about it can, over time, provide a
framework for discernment about which ideas are more or less
faithful to reality.  Without experiencing the shock of clear
disagreement with experiment, scientists might never make the hard
choices needed to obtain theories which are faithful to reality.

In order to obtain a clear confrontation between theory and
experiment, the following conditions must hold:
\begin{itemize}

\item the experimental result be clear and unambiguous,

\item given enough expertise, intelligence, and time, a unique
prediction can be derived from theory, and

\item a confrontation may be made between the experimental result
and the theoretical prediction, obtaining a clear validation or
refutation of the theory.

\end{itemize}
In reality these conditions do not always hold.  A large portion
of the scientific practice is devoted to design and refinement of
both experimental and theoretical techniques which enable a clear
confrontation.

The current era's scientific and technological successes are often
explained with the scientific method.  This account ignores the
crucial role of the scientific community, which is likely more
deserving of credit for progress than the scientific method. A
full understanding of the reasons for scientific progress must
rely on not just philosophical arguments but also historical and
sociological studies of the scientific community\cite{Kuhn70}.

The scientific method (clear confrontation of experiment and
theory) is at risk whenever either theoretical predictions or experimental
results are unreliable, ambiguous, or very complex.  I have already discussed the complexity problem.  It should also be clear that unreproducible calculations are completely ambiguous.  Here I concentrate on the problem of computational unreliability, and consider only the problems it can cause for further scientific research.  I classify these problems into two categories: Small Scale errors and Grand errors.  

Small scale errors require only man-months to man-years to correct. They are confined to research results that are not obtained at great expense and are not widely cited before being discovered and fixed.  If one could not avoid most of them by using basic software disciplines, then they might well be construed as part of the cost of doing physics.

Grand errors occur when an erroneous result is widely adopted, and can require years to decades to correct and have inestimable negative consequences for the scientific process.  I here confine the discussion to grand errors associated with unreliable computations.  These can be perpetuated through a kind of chain reaction. Once an erroneous result is known, people will have incorrect expectations for further results.  As I explained in section \ref{PhysicsReliability}, these incorrect expectations can be expected to subtly determine the results of new calculations.  As long as the physics community relies solely on global tests to check its research results, it has little defense against these chain reactions.  In fact, it is impossible to ascertain whether any of these computation-related grand errors are currently in progress, simply because the physics community does not conduct a systematic search for them using effective testing techniques like those alluded to in section \ref{Bugs}.  Software bugs will not be found unless they are searched for.  It is also very hard to tell whether small scale errors are happening, because there is neither an effective framework for finding them nor a system for reporting them.  When an error is detected an erratum may be published, but errata very rarely document how the error occurred. 
 
If grand errors could not be avoided by adopting a discipline of thoroughly checking every computational result, then perhaps the cycle of their entrenchment and eventual correction could be interpreted as simply a realization of Kuhn's ideas about normal physics vs. paradigm shifts\cite{KarpPrivate, MariansPrivate, Kuhn70}. This hypothetical possibility is purely academic: grand errors can be avoided by taking steps to ensure reproducibility and reliability and to manage complexity; therefore any grand errors that occur would signal collective failures of scientific discipline.  Nonetheless, I want to point out a crucial mistake in such interpretations.  Results obtained using software that has not been tested thoroughly have little connection with either theory or experiment, and permit no confrontation between the two.  Therefore neither the scientific method nor the scientific process as described by Kuhn provide any insurance against grand errors.  Unless, of course, one is correctly managing the computational risks.

Physicists may be tempted to discount the possibility of either small scale or grand errors, and claim the right to be a bit casual about data and computations.   They may justify this attitude of presumed invulnerability with claims that they can tell incorrect data/computations when they see them\cite{MariansPrivate}. Such an ability would have to be intuitive, because it is used to justify avoiding a quantitative or systematic approach to error detection.  Physicists may also claim that the peer review system can be relied on to find errors.  Because the peer review process relies on a subset of the global checks practiced by individual physicists, relying on it alone to detect errors just amounts to a re-assertion that physicists can tell incorrect results when they see them.  Or physicists may claim an elite status and argue that this makes them less vulnerable to error, which again reduces to the same argument.   

A reliance on intuition, at the loss of discipline, is not
fitting in any professional, no matter how experienced.  Moreover, a correct professional intuition does not come from a vacuum, but is built on a
foundation of a lot of training, experience, and discipline.
One must question whether such training and experience are present when the typical physicist
has little or no knowledge and experience of the basics of managing computing risks.
Moreover, even experienced software development professionals are
constantly over-optimistic about software; this is precisely why they have adopted systematic and exacting practices. Human intuition ALWAYS fails when evaluating how unreliable
software is; it is always overoptimistic, and the only way to get
a true knowledge of your software is to methodically test it.
 
I repeat my overall point: the traditional practices for ensuring the success of the physics community, including but not limited to the peer review process, can be expected to fail both on the small scale and on the grand scale if proper steps are not taken to manage computational risk. 

It is possible that by good luck some grand errors have already been detected, but are not widely known. Leo Kadanoff\cite{Kadanoff03} writes about a possible candidate which started in 1992 and lasted about five years: "The whole early history of single bubble luminescence required step by step work to eliminate provocative but incorrect mechanisms.  A set of early experiments by Barber et. al reported very short widths for the emitted pulse of light.  This short width then opened the door to novel mechanisms for explaining the total intensity of the emitted light.  Later developments suggested that the short pulse width was a misstep by the experimentalists.  In contrast to the excellent work on sonoluminescence in the post-1997 period, reported above, this misstep led the simulators and theorists quite astray.  A host of incorrect speculations and mechanisms ran through the field, intended to explain the 'observed' behavior.  Despite one essentially correct simulation, the pre-1997 simulations did almost nothing to weed out these incorrect discussions, undercutting ones[sic] hope that simulations might provide a good tool for such weeding.... Instead the speculations continued unhindered until an experiment by Gomph and coworkers showed that the pulse width was much longer than previously believed.  This implied a lower temperature for the emitting drop.  After this, attention turned away from the incorrect mechanisms so that-- as reported above-- theory, experiment, and simulation began to produce a consensus about what was going on.  The examples of the Oak Ridge paper and some of the earlier sonoluminesce simulations suggest that the models might have been directed toward the wrong goals.  Apparently, rather than being used for a process of checking, criticism, and elimination of incorrect possibilities they were often used to support and exemplify the presumptions of the scientists involved.  A program of modelling should either elucidate new processes or identify wrong directions.  Otherwise, there is no point in carrying it out."

I have not studied the papers which Kadanoff cites, and can only comment on the incident as described in his report.  He does not say whether the original experimental misstep could have been caused by a failure in some computerized experimental control or data analysis.  If so, the whole five year incident would be due to a failure to manage computing risks properly.  Even if the original experimental error had some other cause, one is still left to conclude that proper computing discipline could well have ended the incident much more quickly.

In any case this incident seems relatively small.  It remained confined to one narrow discipline, and lasted only five years.  A citation search on the ISI Web of Science\cite{WebOfScience} indicates that only fifty-six articles cited the original erroneous result before Gomph et al published their correction.  One could easily imagine worse; for example, a computationally caused experimental error in a cosmological experiment like COBE\cite{Hatton94} could profoundly affect our understanding of the universe and the development of particle theory, or a string of erroneous theoretical computations could lead the high $T_{c}$ research community to either erroneously consider an incorrect mechanism for superconductivity or abandon a correct mechanism. 

\section{\label{Consequences}Consequences}
So far I have discussed risks of a rather esoteric and intellectual nature: the risk that a published result may be unreproducible, the risk that a published result may be wrong, the risk that someone's reputation might suffer, the risk that physics may not progress.  Another risk - of ethics and professionalism - may also seem debatable or inconsequential to some, although I would contend that it is the most serious of all.  

There is another another set of risks which few would care to debate, and which I here call consequences to emphasize that they are cannot be considered esoteric or inconsequential in any sense\footnote{I am indebted to Tim Trucano for passing on to me almost all the insights in this section.}.  Here are some of the possible consequences for the physics community as a whole of not taking steps to manage computing risks:
\begin{enumerate}
\item Wasting large sums of other people's money.
\item Difficulty in obtaining increases in funding, or even maintaining present levels of funding.  (Once bit twice shy.)
\item Leading society as a whole or smaller groups in wrong directions.
\item Receiving less respect from others than we might, or even losing others' respect.
\item Losing credibility.
\item Not being able to influence (national) policy makers or society.  (Why should anyone pay attention to advice that comes from unreliable computations?)
\item Failing to speak with authority to issues of grave importance to real people.  (Authority would of course include knowing exactly what the possible errors are and keeping them to an absolute minimum.)  Topping society's priority list (not necessarily ours) are things that can make a life and death difference: global warming and environmental concerns, cryptography, the reliability of potentially deadly machinery like planes and nuclear power plants, advances in technologies that can prevent or cure illnesses, and military technology.  Also at the top of the list are economic concerns, for instance advances in solid state (especially semiconductors), high temperature superconductors, unreliability of power grids, et cetera.  
\end{enumerate}
 We must remember that our honored place in society wasn't given to us as a God-given right: our predecessors earned it by their contributions to World War II and kept it through their continueing contributions to industry, most notably in solid state. We will not retain our status or support if we do not continue to make new contributions of similar importance.  If we can't contribute, we are "unimportant on the overall scale of things (compared to war, pestilence, economics, etc.)"\cite{TrucanoPrivate}

 These are the real consequences of poor choices about computing.  It is interesting that in the software industry the care and expense devoted to getting software right is directly related to the possible severity of retribution from clients.  Probably the most care is given to flight control systems.  Farther down the list comes vendors of enterprise software, where data and services are worth billions of dollars and individual software contracts involve millions of dollars.  Last on the list come the developers of software marketed to the individual user.   
 They issue a notice with each copy of the mass produced software, with text very similar to this text, taken from the GNU copyleft (the caps are not mine): 
 
 "EXCEPT WHEN OTHERWISE STATED IN WRITING THE COPYRIGHT HOLDERS AND/OR OTHER PARTIES PROVIDE THE PROGRAM "AS IS" WITHOUT WARRANTY OF ANY KIND, EITHER EXPRESSED OR IMPLIED, INCLUDING, BUT NOT LIMITED TO, THE IMPLIED WARRANTIES OF MERCHANTABILITY AND FITNESS FOR A PARTICULAR PURPOSE. THE ENTIRE RISK AS TO THE QUALITY AND PERFORMANCE OF THE PROGRAM IS WITH YOU. SHOULD THE PROGRAM PROVE DEFECTIVE, YOU ASSUME THE COST OF ALL NECESSARY SERVICING, REPAIR OR CORRECTION.

12. IN NO EVENT UNLESS REQUIRED BY APPLICABLE LAW OR AGREED TO IN WRITING WILL ANY COPYRIGHT HOLDER, OR ANY OTHER PARTY WHO MAY MODIFY AND/OR REDISTRIBUTE THE PROGRAM AS PERMITTED ABOVE, BE LIABLE TO YOU FOR DAMAGES, INCLUDING ANY GENERAL, SPECIAL, INCIDENTAL OR CONSEQUENTIAL DAMAGES ARISING OUT OF THE USE OR INABILITY TO USE THE PROGRAM (INCLUDING BUT NOT LIMITED TO LOSS OF DATA OR DATA BEING RENDERED INACCURATE OR LOSSES SUSTAINED BY YOU OR THIRD PARTIES OR A FAILURE OF THE PROGRAM TO OPERATE WITH ANY OTHER PROGRAMS), EVEN IF SUCH HOLDER OR OTHER PARTY HAS BEEN ADVISED OF THE POSSIBILITY OF SUCH DAMAGES."\cite{GNUGPL} 
      
Who are the physicists' clients?  Are they other physicists?  Then probably physicists have even less to fear from their clients than do developers of software marketed to the individual user. 
Do physicists want this sort of disclaimer to apply to their research results?  It will be fascinating to see how the physics community chooses to answer this question.

\section{Recommendations and Resources}

\subsection{\label{Professionalism} Professionalism}

I believe that a proper response to the risks of computation starts with recognizing their severity and gravity, choosing to regard computation as an integral part of one's research rather than as an inconvenient add-on, and developing a deeper professionalism and scientific discipline. Each of these steps can be summed up in one word: professionalism.  Professionalism implies the following steps:
\begin{itemize}
\item Choose to recognize and pay very close attention to the full extent of the risks inherent in computing.  This includes cultivating a professional scepticism about the results of computations, whether done by oneself or by others.  Such scepticism should be satisfied only when an author gives cogent and thorough arguments for why her result should be trusted.
\item Consider one's computing activities, whether writing software or using it, to be just another part of one's physics reasoning and research, on par with deriving equations or doing experiments.  If one would prefer to not give the same interest and attention to one's computations that one gives to one's equations or experiments, then the appropriate choice would be to minimize the computing in one's research and to absolutely avoid publishing any results that depend on some computing step for their validity.  For example, this would include avoiding publishing theoretical formulas derived using symbol manipulation software like Mathematica and Maple unless one is willing to check the results thoroughly and to keep exact records of the commands and input that were used to obtain them. The choice to minimize or avoid computing can be a valid choice; it is simply a choice of specialization.  
\item The rest of this list will presume that the reader is not willing to forgo computation.  At the risk of stating the obvious, I remind such readers that publishing a result implies that one is confident that the result is not in error, and also implies that one has obtained this confidence only through a detailed and thorough checking process.  In other words, an author must take full responsibility for her published results. Unfortunately it can be very tempting to make exceptions for computational results; to blame others for having written buggy software, to claim that software risks do not need to be evaluated or minimized because they are implicitly acknowledged whenever a computational result is published, et cetera.   Therefore I encourage the reader to remind herself of her responsibility for the correctness and reproducibility all of her published results, including ones obtained with the help of computational results or software supplied by another party.  This implies choosing methods which have the minimum vulnerability to computational risks and recording all the information required for reproduciblity.  It also implies following the testing guidelines briefly sketched in section \ref{Bugs} and doing detailed cross checking of the results that one publishes.
\item In your papers, give a balanced, professional evaluation of
your results. You should know better than anyone else the things
that could invalidate your results, as well as the uncertainties;
document these in your papers.  A pro and con approach may be helpful.
\item Adopt Buchheit and Donoho's statement\cite{Buchheit95} that
"An article about computational science in a scientific
publication is not the scholarship itself, it is merely
advertising of the scholarship. The actual scholarship is the
complete software development environment and the complete set of
instructions which generated the figures."  Make all
your files freely available to those who want to reproduce your results,
or to build on them.  This can be done either on a per request basis, or on a
charge-free public server like the Computer Physics Communications Program Library\cite{CPC}, the SourceForge web site\cite{SourceForge}, or Mathematica's archive\cite{MathSource}.  One should attempt to make one's scripts, code, and configuration files be well commented and readable.  One should also ensure reproducibility on one's own computer and give some degree of assistance to would-be users. However in many cases one should not feel obligated to write much documentation, provide all the support necessary to make one's configuration files, scripts, or software work on another's computer, or explain how to do calculations.  If there is a real concern on the part of the scientific community about the reproducibility of one's results, that is a different matter and full support for the community's efforts should be supplied.
\item Manage your risks.  Do not feel that you have to achieve 100\% mathematical proof certainty about the validity of a computational result; this is impossible.  Do analyze the risks and build a convincing case for your result's validity.  Risk management is a well developed discipline\cite{Leveson95}; but perhaps it can be boiled down to analyzing both the risks and the costs and benefits of mitigating them, planning and execution of appropriate steps to lessen them, doing the cross-checking necessary to figure out when a risk has become a reality, and fixing failures in a timely fashion.  Obviously the extent of one's efforts to avoid errors will depend on the situation\cite{MariansPrivate, HopkinsPrivate}.  Nonetheless it would be very unprofessional to use situational arguments to justify either avoiding risk analysis and management (perhaps on the grounds that they are too costly) or publishing a result that one is not sure of.
\item  Software always contains bugs; one can expect that even the most conscientious researcher will publish results that contain mistakes and inaccuracies.  If a physicist has worked conscientiously for reproducibility and reliability, then her errors should not be taken to reflect on her quality as a scientist.  Instead of hiding or ignoring one's published mistakes, one should try to find as many as one can and inform as wide an audience as possible about both the mistakes and their corrections. The search for errors will necessarily include supplying contact information with the original article and following up in a timely fashion on any reports of specific errors.  Informing the public will involve both documenting the error on the Internet as soon as it is confirmed and publishing an erratum in the original journal. 
\end{itemize}

\subsection{\label{ATechnicalToolbox}A Technical Toolbox}

\begin{itemize}
\item The way to master complexity is not to throw information away but instead to use software designed for archiving and managing data.  This very powerful strategy allows you to find out the things you need to know when you realize you need to know them.  In particular, source code control software is ideal for keeping track of files that can be read by humans and are subject to revision.  It may also be useful to learn to use data bases, particularly if one's calculations involve many steps or process or generate a lot of data.  One can obtain a reliable and free data base or source code control program on the Internet.
\item Sometimes mathematics programs like Maple, Mathematica, FORM, PARI-GP, Matlab, and Scilab can can tame complexity\cite{GonnetPrivate}.  However, like any other software, the validity of their results should not be assumed, especially in the cases where test suites and bug databases are not freely available to the public. 
\item Wherever possible, simplify.  Given a hard
problem, abstract it into simpler, toy problems, and understand
them thoroughly before attacking the original.
Simple programs are usually better. 
\item Entertain doubts about the practicality of the reductionist
program where everything is a sum of its parts, and if only I
could calculate the problem at infinite resolution I would get
the right answer.  The last fourty years of physics gives a very clear indication that this is often not true.  Moreover, as we have seen the computational approach is
problematic.  Spend extra time looking for other ways of modelling.
\item A computer program should, if at all possible, just
refine a result you can estimate independently.  Or at least it should give known results in some limit.  It is very hard to justify trusting a calculation that one has no way of checking.  
\item Whenever possible, reuse scientific codes written by
experts; many of these are freely available on the Internet. But test them too. Don't believe in magic bullets; to single out one of the most recent ones, open source code that is not thoroughly tested and is not widely and thoroughly used is not likely to be more reliable any other untested software.  
\item If at all possible, use scripts and configuration files to automate all the steps necessary to obtain your computational results.  Most software does allow scripting, whether with batch files, with macros, or with similar technologies. The reason for automating a calculation is that it simplifies reproducibility down to two tasks: (1) keeping archives of all the necessary files, preferably using source code control, and (2) recording the software configuration, including the version of the operating system and of any other software packages that were used.  If unable to record all the necessary information in scripts and files, conscientiously record every detail in a lab notebook, keeping in mind that even the most conscientious human has immense difficulty achieving reproducibility without technical aids like scripts and source code control.  Even when archiving everything, a lab notebook can be very useful anyway. 
\item Upon publishing a result, make a frozen copy of all the files that were used to do the calculation, and leave the copy unchanged as it was upon publication.
\item Keep regular backups, including the the source code control database.
\item Consider putting your files (software, settings, scripts,
tests, and documentation) in an public open source repository.  Consider licensing them with the GNU Public License\cite{GNUGPL}, which will allow them to be freely reused by the public. 
\item Consider practicing the discipline of producing "fully reproducible research." This consists of automating the entire process of creating your
papers (computations, graphing, and type setting) and then making the whole package available to the public. \cite{Buchheit95, Schwab00, Claerbout03, Claerbout94, Claerbout92, deLeeuw01, Sacksteder03}.

\item When checking software or its results, try to do it an automated way, by writing test scripts or test programs.  Then you can test each version of your input files and the software in an automated and reproducible way.  Run tests repeatedly as you make changes (even minor ones) to the calculation.  If unable to automate the tests, create detailed records of each test and its outcome each time that it is performed.
\item Inform yourself about good testing practices\cite{Myers79, Gonnet03}.
\item Inform yourself about the basics of numerical analysis\cite{Golub83, PrycePrivate, NumericalRecipes}.
\item Inform yourself about the field of verification and validation\cite{Oberkampf02, Trucano04, Foundations02, AIAA98}.
\item Do not assume that errors are uncorrelated or governed by a gaussian distribution. 
Rob Easterling\cite{EasterlingPrivate} writes: "Never assume that probability distributions are exactly known.  Include an analysis of the effect of uncertainty in assumed distributions.  Know the difference between random variables and imprecisely estimated unknown constants and perform analyses that reflect this difference."
\item  When one uses a computer to solve a mathematical problem, often the challenge of estimating the accuracy of the computer's solution via an analysis of the solution technique turns out to be even more challenging than solving the original problem.  There is a growing field called uncertainty quantification which is devoted to estimating software accuracy without looking at all its internals\cite{Easterling02, SIAMCSE03, Glimm03, Glimm02, DeVolder01, Glimm01, Glimm99}.  
\end{itemize}

\subsection{For Collaborations and Recipients of Research Grants}
When more than one person is involved in a computing project, or extra funds are awarded to it, its intrinsic challenges are more difficult, and the obligation to not risk delays or failure and waste time and other people's money is even more binding.  Therefore the researchers should go beyond the professional standards outlined in section \ref{Professionalism} and adopt the standard reliability and reproducibility disciplines which are standard in software engineering.  These include many of the technical recommendations in section \ref{ATechnicalToolbox}, but also includes going through a whole software development cycle: 
\begin{enumerate}
\item Finding out the requirements of the prospective users.
\item A design stage which produces a detailed specification of the software's architecture and implementation.
\item Writing software, test suites, and documentation concurrently.  Design changes should be reflected in the specifications. 
\item Making the code available to customers before the final release.
\item A period of testing, client input, and stabilization.
\item Ongoing bug fixing after the release.
\end{enumerate}
Obviously how this process is implemented will change a lot between a two-person low budget project and one involving dozens of people and millions of dollars.  However the rudiments of this process should be there in all cases.

It may be appropriate to adopt an interdisciplinary approach and employ software experts to contribute their expertise to the project.  Remember that in scientific computing both physics expertise and computing expertise are needed; preventing physicsts from coding would be a mistake if they are willing to practice the necessary discipline.   Even in large projects it could be sufficient to let physicists do the whole software cycle, as long they are advised by and held accountable to software professionals who review their work in detail.

\subsection{For Universities and Research Institutions}
\begin{itemize}
\item Consider computing an integral part of physics research. Require that researchers practice the professionalism described in section \ref{Professionalism}.  Reward them for doing so. Strongly encourage reserachers to make their files freely available on the internet; require that at the very least they be made freely available on a per request basis.

\item Educate researchers about the difficulties of computing, professional discipline, and best computing practices. Foster and reward expertise in software testing, managing complexity, numerical analysis, verification and validation, software design, coding skills, scripting, triage, and risk management.  Task particular staff with supporting and educating the researchers who are involved with computing. When possible, create scientific computing interest groups.

\item Provide a source code control server that gives researchers the option of making their projects visible on the Internet or instead keeping them private.  Provide a public bug tracking web server where researchers can promptly document errors and corrections to their published results.   Back up these servers, and provide additional backup services that are easy to use, able to store all of the researcher's files, and can be automated to run regularly.  Allocate staff part time or full time to maintaining and troubleshooting these servers, software, and procedures.

\item Pressure vendors of commercial scientific software to distribute comprehensive test suites with their software, and to publicly  document  both their test suites and each bug report, confirmed or not.  Vendors should also create a web site where users can communicate to each other the bugs they run into and the fixes they find without first verifying these bugs or fixes with the vendor.
\item Ensure that physicists are honored and rewarded when their computing contributions are adopted and reused by others.  Determine whether such reuse is occuring and evaluate the quality of the contribution by soliciting information from users, not by asking the author.  Discourage the practice of trying to sell or barter one's software, or of holding on to it in hopes that someone will eventually want to pay for it.  The physicist should be rewarded by her employer and by the physics community for the software, and should not have to start looking for clients.
\item Reward researchers who quickly check reports of errors in their computational results and quickly post information about confirmed errors on the Internet.
 \end{itemize}

 \subsection{Journals}
\begin{itemize}
\item In so far as possible, require the professionalism described in section \ref{Professionalism}.  Require that peer review no longer rely solely on global qualitative considerations.  Require that each computing result be accompanied within the same paper by a clear and cogent discussion of pros and cons about that result's reliability and reproducibility. Require that the author provide the referees with all the files necessary to reproduce the results, and require that the referees check that the author has in fact taken sufficient steps to be reasonably confident that the result is reliable and could be reproduced if desired.  Encourage referees to comment on every aspect of the quality of the computing.   Encourage the referees to (when practical within a limited time) learn about the computations by running parts of them and experimenting with parameters.
\item Include computational standards in the written description of the journal and its mission\cite{LangtangenPrivate}.  The most stringent example so far may be the ASME Journal of Fluids Engineering\cite{TrucanoPrivate}.  Its Policy Statement on the Control of Numerical Accuracy\cite{ASMEPolicy} states: "The Journal of Fluids Engineering will not consider any paper reporting the numerical solution of a fluids engineering problem that fails to address the task of systematic truncation error testing and accuracy estimation.  Authors should address the following criteria for assessing numerical uncertainty." Ten specific points follow.  Also see the Computer Physics Computations journal, which was discussed in section \ref{JournalPractice}.
 \item Consider requiring the same quality standards from computational results that is required from experiments\cite{LangtangenPrivate}.
\item Require that authors share the files which they used to obtain their results freely on at least a per request basis, and quickly document any errors on the Internet.  If one receives a reliable report that an author has failed to do so, blacklist her.  
\item Restructure the publication process to allow immediate postings of errata, and multiple errata pertaining to the same article.
\item Create a web site for files and code which serves as twin to the journal.  Allow authors to twin their articles with the files and code necessary to reproduce their results.  These files must not be put under the publisher's copyright, but instead made freely available for reuse under a copyleft like the GNU Public License\cite{GNUGPL}.  After initial peer review and publication, allow the author to post fixes to the files, but always retain the original version and make it clear that only the original version was reviewed.  There is some potential for abuse here, but it is likely small if the author wishes to remain in good standing with the scientific community. 
\end{itemize}

\subsection{For Funding Agencies and Professional Organizations}
\begin{itemize}
\item In forming nationwide research agendas, these agencies should
clearly acknowledge the risk that major individual research results could be totally invalidated by computational errors, and also that strings of errors in less important results could also lead to substantial errors on the part of the physics community.  These risks should be analyzed in detail, and detailed plans should be formed to actively control and minimize them.  
\item In section \ref{SoftwareAsDiscourse} I argued that software is better understood as a form of discourse like text or equations than as a machine or mechanism.  If so, software should become part of the subject matter of our scholarly discourse; in our publications we should as ready to discuss the details of our software as we are to discuss details of a theoretical technique or an experiment.
\item A researcher who acts quickly to confirm reports of her published errors and keeps comprehensive information about them on the Internet should be respected and rewarded.  A researcher who is not willing to publish written justifications for her computational results and share the files used to obtain them should be penalized.
\item Require that funding proposals which include plans to use computers analyze the specific problems that could occur, describe the process and resources which will be used for designing, implementing, and maintaining the computation, and give a detailed plan for how risks will be minimized.  Use these as a baseline requirement; i.e. projects will not be considered to have scientific merit unless they give convincing evidence that they have thought through the computation and will take the steps which are necessary for reliability and reproducibility.  In particular, no high profile research project should be funded unless it clearly has adopted a very stringent software development cycle and very stringent verification and validation procedures, comparable to the most exacting software development projects, and has included the expert staff needed to follow through.  Project success and failure should also include evaluations of how the computing turned out.
\item Funding agencies should hire software experts to participate in evaluations of computational aspects of funding proposals and project results.  These same experts should contribute to any formal guidelines within the agencies.
\end{itemize}

\section{Personal Experience}
I wrote the first draft of this article a year and a half ago, and then decided to implement my own recommendations as a test of their validity. Therefore I began keeping all my hand written calculations and observations in notebooks, with each page signed and dated.  This has turned out to be very useful because now I have good records of what I did, and when I want to publish a calculation I just need to go to my notebooks.  I put both the source code for the software I was writing for my research and the texts of the articles I was writing in source code control, and ended my day by recording the day's changes in source code control.  And I did backups of all my files every once in a while, perhaps averaging once a month, perhaps less.  All of these steps were easy to implement; they only required a little conscientiousness.

Other steps were a bit more difficult.  Most of my research over the last year and a half has been devoted to developing software to check whether certain algorithms could be applied to disordered systems, and therefore I had to manage the challenges which face scientists who write software.  I decided that I would not publish a computed result unless there were automated test suites checking each step in producing that result, excepting the steps of printing out and graphing the final numbers.  I also decided that my software should automatically run consistency checks on the results it produces.  Roughly half of my code ended up being testing code; I can estimate that also half of my programming time was spent writing tests.  I consider this cost well worth it because non-automated tests of the same quality are likely impossible and would anyway have consumed far more time because I would have had to repeat them each time I changed my code or configuration. My final published results were the result of months of computer time, and during those months I found it necessary to make numerous changes to the code.  If I had not been able to test my code thoroughly after each change, I would have no reason to be confident in the validity of my final results.  

There was also the challenge of keeping track of the configuration parameters and input data of each individual step in the calculations.  I required that every step in the calculations be controlled by a unique file (not human input, and not a shared file) which contained a complete listing of the configuration parameters and the names of the input files.  I adopted a standard format for these control files, which was a good thing because in the end there were almost twenty thousand of them, associated with the various steps in the several months of calculations.  Perhaps a physicist who is not involved in such intensive computing might not end up managing twenty thousand files, but probably she will redo even a simple calculation many times with different parameters to see how things change, or do many individual steps (for instance, when using Mathematica to manipulate formulas.)  Keeping track of such calculations is about as challenging as what I was doing: the first time is the hardest time.  

There are some decisions about configuration and input management that I would make differently now.  I would store my configuration parameters in a human readable format like the XML (Extensible Markup Language) industry standard instead of a binary format.  I made a poor decision to store configuration data in the same file with the results produced by those configurations, which not only caused a persistent confusion of semantics but also made certain tasks (like throwing away the results and recomputing, or recording a copy of the configuration files) much more difficult than necessary.  And eventually I implemented much of the functionality of a database; i.e. a standard type of software whose sole design goal is to assist with the management of large quantities of data.   Even though many databases might have had some difficulty accomodating my need to develop the software and its configurations incrementally, the most popular databases are pretty well tested and certainly provide better solutions to many of my problems than I was able to implement myself.  Therefore I would consider the possibility of using a database, particularly if I anticipated managing hundreds of files or more.

My investment in reproducibility did serve me well:
\begin{itemize}
\item After an initial study of the matrix step function (density matrix), I decided to study other matrix functions as well.  Because I had full records of how I produced the step function results, I was able to quickly configure my software to study other matrix functions, and then to make detailed comparisons of the matrix functions, knowing that I was comparing apples to apples.
\item On a few occasions, I found reason to doubt the validity of certain results.  For instance, in a particular data set of 33 matrices, twelve or so showed much different results from the rest.  I reran the calculations with the same configuration file as before, and the discrepancy disappeared. This sort of unreproducible, untraceable failure is well known in computing, and has been termed a "Heisenbug."  The only solution is to use all of the following: running automatic consistency checks on computational results, checking the results manually, keeping full records of how the results were produced, and, when a calculation turns out suspiciously, rerunning, examining, and retesting its each individual step.
\item I have been able distribute on the Internet a full copy of all files necessary to reproduce the graphs, numerical results, and typeset copy of my research articles.   This may help others understand what I have done, or adapt my work to obtain new results, or even check the validity of my results.  I too may profit, whether by being able to take up a research project where I left it off some years previously, or by an increased standing with any researchers who use my code.
\end{itemize}

I began development of this code in a Linux emulator named Cygwin which runs on top of Windows. My first publication was accompanied by files which worked in that environment, with the hope that they would work in any other  Linux variant after only minor changes to the compilation process.  Later I converted to Linux and found that simply changing the compilation process was insufficient: subtle differences between the two environments required that certain scripts be altered and file names be changed. This confirms the universal experience that it is impossible to ensure that software will run correctly on a new platform without actually installing and testing it\cite{MariansPrivate}.  I conclude that it is not practical for a physicist to distribute her configuration files and code with a guarantee that they will work for colleagues with only the most minor changes.  I believe that it is still important to distribute these files, but for other reasons: because they are an integral part of one's research publications without which the publications lose their status as scientific works, and because they can be a considerable aid to other physicists who are interested either in reproducing one's own results or else in producing new results of their own.

I originally planned to write detailed specifications of the scientific software and test suites, and to keep these specifications updated to match design changes during implementation.  I also planned to write all of the code very clearly and to add lots of comments. While I believe that most of the current code is concise and clear, there are exceptions, particularly the code which does the final data output and the test suites.  Moreover there are few comments, especially considering that in professional codes every function should be preceded by an explanation of its purpose, inputs, and outputs, and should also contain comments around each logical step of its execution.  I also abandoned the idea of writing specifications at a very early stage, when it became clear that I did not have a precise understanding of how to achieve my design requirements and would have to proceed by incrementally rearchitecting portions of the code.  This is actually a common problem in all software development projects, and in general does not excuse skipping the specification process.  However I felt that in this one-person development project I could rely on detailed test suites and on rereading my own code to be partial substitutes for specifications and comments.  For the most part, this approach seemed to work, but it did exacerbate some subtle semantic problems. I conclude from this experience that single-person software development does not necessarily need to be accompanied by specifications and thorough commenting.  However I believe that once two or three persons are collaborating, these practices become much more important and can not be avoided.

I have discussed several disciplines of varying difficulties which I adopted in order to achieve reproducibility and reliablity.  Yet there was another discipline that was far more difficult and painful than the others even though it saved both time and resources.  This was the triage process.  I had to continually prioritize the things I wanted to achieve, choose to implement new features only if I had already sufficiently tested the existing features and expected to be able to test the proposed new features as well, and work on tasks in a prioritized order instead of according to my whims.  When things went slower than I would have wished (always,) and I did not manage to thoroughly test some features and therefore did not have a good reason to trust the corresponding results, I had to be willing to omit the unchecked results from my publications.  Indeed there was a constant and sometimes almost overpowering temptation to publish the results anyway on the grounds that they were "good enough," despite years of experience in the computing industry and plentiful experience of the negative consequences of such wishful thinking and acting.  I believe that the only thing that saved me was reminding myself that both my personal honesty and my standing within in my professional community were at stake.  Even when I was able to confidently publish a result, I had to acknowledge the uncomfortable fact that the calculation could still be wrong despite my best efforts, and I had to give a somewhat detailed discussion of the risks as I understood them.  Therefore I have a lot of sympathy for the physicist who, less informed about the risks of computing, makes poor choices for managing those risks.

\chapter{\label{LatticeQCD}Unquenched QCD and Disordered Systems}
I here briefly review the current lattice
gauge theory algorithms for simulating fermions and the relations between lattice gauge
theory and mesoscopic physics.

\section{Fermionic Algorithms}
I assume that the reader is familiar with quenched (no dynamical fermions) lattice QCD\cite{Gupta98, Creutz83, Rothe92,Montvay94}, and is also aware that unquenched calculations include fermions by including the
determinant of the Dirac operator in the QCD partition function.  Either this determinant or its derivative must be evaluated quite often, and each evaluation requires a huge amount of computer time.  But even worse, current algorithms show a very rapid growth
in computational time as the fermion mass
is decreased, and therefore the fermion mass is kept at
artificially high values. (See these recent reports from major lattice QCD
collaborations\cite{Kennedy04} for the currently observed behavior.)
In current quenched and unquenched calculations this forces people to use masses of the up and down quarks which are four or five times their physical values\cite{Guadagnoli04, VilladoroPrivate}.  Faster
algorithms would be very useful. 

The most commonly
used lattice QCD algorithm, Hybrid Monte Carlo, evaluates the
fermionic determinant stochastically by introducing new degrees of
freedom which evolve according to forces reflecting the gauge
fields\cite{Kennedy99}. These forces are calculated by solving a
linear system ${Ax}={b}$, and in fact the bulk of the computational
time in modern calculations is spent in repeated solution of this linear
system.

Hybrid Monte Carlo's biggest competitor is probably the Local
Boson Approximation, which follows a similar strategy, using new
degrees of freedom to stochastically evaluate the fermionic
determinant's contribution to the Monte Carlo acceptance
probability\cite{Elser01, Luscher93}. Again computation of the
forces requires solving a linear system. However, the Local Boson
Approximation differs by choosing degrees of freedom which each
correspond to small segments of the Dirac operator's spectrum.
This opens a path for computing large eigenvalues differently than
small eigenvalues. Various efforts to compare Hybrid Monte Carlo
with the Local Boson Approximation suggest that they have similar
performance\cite{Elser01, Peardon02}.

Recently various researchers have been more aware of the
importance of topological features of the gauge field in the
fermion determinant\cite{Edwards01}.  The Atiyah-Singer index
theorem states that fermion zero modes will be created when the
topological index of the gauge field changes\cite{Weinberg96}.  The
resulting zero modes can create serious difficulties for
stochastic unquenched algorithms. But there is also a great deal
of speculation that the behavior of local topological features
("vortex condensation", instantons, etc.) are responsible for
confinement \cite{Kovacs96} and chiral symmetry
breaking\cite{Edwards01, Damgaard99, Verbaarschot00a,
Verbaarschot00b, Gockeler01, Damgaard01}.

Motivated by the computational difficulties caused by zero modes,
recently Duncan et al. have proposed a new strategy for unquenched
calculations \cite{Duncan98}. Rather than evaluate the entire
fermionic determinant stochastically, they evaluate the small
eigenvalues by solving the relevant eigenvalue problem, and then
use a stochastic algorithm (a plaquette expansion, or alternatively
the LBA) to evaluate the rest of the spectrum. This is the first
algorithm that I'm aware of that explicitly exploits the fact that
low energy modes may have much different physics than high energy
modes. As such, it seems to be a step in the right direction.  

Another recent development has been the introduction of
Neuberger's overlap operator\cite{Neuberger99, Luscher01,
Jansen01, Eshof01, Gavai02}, which is special formulation of the Dirac operator.  When this is used instead of the
conventional Dirac operator, chirality is preserved, bringing us
much closer to the real world which is believed to be nearly
chiral.  However, the overlap operator is much more expensive to
compute\cite{Jansen01}.   

\section{Connections With Disordered Systems}
QCD is a disordered system.  In the strong coupling
limit, the lattice gauge links have a correlation length equal to zero; they are independent of each other\cite{Creutz83}.  In modern
lattice calculations, the correlation length is a finite number of lattice spacings. It is expected
that the physically interesting scenario is a weak coupling limit
where the correlation length is physically finite and thus
infinite compared to the lattice spacing.  In this case, the
disorder at any site is small, but the system as a whole is broken
up into uncorrelated domains.

Mesoscopic theory has traditionally concentrated on models where the disordered degrees of freedom are a scalar potential, not a gauge potential like that found in QCD.  Note that scalar potential models may already give a qualitative understanding of certain QCD phenomena. However, some work has also been done on disordered gauge potentials. Most notably, a simplified gauge
model, the random flux model, has been under intense scrutiny
recently because of a conjecture that it is related to a good
model of the integer quantum Hall effect\cite{Ludwig94,
Tsvelik94}. Also, recently Zirnbauer extended the supersymmetric sigma model
technique to create a sigma model for certain disordered gauge potentials\cite{Zirnbauer96, Zirnbauer98, Altland98, Szabo01}.

Mesoscopic theory has traditionally concentrated on disorder with a correlation length equal to zero, because it is easier to handle mathematically than disorder with a finite correlation length.
The step of making a connection to the finite correlation lengths in lattice QCD is not free from difficulty.  One can, however, sweep such difficulties under the rug by assuming scaling in
disordered systems, which means that the microscopic details of
disorder are unimportant at longer length scales, except to set
the initial conditions for renormalization group
equations\cite{Kramer93}. This argument indicates  that
finite-correlation-length systems are very much like
zero-correlation-length systems, and would allow a direct link with QCD.  However, the last decade has seen continued debate about the applicability of the
renormalization group to disordered systems\cite{Kramer93,
Efetov97, Janssen98, Guhr98}.

Notwithstanding these difficulties, the theory of disordered
systems still has a lot to say to certain aspects of lattice QCD.  A number of
researchers have examined the spectra of Dirac operators in gauge field
configurations determined by quenched QCD simulations, and have
found that the low eigenvalues (below the Thouless energy) obey
the level spacing statistics of random matrix
theory\cite{Damgaard99,Verbaarschot00a,Verbaarschot00b,Gockeler01,Damgaard01}.
Sigma model calculations predict the same result analytically.

Disorder induced localization is also observed in lattice QCD. Localization can be
expected in random gauge models as long as no symmetry prohibits
it. The random flux model, for instance, exhibits localization 
throughout its energy band, with the possible (and debated) exception of
its center where there is a discrete symmetry\cite{Altland99}. Localization is seen in quenched simulations of QCD\cite{Baker94, Jansen96, Gattringer01}.  However, unquenching QCD may unlocalize the fermionic modes with energies close to zero that form the chiral
condensate\cite{Verbaarschot00b}. Indeed, Duncan et al. reported
the absence of zero modes in unquenched
calculations\cite{Duncan98}. 

In the last year and a half the lattice QCD community has made its first attempts at understanding unquenched QCD's disorder-related physics. These efforts have been motivated by two considerations:
\begin{enumerate}
\item One of the most popular discretizations of QCD is the staggered discretization.  The staggered approach unfortunately multiplies the number of fermions by four, and therefore one needs to somehow get rid of two of the four, leaving just the up and down quarks.  The traditional prescription for doing this has been to weight the QCD partition function with the square root of the fermion determinant instead of with the whole fermion determinant.  

This prescription is mathematically equivalent to using the square root of the Dirac operator in the original action of lattice QCD.  Recently the lattice community has become painfully aware that an action with a square root in it may not correspond to a local theory.  In a plenary talk at the most recent lattice QCD conference, Kennedy\cite{Kennedy04} described the consequences: 

"If a QFT is local then we are guaranteed that it has the cluster decomposition
property, and that within the context of renormalized perturbation theory it
satisfies the familiar power counting rules, exhibits universality, and is
amenable to a systematic improvement procedure. On the other hand, if it is not
local then there is little we can say about these properties other than that we
we have no \emph{a priori} reason to expect them to hold. In particular, if we
have a non-local lattice theory then we have no good reason to invoke
power-counting arguments to justify taking the na{\"\i}ve continuum limit, or
to expect the lattice theory to be described by continuum perturbation theory
however small the lattice spacing. The fact that a formulation is not manifestly local does not logically imply
that it is not local.... However, in general a non-manifestly local
theory has no reason to be equivalent to a local one. Even if there was a local action corresponding to taking fractional powers of
the fermion determinant in the functional integral, we are still required to
use this local action to measure fermionic quantities.... We should not expect that measuring operators corresponding to a local four taste valence action $M$ on configurations
generated with $\sqrt{\det M}$ to lead to consistent results. Not only might
there be unknown renormalisations of the parameters between the sea and valence
actions (e.g., what is the justification for using the same numerical value for
the quark masses?) but the degrees of freedom are not even the same. "

In summary, the square root trick is an uncontrolled approximation, and can completely change the results, unless one can prove that a particular square root of the Dirac operator is local in some sense and can be used to construct a local field theory.  This question may be hard to decide because there many ways of taking roots of matrices, which are analogous to the many possible choices of branch cuts in scalar roots.  A few researchers have studied the locality of roots of the Dirac operator, and the only major result so far is that the most straightforward choice of root is non-local\cite{Bunk04, Hart04}.
\item Recently there has been some doubt about whether Neuberger's overlap operator can be relied on to always deliver chiral symmetry, which is of course the reason why one would use the operator.  The mechanism of failure would depend on the localization of the Dirac operator's eigenfunctions.  A recent paper by Golterman, Shamir, and Svetitsky\cite{Golterman04} explains this a bit more deeply, discussing first domain wall fermions (which approximate the overlap operator) and then the overlap operator:

"Domain-wall fermions employ an auxiliary, discrete, and (in
practice) finite fifth dimension with spacing $a_5$ and
$N_s$ sites. Finiteness of the fifth dimension ensures locality
but leads to ``residual'' violations of chiral symmetry...." If the wavefunctions of the Dirac operator stretch across the whole fifth dimension and feel its finite size, then the domain wall approach will not deliver chiral symmetry.  Therefore one needs to know that these eigenfunctions are localized in the fifth dimension; that they die off exponentially.  One concludes that understanding the localization physics of QCD is crucial to validating the domain wall approach.

In the limit where the fifth dimension becomes exponentially long, domain-wall fermions are equivalent to the overlap operator; one might expect that the overlap operator's validity  will also be determined by localization physics.  The authors state: "For overlap fermions, chiral symmetry is guaranteed, but not locality [of the overlap operator]...... Deteriorating locality, caused by the proximity of the Aoki phase, will distort physical predictions in an uncontrolled way."  (The Aoki phase corresponds to sponteously breaking the symmetry between the three pions, so that two of them become massless and therefore are even longer ranged than they should be.  It does not correspond to the physical continuum limit, but can nonetheless arise in lattice simulations.)

And now the obvious conclusion: "This brings us to propose that the
range of the overlap operator, as well as the key
spectral quantities of $H_W$ that control it---$\lambda_c$, $\rho(0)$, and
$l_{l}(0)$---should be calculated in any overlap-fermion
simulation,  much as $m_{res}$ is routinely determined in
domain-wall fermion simulations."   In other words, Golterman et al. are suggesting a requirement that the localization physics be understood in all overlap calculations.

\end{enumerate}

\backmatter

\chapter{\label{Bibliography}Bibliography}

\bibliography{Vincent.bib}

\begin{thebibliography}{100}
\bibitem{LinuxSearch}
The search command was: "abs=linux". all physics databases were searched, for
  all years.
\bibitem{MathematicaSearch}
The search command was: "abs=mathematica". all physics databases were searched,
  for all years.
\bibitem{FortranSearch}
The search command was: "abs=fortran". all physics databases were searched, for
  all years.
\bibitem{BugAbstractsSearch}
The search command was: "(abs=bug or abs="software defect")". (ignore the outer
  set of quotes.) all physics databases were searched, for all years.
\bibitem{BugCommentsSearch}
The search command was: "co=bug". all physics databases were searched, for all
  years.
\bibitem{ErrataSearch}
The search command was: "erratum or errata". all physics databases were
  searched, for all years.
\bibitem{SoftwareTestSearch}
The search command was: "(abs=software and abs=test)". all physics databases
  were searched, for all years.
\bibitem{SoftwareVVSearch}
There were three searches, with commands: "(abs=verification and
  abs=validation)", "(abs=verification and abs=software)", and "(abs=validation
  and abs=software)". all physics databases were searched, for all years.
\end{thebibliography}
\bibliographystyle{unsrt}

\newpage
\cleardoublepage
\newpage

\end{document}